%% file: main.tex
\documentclass[a4paper,11pt]{article}

\title{Complexity Theory for Quantum Promise Problems}

\input{macro}

\ifdefined\authorNote
\newcommand{\Knote}[1]{{\color{red} [{K:} #1]}}
\newcommand{\nai}[1]{{\color{blue} [{Nai:} #1]}}
\newcommand{\yc}[1]{{\color{magenta} [{yc:}#1]}}
\newcommand{\HZZ}[1]{{\color{cyan} [{H:}#1]}}
\newcommand{\JW}[1]{{\color{blue} [{JW:}#1]}}

\newcommand{\lch}[1]{{\color{blue} [{LCH:} #1]}}

\else
\newcommand{\Knote}[1]{}%{{\color{red} [{K:} #1]}}
\newcommand{\nai}[1]{}%{{\color{blue} [{Nai:} #1]}}
\newcommand{\yc}[1]{}
\newcommand{\HZZ}[1]{}
\newcommand{\JW}[1]{}

\newcommand{\lch}[1]{}

\fi

\ifdefined\LLNCS
\else
\newcommand{\email}[1]{\href{mailto:#1}{#1}}
\fi

\ifdefined\authorblind
\else
\author{
    Nai-Hui Chia\thanks{Rice University, USA, \email{nc67@rice.edu}.}  \and
	Kai-Min Chung\thanks{Academia Sinica, Taiwan, \email{kmchung@iis.sinica.edu.tw}.}  \and
	Tzu-Hsiang Huang\thanks{Academia Sinica, Taiwan, \email{th57@illinois.edu}} \and
    Chuhan Lu\thanks{Rice University, USA, \email{cl308@rice.edu}}  \and
	Jhih-Wei Shih\thanks{Academia Sinica, Taiwan, \email{g54jo3741953@iis.sinica.edu.tw}.}
}
\fi

\date{}

\begin{document}
\maketitle
\begin{abstract}
\JW{This version contain author note!!.}
Quantum computing introduces many well-motivated problems rooted in physics, asking to compute information from input quantum states. Identifying the computational hardness of these problems yields potential applications with far-reaching impacts across both the realms of computer science and physics. However, these new problems do not neatly fit within the scope of existing complexity theory. The standard classes primarily cater to problems with classical inputs and outputs, leaving a gap to characterize problems involving quantum states as inputs. For instance, breaking new quantum cryptographic primitives involves solving problems with quantum inputs; this significantly changes Impagliazzo's five-world while the complexity classes central to Pessiland, Heuristica, and Algorithmica are grounded in problems with classical inputs and outputs. To bridge these knowledge gaps, we explore the complexity theory for quantum promise problems and potential applications. Quantum promise problems are quantum-input decision problems asking to identify whether input quantum states satisfy specific properties.

%Notably, the emergence of new quantum cryptographic primitives is reshaping our understanding of the minimal assumption in cryptography, which directly affects the relationships between cryptography and complexity theory; in particular, the connection between Minicrypt and Pessiland in Impagliazzo's five worlds view is somewhat unclear --- one reason for this is that breaking new quantum cryptographic primitives involves solving problems with quantum inputs, while the complexity classes central to Pessiland, Heuristica, and Algorithmica are grounded in problems with classical inputs and outputs.

%Notably, the emergence of new quantum cryptographic primitives is reshaping our understanding of the minimal assumption in cryptography, which directly affects the relationships between cryptography and complexity theory; in particular, the connection between Minicrypt and Pessiland in Impagliazzo's five worlds view is somewhat unclear --- one reason for this is that breaking new quantum cryptographic primitives involves solving problems with quantum inputs, while the complexity classes central to Pessiland, Heuristica, and Algorithmica are grounded in problems with classical inputs and outputs.

%To bridge these knowledge gaps, we explore the complexity theory for quantum promise problems and potential applications. Quantum promise problems are quantum-input decision problems asking to identify whether input quantum states satisfy specific properties. 

We begin by establishing structural results for several fundamental quantum complexity classes: p/mBQP, p/mQ(C)MA, $\text{p/mQSZK}_{\text{hv}}$, p/mQIP, p/mBQP/qpoly, p/mBQP/poly, and p/mPSPACE. This includes identifying complete problems, as well as proving containment and separation results among these classes. Here, p/mC denotes the corresponding quantum promise complexity class with pure (p) or mixed (m) quantum input states for any classical complexity class C. Surprisingly, our findings uncover relationships that diverge from their classical analogues --- specifically, we show unconditionally that p/mQIP$\neq$p/mPSPACE and p/mBQP/qpoly$\neq$p/mBQP/poly. This starkly contrasts the classical setting, where QIP$=$PSPACE and separations such as BQP/qpoly$\neq$BQP/poly are only known relative to oracles. More interestingly, these separation results further connected to the topic of for both quantum property testing and unitary synthesis.

This new framework has numerous applications in quantum cryptography, particularly in the contexts of Microcrypt. We provide a better characterization of its primitives; for example, we show that OWSG and PRS can be broken by a p/mQCMA oracle, leading to a natural quantum analogue of Impagliazzo’s five worlds by substituting the classical complexity classes in Pessiland, Heuristica, and Algorithmica with mBQP and mQCMA. Moreover, we establish the relativization barrier for proving the existence of EFI, noting that no such barrier currently exists within traditional complexity theory.

\end{abstract}

%% motivation -> classical complexity can't describe -> implication (better understanding on crypto hardness), further study. Why new definition

\pagenumbering{gobble}

\tableofcontents
\newpage

\pagenumbering{arabic}
%\setcounter{page}{1}

%\input{0_Extended_abstract}

\input{1_Intro}
\input{2_Prelim}

\input{3_Definition}
\input{4_qma}

\input{7_Uncond_separation}
\input{6_App}

\bibliographystyle{alphaurl}
\bibliography{References.bib,nai.bib}

\appendix
\input{Appendix}
\input{Appendix_pQSZK}

\input{Appendix_rep_KA04}

\end{document}

%% file: macro.tex
\usepackage[utf8]{inputenc}
\usepackage{amsmath,amssymb}
\usepackage{framed,caption,color,url}
\usepackage{standalone}
\usepackage{times}
\usepackage{mdframed}
\usepackage[noend]{algorithmic}
\usepackage{algorithm}
\usepackage{graphicx}               % Necessary to use \scalebox
\usepackage{cite} % Makes multiple citations look better
\usepackage{nicefrac} % with \nicefrac coomant one can produce dioginal divigin which looks better than a/b
\usepackage{paralist} % use \begin{inparaenum} or \begin{asparaenum} (or inparaitem or asparaitem) to enumerate inside a paragraph.
\usepackage{boxedminipage} % to have miniboxes to emphasize something
\usepackage[pdfstartview=FitH,colorlinks,linkcolor=blue,filecolor=blue,citecolor=blue,urlcolor=blue,pagebackref=true]{hyperref}
\usepackage{upgreek}
\usepackage{centernot}
\usepackage{mathtools}
\usepackage{bbm}
\usepackage{cleveref} %% for referencing use \cref
\usepackage{subcaption}
\usepackage{stmaryrd} % for quantum transition capacities
\usepackage{hyperref}
\usepackage{mathrsfs}
\usepackage{enumitem}

\ifdefined\LLNCS
\else
\usepackage{fullpage}
\usepackage{amsthm}
\fi

\newcommand{\fhat}[2]{\ifthenelse{\equal{#2}{}}{\hat{f}[#1]}{\ifthenelse{\equal{#2}{0}}{\hat{f}[\emptyset]}{\hat{f}[#1_{\leq #2}]}}}
\newcommand{\gain}[2]{\ifthenelse{\equal{#2}{}}{g[#1]}{g[#1_{\leq #2}]}}

\newcommand{\pr}[2][]{\Pr_{\ifthenelse{\isempty{#1}}{}{{#1}}}\left[{#2}\right]}

\usepackage{dsfont}

\newcommand{\E}{\Exp}

\newcommand{\remove}[1]{}

\newcommand{\nin}{\not \in}

\newcommand{\pad}{\quad\hspace{0.15em}}

\newcommand{\ie}{i.e.,\ }

\newcommand{\ceil}[1]{\lceil #1 \rceil}

\newcommand{\bra}[1]{\langle#1\rvert}
\newcommand{\braket}[1]{\langle #1 \rangle}
\newcommand{\ket}[1]{\lvert#1\rangle}

\newcommand{\ketbra}[1]{\lvert #1\rangle\!\langle #1\rvert}

\newcommand{\half}{\tfrac{1}{2}}

\newcommand{\C}{\mathbb{C}}

\newcommand{\tp}{\otimes}

\newcommand*{\fnref}[1]{\textsuperscript{\normalfont\ref{#1}}}

\newcommand{\R}{{\mathbb R}}
\newcommand{\N}{{\mathbb N}}

\newcommand{\cA}{{\mathcal A}}
\newcommand{\cB}{{\mathcal B}}
\newcommand{\cC}{{\mathcal C}}
\newcommand{\cD}{{\mathcal D}}

\newcommand{\cF}{{\mathcal F}}

\newcommand{\cH}{{\mathcal H}}

\newcommand{\cL}{{\mathcal L}}
\newcommand{\cM}{{\mathcal M}}

\newcommand{\cO}{{\mathcal O}}
\newcommand{\cP}{{\mathcal P}}

\newcommand{\cR}{{\mathcal R}}

\newcommand{\cT}{{\mathcal T}}
\newcommand{\cU}{{\mathcal U}}
\newcommand{\cV}{{\mathcal V}}

\usepackage{bm}
\newcommand{\eps}{\varepsilon}
\newcommand{\poly}{\operatorname{poly}}

\newcommand{\Exp}{\operatorname*{\mathbb{E}}}

\newcommand{\view}{\operatorname{view}}

\newcommand{\Com}{\operatorname{Com}}

\newcommand{\negl}{\operatorname{negl}}

\newcommand{\prover}{\operatorname{prover}}
\newcommand{\sample}{\operatorname{sample}}

\newcommand{\epr}{\operatorname{EPR}}

\newcommand{\TD}{\operatorname{TD}}
\newcommand{\halfs}{\operatorname{HALF}}

\newcommand{\class}[1]{\ensuremath{\mathbf{#1}}}
\newcommand{\NP}{\ensuremath{\class{NP}}}

\renewcommand{\P}{\class{P}}

\newcommand{\BQPpoly}{\class{BQP/poly}}
\newcommand{\BQPqpoly}{\class{BQP/qpoly}}
\newcommand{\pBQPpoly}{\class{pBQP/poly}}
\newcommand{\pBQPqpoly}{\class{pBQP/qpoly}}
\newcommand{\mBQPpoly}{\class{mBQP/poly}}

\newcommand{\mBQPqpoly}{\class{mBQP/qpoly}}

\newcommand{\PSPACE}{\class{PSPACE}}
\newcommand{\QPSPACE}{\class{QPSPACE}}
\newcommand{\pp}{\class{PP}}
\newcommand{\BQP}{\class{BQP}}
\newcommand{\pQCMMA}{\class{pQ(C)MA}}
\newcommand{\mQCMMA}{\class{mQ(C)MA}}
\newcommand{\pmQCMMA}{\class{p/mQ(C)MA}}
\newcommand{\pmclass}{\class{p/m}}
\newcommand{\pmBQP}{\class{p/mBQP}}
\newcommand{\pBQP}{\class{pBQP}}
\newcommand{\mBQP}{\class{mBQP}}
\newcommand{\QCMA}{\class{QCMA}}
\newcommand{\pmQCMA}{\class{p/mQCMA}}
\newcommand{\pQCMA}{\class{pQCMA}}
\newcommand{\mQCMA}{\class{mQCMA}}
\newcommand{\pmQMA}{\class{p/mQMA}}
\newcommand{\pmPSPACE}{\class{p/mPSPACE}}
\newcommand{\QMA}{\class{QMA}}
\newcommand{\pQMA}{\class{pQMA}}
\newcommand{\pPSPACE}{\class{pPSPACE}}
\newcommand{\mQMA}{\class{mQMA}}
\newcommand{\mPSPACE}{\class{mPSPACE}}
\newcommand{\QIP}{\class{QIP}}
\newcommand{\pQIP}{\class{pQIP}}
\newcommand{\mQIP}{\class{mQIP}}

\newcommand{\pQIPsec}{\class{pQIP[2]}}
\newcommand{\mQIPsec}{\class{mQIP[2]}}
\newcommand{\pmQIP}{\class{p/mQIP}}
\newcommand{\mQSZKhv}{\ensuremath \class{mQSZK}_{\mathrm{hv}}}

\newcommand{\mQSZKhvsec}{\ensuremath \class{mQSZK[2]}_{\mathrm{hv}}}
\newcommand{\mpolyQSZKhv}{\ensuremath \class{mQSZK}_{\mathrm{hv}}^{\poly}}
\newcommand{\pQSZKhv}{\ensuremath \class{pQSZK}_{\mathrm{hv}}}

\newcommand{\pQSZKhvsec}{\ensuremath \class{pQSZK[2]}_{\mathrm{hv}}}
\newcommand{\pQSZK}{\class{pQSZK}}
\newcommand{\QSZKhv}{\ensuremath \class{QSZK}_{\mathrm{hv}}}
\newcommand{\pQCZKhv}{\class{pQCZK_{hv}}}
\newcommand{\QCZKhv}{\class{QCZK_{hv}}}
\newcommand{\QSZK}{\class{QSZK}}

\newcommand{\pmINF}{\class{p/mALL}^{\poly}}
\newcommand{\pINF}{\class{pALL}^{\poly}}
\newcommand{\mINF}{\class{mALL}^{\poly}}

\ifdefined\LLNCS
\newtheorem{observation}[theorem]{Observation}
\newtheorem{fact}[theorem]{Fact}

\newtheorem{construction}[theorem]{Construction}
\newtheorem{algorithm1}[theorem]{Algorithm}
\newtheorem{assumption}[theorem]{Assumption}

\makeatletter
\renewcommand{\@Opargbegintheorem}[4]{%
  #4\trivlist\item[\hskip\labelsep{#3#2\@thmcounterend}]}
\makeatother

\else

\newtheorem{theorem}{Theorem}[section]
\theoremstyle{plain}
\newtheorem{claim}[theorem]{Claim}

\newtheorem{lemma}[theorem]{Lemma}
\newtheorem{corollary}[theorem]{Corollary}

\theoremstyle{definition}

\newtheorem{definition}[theorem]{Definition}

\newtheorem{example}[theorem]{Example}

\newtheorem{remark}[theorem]{Remark}

\theoremstyle{definition}

\fi

\ifdefined\LLNCS
\else

\newcommand{\sdotfill}{\textcolor[rgb]{0.8,0.8,0.8}{\dotfill}} %change to \cdotfill later

\makeatletter
\def\th@protocol{%
\normalfont % body font
\setbeamercolor{block title example}{bg=orange,fg=white}
\setbeamercolor{block body example}{bg=orange!20,fg=black}
\def\inserttheoremblockenv{exampleblock}
}
\makeatother
\theoremstyle{protocol}
\newtheorem{proto}[theorem]{Protocol}
\newtheorem{protoc}[theorem]{Protocol}

\fi

\newcommand{\namedref}[2]{#1~\ref{#2}}

\newcommand{\torestate}[3]{%
\expandafter \def \csname BBRESTATE #2 \endcsname{#3}
\theoremstyle{plain}
\newtheorem{BBRESTATETHMNUM#2}[theorem]{#1}
\begin{BBRESTATETHMNUM#2}\label{#2}\csname BBRESTATE #2 \endcsname   \end{BBRESTATETHMNUM#2}
\newtheorem*{BBRESTATETHMNONNUM#2}{\namedref{#1}{#2}}
}

\newcommand{\restate}[1]{\begin{BBRESTATETHMNONNUM#1}[Restated] \csname BBRESTATE #1 \endcsname
\end{BBRESTATETHMNONNUM#1}}

\DeclareMathOperator{\Tr}{Tr}

\newcommand{\aprov}{\ensuremath{\alpha_{\mathrm{prover}}}}

\newcommand{\ans}{\ensuremath{\mathrm{ans}}}
\newcommand{\aans}{\ensuremath{A_i^{\ans}}}

\newcommand{\KeyGen}{\mathsf{KeyGen}}
\newcommand{\StateGen}{\mathsf{StateGen}}
\newcommand{\Ver}{\mathsf{Ver}}

\newenvironment{protocal}[1]
{
\begin{mdframed}
    \begin{center}\textbf{#1}\end{center}
}
{
\end{mdframed}
}

%% file: 1_Intro.tex
% orginal order
%\input{1-1_Intro}
%\input{1-2_Results}
%\input{1-3_Compare}
%\input{1-4_Open}

% STOC order
\input{1-1_Intro}

\input{1-2_Results_focs}

\input{1-3_tech_overview}
\input{1-4_Open}
\input{1-5_Compare}
\ifdefined\authorblind
\else
\subsection{Acknolwedgement}
We would like to thank Yu-Hsuan Huang, Yu-Ching Shen, and Shota Yamada for valuable discussions. We also appreciate helpful comments from anonymous reviewers. Nai-Hui Chia is supported by NSF Award FET-2243659, NSF Career Award FET-2339116, Google Scholar Award, and DOE Quantum Testbed Finder Award DE-SC0024301. Kai-Min Chung is supported in part by the Air Force Office of Scientific Research under award number FA2386-23-1-4107 and the 2024 Academia Sinica Seed Grant for Grand Challenge Program. (AS-GCS-113-M06). Tzu-Hsiang Huang and Jhih-Wei Shih are supported in part by the Air Force Office of Scientific Research under award number FA2386-23-1-4107 and NSTC QC project, under Grant no. NSTC 113-2119-M-001 -009 -
\fi
.

%% file: 1-1_Intro.tex
\section{Introduction}
%Complexity theory characterizes the computational resources required to solve decision problems. Both practical applications and theoretical advancements have driven the exploration of complexity classes. \HZZ{For example, proving that a problem is NP-hard motivates the development of approximation algorithms for practical use, while investigating relationships between complexity classes provides insights into computational resources. Beyond these areas, complexity theory has shaped fields such as cryptography, optimization, and formal verification, making it a foundational area of computer science.} However, traditional complexity theory, which focuses on classical input and output, fails to address the problems posed by quantum computing. Quantum computing introduces a wealth of intriguing problems rooted in physics, many of which involve extracting classical information from quantum input states. Understanding the computational complexity of these novel problems has the potential to unlock groundbreaking applications, highlighting the need to extend traditional complexity theory.

Complexity theory characterizes the computational resources required for computational problems. The study of complexity theory guides the development of many areas in computer science. For example, proving that a problem is \class{NP}-hard motivates the development of approximation algorithms for practical use, while investigating relationships between complexity classes provides insights into computational resources. Beyond these areas, complexity theory has shaped fields such as cryptography, optimization, and formal verification, making it a foundational area of computer science. Meanwhile, the rise of quantum computation and information introduces a new set of problems, many of which are rooted in the processing of quantum information. Understanding the computational complexity of these new problems has the potential to unlock groundbreaking applications of quantum computing. However, these new problems do not fit neatly within the scope of existing complexity classes, which primarily address problems involving classical information. Notably, classical complexity theory also includes various complexity classes for different types of computational tasks such as decision problems, relation problems~\cite{Aar14,ABK24}, sampling problems~\cite{Aar14}, and distribution testing problems~\cite{CG18}. This further suggests the need for a new quantum complexity theory.
%Notably, classical complexity theory also includes various types of computational tasks such as decision problems, relation problems~\cite{Aar14,ABK24}, sampling problems~\cite{Aar14}, and distribution testing problems~\cite{CG18}. This gap highlights the need for a new complexity theory. 

Indeed, new complexity classes have been introduced for quantum states and unitary matrices synthesis. Leading this effort, Rosenthal and Yuen~\cite{rosenthal2022interactive} introduced complexity classes for state synthesis, studying the complexity of generating quantum states indexed by classical inputs. Bostanci et al.\cite{bostanci2023unitary} conducted an exhaustive study on complexity classes for unitary synthesis, targeted at generating unitary matrices indexed by classical input strings. It is worth noting that both frameworks aim to characterize problems for \emph{generating quantum objects}.%\footnote{See the related work in Section \ref{sec:related} for a brief survey and comparison with different frameworks.} 

In addition to synthesizing quantum objects, many problems ask to extract classical information from quantum states. Briefly, given multiple copies of unknown quantum states (either pure or mixed states), these problems ask to identify whether the input states satisfy some properties A or B (properties A and B should be far apart). We call these problems as \emph{quantum decision promise problems}, as they are decision problems that generalize promise problems in complexity theory. For simplicity, in the remainder of this article, we abbreviate \emph{quantum decision promise problems} as \emph{quantum promise problems} (QPPs). Unlike synthesizing quantum states and unitaries, which are problems for \emph{generating quantum objects}, quantum promise problems have quantum inputs and classical outputs. Analogous to the classical setting, various categories of computational problems are classified into distinct complexity classes, including counting classes (e.g., \class{\#P}~\cite{Val79}) and sampling classes (e.g., \class{SampleBQP}~\cite{Aar14}), each with its own defining features. The same applies to unitary synthesis, state synthesis, and quantum promise problems.\footnote{See Section~\ref{sec:related} for more discussion.}

%Just as in the classical setting, different kinds of computational problems has distinct complexity theory counting compleixty class (#p) or sampling compleixty class (SampleBQP), unitary synthesizing probelm, state symthesizeing problm and quantum promise problem capturing their own unique features\footnote{See Section~\ref{sec:related} for more discussion}.
%% note add how to compare ..

One example of a quantum promise problem is the product test for states, which aims to test whether a given quantum state is a product state or far from a product state~\cite{HM13}. Product testing is one example of quantum property testing, and other interesting properties~\cite{MdeW16} can also be described by quantum promise problems. Questions in quantum metrology and learning often aim to identify unknown parameters encoded within a quantum state~\cite{meyer2023quantum,simon2017quantum,anshu2023survey}. Furthermore, new quantum cryptographic primitives~\cite{pqs,pqs1,pqs2,morimae2022one,brakerski2022computational,cao2022constructing} are proven valuable in constructing complex cryptographic applications without one-way functions, which securities are based on solving problems with quantum state inputs.   
Intriguingly, recent research has ventured into areas that bridge the disciplinary divide between computer science and physics, such as learning quantum states~\cite{anshu2023survey}, pseudorandomness and wormhole volume~\cite{bouland_et_al:LIPIcs:2020:11748}, blackhole decoding tasks~\cite{scottlecture16,HH13,brakerski2023blackhole}, the state minimum circuit size problems~\cite{chia2021mcsp}, etc. All these problems can be viewed as quantum promise problems or variants. Additionally, the emergence of new quantum problems and cryptographic primitives is reshaping our understanding of the interplay between quantum cryptography and complexity theory. Impagliazzo's ``five-worlds'' view~\cite{impagliazzo1995personal}, each world representing different fundamental cryptography and complexity assumptions, characterizes the variable powers of algorithms or cryptography based on which world we are in. The introduction of quantum promise problems, combined with recent advancements in quantum cryptography, could remarkably shift our understanding of these fields.\footnote{See Section \ref{section:open_five_world} for more discussion.}

Along this line, we might need complexity classes and corresponding theory for quantum promise problems to better study the computational complexity of these novel problems. Kashefi and Alves~\cite{kashefi2004complexity} introduced $\class{QMA}$ and $\class{BQP}$ for quantum promise problems involving pure states and identified some problems within these complexity classes.\footnote{In their work~\cite{kashefi2004complexity}, quantum promise problems are referred to as quantum languages.} However, many fundamental aspects and variations of these complexity classes remain largely unexplored. These uncharted territories include issues like complete problems, inclusions and exclusions, considerations involving mixed states\nai{Check}, numbers of copies of quantum states, space complexity, etc. Crucially, quantum promise complexity classes provide a direct and natural framework for assessing the computational hardness of quantum promise problems in comparison to other types of complexity classes and thus offer a promising avenue for achieving a more comprehensive characterization of novel quantum cryptographic primitives and related problems. Additionally, quantum promise problems naturally extend the concept of promise problems, suggesting that the corresponding complexity classes may preserve several well-established properties of promise complexity classes. These insights suggest that quantum promise complexity classes can be a useful framework for studying the computational hardness of quantum promise problems. Inspiring by all these challenges, in this work, we aim to do the following: 
\begin{center}
    \emph{Establish the proper theoretical framework to study the hardness of quantum promise problems.}
\end{center}

%% file: 1-2_Results_focs.tex
\subsection{Our results}

This work introduces and revisits complexity classes for quantum promise problems, highlights their structural theorems, and demonstrates applications in property testing, unitary synthesis, and cryptography. We provide a brief overview of our findings below.

\paragraph{Quantum promise complexity classes} 
Consider two sets of quantum states, $\cL_Y$ and $\cL_N$, where any state in $\cL_Y$ is far from any state in $\cL_N$. A quantum promise problem $(\cL_Y,\cL_N)$ is to determine whether a given quantum state $\rho$, with multiple copies available, belongs to $\cL_Y$ or $\cL_N$. When $\cL_Y$ and $\cL_N$ only contain pure quantum states, we call $(\cL_Y,\cL_N)$ a \emph{pure-state quantum promise problem}; otherwise, we call $(\cL_Y,\cL_N)$ a \emph{mixed-state quantum promise problem}.

We define complexity classes for pure-state and mixed-state quantum promise problems under various computational resources. Take some examples as follows: 

\begin{definition}[$\pBQP$ and $\mBQP$ (Informal)]\label{intro_def:bqp}
$\pBQP$ ($\mBQP$) represents the set of pure-state (mixed-state) quantum promise problems, denoted by $(\cL_Y, \cL_N)$. For each problem in $\pBQP$ or $\mBQP$, there exists a polynomial-time uniform quantum algorithm $\cA$. Given a polynomial number of inputs $\rho$, the algorithm accepts with a probability of at least $\frac{2}{3}$ if $\rho \in \cL_Y$. Conversely, if $\rho \in \cL_N$, the algorithm accepts with a probability of at most $\frac{1}{3}$.
\end{definition}

The definition of $\pPSPACE^{\poly}$ ($\mPSPACE^{\poly}$) is analogous to Definition~\ref{intro_def:bqp}, except that the algorithm $\cA$ is required to run uniformly in polynomial space and receives, at the beginning, only polynomially many copies of the input state — a natural input model for polynomial-space computation. Another model considers the input state to be provided via a query oracle, allowing a polynomial-space algorithm to access more than a polynomial number of copies. \footnote{In this model, new copies of the input state can be requested during the computation, allowing a polynomial-space algorithm to potentially access more than a polynomial number of copies.} In this paper, we primarily focus on the model in which the input state is provided only at the beginning, restricting polynomial-space algorithms to polynomially many copies. For simplicity, we abbreviate $\pPSPACE^{\poly}$ ($\mPSPACE^{\poly}$) as $\pPSPACE$ ($\mPSPACE$).

%\footnote{An alternative notation would be $\pPSPACE^{poly}$ to indicate that the number of copies is at most polynomial. However, since our work only considers the case where the polynomial-space algorithm $\cA$ has access to only a polynomial number of copies, we omit the superscript “poly” for notational simplicity.}

%\footnote{There can be two ways to access quantum states: The first one gets all copies of quantum states prior to the computation, and the second one can request new copies during the computation. A $t$-time algorithm can access up to $t$ quantum states in both models. Conversely, an $s$-space algorithm in the first model can only use at most $s$ copies, while the second model could potentially utilize more copies. In this paper, we mainly focus on the first model.} 

\begin{definition}[$\pQIP$ and $\mQIP$ (Informal)\footnote{We reserve the notation of $\class{p/mIP}$ as interactive proofs with classical communication.}]\label{intro_def:qip}
$\pQIP$ ($\mQIP$) represents the set of pure-state (mixed-state) quantum promise problems, denoted by $(\cL_Y, \cL_N)$. For each problem in $\pQIP$ or $\mQIP$, there exists interactive algorithms $\cP$ and $\cV$ satisfying the following properties: 
\begin{itemize}
    \item \textbf{Resources:}  $\cP$ is an unbounded time quantum algorithm with \emph{unbounded} copies of inputs. $\cV$ is a polynomial-time uniform quantum algorithm with polynomial copies of inputs.
    \item \textbf{Completeness:} If $\rho \in \cL_Y$, a prover $\cP$ exists to convince $\cV$ that $\rho \in \cL_Y$ with probability at least $2/3$. 
    \item \textbf{Soundness:} If $\rho \in \cL_N$, no prover $\cP^*$ can convince $\cV$ that $\rho \in \cL_Y$ with probability greater than $1/3$.  
\end{itemize}
\end{definition}
We also consider alternative definitions of interactive proofs with restricted prover's resources.
\begin{definition}[$\pQIP^{\poly}$ and $\mQIP^{\poly}$ (Informal)]
    $\pQIP^{\poly}$ ($\mQIP^{\poly}$) is similar to $\pQIP$ ($\mQIP$), with the following difference: (i) For completeness, an honest prover receives only a polynomial number of copies of the input state. (ii) For soundness, the verifier still competes against a malicious prover with unlimited access to the input state.
\end{definition}
As a correspondence, we have $\pQIP^{\poly} \subseteq \pQIP$ and $\mQIP^{\poly} \subseteq \mQIP$, as we only limit the power of honest prover. This alternative definition gives us a different view when defining new Impagliazzo’s ``five-worlds” (See Section \ref{section:open_problem} for more discussion).  

In addition, $\pQCMMA$ and $\mQCMMA$ follow Definition \ref{intro_def:qip} with the prover only sending a single quantum (classical) message. In general, let $\mathcal{C}$ be a classical complexity class. We define $\textbf{p}\mathcal{C}$ and $\textbf{m}\mathcal{C}$ as the pure-state and mixed-state quantum promise complexity classes, respectively, both imitating the definition of $\mathcal{C}$. Furthermore, we abbreviate ``$\textbf{p}\mathcal{C}$ and $\textbf{m}\mathcal{C}$" as $\textbf{p/m}\mathcal{C}$.
%\footnote{There can be two ways to access quantum states: The first one gets all copies of quantum states prior to the computation, and the second one can request new copies during the computation. A $t$-time algorithm can access up to $t$ quantum states in both models. Conversely, an $s$-space algorithm in the first model can only use at most $s$ copies, while the second model could potentially utilize more copies. In this paper, we mainly focus on the first model.} 

In this study, we focus on the complexity classes \class{p/mBQP}, \class{p/mQMA}, \class{p/mQCMA}, \class{p/mQSZK_{hv}}, \class{p/mQIP},and \class{p/mPSPACE}. We have chosen these specific complexity classes due to our observations of their close relevance to many fundamental problems in quantum information.\footnote{See Section \ref{section:app} for more discussion on application.} In this paper, we call these complexity classes for quantum promise problems as \emph{quantum promise complexity classes}.

\subsubsection{Important properties of quantum promise complexity classes}\label{sec:properties}

Here, we outline the properties that distinguish the study of quantum promise complexity classes from that of standard complexity classes for languages and promise problems. %These properties either post new challenges to studying quantum promise complexity classes or enable  
%There are two properties worth to be noted for quantum promise complexity classes:

\paragraph{The number of input copies matters} A quantum promise problem can only be solved with a sufficient number of copies of input states.\footnote{See Section~\ref{section:open_copy} for more discussion on different numbers of input copies.} This requirement significantly affects the relationships between different complexity classes. 
For example, it is straightforward to show that $\class{QIP}$ can be simulated by a double exponential time algorithm by trying all possible responses from the prover. However, the same approach cannot show similar results for quantum promise complexity classes. In particular, it cannot show that an unbounded algorithm with a limited number of input states solves problems in $\pQIP$ and $\mQIP$. This is because, in $\pQIP$ and $\mQIP$, the prover can generate responses using super-polynomially many input states, and an algorithm with polynomial input states cannot simulate the prover in general. 
%In contrast, only a polynomial number of copies are available in a single-party algorithm. 
In fact, we are able to show that non-interactive protocols (including $\pmQCMA$ and $\pmQMA$) can still be simulated by a a polynomial-space algorithm given polynomially many copies of the input state.
Conversely, interactive protocols cannot, in general, be simulated by any algorithm that has access to only polynomially many copies of the input. That is, $\pmQIP[2]\not\subseteq \pmINF$, where $\pmINF$ denotes the class of problems solvable in unbounded time with arbitrary advice, given access to only polynomially many copies of the input. This result is different from the well-known equality $\QIP = \PSPACE$ (See Theorem \ref{thm:informal_up} and Theorem \ref{thm:mpQIP_2_not_in_mpINF}).

%\footnote{The reason for the notation $\class{UNBOUND}$, instead of $\class{ALL}$, is that the algorithm is restricted to having access to only polynomially many copies of the input.}

%For example, it is straightforward to show that $\class{QIP}$ is in $\class{PSPACE}$ by having a polynomial-space algorithm simulate all possible responses from the prover. However, the same strategy cannot be applied to show that $\pQIP$ is in $\pPSPACE$. This is because, in $\pQIP$ and $\mQIP$, the prover can generate responses using super-polynomially many copies of the input state. In contrast, only a polynomial number of copies are available in $\pPSPACE$. In fact, we can show that $\pQIP \neq \pPSPACE$ under this observation. 

%\paragraph{$\mathcal{C}$ is a subset of $\textbf{p}\mathcal{C}\subset \textbf{m}\mathcal{C}$.}  

%\paragraph{Quantum promise complexity classes contain their classical counterparts ($\class{C}\subseteq \textbf{p/m}\class{C}$)}

%\paragraph{Quantum promise complexity classes contain their classical counterparts ($\class{C}\subseteq \textbf{p/m}\class{C}$)} 
\paragraph{Containment and separation of quantum promise complexity classes}
 
Our definition of quantum promise problems generalizes classical problems. Indeed, all classical problems can be regarded as special cases of quantum promise problems. This fundamental insight leads to the following remark. 

\begin{remark}\label{thm:informal_sep}
    If a separation exists between two classical complexity classes $\cC_1$ and $\cC_2$, the separation extends to $\textbf{p/m}\cC_1$ and $\textbf{p/m}\cC_2$ (Claim~$\ref{Def:prop3}$). Additionally, if a quantum promise problem $\cL$ belongs to $\textbf{p/m}\cC$, the subset of $\cL$ containing only classical inputs belongs to $\cC$ (Claim $\ref{Def:prop2}$).
\end{remark}
Moreover, it is possible to show that $\textbf{p/m}\cC_1 \neq \textbf{p/m}\cC_2$ unconditionally, even though $\cC_1 = \cC_2$ remains an open question. Intuitively, this is because quantum promise complexity classes can encompass problems with quantum inputs that naturally reflect differences between models with varying quantum resources. For instance, we are able to show that $\textbf{p}\class{BQP/poly} \subsetneq \textbf{p}\class{BQP/qpoly}$ unconditionally. Notably, an unconditional separation between $\BQPpoly$ and $\BQPqpoly$ is still considered beyond the reach of current techniques.

However, showing any separation in the quantum promise complexity classes should not be taken for granted, as classical relativized barriers often carry over to quantum promise complexity classes. For example, $\pBQP^{\pPSPACE}=\pPSPACE^{\pPSPACE}$. This implies that, to uncondionally separate $\pBQP$ and $\pPSPACE$, a non-relativizing technique will be required. We propose that Open Question 1 in Section~\ref{section:open_strcut_result} may inspired a potential non-relativize technique for achieving a separation between $\pBQP$ and $\pPSPACE$.

We conclude that although quantum promise complexity classes provide a more refined characterization of quantum computational resources than classical decision problems, significant barriers remain when attempting to establish unconditional separations or containments between such classes.

%Likewise, assuming the existence of hard problems in a quantum promise complexity class is a ``weaker assumption'' than assuming the existence of hard problems in the corresponding classical complexity classes.

\paragraph{Difference between pure and mixed states makes a difference}
A quantum promise problem that includes mixed states is generally more challenging to decide. A key difference between pure-state and mixed-state quantum promise problems is the ability to identify the equality between two states. For pure-state quantum promise problems, the ``swap test" technique is sufficient. However, this technique is ineffective when dealing with mixed states, as even two identical mixed states cannot pass the swap test. For example, by introducing the swap test technique, we can readily prove that a variant of the local Hamiltonian problem is $\pQMA$-complete. However, this finding does not extend to the $\mQMA$-complete problem. See Theorem $\ref{Thm:completeProblem}$ for the sketch of a proof.

\subsubsection{Structural results for quantum promise problem}

%We delve deeper into the investigation of whether these quantum promise complexity classes maintain key properties akin to their classical counterparts. This exploration encompasses various facets, including the notions of reductions, finding natural complete problems within the classes, containments of complexity classes, the existence of the amplification lemma, and search-to-decision reductions. 

\begin{figure}
\centering
\includegraphics[width=0.8\textwidth]{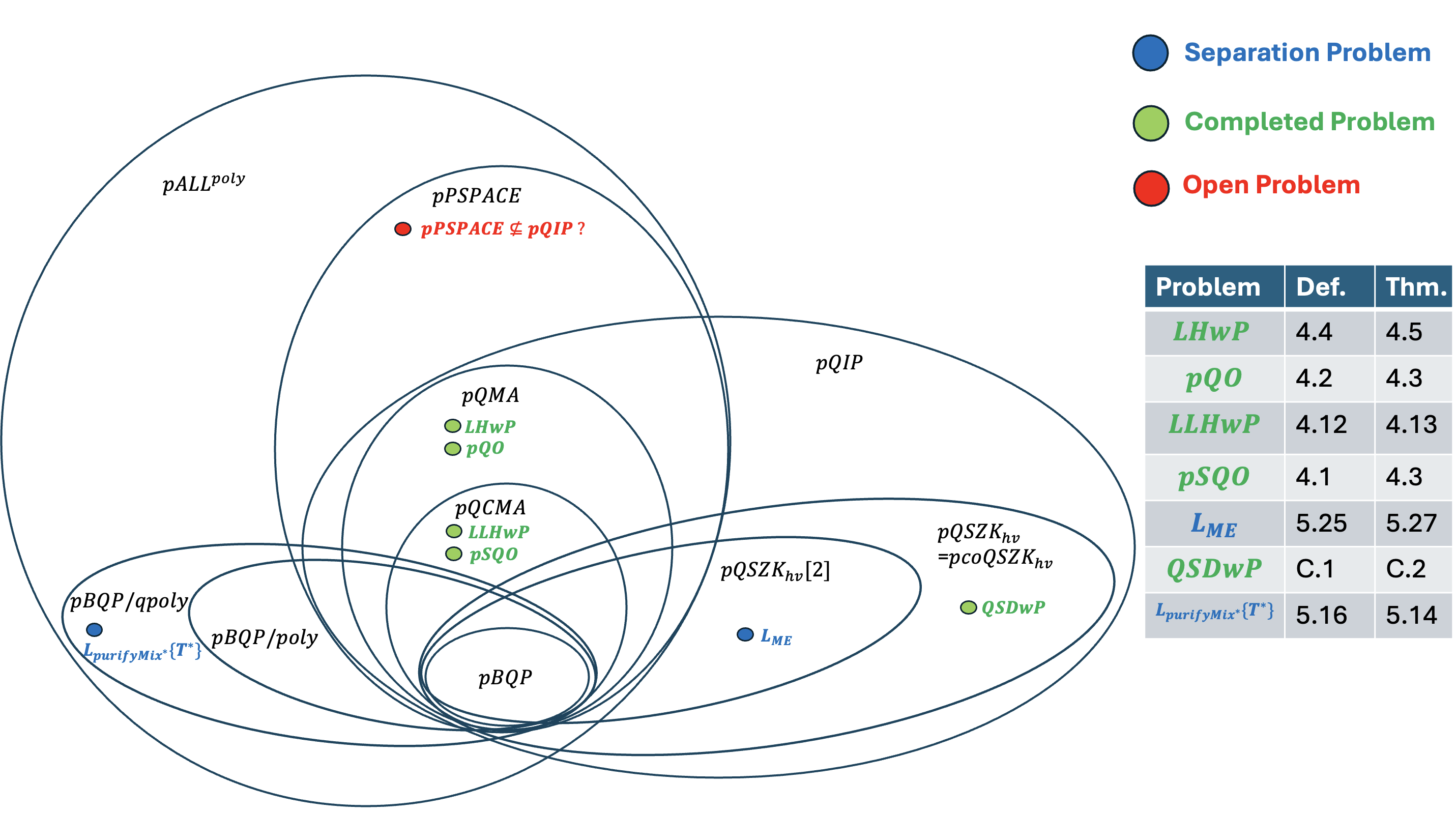}
\caption{\label{fig:pure_class_relation} We include the inclusion (Theorem \ref{thm:informal_up}), separation (Theorem \ref{thm:mpQIP_2_not_in_mpINF}, Theorem \ref{thm:intro_sep_quantum_property} and Theorem~\ref{thm:pure_BQPpolyneqBQPqpoly}), complete problems (Theorem \ref{Thm:completeProblem} and Theorem \ref{thm:qob_complete_qma}), and complement (Theorem \ref{thm:intro_pQSZKhv_close}) results for the pure complexity class. }
\end{figure}

We start by presenting our results for the pure complexity class. The results can be categorized into three parts: complete problems, containments, and separations of complexity classes. We provide a summary in Figure~\ref{fig:pure_class_relation} and give a more detailed explanation below.

\paragraph{First QMA- and QCMA-complete problems from Quantum OR problems} First, we show that the famous \emph{Quantum Or} problems \cite{harrow2017sequential,Aar20} are complete for $\pmQCMA$, $\pmQMA$. Briefly, given polynomial copies of quantum states $\rho$ and a set of observables with the promise that either there exists an observable $O$ such that $\Tr(O \rho)$ close to $1$ or $\Tr(O \rho)$ is small for all $O$, Quantum OR problems are to decide which case it is. See Definition \ref{qcma:qob} and Definition \ref{qma:qob} for formal definitions. 

\begin{theorem}[Theorem $\ref{thm:QOB_complete}$] Quantum OR problems (Definition \ref{qcma:qob} and Definition \ref{qma:qob}) are complete for $\pmQCMA$, $\pmQMA$.
\label{thm:qob_complete_qma}
\end{theorem}
%Roughly, the proof of Theorem~\ref{thm:qob_complete_qma} uses the observations that one can view the verification process as a set of efficient measurements indexed by the witness.

The Quantum OR problems have various applications in quantum input problems, such as shadow tomography \cite{Aar20}, black-box separation of quantum cryptographic primitives \cite{chen2024power}, and quantum property testing \cite{harrow2017sequential}. By Theorem \ref{thm:qob_complete_qma}, we know it is even more powerful, capable of solving all quantum promise problems in $\pmQCMA$ and $\pmQMA$.

\paragraph{Other complete problems inspired by classical counterparts}In addition, we identify complete problems for $\pmQMA$, $\pmQCMA$, and $\pQSZKhv$ that naturally extend classical complete problems. We show that variants of the local Hamiltonian problem or quantum state distinguishability are complete for $\pQCMA$, $\pQMA$, $\pQSZKhv$, $\mQCMA$, or $\mQMA$.

\begin{theorem} Variants of the local Hamiltonian problem is complete for $\pQCMA$, $\pQMA$, $\pQSZKhv$, $\mQCMA$, and $\mQMA$. Specifically,
\begin{itemize}
    \item 5-\textbf{LHwP} (Definition \ref{qma:p_problem}) is \class{pQMA}-complete (Theorem \ref{thm:pqma_complete}).
    \item  5-\textbf{LLHwP} (Definition \ref{qcma:p_problem}) is \class{pQCMA}-complete (Theorem \ref{qcma:p_complete}).
    \item \textbf{QSDwP} (Definition \ref{definition:QSDwP}) is $\pQSZKhv$-complete (Theorem \ref{thm:UNQSD_QSZKcomplete}).
    \item 5-\textbf{LHwM} (Definition \ref{qma:m_problem}) is \class{mQMA}-complete (Theorem \ref{thm:mqma_complete}).
    \item 5-\textbf{LLHwM} (Definition \ref{qcma:m_problem}) is \class{mQCMA}- complete (Theorem \ref{qcma:m_complete}).
\end{itemize}
\label{Thm:completeProblem}
\end{theorem}

One might initially think that complete problems for $\QCMA$, $\QMA$, and $\QSZK$ could serve as potential candidates for complete problems in their corresponding quantum promise complexity classes, since classical complexity classes are subsets of these quantum promise classes. For example, Kitaev, Shen, and Vyalyi \cite{kitaev2002classical} show that the local Hamiltonian problem is $\QMA$ complete. The problem input is a Hamiltonian given as a sum of local terms, $H := \sum\limits_{s \in S} H_s$. The task is to determine whether the ground state energy of $H$ (i.e., its minimum eigenvalue) is low or high. However, it is unclear how problems with unknown quantum inputs can be reduced to the local Hamiltonian problems. We introduce a natural variant of the local Hamiltonian problem (5-\textbf{LHwP} (Definition \ref{qma:p_problem})) where the problem input is defined as $H_{\ket{\psi}}:= \sum\limits_sH_s-\sum\limits_\ell\big(\ket{\psi}\bra{\psi} \otimes H_\ell\big)$. At a high level, the above problem can be viewed as a variant of the local Hamiltonian problem that involves an unknown quantum state. We highlight that $\ket{\psi}\bra{\psi}$ is a non-local term, whereas $H_s$ and $H_\ell$ remain local Hamiltonians.

This type of Hamiltonian, one that includes an unknown state, has appeared in the literature in the context of quantum PCA~\cite{LMR14}, although prior work has primarily focused on Hamiltonian simulation. In that work, the authors also provide an algorithm that is efficient in both sample complexity and runtime for simulating Hamiltonians involving unknown quantum states. On the other hand, we show that deciding whether the ground state energy of Hamiltonians involving unknown quantum states is low or high is $\pQMA$-complete. This mirrors the classical case, where the Local Hamiltonian problem is $\QMA$-complete, yet Hamiltonian simulation has an efficient quantum algorithm. To obtain a complete problem for $\mQMA$, we cannot simply modify the pure-state problem to the form $H_{\rho}:= \sum\limits_sH_s-\sum\limits_\ell\big(\rho \otimes H_\ell\big)$. For an explanation of why this approach fails and how we construct a $\mQMA$-complete problem, we refer the reader to the technical overview in Section~\ref{sec:tech_pmQMA_complete}.

\paragraph{Unconditional separation between quantum promise complexity classes and its implication.}

Next, we investigate the separation between quantum promise complexity classes. We study the separation between classical and quantum advice in the context of our quantum promise problem. In complexity-theoretic terms, one example is examining the relationship between $\pBQPpoly$ and $\pBQPqpoly$.
\begin{theorem}[Theorem~\ref{thm:sep_pure_BQPpoly_BQPqpoly}, Theorem~\ref{thm:sep_mix_BQPpoly_BQPqpoly}]\label{thm:pure_BQPpolyneqBQPqpoly}
    $\pBQPpoly \subsetneq \pBQPqpoly$ and $\mBQPpoly \subsetneq \mBQPqpoly$.
\end{theorem}

Theorem~\ref{thm:pure_BQPpolyneqBQPqpoly} shows that we can separate $\pBQPpoly$ and $\pBQPqpoly$ unconditionally for our quantum promise problems. This is surprising given the long history of studying $\BQPpoly$ and $\BQPqpoly$ in complexity theory~\cite{AK07,liu23,NN23,LLPY24}. The construction of a classical oracle $\cO$ that separates $\BQPpoly$ and $\BQPqpoly$ remains an open major question in complexity theory, and demonstrating an unconditional separation between these classes is still considered beyond the reach of current techniques. 

\begin{remark}
    The unconditional separations in Theorem~\ref{thm:pure_BQPpolyneqBQPqpoly} show that our quantum promise complexity classes provide a more fine-grained characterization of quantum resources compared to classical decision problems.\HZZ{I forget what does this sentence means (although we discussed this)?} To the best of our knowledge, the only other relevant unconditional separation is $\class{FBQP/qpoly} \neq \class{FBQP/poly}$ as shown in~\cite{ABK24}, which addresses classical relation problems rather than decision problems.
\end{remark}

%This result arises as a byproduct of our construction of an unconditionally secure commitment scheme. Technically, we leverage lower bounds on sample complexity from quantum property testing, together with the concentration properties of the Haar measure. For more details, see Theorem~\ref{thm:info_un_shcsb_com} and Section~\ref{sec:tech_un_shcb_com}.

%We showed that some containment results still hold for quantum promise complexity classes. Surprisingly, we can also show that $\pQIP[2]$ and $\mQIP[2]$ are not contained in $\pPSPACE$ and $\mPSPACE$; however, $\pmQMA$ and $\pmQCMA$ are in $\pmPSPACE$.

%The following theorem says that the $\PSPACE$ algorithm with polynomial copies of input states is more powerful than quantum Merlin-Arthur, where Merlin receives unbounded copies of input.

Next, we present two separation results that differ from the corresponding results in classical complexity theory. First, we show that $\pQIP$ and $\mQIP$ are not contained in $\pPSPACE$ and $\mPSPACE$, respectively. We actually show something stronger (Theorem \ref{thm:mpQIP_2_not_in_mpINF}): even $\class{p/mQSZK_{hv}[2]}$ is not contained in $\pmINF$, where the latter refers to any single-party algorithm that has access to polynomial copies of the input (Definition~\ref{Def:physically_realizable}). Second, we show that $\mQIP$ and $\mQSZKhv$ are not closed under complement. In contrast, in classical complexity theory, $\QIP$ is closed under complement because $\QIP = \PSPACE$, and Watrous~\cite{Wat02} showed that $\QSZKhv$ is also closed under complement. We state both results as follows:

\begin{theorem}[Theorem \ref{thm:mQSZKhv_not_inside_mINF}, Theorem \ref{thm:pQSZKhv_not_inside_pINF}]\label{thm:mpQIP_2_not_in_mpINF}
    $\class{p/mQSZK_{hv}[2]} \not \subseteq \pmINF$.
\end{theorem}

\begin{theorem}[Theorem \ref{thm:mQSZKhv_and_mQIP_not_close_under_complement}]\label{thm:intro_mQSZKhv_and_mQIP_not_close_under_complement}
$\class{mQIP}$ and $\class{mQSZK_{hv}}$ are not closed under complement.
\end{theorem}

\begin{remark}
 In our framework, we have two types of separation results (i) sample complexity (ii) computational power. We say that a separation result $\class{p/mC_1} \nsubseteq \class{p/mC_2}$ is due to sample complexity if the condition $\class{p/mC_1} \nsubseteq \pmINF$ holds. Note that $\pmINF$ denotes the class of problems solvable in unbounded time with arbitrary advice, given access to only polynomially many copies of the input. In this case, the separation arises from limitations in sample complexity rather than computational power. For example, the separation result $\pmQIP \nsubseteq \pmPSPACE$ is attributable to sample complexity, since we also obtain $\pmQIP \nsubseteq \pmINF$. On the other hand, if $\class{p/mC_1} \nsubseteq \class{p/mC_2}$ and $\class{p/mC_1} \subseteq \pmINF$, then we say that the separation result is computational. Indeed, the separation result $\pBQPpoly \subsetneq \pBQPqpoly$, which equivalent to $\pBQPqpoly \not\subseteq \pBQPpoly$, arises from computational power, as $\pBQPqpoly \subseteq \pINF$. \HZZ{Why? I feel this is wrong?}   
\end{remark}

%Now, we provide a high-level idea for proving Theorem \ref{thm:mpQIP_2_not_in_mpINF}. We first consider the mixed-state version. To show a separation between the two classes, we need to define a mixed-state quantum promise problem $\cL$ satisfying that 
%a) Any single-party algorithm with access to polynomial copies of input cannot decide $\cL$, and b) there exists an interactive proof where the prover, with access to unbounded copies of the input, can decide $\cL$. We consider the mixedness testing problem $\cL_{mix}$ as follows:

Even more surprisingly, we are able to connect both the separation results Theorem~\ref{thm:mpQIP_2_not_in_mpINF} and Theorem~\ref{thm:intro_mQSZKhv_and_mQIP_not_close_under_complement} to two different topics: quantum property testing and quantum input unitary synthesize problems. 

\paragraph{Implication for quantum property testing.}

First, we show that we can extend Theorem~\ref{thm:mpQIP_2_not_in_mpINF} and Theorem~\ref{thm:intro_mQSZKhv_and_mQIP_not_close_under_complement} to some well-studied quantum property testing problems.\footnote{Quantum property testing is a special case of quantum promise problems, where the no case includes all instances which are $\epsilon$-far from the yes case.} We state the results as follows:

\begin{theorem}[Informal, separation in quantum property testing]\label{thm:intro_sep_quantum_property} Let $\cL_{mix}$ (Definition~\ref{def:L_maximall_mixed_property_problem}) be the mixed property of testing a state is a maximally mixed state or $\epsilon$-far from it and $\cL_{ME}$ (Definition~\ref{def:purifyMix_property}) be the pure property of testing a state is a maximally entangle state or $\epsilon$-far from it.\footnote{Here, $\epsilon(\cdot)$ is fixed to be some $1/\poly$.} Then the following holds,
\begin{enumerate}
    \item $\overline{\cL_{mix}} \notin \mINF \text{ and } \overline{\cL_{mix}} \in \mQSZKhvsec$. (Corollary~\ref{corollary:mixness_cpmlement_in_mQSZK})
    \item $\cL_{ME} \notin \pINF \text{ and } \cL_{ME} \in \pQSZKhvsec$. (Theorem~\ref{thm:purifymixPorperty_in_pQSZKhv})
    \item $\overline{\cL_{mix}} \in \mQIP (\mQSZKhv) \text{ and } \cL_{mix} \notin \mQIP (\mQSZKhv)$. (Corollary~\ref{cor:mixed_prpoery_hard_for_single_and_interactive})
\end{enumerate}
\end{theorem}

Previous work on quantum property testing has focused primarily on the sample complexity of single-party algorithms. Recently, several works~\cite{CG18,HR22,HR24,HR25,BBCS26} have studied interactive models for distribution testing, which can be considered as a special case of mixed properties. To the best of our knowledge, such a model has not been previously studied in quantum property testing, and Theorem~\ref{thm:intro_sep_quantum_property} shows how interaction fundamentally changes the landscape of quantum property testing. In the following, we provide an interpretation of Theorem~\ref{thm:intro_sep_quantum_property}. 

The first two results demonstrate the power of interactive proofs. It is known that deciding $\cL_{mix}$ or $\cL_{ME}$ requires exponentially many copies of the input state for any single-party algorithm~\cite{CHW07,o2021quantum,CWZ24}. In contrast, with the assistance of a prover, a verifier can decide these properties using only polynomially many copies of the input, polynomial communication, and polynomial verification time. Moreover, these protocols can be made zero-knowledge.

On the other hand, our final result reveals a fundamental limitation of the interactive model. Even with the help of a prover, if the verifier is restricted to polynomially many copies of the input state, then $\cL_{mix}$ cannot be verified. This suggests that the interaction model is insufficient to overcome the sample complexity barrier for all quantum properties. An interesting direction for future work is to identify other natural quantum properties for which interaction either does or does not provide a similar advantage.

\paragraph{Implication for quantum input unitary synthesize problems}

Interestingly, as a byproduct of the separation results in Theorem~\ref{thm:mpQIP_2_not_in_mpINF}, we also obtain the impossibility results of the quantum input unitary synthesize problem. We state the results as follows. 
\begin{theorem}[Informal, Theorem \ref{thm:lower_bound_uhl}]\label{thm:intro_lower_bound_uhl}
    Fix any two bipartite state $\ket{\phi}_{AB}$ and $\ket{\psi}_{AB}$. Let $U^{Uhl}_{\ket{\phi},\ket{\psi}}$ be the Uhlmann's transformation (Definition~\ref{def:unitary_Uhl}):  which map $\ket{\phi}$ as close as possible to $\ket{\psi}$ by acting only on the $B$-regitser of $\ket{\phi}$. Then, given polynomially many copies of input states $\ket{\phi}$ and $\ket{\psi}$, no unbounded-time algorithm, even one equipped with arbitrary-size advice\HZZ{Is this quantum or classical advice?}, can synthesize unitary $U^{Uhl}_{\ket{\phi},\ket{\psi}}$ within error $1 - \frac{1}{p(\cdot)}$ for any polynomial $p(\cdot)$.
\end{theorem}

\begin{theorem}[Informal, Theorem \ref{thm:lower_bound_dis}]\label{thm:intro_lower_bound_dis}
    Fix any two mixed state $\rho$ and $\sigma$ promising that the fidelity of $\rho$ and $\sigma$ is at most $\frac{1}{\alpha(\cdot)}$, where $\alpha(\cdot)$ is some polynomial.  Let $U^{dis}_{\rho,\sigma,p(\cdot)}$ be the state discrimination unitary (Definition~\ref{Def:unitary_dis}):  which distinguish $\sigma$ and $\rho$ with advantage at least $\frac{1}{p(\cdot)}$. Then, given polynomially many copies of input states $\rho$ and $\sigma$, no unbounded-time algorithm, even one equipped with arbitrary-size advice, can synthesize unitary $U^{dis}_{\sigma,\rho,p(\cdot)}$ for any polynomial $p(\cdot)$.\footnote{Note that \cite{Qia23, MNY24} also implicitly gave the same result as Theorem~\ref{thm:intro_lower_bound_dis}. However, we provide an alternative proof: while their approach relies on the multi-instance games technique, ours reduces the problem to an interactive protocol for a quantum promise problem.}\HZZ{Is this quantum or classical advice?}
\end{theorem}

In fact, any separation between single-party algorithm ($\pINF$ or $\mINF$) and interactive proof model ($\pQIP$ or $\mQIP$) for QPPs immediately yields an impossibility result for certain quantum-input unitary synthesis problems. To see why such connection exists, interested readers are referred to Section~\ref{subsec:imposs_unitary_synth} for a more detailed discussion.

\paragraph{Other structure results for QPPs complexity class: inclusion, complement and search-to-decision. }

Next, we examine the inclusion relationships among various quantum promise complexity classes, yielding results analogous to those in classical complexity theory.

\begin{theorem}[Theorem $\ref{thm:QMA_Upperbound}$]
\label{thm:informal_up}
Pure-state complexity classes have the following relation: $\pBQP \subseteq \pQCMA \subseteq \pQMA \subseteq \pPSPACE$. Also, mixed-state complexity classes have the following relation: $\mBQP \subseteq \mQCMA \subseteq \mQMA \subseteq \mPSPACE$.
\end{theorem}

The first two inclusions follow from the same reasons as $\BQP\subseteq \QCMA \subseteq \QMA $. Along this line, one might expect that the last inclusion, $\pmQMA \subseteq \pmPSPACE$ also holds by using the same idea for proving $\QMA \subseteq \PSPACE$. However, as we discuss in Section~\ref{sec:properties}, since $\pmQMA$ allows the prover to use unbounded copies of quantum inputs, it is unclear how a polynomial-space quantum machine with at most polynomial copies of quantum inputs can simulate the prover in general. Therefore, the original proof for showing $\QMA \subseteq \PSPACE$ does not work for $\pmQMA \subseteq \pmPSPACE$. We show that the $\pmQMA$ complete problem, Quantum Or problems, belongs to $\pmPSPACE$ by adapting Aaronson's de-merlinization protocol \cite{aaronson2006qma, harrow2017sequential}. Combining with Theorem~\ref{thm:mpQIP_2_not_in_mpINF}, we observe an unconditional separation between non-interactive and interactive proof, i.e., $\pmQMA \subsetneq \pmQIP[2]$.

Last, we also study the complement property for $\pQSZKhv$. In classical complexity theory, $\class{QSZK_{hv}}$ is closed under complement~\cite{Wat02}. Surprisingly, we can show that $\pQSZKhv$ is also closed under complement; 

\begin{theorem}[Theorem \ref{thm:pQSZKhv_close}] \label{thm:intro_pQSZKhv_close}
    $\class{pQSZK_{hv}}$ is closed under complement. 
\end{theorem}
% \begin{theorem}[Theorem \ref{thm:mQSZKhv_and_mQIP_not_close_under_complement}]\label{thm:intro_mQSZKhv_and_mQIP_not_close_under_complement}
% $\class{mQIP}$ and $\class{mQSZK_{hv}}$ are not closed under complement.
% \end{theorem}
Combined with Theorem~\ref{thm:intro_mQSZKhv_and_mQIP_not_close_under_complement}, the above results demonstrate that the complexity classes for pure-state and mixed-state quantum promise problems can behave differently.

Analogous to $\QCMA$, $\pQCMA$ and $\mQCMA$ also have the search-to-decision reductions (Theorem \ref{intro_thm:StoD}). On the other hand, we do not know how to obtain search-to-decision reduction for $\pmQMA$. Note that whether $\QMA$ has a search-to-decision reduction is a long-standing open question. See Open problem 1 in Section \ref{section:open_problem} for more discussion. 
\begin{theorem}[Search-to-decision reduction \ref{lem:searchToDecision_version2}, informal]\label{intro_thm:StoD}
    Consider a $\pmQCMA$-complete problem $\cL$. Then, there exists a polynomial-time oracle algorithm $\mathcal{A}^{\cL}$ that finds the witness for a ``yes" instance of $\cL$. 
\end{theorem}
The statement of Theorem \ref{intro_thm:StoD} is not well-defined until we formally define oracle access to a quantum promise complexity class. When an algorithm $\cA$ queries a quantum oracle $\cO$, $\cA$ is allowed to send only a polynomial number of copies of a quantum state to the oracle.\footnote{All of our work deals with polynomial copies of input when considering a single-party algorithm. Therefore, in our definition, we require that the oracle only receive polynomial copies of input.} We generally require that $\cO$ must operate linearly, according to the postulates of quantum physics. More formally, we consider physically realizable $\cO$, where a CPTP map exists as an instantiation of $\cO$ (See Section $\ref{Def:Oracle}$ for a formal definition). With this definition, $\pmQCMA$ can serve as an oracle because it is known that $\pmQCMA$ is upper-bounded by $\pmPSPACE$ by Theorem \ref{thm:informal_up}. We also observe that not all quantum promise complexity classes can serve as oracles (e.g., $\pQIP$). Specifically, when $\cA$ attempts to query $\pQIP$, it cannot send a sufficient number of copies, and thus no physically realizable oracle $\cO$ can solve $\pQIP$ with only polynomially many copies of the input.

%Let us come back to search-to-decision reduction. $\mathcal{A}^{\pmQCMA}$ can be view as $\cA$ solving all $\pmQCMA$ problems in a single step. The proof is analogous to the classical version. However, $\NP$-type search-to-decision reduction deals with languages, not promise problems. Therefore, input instances never ``drop'' outside the promise, but this is not true for $\pmQCMA$. To address this issue, we modify the size of $\cL_Y$ and $\cL_N$ region for queries with different lengths of prefixes. Precisely, the yes region of longer prefix lengths always encompasses the complement of no region of shorter prefix lengths \footnote{\cite{Aar20} also use the similar technique.}.

By applying the search-to-decision reduction, the prover only needs a polynomial number of input states to find a classical witness.
\begin{corollary}[Corollary~\ref{thm:qcmaR}]\label{intro:qcmaR}
    $\pmQCMA^{\poly} = \pmQCMA$.
\end{corollary}

\subsubsection{Characterize quantum cryptography using QPP complexity theory}

Traditional complexity theory plays an important role in understanding the computational hardness cryptography. One direction is to determine what kinds of hardness assumption are necessary to construct cryptographic primitives. A notable example is that the existence of one-way functions (OWFs) implies that $\class{P} \neq \class{NP}$. This result suggests that proving the existence of OWFs is beyond current techniques, since establishing $\class{P} \neq \class{NP}$ faces numerous barrier results.
The other direction is to understand what kinds of hardness assumption are sufficient to construct cryptographic primitives—for instance, constructing OWFs based on worst-case traditional complexity assumptions~\cite{HN24, LMP24}. We demonstrate that QPP complexity theory characterizes many primitives in the field of quantum cryptography. We focus on the recently proposed new cryptographic world: Microcrypt. 

%Recently, two distinct cryptographic worlds have been introduced in the field of quantum cryptography: Microcrypt and unconditional cryptography. We demonstrate that QPP complexity theory characterizes many primitives in these two new worlds.

%\paragraph{Application to Microcrypt:} 

The world of Microcrypt encompasses set of cryptographic primitives that are potentially weaker than OWFs. Examples include the one-way state generator (OWSG)~\cite{morimae2022one}, which is the quantum analogue of OWF with its output replaced by a quantum state. Other examples, such as pseudorandom states (PRS)~\cite{pqs} and EFI~\cite{brakerski2022computational}, can be viewed as quantum analogues of pseudorandom generators and EFID~\cite{Gol90}, respectively, again with outputs changed to quantum states. The security of these primitives are quantum-input problems, and notably, none of their hardness are known to imply $\class{P} \neq \class{NP}$. Currently, the smallest known traditional complexity separation arises from the PRS primitive, whose existence implies  $\class{BQP}\neq \class{PP}$~\cite{Kretschmer2021QuantumPA}; for EFI, no known traditional complexity separation exist. Hence, constructing EFI unconditionally does not encounter any of the barrier results from traditional complexity theory.
The central reason is that the security of new primitives are inherently quantum-input problems, whereas traditional complexity theory primarily deals with classical-input decision problems.
However, this perspective shifts considerably when viewed through the lens of QPP complexity theory. 

We briefly summarize our results related to Microcrypt. First, we present a necessary hardness assumption for EFI:
\begin{theorem}\label{app:crypo_EFI}
If EFI exists, then $\mBQP \subsetneq \mPSPACE$ and $\mBQP \subsetneq \mpolyQSZKhv$.\footnote{The definition of $\mpolyQSZKhv$ is similar to $\mQSZKhv$, with the only difference being in completeness. The honest prover is limited to a polynomial number of copies of input to run the protocol. This limitation subtly stems from the fact that $\mQSZKhv \not\subseteq \mINF$.}   
\end{theorem}
Currently, the relationship between $\mPSPACE$ and $\mpolyQSZKhv$ remains unknown, and thus the corresponding results are incomparable. Notably, the separation $\mBQP \subsetneq \mPSPACE$ implies the relativization barrier for constructing EFI, since even establishing a separation between the mixed-state classes $\mBQP$ and $\mPSPACE$ is known to encounter such a barrier.\footnote{See ``Containment and separation of quantum promise complexity classes'', in Section~\ref{sec:properties}, for further discussion.}

Next, we present a natural necessary hardness assumption for OWSG and PRS:
\begin{theorem}\label{app:crypo_OWSG_PRS} If a pure-state OWSG or PRS exists, then $\pBQP \subsetneq \pQCMA$. Moreover, if a mixed-state OWSG exists, then $\mBQP \subsetneq \mQCMA$. \end{theorem}

Theorem~\ref{app:crypo_EFI} and Theorem~\ref{app:crypo_OWSG_PRS} also provide new insights that motivate a quantum analogue of Impagliazzo’s five worlds.\footnote{See Section~\ref{section:open_five_world} for more discussion.}
Lastly, we show that quantum promise complexity theory also serves as a useful hardness assumption for constructing primitives in Microcrypt. Here is our result:
\begin{theorem}[Informal, Theroem \ref{Thm:pQSZK_hv_imply_EFI}]\label{thm:EFI_construction}
        If there is a quantum promise problem $\mathcal{L}$ in $\pQCZKhv$ that is hard on average, then EFI pairs exist. 
\end{theorem}
Our result generalizes~\cite{brakerski2022computational}, which constructs EFI from the average-case hardness of $\QCZKhv$. As $\QCZKhv$ has classical inputs, prior EFI constructions were limited to classical-input problems. We extend this to the average-case hardness of $\pQCZKhv$, yielding a weaker assumption since $\QCZKhv \subseteq \pQCZKhv$.

%% file: 1-3_tech_overview.tex
\subsection{Technical overview}

In the technical overview section, we focus on two main results. The first is a $\mQMA$ complete problem. The second is about our three unconditionally separation results (i) $\mQSZKhv \neq \mINF$, (ii) $\pQSZKhv \neq \pINF$ and (iii) $\pBQPpoly \neq \pBQPqpoly$.

\subsubsection{$\pmQMA$ complete problem}\label{sec:tech_pmQMA_complete}

%One might initially think that complete problems for $\QCMA$, $\QMA$, and $\QSZK$ could serve as potential candidates for complete problems in their corresponding quantum promise complexity classes, since classical complexity classes are subsets of these quantum promise classes. For example, 
We first recall the result that the Local Hamiltonian problem is $\QMA$-complete~\cite{kitaev2002classical}, and show that by slightly modifying the problem, we can obtain a pure-promise problem that is $\pQMA$-complete. Finally, we explain the challenges in obtaining a $\mQMA$-complete problem and how we overcome them.
Kitaev, Shen, and Vyalyi \cite{kitaev2002classical} show that the local Hamiltonian problem is $\QMA$ complete, where the problem input is the sum of local Hamiltonians, $H:=\sum\limits_{s \in S}H_s$. To prove this problem is $\QMA$-hard, the reduction reduces any instance $x$ to a local Hamiltonian $H:=H_{in}^I+H_{in}^A+H_{out}+H_{prop}+H_{stab}$, where each term in $H$ imposes a specific constraint on the witness state.\footnote{For a good introduction, the reader is referred to \cite{kitaev2002classical}. $H_{in}^I$ requires the input register of the witness (with the clock register at the starting time) to be $\ket{x}$; $H_{in}^A$ requires the ancilla register of the witness (with the clock register at the starting time) to be $\ket{0}$; $H_{out}$ requires the answer register of the witness (with the clock register at the ending time) to be $\ket{1}$ (which means accept); $H_{prop}$ requires the witness to evolve correctly based on the $\QMA$ verifier; and $H_{stab}$ requires the witness encoded the clock register correctly.} If $x \in \cL$, then $H$ has a low-energy state. Otherwise, $H$ has no low-energy state. However, it is unclear how problems with unknown quantum inputs can be reduced to the local Hamiltonian problems. This challenge arises because such reductions must map quantum inputs of quantum promise problems to the classical inputs of these classical complete problems.

We overcome this difficulty by the following approach: embedding an unknown state into the inputs so that the modified problem has the proper interface for reductions from quantum promise problems in the complexity class. Specifically, consider a variant of the local Hamiltonian problem where the problem input is defined as $H_{|\psi\rangle}:= \sum\limits_sH_s-\sum\limits_\ell\big(\ket{\psi}\bra{\psi} \otimes H_\ell\big)$. Note that $H_s$ and $H_\ell$ are local Hamiltonians, but $\ket{\psi}\bra{\psi}$ is a non-local term. To show that this problem is $\pQMA$-hard, we reduce any instance $\ket{\psi}$ to $H:=H_{in}^I+H_{in}^A+H_{out}+H_{prop}+H_{stab}$, where $H_{in}^I:=(I-\ket{\psi}\bra{\psi})\otimes H_\ell$ verifies that the witness has the correct input state, and the other terms of $H$ are the same as in the classical cases. To show that this problem belongs to $\pQMA$, we can combine the original method with the swap test technique to measure the overlap between the witness and $\ket{\psi}\bra{\psi}\otimes H_\ell$.

Unfortunately, we cannot use a similar method to define candidates for $\mQMA$- or $\mQCMA$-complete problems. Indeed, when embedding an unknown mixed state into local Hamiltonian problems, two issues arise: (i) when proving containment results, the verifier cannot use the swap test technique to estimate the overlap between the witness and $\rho \otimes H_{\ell}$. Even checking the fidelity between two mixed states is impossible, as shown in \cite{o2021quantum}. (ii) When proving hardness results, ``yes" instances cannot be reduced to a Hamiltonian $H$ that has a low-energy state. Indeed, the input $\rho$ may be far from any pure state. 

To address this, the $\mQMA$-complete problem involves a distribution of $\{H_{\ket{\psi}}\}$, where the distribution of $\ket{\psi}$ corresponds to the input mixed state $\rho$. The task is to determine whether $\{H_{\ket{\psi}}\}$ has an expected low-energy state. That is, suppose $\rho:= \sum\limits_i \lambda_i\ket{\psi_i}\bra{\psi_i}$. Then, the problem asks whether $\sum\limits_i \lambda_i \min\limits_{\eta} \bra{\eta} H_{\ket{\psi_i}} \ket{\eta}$ is small. We can prove that the distributional local Hamiltonian problem is $\mQMA$-hard by adapting the technique in \cite{kitaev2002classical} and the average argument.

\paragraph{Challenges of proving mQMA-complete} This modified problem is likely too hard to fall within the $\mQMA$ complexity class, as we explain in the following paragraphs. We will also describe how to introduce a properly defined problem that is $\mQMA$-complete. Specifically, we add two additional constraints in the following Steps 1 and 2 to ``simplify'' the problem, and design a verification procedure in Step 3, so that the problem is in $\mQMA$.

\paragraph{Step 1: State-dependent witness} To begin with, we can view the input mixed state $\rho$ as a distribution over pure states. Additionally, for simplicity, we assume $H_{\ket{\psi}}:=H_s-\ket{\psi}\bra{\psi} \otimes I$.
The verifier should use one copy of such a pure state and a witness to estimate the energy. However, the prover is unaware of the low-energy state, as the prover does not know the input sample, $\ket{\psi}$, obtained by the verifier. Hence, the low-energy state, denoted as $\ket{\eta_{\psi,\phi}}$, is constructed by the verifier itself, with the help of the prover. Specifically, the prover sends a polynomial-size circuit $\cC$ and a witness $\ket{\phi}$ such that, for any state $\ket{\psi}$ sampled from the input mixed state, the verifier uses $\cC$, $\ket{\phi}$, and $\ket{\psi}$ to construct a low-energy state $\ket{\eta_{\psi,\phi}}$. Hence, we additionally assume that in a yes-instance, the Hamiltonian $H$ satisfies the following promise: there exists a circuit $\cC$ and a witness $\ket{\phi}$ such that
\begin{equation*}
    \sum\limits_i \lambda_i  \bra{\eta_{\psi_i,\phi}} H_{\ket{\psi_i}} \ket{\eta_{\psi_i,\phi}}
\end{equation*}
is small. Additionally, we split the expected energy into two terms, $I$ and $D$: the first term corresponds to the part where the Hamiltonian is independent of the input state $\ket{\psi}$, while the second term corresponds the part that is dependent on the input state $\ket{\psi}$. That is,
\begin{equation*}
\begin{cases}
    I := \sum\limits_i \lambda_i  \bra{\eta_{\psi_i,\phi}} H_s \ket{\eta_{\psi_i,\phi}} \\
    D := \sum\limits_i \lambda_i D_{\psi_i} \text{, where } D_{\psi_i}:= \bra{\eta_{\psi_i,\phi}} \cdot (\ket{\psi_i}\bra{\psi_i} \otimes I) \cdot \ket{\eta_{\psi_i,\phi}}. 
\end{cases}
\end{equation*}
Also, we note that the energy $\sum\limits_i \lambda_i  \bra{\eta_{\psi_i,\phi}} H_{\ket{\psi_i}} \ket{\eta_{\psi_i,\phi}}$ is equal to the difference of the two components, namely $I-D$.

%\lch{I is overloaded: also denote identity and input register}

\paragraph{Step 2: Estimate the expected energy using a single-copy state with the ``Hadamard test"} 

The first challenge is to estimate $D_\psi$ using only a single copy of the state $\ket{\psi}$ (We suppress the subscript $i$ for simplicity). We are no longer allowed to use the swap-test technique. This is because the state $\ket{\eta_{\psi,\phi}}$ is constructed from $\ket{\psi}$, and the swap test would need to be applied to the joint state $\ket{\eta_{\psi,\phi}} \otimes \ket{\psi}$. Instead, the verifier uses ``Hadamard testing" to estimate the energy. Specifically, suppose we use a single copy of $\ket{\psi}$, a single copy of witness $\ket{\phi}$, and the circuit $\cC$ to construct the following state
\begin{equation}\label{intro:qma_complete_hadamardtest}
    \frac{1}{\sqrt{2}}\ket{0} \otimes \ket{\psi}\ket{\phi}\ket{0}+\frac{1}{\sqrt{2}}\ket{1}\otimes \ket{\eta_{\psi,\phi}},
\end{equation}
To estimate the energy $D_\psi$, we further rewrite $\ket{\eta_{\psi,\phi}}$ as follows:
\begin{equation*}
    \ket{\eta_{\psi,\phi}} := \alpha\ket{\psi}\ket{G} + \sum\limits_{\psi^\perp} \beta_{\psi^\perp}\ket{\psi^\perp}\ket{G_{\psi^\perp}}.
\end{equation*}
Furthermore, for simplicity, we ignore the terms involving $\ket{\psi^\perp}$, so that $\ket{\eta_{\psi,\phi}}$ becomes a non-normalized state. Then, we have
\begin{equation*}
\begin{aligned}
    &\,\text{Equation}~(\ref{intro:qma_complete_hadamardtest}) \\
    := & \frac{1}{\sqrt{2}}\ket{0} \otimes \left(\ket{\psi}\ket{\phi}\ket{0}\right) +\frac{1}{\sqrt{2} }\ket{1} \otimes \alpha\ket{\psi}\ket{G},
\end{aligned}
\end{equation*}
If we measure the first qubit in the Hadamard basis, the quantity $-\textbf{Re}(\alpha \cdot \braket{\phi, 0|G})$ is encoded in the success probability. More formally, if we define a random variable $X$ being the outcome of the above "Hardmard test", we get $\E[X] = -\textbf{Re}(\alpha \cdot \braket{\phi, 0|G})$. However, the goal we want to estimate is $-D_{\psi}=-|\alpha|^2$.

Hence, we additionally promise that the yes-instance satisfies the following properties: (i) $\alpha$ is a real number (ii) $\braket{\phi, 0|G}=1$. Also, for the sake of clarity, we additionally assume that $\alpha$ is a positive number and take this as given.\footnote{In our actual promise problem, this assumption is not required.}   Hence, when the prover is honest, we can estimate $-\alpha$, which equals $-\sqrt{D_{\psi}}$. Nevertheless, this already brings us one step closer. To analyze soundness, observe that when $\braket{G_0|G_1} < 1$, or when $\alpha \cdot \braket{\phi, 0|G}$ is not a positive real number, the energy will be overestimated. Because a malicious prover aims to present a state with energy as low as possible in order to pass the verification, the optimal cheating strategy must still satisfy the same promise conditions as in the ``yes" case.

% Therefore, at a high level, we must promise that in the ``yes'' instance, the following condition holds: there exists a low-energy state $\ket{\eta_{\psi,\phi}}$ such that, after performing Hadamard testing, we recover the value $|\alpha_0\alpha_1|$. This can be achieved by requiring that $\braket{G_0|G_1} = 1$ and $\alpha_0\alpha_1 \geq 0$.\footnote{In our actual promise problem, we do not require the condition $\alpha_0\alpha_1 \geq 0$. However, for the sake of clarity and simplicity, we impose this restriction in the technical overview section.} Moreover, a malicious prover will attempt to present a state with energy as low as possible in order to pass the verification. Hence, even in the presence of a dishonest prover, the optimal cheating strategy must still satisfy the same promise conditions as in the ``yes" case. \nai{However, this approach can only obtain $\sqrt{B}$.}

%This brings us to our second problem: \textbf{Bias from Entangled Quantum Witnesses}.

\paragraph{Step 3: Replace estimation with verification using a ``hint'' from the prover} 
The Hadamard test estimates only $\sqrt{D_\psi}$ rather than the needed $D_\psi$. For completeness, a natural approach is to run the test on $N$ separate blocks of the witness, obtaining outcomes $X_1, \dots, X_N \in \{-1, 0, 1\}$, average them to estimate $\sqrt{D_\psi}$, and then square the result. This works for honest prover where each block is mutually independent. However, a cheating prover may send an entangled ``witness'', resulting in correlated outcomes whose square of average may far to $D_\psi$.

We circumvent this issue by replacing estimation with verification. 
%Instead of estimating $\sqrt{D_\psi}$ from the measurements, the verifier asks the prover to declare its value in advance and uses the measurements only to check this declaration.
Along with the circuit and the witness state, the prover declares a real number $\aprov\in [-1,1]$ claimed to be $\sqrt{D_\psi}$. The verifier checks this declaration against the measurement outcomes and reject unless $\left|\aprov -\overline{X}\right| > \eps$ where $\overline{X}\coloneqq \frac1N \sum_j X_j$. If the check passes, the verifier uses $\aprov^2$ in place of $D_\psi$.

The reason why this bypasses the correlation problem is that the cheating prover must send a large value of $\aprov$ to claim that the energy is small. If $\aprov$ is smaller than $\E[\overline{X}] + \epsilon$, the final energy estimation will still be larger than the threshold. We only need to catch the case where $\aprov$ is larger than $\E[\overline{X}] + \epsilon$. We don't need to deal with the correlation in such case because we only care about the distance between $\aprov$ and $\E[\overline{X}] + \epsilon$. Finally, we need to argue how the verification catch the case of $\aprov > \E[\overline{X}] + \epsilon$. We can not directly apply Chernoff bound in such case because the cheating prover can make the $\{X_i\}$ arbitrary correlated. Fortunately, we can apply Markov style like argument to say that if $\aprov > \E[\overline{X}] + \epsilon$, the verification fails $\left(|\aprov - \overline{X}| > \epsilon\right)$ with probability with at least $\epsilon/3$.

\subsubsection{Three unconditional separations}

In this subsection, we will provide a technical overview of our three unconditional separations (i) $\mQSZKhv \neq \mINF$, (ii) $\pQSZKhv \neq \pINF$ and (iii) $\pBQPpoly \neq \pBQPqpoly$.  

\paragraph{Separation between $\mQSZKhv$ and $\mINF$.}
We first define our QPPs $\cL_{mix^*}$as follows
\begin{equation*}
    \cL_{mix^*,Y}:=\{T\rho_{\halfs}T^{\dagger}: T \text{ is arbitrary } \lambda \text{-qubit unitary}\}\text{, } \cL_{mix^*,N}:=\{\frac{I}{2^\lambda}\},
\end{equation*}
where \(\rho_{\halfs}:= \frac{1}{2^{\lambda-1}} \sum_{i \in \{0,1\}^{\lambda-1}}\ket{i||0}\bra{i||0}\) and $a||b$ mean the string concatenation between $a$ and $b$. We first observe that $\rho_{in} \in \cL_{mix,Y}$ is just the maximally mixed state on some subspace with dimension $2^{\lambda-1}$, and $\cL_{mix,N}$ contains only one mixed state, which is maximally mixed state. The sample complexity of $\cL_{mix^*}$ for the single-part algorithm is a well-studied problem in the field of quantum property testing (related to testing a state is a maximally mixed state or $\epsilon$-far from it). We state the hardness result for $\cL_{mix^*}$~\cite{CHW07,MdeW16,o2021quantum} as follows: for any polynomial $p(\cdot)$, and any algorithm $\cA$
\begin{equation}\label{equation:mixed_hard}
        \left| \Pr\limits_{T \leftarrow Haar}\left[ \cA( (T\rho_{\halfs}T^{\dagger})^{\otimes q})=1 \right] - \Pr\left[ \cA( (\frac{I}{2^\lambda})^{\otimes q})=1 \right] \right| = \negl(\lambda),  
\end{equation}
where $\text{Haar}$ is the distribution of $\lambda$-qubit Haar random unitaries. This implies $\cL_{mix^*} \notin \mINF$. To obtain the  separation, it remains to show $\cL_{mix^*} \in \mQSZKhv$. Our protocol mimics the classic graph non-isomorphism protocol and proceeds as follows. The verifier first samples $b \leftarrow \{0,1\}$. If $b=0$, the verifier sends  $(\rho_{in})^{\otimes \lambda}$ to the prover. Otherwise, the verifier sends  $(\frac{I}{2^\lambda})^{\otimes \lambda}$. The prover is then asked to output a bit $b'$ such that $b'=b$.

If $\rho_{in} \in \cL_{mix^*,Y}$, the prover can perform the optimal measurement to distinguish $\rho_{in}^{\otimes \lambda}$ from $\left(\frac{I}{2^\lambda}\right)^{\otimes \lambda}$. We claim that the error can only be negligible. We observe that the trace distance between $\rho_{in}$ and $\frac{1}{2^\lambda}$ is equal to $\half$ because $\rho_{in}$ is a maximally mixed state on some subspace with dimension $2^{\lambda-1}$. Then, trace distance can be further amplified to $1-\negl(\lambda)$ given $\lambda$ copies of the state. Hence, the optimal measurement can only cause a negligible error.  On the other hand, if $\rho_{in} \in \cL_{mix^*,N}$, no algorithm can guess $b$ with probability better than $\frac{1}{2}$ because $\rho_{in}$ is exactly equal to $\frac{I}{2^\lambda}$. Finally, we apply the standard parallel repetition to reduce the soundness error to negligible. The protocol is honest-verifier zero knowledge because the simulator already knows the challenge bit $b$ and simply outputs $b'=b$. Therefore, the simulation error is negligible.  

\paragraph{Separation between $\pQSZKhv$ and $\pINF$.} We define our QPPs $\cL_{purifyMix^*}$ as follows
\begin{equation*}
    \begin{aligned}
        &\cL_{purifyMix^*,Y}:=\left\{ (T_1 \otimes T_2)\ket{\halfs}:T_1,T_2 \text{ are arbitrary $\lambda$-qubit unitaries} \right\},\\
        &\cL_{purifyMix^*,N}:=\left\{(I \otimes T)\ket{\epr)} : T \text{ is arbitrary $\lambda$-qubit unitary}\right\},
    \end{aligned}
\end{equation*}
where $\ket{\epr}:=\frac{1}{\sqrt{2^\lambda}}\sum\limits_{i \in \{0,1\}^\lambda}\ket{i}\ket{i}$ and $\ket{\halfs}:=\frac{1}{\sqrt{2^{\lambda-1}}}\sum\limits_{i \in \{0,1\}^{\lambda - 1}}\ket{i\|0}\ket{i\|0}$. It is easy to see that $\cL_{purifyMix^*}$ is a purify version of $\cL_{mix^*}$. Then, we can apply the same protocol as in the previous section to decide $\cL_{purifyMix^*}$ because we can simply discard the purification part of the input state. This implies $\cL_{purifyMix^*} \in \pQSZKhv$. It remains to show $\cL_{purifiyMix^*} \notin \pINF$.

To prove the hardness result, we apply a technique from~\cite{CWZ24}. The authors of~\cite{CWZ24} study how purification affects the sample complexity of mixed-state properties. This is exactly the setting of $\cL_{mix^*}$ and $\cL_{purifyMix^*}$. Informally, they show that, under some conditions (Definition~\ref{def:unitarily-invariant}), changing from a mixed-state property to its purification does not reduce the sample complexity. Moreover, the same conclusion holds in the average-case setting, provided that the underlying distribution satisfies the required conditions (Definition~\ref{def:unitarily-invariant-dist}).

In order to transfer the average-case hardness of $\cL_{mix^*}$ (Equation~(\ref{equation:mixed_hard})) to the average-case hardness of $\cL_{purifyMix^*}$ by their results. We design the distribution $T,T_1,T_2$, defined in $\cL_{purifyMix^*}$, are all sampled from Haar-random unitary.
It is easy to see that $\cL_{purifyMix^*}$ satisfy Definition~\ref{def:unitarily-invariant} and the distribution satisfies Definition~\ref{def:unitarily-invariant-dist}. Therefore, by applying their result together with Equation~(\ref{equation:mixed_hard}), we obtain the following average-case hardness result of $\cL_{purifyMix^*}$: for any polynomial $p(\cdot)$ and any algorithm $\cA$
\begin{equation}\label{equation:pure_hard}
        \left| \Pr\limits_{T \leftarrow Haar}\left[ \cA (((I \otimes T)\ket{\epr})^{\otimes q})=1 \right] - \Pr\limits_{T_1,T_2 \leftarrow Haar}\left[ \cA (((T_1 \otimes T_2)\ket{\halfs})^{\otimes q})=1 \right] \right|  
        = \negl(\lambda).
\end{equation} The average-case hardness trivially implies the worst-case hardness, then we get $\cL_{purifyMix^*}\notin \pINF.$

\paragraph{Separation between $\pBQPqpoly$ and $\pBQPpoly$.} We define $\cL_{purifyMix^*}(\{T^*\})$ as a parameterized version of $\cL_{purifyMix^*}$ as follows
\begin{equation*}
    \begin{aligned}
        &\cL_{purifyMix^*,Y}:=\left\{ (T_1 \otimes T_2)\ket{\halfs}:T_1,T_2 \text{ are arbitrary $\lambda$-qubit unitaries} \right\},\\
        &\cL_{purifyMix^*,N}:=\left\{(I \otimes T^*)\ket{\epr)} \right\}.
    \end{aligned}
\end{equation*}
The only difference between $\cL_{purifyMix^*}(\{T^*\})$ and $\cL_{purifyMix^*}$ is that the no-instances of the former consist of the single state $(I \otimes T^*)\ket{\epr}$, whereas the no-instances of the latter consist of all states of the form $(I \otimes T)\ket{\epr}$, where $T$ ranges over all $\lambda$-qubit unitaries. Since the no-case of $\cL_{purifyMix^*}(\{T^*\})$ contains only a single pure state, we can decide QPP by performing swap-test between $\ket{\phi_{in}}$ and $(I \otimes T^*)\ket{\epr}$. The only issue is that the state $(I \otimes T^*)\ket{\epr}$ may not be efficiently generated. We simply take the quantum advice to be $(I \otimes T^*)\ket{\epr}$. Then we get the following: for any family of $\{T^*\}$,  $\cL_{purifyMix^*}(\{T^*\}) \in \pBQPqpoly$. It remains to show that there exists a family of $\{T^*\}$ such that $\cL_{purifyMix^*}(\{T^*\}) \not\in \pBQPpoly$. We require the following two notation
\begin{equation*}
\begin{aligned}
    &\beta_{\cA,q} : =  \Pr\limits_{T_1,T_2 \leftarrow Haar}\left[ \cA (((T_1 \otimes T_2)\ket{\halfs})^{\otimes q})=1 \right], \\
    & f_{\cA,q}(T): = \Pr\left[ \cA (((I \otimes T)\ket{\epr})^{\otimes q})=1 \right] - \beta_{\cA,q}.
\end{aligned}
\end{equation*} We observe that$|f_{A,q}(T^*)|$ is the advantage of $\cA$ to decide an average case of $\cL_{purifyMix^*}(\{T^*\})$ with distributions $T_1,T_2$ being sampled from Haar unitaries. We can also restate our goal to show that there exists $T^*$ such that for all efficient quantum circuits $C$, $|f_{C,q}(T^*)| = \negl(\lambda)$\footnote{Here we treat classical advice as a circuit, it is sufficient to consider all efficient quantum circuits.}.

We first rewrite Equation~(\ref{equation:pure_hard}) as follows: for any polynomial $q(\cdot)$ and any adversary $\cA$
\begin{equation} \label{equation:pure_hard_rewrite}
  |\E_{T \leftarrow Haar}[f_{\cA,q}(T)]| = \negl(\lambda).   
\end{equation}
If the Lipschitz value of function $f_{\cA,q}(T)$ is at most polynomial in $\lambda$, we can apply the Haar random concentration~\cite{Mec19} technique to get a double exponential concentration as follows: for all polynomial $q(\cdot)$ and any adversary $\cA$
\begin{equation*}
               \Pr_{T\leftarrow Haar} [|f_{\cA,q}(T)| \leq \negl(\lambda) + \E_{V \leftarrow Haar}[f_{\cA,q}(V)]] >  1 - \exp(-2^{\frac{\lambda}{4}}).
\end{equation*} By re-arranging the above equation and fitting the value in Equation~(\ref{equation:pure_hard_rewrite}), we get the following: for all polynomial $q(\cdot)$ and any adversary $\cA$
\begin{equation}\label{equation:double_exp_concen}
               \Pr_{T\leftarrow Haar} [|f_{\cA,q}(T)| \text{ is notictable}]] \leq \exp(-2^{\frac{\lambda}{4}}).
\end{equation}

Since we know the number of distinct efficient quantum circuit is at most exponential. Then, by the double exponential concentration in equation~(\ref{equation:double_exp_concen}) and union bound, we know that there must exist $T^*$ such that for all polynomial $q(\cdot)$ and all efficient quantum circuits $C$, $|f_{C,q}| = \negl(\lambda)$. 

Lastly, it remains to show that the Lipschitz value of the function $f_{\cA,q}(T)$ is at most polynomial in $\lambda$. Since the value $\beta_{\cA,q(\cdot)}$ is constant (independent of $T$), it suffices to consider the term $\Pr\left[\cA\left(((I\otimes T)\ket{\epr})^{\otimes q}\right)=1\right]$. We apply the technique in~\cite{Kretschmer2021QuantumPA}, which states that if $A^{U}$ is a $t$-query algorithm with oracle access to a $\lambda$-qubit unitary $U$, then the function $f'(U):=\Pr[A^{U}=1]$
is $t$-Lipschitz. Therefore, it suffices to show that $\cA(((I\otimes T)\ket{\epr})^{\otimes q})$ can be simulated by a $q$-query quantum algorithm. Because we know that $q$ is a polynomial, the Lipschitz value of $f_{\cA,q}$ is at most polynomial in $\lambda$.

To see why $\cA(((I\otimes T)\ket{\epr})^{\otimes q})$ can be simulated by a $q$-query quantum algorithm $\cA^{T}$. This is immediate because the only dependence on the unitary $T$ is through the input state $(I\otimes T)\ket{\epr}$. The simulator first prepares $q$ copies of $\ket{\epr}$ and then applies the oracle $T$ to the second register of each copy, requiring exactly $q$ oracle queries. The resulting state is precisely $((I\otimes T)\ket{\epr})^{\otimes q}$, after which the simulator $\cA^{T}$ runs $\cA$ unchanged. Hence, the simulation is perfect and uses exactly $q$ queries to $T$.

%% file: 1-4_Open.tex
\subsection{Discussion and Open questions}\label{section:open_problem}

We have already obtained several interesting results in the study of quantum promise complexity theory, but there remains much to explore. One direction is to further investigate structural results for the complexity classes defined in this paper (see Section~\ref{section:open_strcut_result}). Another direction is to study how different settings for the number of input copies give rise to distinct complexity classes, and to explore the relationships among these classes (see Section~\ref{section:open_copy}).  Finally, we discuss possible quantum analogues of Impagliazzo’s five worlds and outline several related open questions (see Section~\ref{section:open_five_world}).

\subsubsection{Open questions related to quantum promise complexity theory}\label{section:open_strcut_result}

There are many open questions in quantum promise complexity theory. We list some of them below:

\begin{itemize}
    \item \textbf{Open question 1: Can the prover find a witness of $\pmQMA$, given polynomial copies of input?}\label{open:find_qma_witness} In classical $\QMA$, the prover can always find a witness (in a yes instance) by exhausting all quantum states. Poulin and Wocjan \cite{poulin2009preparing} give an exponential time method to find a ground state of local Hamiltonian by using phase estimation and Grover’s algorithm. Subsequently, Irani et al. \cite{irani2021quantum} gives an algorithm with only a single query of $\pp$ oracles. However, we do not know whether the prover can find a witness of $\pmQMA$ within only polynomial copies of yes instances. We can still say something more that bridges the relation between finding witnesses of $\pmQMA$ and separating complexity classes. Combine Theorem~\ref{thm:informal_up} with Corollary~\ref{intro:qcmaR} (for the pure-state version only, since the mixed-state version has the same argument here), we have:
\begin{center}\label{thm:relation}
    $\pBQP \subseteq \pQCMA = \pQCMA^{\poly} \subseteq \pQMA^{\poly} \subseteq \pQMA \subseteq \pPSPACE$.
\end{center}
Now, we can obtain a ``win-win" result. Either (i) the prover can find a witness of $\pQMA$ within only polynomial copies of yes instances or (ii) $\pBQP \subsetneq \pPSPACE$ (or we can change (ii) to $\pQCMA \subsetneq \pQMA$). Indeed, if (i) does not hold, then $\pQMA^{\poly} \subsetneq \pQMA$. Note that proving the classical analogs, such as $\BQP \subsetneq \PSPACE$ or $\QCMA \subsetneq \QMA$, are considered Holy Grails in computer science. %However, it remains unclear whether the quantum counterparts are tractable.
    \item \textbf{Open question 2: Can we show $\pQSZKhv = \pQSZK$?} It is straightforward to extend their definition to define $\pQSZK$. In classical complexity $\QSZKhv = \QSZK$ \cite{Wat06}. The proof first shows that $\QSZKhv$ has a public coin protocol. Then, they use a rewinding technique to get $\QSZKhv = \QSZK$. We can extend the first part of the proof to $\pQSZKhv$ with slightly worse soundness (See Appendix \ref{appthm:pQSZKhv_public_coin} for more detail). For the rewinding part, recall in \cite{Wat06}, they apply $\Pi:=\ket{0^{q(n)}}\bra{0^{q(n)}}\otimes I$ to project states to the initial state. In $\pQSZK$, our initial state contains some unknown state $\ket{\phi}$. Since the simulator can only have polynomial copies of $\ket{\phi}$, it is unclear how to do a projection on the unknown states with negligible error.
    \item \textbf{Open question 3: Can we show $\pmPSPACE \subsetneq \pmQIP$? } In traditional complexity, it is known that $\class{PSPACE}= \class{QIP}$. Besides, Theorem~\ref{thm:mpQIP_2_not_in_mpINF} implies that $\pmQIP \not\subseteq \pmPSPACE$. Hence, it is intriguing to ask whether $\pmPSPACE \subsetneq \pmQIP$? The original proof of $\class{PSPACE} \subseteq \class{IP}$ relies on the algebraic structure of a $\class{PSPACE}$-complete problem. It is not clear whether our $\pmPSPACE$-complete problem has the same structure.
    \item \textbf{Open question 4: Applications to quantum state learning} It can be shown that the task of learning quantum states prepared by polynomial-size quantum circuits can be solved using a $\mQCMA$ oracle through search-to-decision reductions. An interesting question is whether $\mBQP \neq \mQCMA$ would imply the existence of a class of states that are efficiently preparable but remain hard to learn.
\end{itemize}

\subsubsection{Open questions related to different input copies of quantum promise complexity theory}\label{section:open_copy}

In discussing the structural results of quantum promise problems, we focus on the following setting: for any single-party algorithm, access is limited to a polynomial number of input copies, even if the computational power exceeds polynomial time; whereas for any two-party algorithm (interactive protocol), both honest and malicious provers are allowed access to an unbounded number of input copies. We believe this setting is the most natural one, although it remains interesting to explore alternative settings. 

We list several different settings, each interesting in its own right
\begin{itemize}
    \item \textbf{More than polynomial copies (single party)}: It is natural to consider that any algorithm running in more than polynomial time may also have access to more than a polynomial number of copies of the input. In this section, we focus on the class $\class{PSPACE}$. We introduce some notation that will be used only here. The notation $\pmPSPACE^{\poly}$ (respectively, $\pmPSPACE^{\infty}$) denotes that a polynomial-space algorithm $\cA$ has access to only polynomially many (respectively, an unbounded number of) copies of the input. Note that even though $\pmPSPACE^{\infty}$ allows access to an unlimited number of copies, the total space used remains polynomial.
    \item \textbf{Limit to polynomial copies (two parties):} For interactive proof systems, it is also natural to consider provers that are limited to a polynomial number of copies of the input. We introduce some notation that will be used only in this section. The notation $\pmQIP^{\poly}$ (respectively, $\pmQIP^{\infty}$) indicates that both the honest and malicious provers have access to only polynomially many (respectively, an unlimited number of) copies of the input, while the verifier has access to only polynomially many copies.
    \item \textbf{Bounded copies}: In this setting, the algorithm is allowed to access only a prior bounded number of copies of the input state. This scenario is particularly interesting because certain quantum input problems are inherently bounded copies. A notable example is unclonable cryptography. For example, in quantum money, an adversary receives one valid money state and is challenged to generate two valid ones. This task is inherently a bounded-copies setting. Recently, Cavalar et al.~\cite{CCCGHJL25} introduced a new quantum input meta-complexity problem that also involves only a single copy of the input state.
\end{itemize}

Here is some open questions related to different input copies of quantum promise complexity theory. 
\begin{itemize}
    \item \textbf{Whether $\pmQIP^{\poly}$ is equal to $\pmPSPACE^{\poly}$?} The original proof of $\QIP \subseteq \PSPACE$ relies on expressing a $\QIP$ protocol as a semidefinite program (SDP) and then applying the multiplicative weights algorithm in $\PSPACE$ to solve the SDP. This proof no longer applies in our setting, since the SDP constraints involve an unknown input state, and it is unclear how to apply the multiplicative weights algorithm under such constraints. The reverse direction follows for the same reason discussed in Open Question 3 of the previous section.

    \item \textbf{Whether $\pmQIP^{\infty}$ is equal to $\pmPSPACE^{\infty}$?} The original proof of $\QIP \subseteq \PSPACE$ may appear applicable when considering $\pmPSPACE^{\infty}$. A natural attempt would be for $\pmPSPACE^{\infty}$ to perform tomography on the input state in order to obtain complete information, and then apply the multiplicative weights algorithm as in the original proof. However, this approach fails because, even if full tomography on the input state were possible, it would be impossible to store the resulting exponentially large density matrix within polynomial space. Moreover, it remains unclear how to apply the multiplicative weights algorithm in such a setting. The reverse direction follows for the same reason discussed in Open Question 3 of the previous section.

    %\item \textbf{Can the hardness of unclonable cryptography and bounded-copy quantum input state problem (e.g., ~\cite{CCCGHJL25}) be captured by quantum promise complexity class?} These problems are inherently formulated in the bounded-copies setting, it is not clear how to capture them within polynomial-copies quantum promise complexity classes. It is therefore natural to consider bounded-copies quantum promise complexity classes instead. However, even in the bounded-copies setting, the hardness of unclonable cryptography is not directly captured, because the security game is closer to a quantum input state-synthesis problem than to a quantum input decision problem.
    \item \textbf{Can the hardness of unclonable cryptography and bounded-copy quantum input problems (e.g.,~\cite{CCCGHJL25}) be captured by quantum promise complexity classes?} These problems are inherently formulated in a bounded-copies model, and it is unclear how to express them within polynomial-copies quantum promise classes. 
    One possibility is that certain bounded-copies problems collapse to their polynomial-copies counterparts, analogous to the situation for the EFI security problem. For example, the Kolmogorov-complexity problem in~\cite{CCCGHJL25} may exhibit such a collapse, or be closely related to the state minimum circuit size problem (which allows polynomially many copies) studied in~\cite{chia2021mcsp}. Another direction is studying \emph{bounded-copies} quantum promise classes instead. However, even in that setting, the hardness of unclonable cryptography is not directly captured, since its security game more closely resembles a \emph{state-synthesis} task with quantum inputs than a standard quantum-input \emph{decision} problem.

\end{itemize}

\subsubsection{Open question related to new Impagliazzo's five worlds}\label{section:open_five_world}

Our framework can provide some new insights into the following question: 
\begin{center}
    \emph{What will Impagliazzo's five worlds look like in the quantum setting?}
\end{center}
%\paragraph{Discussion:  What will be the new Impagliazzo's five worlds view in the quantum setting?\newline\newline} 
Impagliazzo's five worlds view~\cite{impagliazzo1995personal} presents different cryptographic and complexity assumptions, where each world reflects a different level of computational hardness. For example, in the world of \emph{Algorithmica}, where $\class{P} = \class{NP}$, efficient algorithms exist for $\class{NP}$-complete problems, meaning there are no hard challenges in this world. In the world of \emph{Heuristica}, where $\class{P} \neq \class{NP}$ and $\class{DistNP} \subseteq \class{AvgP}$, some hard challenges, but we do not know how to generate them efficiently. In the world of \emph{Pessiland}, where $\class{DistNP} \not\subseteq \class{AvgP}$ and one-way functions do not exist, we can efficiently sample hard challenges, but we do not know how to generate their solutions efficiently. Finally, in worlds like \emph{Minicrypt} or \emph{Cryptomania}, the existence of one-way functions or public-key encryption guarantees secure protocols but implies that $\class{P} \neq \class{NP}$, preventing efficient solutions for $\class{NP}$-complete problems.

%What will the new Impagliazzo's five worlds view look like in the quantum setting? 
Answering the above question involves a new characterization of worlds based on different fundamental assumptions in quantum complexity theory and cryptography.
There are some implicit guiding features in Impagliazzo's five worlds: (i) The assumption in \emph{Minicrypt} implies the average-case hardness of the complexity class defined in \emph{Heuristica}.\footnote{If the complexity class defined in \emph{Algorithmica} has a worst-to-average-case reduction, then the assumption in \emph{Minicrypt} should imply the worst-case hardness of that complexity class.} This gives us a necessary assumption for the existence of cryptographic primitives. (ii) The complexity class defined in \emph{Algorithmica} is the smallest one sufficient to break most cryptographic assumptions, assuming it is easy to solve, and it should also play a central role in other fields of computer science.

We can view Impagliazzo's five worlds in two parts: \emph{Algorithmica} and \emph{Heuristica} are complexity-related assumptions, while \emph{Minicrypt} and \emph{Cryptomania} pertain to cryptography-related assumptions. In recent developments in quantum cryptography, researchers have introduced several new quantum primitives, such as EFI, OWSG, and PRS, along with their relationships.\footnote{For more example and relation, we refer to reader to the graph \url{https://sattath.github.io/qcrypto-graph/}} These primitives could form the basis of cryptography-related assumptions in the new Impagliazzo's five worlds. On the other hand, using traditional language for complexity-related assumptions presents a limitation. Indeed, one possibility is to replace $\class{NP}$ with $\class{QCZK}$ or $\class{P}^\class{\#P}$ \HZZ{Check}, since the average-case or worst-case hardness of these complexity classes implies EFI \cite{brakerski2022computational,KT24}. However, these complexity classes do not satisfy the first feature (i.e., the other direction). Another possibility is to replace $\class{NP}$ with $\class{PP}$, as \cite{Kretschmer2021QuantumPA} shows that $\class{PP}$ satisfies the first feature. However, $\class{PP}$ does not satisfy the second feature because $\class{PP}$ is a very large complexity class.\footnote{For example, we know that $\class{PH}\subseteq\class{P^{PP}}$ by 
Toda's Theorem} Fortunately, our quantum complexity framework could serve as a potential language for expressing complexity-related assumptions. Based on this, we propose three different versions of Impagliazzo's five worlds. The first version, based on $\mQSZKhv$ and EFI pair, is as follows

\begin{itemize}
    \item \textbf{Heuristica:} $\textbf{Dist-}\mQSZKhv \subseteq \textbf{Avg-}\mBQP$.\footnote{We use Dist/Avg only informally for high-level discussion. A formal definition could be given following the survey on average-case complexity by Bogdanov and Trevisan.}
    \item \textbf{Pessiland:} $\textbf{Dist-}\mQSZKhv \not\subseteq \textbf{Avg-}\mBQP$ and EFI pair does not exist.
    \item \textbf{MiniQcrypt:} EFI pair, OWSG, PRS, or post-quantum OWF exists. \HZZ{Check}
    \item \textbf{Cryptomania:} Public-key or public unclonable quantum cryptography exists.\footnote{Public unclonable cryptography include public key quantum money, public verifiable secure software leasing, public verifiable certified deletion, public verifiable software copy protection, etc.}
\end{itemize}
Note that Morimae~\cite{Mor24} introduced a new world called \emph{Microcrypt} based on EFI, OWSG, and PRS, but without OWF. Hence, the world \emph{MiniQcrypt} refers to \emph{Microcrypt} combined with \emph{Minicrypt}.
Since $\mBQP \subsetneq \mQSZKhv$ (Theorem \ref{thm:mQSZKhv_not_inside_mINF}), we rule out the possibility of \emph{Algorithmica}. Ruling out \emph{Algorithmica} is based on an information-theoretic statement. One might argue that the class $\mQSZKhv$ is too large, and thus it can be relaxed. The second version, based on $\mpolyQSZKhv$ and EFI pairs, is as follows:

\begin{itemize}
    \item \textbf{Algorithmica:} $\mBQP = \mpolyQSZKhv$.
    \item \textbf{Heuristica:} $\mBQP \subsetneq \mpolyQSZKhv$ and $\textbf{Dist-}\mpolyQSZKhv \subseteq \textbf{Avg-}\mBQP$.
    \item \textbf{Pessiland:} $\textbf{Dist-}\mpolyQSZKhv \not\subseteq \textbf{Avg-}\mBQP$ and EFI pair does not exist.
    \item \textbf{MiniQcrypt:} EFI pair, OWSG, PRS, or post-quantum OWF exists.
    \item \textbf{Cryptomania:} Public-key or public unclonable quantum cryptography exists.
\end{itemize}
However, Impagliazzo’s original five worlds are based on complexity class $\class{NP}$. A natural extension to our quantum complexity class is $\mQCMA$. Therefore, the third version, based on $\mQCMA$ and mixed-state OWSG, is as follows:
\begin{itemize}
    \item \textbf{Algorithmica:} $\mBQP = \mQCMA$.
    \item \textbf{Heuristica:} $\mBQP \subsetneq \mQCMA$ and $\textbf{Dist-}\mQCMA \subseteq \textbf{Avg-}\mBQP$.
    \item \textbf{Pessiland:} $\textbf{Dist-}\mQCMA \not\subseteq \textbf{Avg-}\mBQP$ and mixed-state OWSG does not exist.
    \item \textbf{MiniQcrypt:} Mixed-state OWSG, PRS, or post-quantum OWF exists.
    \item \textbf{Cryptomania:} Public-key or public unclonable quantum cryptography exists.
\end{itemize}
In the above world, we do not include EFI in \emph{MiniQcrypt} since it is unclear whether the existence of EFI implies $\mBQP \subsetneq \mQCMA$, which is an intriguing open problem. If the answer is positive, our framework creates a similar view of the original Impagliazzo’s five worlds.

Different complexity frameworks give rise to different versions of Impagliazzo's worlds. In contrast to our framework, another new world can be obtained based on $\class{avgUnitarySZK}_{\text{HV}}$ (in unitary complexity) and EFI pairs. In their worlds, they do not have \emph{Pessiland} since $\class{avgUnitarySZK}_{\text{HV}} \not\subseteq \class{avgUnitaryBQP}$ if and only if EFI pairs exist \cite{bostanci2023unitary, bostanci2024efficient}. %Last, we summarize the open problem as follows.
%However, within the framework of unitary complexity, it remains unclear whether Impagliazzo's worlds can be constructed with purely complexity language without relying on the zero-knowledge property.

%\paragraph{Open question related to new Impagliazzo's five worlds.}\label{open:Impga_five} Here, we summarize open problems related to the new five worlds as follows: What are the relationships between Public Key Encryption (PKE) and other quantum primitives? What are the relationships between \emph{Cryptomania} and \emph{MiniQcrypt}? Would average-case hardness of $\mpolyQSZKhv$ be equivalent to the existence of EFI? Do we have the worst to average-case reduction for $\mpolyQSZKhv$, $\mQSZKhv$, and $\mQCMA$? Does the existence of EFI imply $\mBQP \subsetneq \mQCMA$? \HZZ{What is implications to 5 world?}

%\nai{I revise the open question for five worlds as follows. The old one is commented.}
%\paragraph{Open question related to new Impagliazzo's five worlds.}\label{open:Impga_five}

Here, we summarize open problems related to the new five worlds as follows: 
\begin{itemize}
    \item \textbf{What are the relationships between \emph{Cryptomania} and \emph{MiniQcrypt}?} In particular, would it be possible that one can obtain unclonable cryptography from OWSG or other primitives in MiniQcrypt? Does the existence of unclonable cryptography imply MiniQcrypt? Or, do quantum primitives, such as EFI and OWSG, imply public-key encryption with a classical public key? \nai{Check.}
    \item \textbf{What are the relationships between \emph{Pessiland} and \emph{MiniQcrypt}?} Specifically, in the first and second versions of the five-world view, if we can show that the average-case hardness of $\mpolyQSZKhv$ implies EFI, then \emph{Pessiland} and \emph{MiniQcrypt} are equivalent. For the third version, one might consider whether barrier results related to Impagliazzo's five worlds extend to this context.
    \item \textbf{Do we have worst- to average-case reductions for $\mpolyQSZKhv$, $\mQSZKhv$, and $\mQCMA$?} Note that it is open whether average-case $\mQSZKhv$ can be solved with only polynomial copies. If this is true, it directly implies that worst- to average-case reduction for $\mQSZKhv$ does not exist; otherwise, it will rule out the existence of Heuristica in the first version of the five worlds.
    \item \textbf{Which complexity classes shall we consider for Algorithmica and Heuristica?} In principle, we want a class that is the tightest upper bound for breaking cryptographic primitives. So, it is worth investigating whether $\mQCMA$ suffices to break EFI and exploring the relationship between $\mQCMA$ and $\mpolyQSZKhv$.
\end{itemize}

%% file: 1-5_Compare.tex
\subsection{Discussion on complexity theory for different computational problem}\label{sec:related}
%\HZZ{qunatum complexity型別上的比較，state/unitary synthesis 都是input classical input, 造東西}

In computational complexity theory, problems arise in fundamentally different forms, such as decision problems, relation problems, counting problems, sampling problems, and distribution problems. These categories are not naturally reducible to one another—for instance, the task of determining whether a statement is true (decision) differs essentially from generating a sample from a distribution (sampling), or from identifying a valid solution among multiple possible outputs (relation). Since these problem types capture distinct notions of computation, a single complexity class cannot adequately characterize all of them. For example, \emph{finding a Nash equilibrium} is complete for the class \class{PPAD}\cite{DWP09}. This problem can be naturally captured as a relational problem. However, we do not yet know how to formulate it as a decision problem. As another example, computing the \emph{gaussian permanent estimation} is in $\class{FBPP^{NP}}$ if $\class{SampP} = \class{SampBQP}$\cite{Aar14}. No analogous result is currently known for the corresponding decision problem. Consequently, after decision problems, separate complexity classes have been introduced for counting problem~\cite{Val79}, relation problems~\cite{Aar14,ABK24}, sampling problems~\cite{Aar14}, and distribution testing problems~\cite{CG18}.

Following the same reasoning, several new complexity classes have also been proposed to study quantum states and operators. Notably, complexity classes for \emph{state synthesis problem}, introduced by Rosenthal and Yuen \cite{rosenthal2022interactive}, are analogous to sampling complexity classes~\cite{aaronson2010computational}.
Bostanci et al.\cite{bostanci2023unitary} conducted an exhaustive study on complexity for \emph{unitary synthesis problems},  identifying complete problems for specific classes, formalizing a series of fundamental problems within these complexity classes, exploring their implications for quantum cryptography, and determining equivalences between certain unitary complexity classes. The complexity classes for state synthesis problems, unitary synthesis problems, and quantum promise problems each focus on different quantum tasks. For instance, in our framework, the outputs are decisional, yet they consist of quantum states.

\paragraph{Unitary synthesis problem} Bostanci et al. \cite{bostanci2023unitary} define a new notion of 
\textit{unitary complexity classes}, which describe the computational resources needed to perform unitary, also called state transformation. The unitary complexity class captures the difficulty of tasks involving both quantum inputs and outputs. In contrast, our framework provides a different perspective by focusing on decisional problems with quantum input. Before comparing their results with ours, let us recall the definitions in their paper.

A unitary synthesis problem is a sequence of unitary $\cU:=(U_x)_{x\in\{0,1\}^*}$, and a unitary complexity class is a collection of unitary synthesis problems. They define classes such as $\textbf{unitaryBQP}$, $\textbf{unitaryPSPACE}$. Informally, $(U_x)_{x\in\{0,1\}^*} \in \textbf{unitaryBQP}$ if there exists a uniform polynomial-time quantum algorithm that on input $x$ implement $U_x$. Besides, they also explore interactive proofs for unitary synthesis problems. Informally, $(U_x)_{x\in\{0,1\}^*} \in \textbf{unitaryQIP}$ is defined as follows: The verifier receives an instance $x$ and a register $A$, where the quantum state in the register $A$ may be entangled with a larger quantum system. The dishonest prover receives a classical instance $x$ only. The goal of the verifier is to apply $U_x$ on the register $A$. The prover must not know the state in register $A$ beforehand.
Besides, they also define average-case complexity classes, such as $\textbf{avgUnitaryBQP}$, $\textbf{avgUnitaryPSPACE}$, $\textbf{avgUnitaryQIP}$, and $\textbf{avgUnitarySZK}_{\textbf{HV}}$.

Their framework aims to synthesize a unitary, which captures the quantum input and output problems. For example, they show that 
\begin{equation*}
    \textit{UHLMANN}_{1-\epsilon} \text{ is complete for } \textbf{avgUnitarySZK}_{\text{HV}},
\end{equation*}
where $\textit{UHLMANN}_{1-\epsilon}$ is a unitary synthesis problem related to Uhlmann transformation. On the contrary, our framework focuses on decisional problems with quantum input. For instance, the property testing problems, specifically the Quantum Or problems (Definition \ref{qcma:qob} and Definition \ref{qma:qob}), are complete for $\class{pQ(C)MA}$. 
    
Furthermore, the complexity class involving interactive proofs fundamentally differs in our work compared to theirs (e.g., $\pQIP$ versus $\textbf{unitaryQIP}$). The prover in our setting knows the complete information about the quantum inputs, while the prover in their setting knows the unitary but not the quantum states to which it should be applied. For example, the definition of $\textbf{unitaryQMA}$ currently has no known applications. In their framework, the prover can send only one message, independent of register $A$. The verifier, upon receiving this message, must decide whether to accept. If it accepts, the verifier synthesizes $U_x$ and applies it to register $A$, implying that $U_x$ consists of polynomial quantum gates. However, it remains unclear whether polynomial-sized gates with additional advice and witnesses are useful for solving tasks involving quantum inputs and outputs. On the other hand, in our framework, the witness plays a fundamental role in solving decisional quantum tasks. For instance, using $\QCMA$ oracle is sufficient to break OWSG, which mirrors the classical results.

%This makes the definition of $\textbf{unitaryQMA}$ doesn't describe the class we imagine. Indeed, the prover could only send one message independent of the register $A$. The verifier who receives this message should decide whether to accept. If the verifier accepts, it needs to synthesize $U_x$ and apply to register $A$. This implies that $U_x$ consists of polynomial quantum gates. Specifically, consider a sequence of unitary $\cU \in (U_n)_{n\in \mathbf{N}}$ (Each security parameter only has a single unitary). Then, $\cU \in \textbf{unitaryQMA} \implies \cU \in  \textbf{unitaryBQP}\backslash \textbf{qpoly}$. \HZZ{I don't know if the qma part explains correctly.}

Besides, consider the quantum complexity classes involving only a single party (e.g., $\pBQP$ compare to $\textbf{unitaryBQP}$). Traditional complexity class can be reduced to both definitions (See Section \ref{Def:Oracle} and \cite{bostanci2023unitary} (Section 3.1, unitary synthesis problem). However, neither definition of quantum complexity classes can trivially reduce to the other. Specifically, the unitary synthesis problem $\cU$ reduces to a quantum promise problem $\cL$ if the following holds: Let $\cC$ be some computational resources (e.g., polynomial time or polynomial space). Given $\cU$, we can define a quantum promise problem $\cL$ such that $\cU \in \textbf{unitary}\cC$ if and only if $\cL \in \class{p\cC}$. This seems impossible since the promise problem is a decision problem, while unitary synthesis describes a quantum output problem. On the other hand, a quantum promise problem cannot trivially be reduced to a synthesis problem. One might think that given a quantum promise problem $\cL$, we can define the unitary synthesis problem $\mathcal{U}$ as follows: Let $\cU:=(U_n)_{n\in \mathbb{N}}$ (Each security parameter only have one unitary). Define the unitary $U_n$ such that for all $\ket{\psi} \in \cL_Y \cup \cL_N$, $b \in \{0,1\}$, $\ket{\psi}^{\otimes t}\ket{b}\ket{0^*} \xmapsto{U_n} \ket{\psi}^{\otimes t}\ket{b \oplus \cL(\psi)}\ket{0^*}$, where $\cL(\psi) =1 $ iff $\ket{\psi} \in \cL_Y$. However, this definition has plenty of issues. First and foremost, $U_n$ may not be unitary. Indeed, suppose $\ket{\psi_Y} \in \cL_Y$ and $\ket{\psi_N} \in \cL_N$. Then, $\ket{\psi_Y}^{\otimes t}$ and $\ket{\psi_N}^{\otimes t}$ are not orthogonal but rather ``almost" orthogonal. Second, any algorithm that decides $\cL$ may produce some garbage states that need to be specified. Third, we must know the number of copies $t$ and the size of the ancilla before defining $U_n$. Finally, we must define $U_n$ on the subspace $span\{ \ket{\psi}^{\otimes t}: \ket{\psi} \in \cL_Y \cup \cL_N\}$, as required by the definition of the unitary synthesis problem (see ``partial isometry" in \cite{bostanci2023unitary}). Hence, the second attempt is to define the unitary $U_n$ such that for all $\ket{\psi} \in \cL_Y \cup \cL_N$, $b \in \{0,1\}$, 
\begin{equation*}
    \ket{\psi}^{\otimes t}\ket{b}\ket{0^*} \xmapsto{U_n} \alpha_{\psi}\ket{G_{\psi}}\ket{b \oplus \cL(\psi)} + (1-\alpha_{\psi}^2)\ket{G'_{\psi}}\ket{b \oplus \overline{\cL(\psi)}},
\end{equation*}
where $\ket{G_\psi}$ and $\ket{G'_\psi}$ are some arbitrary garbage states, and $\alpha_\psi \geq \frac{2}{3}$. Note that this method solves all of the issues but the last one. To solve the last issue, we can choose a basis of $span\{ \ket{\psi} \in \cL_Y \cup \cL_N\}$ and define the map for the basis. Unfortunately, in this method, the reduction is stated as:
\begin{center}
    \textit{$\cL \in \class{p\cC}$ if and only if there exists a sequence of unitary $\cU$ (from infinitely many candidates, since we must fix parameter $t$, $\ket{G_\psi}$, $\ket{G'_\psi}$, $\alpha_\psi$, and basis) belongs to \textbf{unitaryC}}. 
\end{center}
We find the quantifier ``there exists" in the statement undesirable. Worse, considering mixed-state language makes it even more complicated. 

In conclusion, we find both definitions of quantum complexity classes offer valuable perspectives for characterizing the hardness of different types of quantum problems, and currently, there is no unified method to approach them.

\paragraph{State synthesis problem} Rosenthal and Yuen \cite{rosenthal2022interactive} defined the complexity class of state synthesis problem. For example, \class{statePSPACE} and \class{stateQIP}~\cite{rosenthal2022interactive} contain all sequence of quantum states $\{\ket{\psi}_x\}_{x\in \{0,1\}^*}$ generated in polynomial space or by a polynomial-time quantum verifier interacting with an all-powerful dishonest prover. They~\cite{metger2023stateqip,rosenthal2022interactive,rosenthal2024efficient} demonstrated the equivalence of the two classes as $\class{statePSPACE}=\class{stateQIP}$, mirroring the $\class{QIP}=\class{PSPACE}$ relationship~\cite{qippspace11}. The state synthesis problem captures the computational resources needed to construct a state. In contrast, our framework focuses on the resources required to test a property. Surprisingly, the state synthesis problem is related to QPP complexity theory. Specifically, one could break EFI pairs with $\mPSPACE$ algorithm by using the algorithmic Uhlmann’s theorem~\cite{metger2023stateqip}
(Theorem~\ref{thm:EFI_PSPACE}).

%% file: 2_Prelim.tex
\section{Preliminaries}

%%%%%%%%%%%%%%%%%%%%%%%%%%%%%%%%%%%
%%%%%%%%%% Notation 5%%%%%%%%%%%%%%
%%%%%%%%%%%%%%%%%%%%%%%%%%%%%%%%%%%

\subsection{Notation}\label{Notation}
\begin{enumerate}
\item\label{notation_q} Notation related to quantum states and unitary:
    \begin{enumerate}
         \item A $n$-qubit pure state $\ket{\psi}$ is a vector in $\C^{2^n}$ whose 2-norm is 1. A $n$-qubit mix state $\rho$ is a $2^n \times 2^n$ positive semidefinite matrix (and hence self-adjoint or Hermitian) with $\Tr(\rho)=1$. If $\rho$ has rank $1$, then we also view $\rho$ as a pure state. Let $\mathcal{H}(n)$ and $\mathcal{D}\big(\mathcal{H}(n)\big)$ represent the sets of all $n$-qubit pure states and mixed states, respectively. Also, let $\mathcal{H}(\N) = \bigcup\limits_{\lambda \in \N}\mathcal{H}(\lambda)$ and $\mathcal{D}\big(\mathcal{H}(\N)\big) = \bigcup\limits_{\lambda \in \N}\mathcal{D}\big(\mathcal{H}(\lambda)\big)$. Additionally, define $qubit(\rho)$ as the number of qubits of $\rho$. Let $qubit(\rho)=n$, if $\rho \in \mathcal{D}\big(\mathcal{H}(n)\big)$ and let $\dim \cH(n) = \dim \cD(\cH(n)) := 2^n$.  \nai{Is $\mathcal{H}$ the same notation as Hilbert space?}
        \item Let $\rho$ be a mixed state. We define the decomposition set $\mathcal{D}_{\rho}$ as the collection of all possible eigenbasis of $\rho$. Let $\mathcal{D}\in \mathcal{D}_{\rho}$, where we refer to $D$ as an eigenbasis decomposition of $\rho$. Suppose $\mathcal{D}=\{\ket{\psi^i}\}$ and $\rho = \sum\limits_i\lambda_i \ket{\psi^i}\bra{\psi^i}$. We overload the meaning of $\mathcal{D}$ to represent an ensemble of states and their corresponding probability, i.e., $\mathcal{D}=\{(\ket{\psi^i}, \lambda_i)\}$. Besides, we define $\ket{\psi} \leftarrow \mathcal{D}$ as a sampling process that we get $\ket{\psi}=\ket{\psi^i}$ with probability $\lambda_i$. Additionally, we will simplify $\ket{\psi} \leftarrow \cD$ to $\ket{\psi} \leftarrow \rho$ only if the choice of $\cD$ does not affect the statement.  [Used in Section \ref{qma:section_m}.] 
        \item Let $\ket{0}^t$ be an abbreviation of $\ket{0}^{\otimes t}$. Moreover, $t$ will be omitted if it is unimportant; in such cases, we denote it as $\ket{0^*}$ or even $\ket{0}$. Additionally, if a state is unimportant or can be arbitrary, we denote it as $\ket{*}$.
        %\item Let $I_S$ denote the uniformly mixed state on a subspace $S$. Formally, we define $I_S:=\frac{1}{\text{dim}(S)}\sum_{i= 1}^{\text{dim}(S)} \ket{v_i}\bra{v_i}$, where $\ket{v_1}, \dots ,\ket{v_{\text{dim}(S)}}$ form any orthonormal basis of $S$. [Used in Section \ref{sec:mQIP_not_in_mPSPACE}.]
        \item A quantum process that maps pure quantum states to pure quantum states is defined as a unitary transformation. To implement any arbitrary unitary transformation, we can choose a set of local unitary operations capable of approximating any unitary transformation with arbitrary precision. These local unitary operations are called \emph{quantum gates}, and a collection of them is called a \emph{universal gate set}, denoted as $\mathcal{U}$, if any unitary operation can be approximated by a finite sequence of gates from this set. For instance, ${T, H}$ is a universal quantum gate set, where $T$ is the Toffoli gate and $H$ is the Hadamard gate. %The elements of the universal quantum gate set are called elementary gates. We assume that a circuit consists of a sequence of these elementary gates. 
        \item A quantum channel is a completely positive trace-preserving (CPTP) linear map between spaces of mixed states. 
    \end{enumerate}
\item Notation related to register:
\begin{enumerate}
    \item Let $A$ be a register, and let $U$ be a unitary operation acting on register $A$. Define $qubit(A)$ as the number of qubits in register $A$. Besides, we define $qubit(U_A):=qubit(A)$.
    %Let $U_{AB}$ and $S_{BC}$ be unitaries. Define $qubit(U):=qubit(AB)$ and $qubit(U\cap S):=qubit(B)$, i.e., the number of qubits of the register that is the intersection of $U$ and $S$. \nai{Why do we need this notation? Intersection of two unitaries looks strange. }
    \item \label{pre:register} Usually, register name has it meaning. For example: $T$ for test or time, $I$ for input, $W$ for witness, $A$ for ancilla, $A^{ans}$ for the answer, $P$ for purification, $C$ for commit, and $R$ for reveal. %\nai{I think this is redundant. If the name really matters, state it when you mention it. reader will not go back here to check the meaning of the name.} 
    \item  Let $U$ be a unitary operation acting on register $A$ and let $(S \otimes U_A)$ act on register $AB$. Then, $S$ implicitly represents a unitary operation acting on register $B$. %Sometimes, the whole space is defined implicitly. For example, let $T$ be another unitary act on register $AB$. Then, the unitary $S$ of $(S \otimes U_A)T_{AB}$ acts on register $B$.
    Besides, in this form, we emphasize that $S$ does not act on register $A$.
    \item If $U$ is a unitary operation acting on register $AB$ and $U = S_A \otimes S_B$, we define $U|_{-B}:=S_A$.
\end{enumerate}
\item\label{notation_r} Notation related to random variables:
\begin{enumerate}
    \item Let $X$ be a random variable and $A \subseteq \R$ a set. The indicator function, denoted by $\mathbbm{1}$, is defined as follow:
    \begin{equation*}
        \mathbbm{1}_{A}(X)= 
        \begin{cases}
            1,& \text{if } X \in  A\\
            0,              & \text{otherwise}.
        \end{cases}
    \end{equation*}
    Additionally, we have overloaded the indicator function $\mathbbm{1}$. Let $S$ be a statement. We define as follows: $\mathbbm{1}(S)=1$, if $S$ is true. Otherwise,  $\mathbbm{1}(S)=0$.
    \item A Rademacher random variable $R$ has the following probability mass function: 
    \begin{equation*}
        f_R(k)= 
        \begin{cases}
            \frac{1}{2},& \text{if } k \in \{1,-1\}\\
            0,              & \text{otherwise}.
        \end{cases}
    \end{equation*}
\end{enumerate}

\item Notation related to trace, distance, and fidelity:
    \begin{enumerate}
        \item For any square matrix $X$, the trace norm $\|X\|_{tr}$ is defined as $\|X\|_{tr}:= \frac{1}{2}\Tr\sqrt{X^*X}$. The trace distance between two states $\rho$ and $\sigma$ is defined as $\TD(\rho,\sigma):=\|\rho-\sigma\|_{tr}$. If $\rho:=\ket{\phi}\bra{\phi}$ and $\sigma:=\ket{\psi}\bra{\psi}$ are pure states, we commonly abbreviate $\TD(\ket{\phi}\bra{\phi}, \ket{\psi}\bra{\psi})$ as $\TD(\ket{\phi}, \ket{\psi}):=\|\ket{\phi} - \ket{\psi}\|_{tr}$.
        \item For any two mixed states $\sigma$ and $\rho$, the  fidelity between $\sigma$ and $\rho$ is defined as $F(\sigma,\rho):= ||\sqrt{\sigma}\sqrt{\rho}||_{1}$. For two pure states $\ket{\phi}$ and $\ket{\psi}$, we abbreviate $F(\ket{\phi}\bra{\phi}, \ket{\psi}\bra{\psi})$ as $F(\ket{\phi},\ket{\psi})=|\braket{\phi|\psi}|$. If only one of them is a pure state, we also abbreviate $F(\ket{\phi}\bra{\phi}, \sigma)$ as $F(\ket{\phi}, \sigma)$.
        \item $\Tr_{A}(\rho_{AB})$ denotes the reduced state of subsystem (or register) $B$ from $\rho_{AB}$. We say that register $A$ of $\rho_{AB}$ has been traced out. If $\rho:=\ket{\phi}\bra{\phi}_{AB}$ is a pure state, we commonly abbreviate $\Tr_{A}(\ket{\phi}\bra{\phi}_{AB})$ as $\Tr_{A}(\ket{\phi}_{AB})$. In some context, we denote $\Tr_{> n}(\rho)$ as the partial trace over all but the first $n$ qubits. $\Tr_{\geq n}(\cdot)$, $\Tr_{\leq n}(\cdot)$, and $\Tr_{\leq n, >m}(\cdot)$ are defined in a similar manner.
    \end{enumerate}
\item Notation related to string, set, and matrix:
    \begin{enumerate}
        \item Let $x,y \in \{0,1\}^n$, we write $x||y$ to denote the string concatenation of $x$ and $y$. Besides, $|\cdot|$ denotes the length of a string, $x_i$ refers to the $i$-th bit of $x$, and $x_{i,j}$ refers to the substring of $x$ from index $i$ to index $j$.
        \item $1^n$ and $0^n$ denote strings of length $n$ consisting entirely of 1s and 0s, respectively.
        \item  We use the special symbol $\#$ for padding.
        \item Let $n\in\N$. Define $[n]:=\{1,2,\dots,n\}$ and $[n]_0:=\{0,1,2,\dots,n\}$.
        \item $\R^+$ refers to the interval $(0,\infty)$ and $\R^+_0$ refers to the interval $[0,\infty)$.
        \item $\mathcal{P}(\N)$ denotes the set of all integer-valued polynomials with non-negative leading coefficients.
        \item The \emph{direct sum} of two sets $A$ and $B$, denoted as $A \oplus B$, consists of the ordered pairs $(a,b)$, where $a \in A$ and $b \in B$.
        \item Let $A$ and $B$ be two square matrices. We say that $A \preceq B$ if $B-A$ is positive semi-definite.
        \item Define $\lambda_{min}$ as a function that maps a Hermitian (self-adjoint) matrix to its minimum eigenvalue.
    \end{enumerate}
\item Asymptotic notation: Unless otherwise specified, all functions in this paper are all non-negative value functions with domain $\mathbb{N}$.
    \begin{enumerate}
        \item A function $f(n)$ is called negligible if, for every constant $c > 0$ and for all sufficiently large $n$, $f(n) \leq \frac{1}{n^c}$. Besides, $\negl(n)$ denotes an arbitrary negligible function.
        \item A function $f(n)$ is called polynomial bounded if there exists a constant $c > 0$ such that for all sufficiently large $n$, $f(n) \leq {n^c}$. Besides, $\poly(n)$ denotes an arbitrary polynomial bounded function.
        \item Consider two functions $f$ and $g$. We say that $f \geq g$ if, for sufficiently large $n$, $f(n) \geq g(n)$.
        \item QPT stands for quantum polynomial times algorithm.
    \end{enumerate}
    \HZZ{Need to define notation psd}
\end{enumerate}

\subsection{Useful technical lemmas for testing states}
\begin{theorem}[{\cite[Section 4,\:restated]{harrow2017sequential}} \label{thm:QOR}]
Consider a bipartite Hilbert space $\mathscr{H}_A\otimes \mathscr{H}_B$, where $dim\: \mathscr{H}_B = N$. Let $\rho$ be an unknown mixed state and $\Lambda$ be an orthogonal projector on Hilbert space $\mathscr{H}_A\otimes \mathscr{H}_B$.
Supposed we're promised that either

\begin{enumerate}
    \item[(i)] There exists $\sigma \in \mathscr{H}_B$ such that $\Tr\left(\Lambda (\rho \otimes \sigma)\right) \geq \eta \geq \frac{1}{2}$, or else
    \item[(ii)] 
    For all states $\sigma \in \mathscr{H}_B$, $\Tr\left(\Lambda (\rho \otimes \sigma)\right) \leq \delta$.
\end{enumerate}
There is an algorithm that uses one copy of $\rho$, and that accepts with probability at least $\frac{\eta^2}{7}$ in case $(i)$ and accepts with probability at most $4N \cdot \delta$ in case $(ii)$. 
\end{theorem}
We briefly explain the algorithm for Theorem~\ref{thm:QOR}. For $i \in [N-1]_0$, define a 2-outcome projective measurement $\Lambda_i$ on Hilbert space $\mathscr{H}_A \otimes \mathscr{H}_B$ as follow: (i) Apply $I_A \otimes ({X_i})_B$, where $X_i: \ket{a} \rightarrow \ket{a \oplus i}$, (ii) Apply $\Lambda$, and (iii) Apply $I_A \otimes ({X_i})^\dagger_B$. Define $\Pi:= \sum\limits_{i=0}^{N-1} (\Lambda_i)_{AB} \otimes Q\ket{i}\bra{i}Q^\dagger_C$, where $Q$ is the Fourier transform. Also, define $\Delta := I_{AB} \otimes \ket{0}\bra{0}_C$. The algorithm works as follows:
\begin{enumerate}
    \item Create the state $\rho_A \otimes \ket{0}\bra{0}_B \otimes \ket{0}\bra{0}_C$.
    \item Repeat $\ceil{\frac{N}{\eta}}$ times or until the algorithm accepts:
    \begin{enumerate}
        \item Perform the projective measurement $\{\Pi, I-\Pi\}$. If the first result is returned, accept.
        \item Perform the projective measurement $\{\Delta, I-\Delta\}$. If the second result is returned, accept.
    \end{enumerate}
    \item Reject.    
\end{enumerate}

\begin{lemma}[Swap test\cite{KMY03}]\label{lemma:swap-test}For any $n \in \N$ qubits mixed-state $\sigma$ and $\rho$, consider the following state: %\HZZ{Register is not clear} 
    \begin{align*}
        \Bigl((H_A\otimes I_{BC})(c-\mathbf{SWAP})(H_A\otimes I_{BC})\Bigr)\ket{0}\bra{0}_A\otimes \sigma_B \otimes \rho_C \Bigl((H_A\otimes I_{BC}) (c-\mathbf{SWAP})^{*}(H_A\otimes I_{BC})\Bigr).
    \end{align*}
The $(c-\mathbf{SWAP})$ is a controlled-swap gate, where the first qubit in register $A$ controls the swapping of states between $B$ and $C$. $H$ is the Hadamard gate. 
Measuring the first qubit gives outcome $0$ with probability $\frac{1}{2} + \frac{1}{2}\Tr(\sigma\rho)$. We say $(\sigma, \rho)$ passes the swap test if the outcome gets $0$. If one of the states becomes a pure state, $\ie \sigma = \ket{\phi}\bra{\phi}$, the probability of passing the swap test can be written as $\frac{1}{2} + \frac{1}{2}F(\rho, \ket{\phi})^2$.
\end{lemma}
When the states in registers $B$ and $C$ are pure, this test can test whether the two states are close. Conversely, this test fails to test the closeness of two mixed states in general. Additionally, We can derive the following lemma by the swap test.  
\begin{lemma}[Partial swap test, Appendix \ref{App:swap}]\label{pre:sqap}
    Consider two states $\ket{\phi}_{BC}$ and $\ket{\psi}_D$, where $qubit(B)=qubit(D)$. 
    We write $\ket{\phi}$ as the following.
    \begin{equation*}
        \ket{\phi} = \alpha\ket{\psi}_B\ket{G}_C + \sum\limits_j \beta_j \ket{\psi_j^\bot}_B\ket{G_j}_C,
    \end{equation*}
    where $\ket{G}$ and $\{\ket{G_j}\}$ are unimportant garbage state and $\{|\psi^{\perp}_j\rangle\}$ is a basis of subspace $\big\{\ket{\eta} \::\: \bra{\eta}\cdot \ket{\psi}=0\big\}$. Consider the following state
    \begin{equation*}
        H_A(c_A-\mathbf{SWAP}_{BD})H_A \ket{0}_A\ket{\phi}_{BC}\ket{\psi}_D.
    \end{equation*}
    Measuring the register $A$ gives outcome 0 (which we call accept) with probability $|\alpha|^2 + \frac{1}{2}\sum\limits_j |\beta_j|^2$.
\end{lemma}

\begin{lemma}[Quantum union bound \label{pre:qub} \cite{o2022quantum},\cite{gao2015quantum},\cite{sen2012achieving}]
    Let $\rho$ be a mixed state, and let $E_1,\dots,E_m$ be two-outcome measurements such that for all $i\in [m]$, $\Tr(E_i\rho) \geq 1-\epsilon_i$. Suppose we sequentially measure $\rho$ with $E_1,\dots,E_M$, then they all accept with probability at least 
    \begin{equation}
        \label{pre:qub_eq}
         1-4\sum\limits_{i=1}^M\epsilon_i.
    \end{equation}
    Condition on all measurements accept, let $\Tilde{\rho}$ be the resulting state from $\rho$. Then,
    \begin{equation*}
        \|\Tilde{\rho}-\rho\|_{tr} \leq  \sqrt{\sum\limits_{i=1}^M\epsilon_i}.
    \end{equation*}
\end{lemma}
As a historical note, Sen \cite{sen2012achieving} proves the lower bound $1-2\sqrt{\sum\limits_{i=1}^M\epsilon_i}$ instead of Equation (\ref{pre:qub_eq}). Subsequently, Gao \cite{gao2015quantum} obtained the square of Sen's error. After that, O'Donnell and Venkateswaran \cite{o2022quantum} obtained a remarkably simpler proof with the same lower bound. Both lower bounds worked fine for us.

\begin{theorem}[Uhlmann's Theorem\cite{NC10}]\label{theorem:Ulman} Let $\rho$ and $\sigma$ $\in \mathcal{D(H)}$ and $\ket{\phi}$ and  $\ket{\psi} \in \mathcal{H}\otimes\mathcal{H}$ be the purification of $\sigma$ and $\rho$. Then  Uhlmann's theorem states that there exists a $U$, working on only the purification part, satisfies the following
\begin{equation*}
    F(\sigma,\rho) = \max\limits_{U}|\bra{\psi}U\otimes I\ket{\phi}|.
\end{equation*}
\end{theorem}

\HZZ{What name to choose}
\begin{lemma}[Non interference Lemma]\label{lemma:no_interference}
    Consider a collection of inputs $\{\ket{\psi_i}_A\}_i$ and any quantum state $\sum\limits_i \alpha_i \ket{\psi_i}$. Let $F$ be an isometry acting on the register $A$ and output non-zero qubits registers $B$ and $C$. Let $p_i^x$ and $p^x$ be the probability of obtaining outcome $x$ when measuring the register $B$ of $F(\ket{\psi_i})$ and $F\left(\sum\limits_i \alpha_i \ket{\psi_i}\right)$ in computational basis, respectively; and let $\ket{x}\ket{\phi_i^x}$ be the post-measurement state after measuring $F(\ket{\psi_i})$. If $\{\phi_i^x\}_{i,x}$ is an orthonormal set, then
    \begin{equation*}
        p^x = \sum\limits_i \alpha_i^2p^x_i.
    \end{equation*}
\end{lemma}
\begin{proof}
    \begin{equation*}
        F\left(\sum\limits_i \alpha_i \ket{\psi_i}\right) := \sum\limits_{i,x} \alpha_i \beta_i^x  \ket{x}\ket{\phi_i^x},
    \end{equation*}
    where $(\beta_i^x)^2 = p_i^x$. It is clear that $ p^x = \sum\limits_i \alpha_i^2p^x_i$.
\end{proof}

\iffalse
The following lemma provides a lower bound for the sum of the squares of expectations. We defer the following proof to Appendix \ref{Appendix:RV-2}. 
\begin{lemma}\label{Pre:RV}
    Fixed some $n \geq 2$. Given $n$ independent but not identical random variables $X_1, X_2,\dots, X_n$. Suppose, for all $i$, $E[X_i]=\alpha_i$ and has support $\{-1,1\}$. Define $\Bar{X}=\frac{1}{n}\sum\limits_i^nX_i$ and $W=\frac{n}{1-n}\sum\limits_i^n(X_i-\Bar{X})^2 + n$. Then, 
    \begin{equation*}
        \E[W] \leq \sum\limits_i^n \alpha_i^2.
    \end{equation*}
    The equality holds if and only if $X_i$'s are all identical. Besides the range of $\,W$ is $[-7n,n]$.
\end{lemma}
\fi

\subsubsection{Concentration inequalities}
Here, we introduce some concentration inequalities that will be used in our analysis. 
\begin{lemma}[Chernoff Hoeffding's bound] \label{lemma:chernoff_bound}
    If $X_1,\cdots,X_n$ are independent random variables such that $a_i \leq X_i \leq b_i$. Let $\bar{X} = \frac{1}{n}\sum\limits_{i = 1}^n X_i$. Then,
    \begin{equation*}
        \Pr\left[ \left|\bar{X} - E[\bar{X}] \right| \geq t \right] \leq 2\cdot \exp\left(\frac{-2n^2t^2}{\sum_{i = 1}^n (a_i - b_i)^2}\right). 
    \end{equation*}    
\end{lemma}

\begin{lemma}[Azuma--Hoeffding's bound]\label{pre:azuma}
    Let $X_1, \cdots, X_n$ be random variables produced by a process in $n$ steps, such that $a_i \leq X_i \leq b_i$,
    and let $\cF_{i-1}$ denote the outcomes of the first $i-1$ steps, on which $X_i$ may depend arbitrarily.\footnote{Formally, $\cF_{i-1}$ is the $\sigma$-algebra generated by the first $i-1$ steps and $X_i$ is $\cF_i$-measurable; over a finite set of outcomes the two formulations agree.}
    Let $\bar X = \frac1n \sum_{i=1}^n X_i$.  If there is a
    $\mu \in \R$ with 
    \begin{equation*}
        \E\left[X_i \mid \cF_{i-1}\right] \;\geq\; \mu
        \qquad \text{for every $i$ and every history $\cF_{i-1}$,}
    \end{equation*}
    then for every $t>0$,
    \begin{equation*}
        \Pr\left[\bar X \leq \mu - t\right] \;\leq\;
        \exp\left(\frac{-2n^2t^2}{\sum_{i=1}^n (a_i-b_i)^2}\right).
    \end{equation*}
\end{lemma}

\begin{lemma}[\cite{Kretschmer2021QuantumPA}]\label{lemma:Haar}
Let $\ket{\phi} \in \mathcal{H}(m)$, and let $\epsilon > 0$. Then:
\begin{equation*}
    \Pr\Big[\big|\braket{\phi |\psi}\big|^2 \geq \epsilon : \ket{\psi} \leftarrow \mu_m\Big] \leq e^{-\epsilon (2^m-1)},
\end{equation*}
where $\mu_m$ is the Haar measure on m-qubit states.
\end{lemma}

\begin{theorem}[\cite{Mec19}, Theorem 5.17]\label{thm:concentration_on_unitary}
    Let $\mathbb{U}(n)$ denote the set of unitaries acting on $n$-qubit systems. Let $\mu_n$ be the Haar measure over $\mathbb{U}(n)$. Suppose that $f:\mathbb{U}(n) \rightarrow \R$ is $L$-Lipschitz with respect to the Frobenius norm. Then, for all $t>0$, we have:
    \begin{equation*}
        \Pr\limits_{U \leftarrow \mu_n}\left[ \left |f(U) - \E\limits_{V \leftarrow \mu_n}[f(V)] \right| \geq t  \right] \leq 2\cdot \exp\left(-\frac{(2^n-2)t^2}{24L^2}\right).
    \end{equation*}
\end{theorem}

\begin{lemma}[\cite{Kretschmer2021QuantumPA}, Lemma 28]\label{lemma:query_Lipschitz}
 Let $\mathbb{U}(n)$ denote the set of unitaries acting on $n$-qubit systems. Let $\cA^U$ be a quantum algorithm that makes at most $L$ queries to a unitary $U \in \mathbb{U}(n)$. Define $f:\mathbb{U}(n) \rightarrow \R$ by $f(U):=\Pr\left[\cA^U=1\right]$. Then $f$ is $L$-Lipschitz with respect to the Frobenius norm.
\end{lemma}

\subsubsection{Results from \cite{CWZ24}}
We will require some results from~\cite{CWZ24}. In order to state their results, we require the following definitions. 

\begin{definition}[Local testers for bipartite states (Definition 1.1 in~\cite{CWZ24})]
\label{def:local-testers}
Let $\mathscr{H}_{AB} = \mathscr{H}_A \otimes \mathscr{H}_B$ be a bipartite Hilbert space. 
A \emph{(global) tester} $\mathcal{T}$ with sample complexity $N$ for bipartite states in 
$\mathscr{H}_{AB}$ acts on $\mathscr{H}_{AB}^{\otimes N}$. 
The tester $\mathcal{T}$ is \emph{local on $\mathscr{H}_A$} if it acts non-trivially only on 
$\mathscr{H}_A^{\otimes N}$.
\end{definition}

\begin{definition}[Unitarily invariant properties (Definition 1.2 in~\cite{CWZ24})]
\label{def:unitarily-invariant}
A property $\cP = \bigl(\cP^{\mathrm{yes}}, \cP^{\mathrm{no}}\bigr)$
of bipartite \emph{pure} states in $\mathscr{H}_{AB}$ 
is said to be \emph{unitarily invariant on $\mathscr{H}_A$} if
$(U_A \otimes I_B)\,\ket{\psi}_{AB} \in \cP^X$ for every $\ket{\psi}_{AB} \in \cP^X,$
where $U_A$ is a unitary operator on $\mathscr{H}_A$, $X \in \{\mathrm{yes},\mathrm{no}\}$,
and $I_B$ denotes the identity operator on $\mathscr{H}_B$.
\end{definition}

\begin{definition}[Unitarily invariant distributions (Definition 3.1 in~\cite{CWZ24})]
\label{def:unitarily-invariant-dist}
Let $D$ be a probability distribution on bipartite pure states in $\mathscr{H}_{AB}$. 
Then, $D$ is said to be \emph{unitarily invariant on~$\mathscr{H}_B$} if, 
for any measurable subset $S$ of bipartite pure states in $\mathscr{H}_{AB}$ 
and any $U \in \mathcal{U}_d$,
$$
\Pr_{\ket{\psi}_{AB}\sim D}\bigl[\ket{\psi}_{AB}\in S\bigr]
\;=\;
\Pr_{\ket{\psi}_{AB}\sim D}\bigl[\ket{\psi}_{AB}\in (I_A \otimes U_B)\,S\bigr],
$$
where $(I_A \otimes U_B)\,S \;=\; 
\{\,(I_A \otimes U_B)\ket{\psi}_{AB} : \ket{\psi}_{AB}\in S\}.
$
\end{definition}

\begin{definition}[Average-case testers (Definition 3.2 in~\cite{CWZ24})]
\label{def:avg-case-testers}
Let $\cP = \bigl(\cP^{\mathrm{yes}}, \cP^{\mathrm{no}}\bigr)$ be a property of bipartite pure states, 
and let $D^{\mathrm{yes}}$ and $D^{\mathrm{no}}$ be probability distributions on $\cP^{\mathrm{yes}}$ and 
$\cP^{\mathrm{no}}$, respectively. For $0 \le s < c \le 1$, a tester $\mathcal{T}$ is called an 
\emph{average-case $(c,s)$-tester} for $\cP$ with respect to $(D^{\mathrm{yes}},D^{\mathrm{no}})$ 
\emph{with sample complexity~$N$}, if
$$
\mathbb{E}_{\ket{\psi}_{AB}\sim D^{\mathrm{yes}}} 
  \Bigl[\Tr\!\bigl(\mathcal{T}\,\ket{\psi}\bra{\psi}^{\otimes N}_{AB}\bigr)\Bigr] 
  = c,
\quad
\mathbb{E}_{\ket{\phi}_{AB}\sim D^{\mathrm{no}}} 
  \Bigl[\Tr\!\bigl(\mathcal{T}\,\ket{\phi}\bra{\phi}^{\otimes N}_{AB}\bigr)\Bigr] 
  = s.
$$
Moreover, if $c = \tfrac{2}{3}$ and $s = \tfrac{1}{3}$, we simply call $\mathcal{T}$ an \emph{average-case tester}.
\end{definition}

We will require the following theorem in Section~\ref{sec:un_shcb_commit}.

\begin{theorem}[Optimal average-case local tester (Theorem 3.7 in~\cite{CWZ24})]
\label{thm:optimal-average-case-local-tester}
Suppose $\cP = \bigl(\cP^{\mathrm{yes}}, \cP^{\mathrm{no}}\bigr)$ 
is a property of bipartite pure states in $\mathscr{H}_{AB}$ that is unitarily invariant 
on $\mathscr{H}_B$, and suppose $D^{\mathrm{yes}}, D^{\mathrm{no}}$ are probability 
distributions on $\cP^{\mathrm{yes}}, \cP^{\mathrm{no}}$ that are 
unitarily invariant on $\mathscr{H}_B$. Let $\mathcal{T}$ be an average-case 
$(c,s)$-tester for $\cP$ with respect to $(D^{\mathrm{yes}}, D^{\mathrm{no}})$ 
with sample complexity $N$, for some parameters $0 \le s < c \le 1$. Then, 
there is a \emph{local tester} $\widehat{\mathcal{T}}$ on $\mathscr{H}_A$ with sample complexity $N$ that is also an average-case $(c,s)$-tester for $\cP$ 
with respect to $(D^{\mathrm{yes}}, D^{\mathrm{no}})$.
\end{theorem}

\subsubsection{Canonical quantum bit commitments and the flavor of conversion lemma}

We define canonical quantum bit commitments as defined in~\cite{yan2022general}.

\begin{definition}[Canonical quantum bit commitments]\label{definiton:can_com}
A \emph{canonical quantum bit commitment scheme} is represented by a family 
$\{Q_0(\lambda), Q_1(\lambda)\}_{\lambda \in \mathbb{N}}$ of polynomial-time computable unitaries 
over two registers $\mathbf{C}$ (called the \emph{commitment register}) and $\mathbf{R}$ 
(called the \emph{reveal register}). In the rest of the paper, we often omit $\lambda$ 
and simply write $Q_0$ and $Q_1$ to mean $Q_0(\lambda)$ and $Q_1(\lambda)$.
\end{definition}

\begin{definition}[Hiding]
We say that a canonical quantum bit commitment scheme $\{Q_0, Q_1\}$ is 
\emph{computationally} (resp. \emph{statistically}) \emph{hiding} if 
$\mathrm{Tr}_{\mathbf{R}}\!\left(Q_0 ( \ket{0}\!\bra{0} )_{\mathbf{C}, \mathbf{R}} Q_0^{\dagger}\right)$ 
is computationally (resp. statistically) indistinguishable from 
$\mathrm{Tr}_{\mathbf{R}}\!\left(Q_1 ( \ket{0}\!\bra{0} )_{\mathbf{C}, \mathbf{R}} Q_1^{\dagger}\right)$. 
We say that it is \emph{perfectly hiding} if they are identical states.
\end{definition}

\begin{definition}[Honest Binding]
We say that a canonical quantum bit commitment scheme $\{Q_0, Q_1\}$ is 
\emph{computationally} (resp. \emph{statistically}) \emph{binding} 
if for any polynomial-time computable (resp. unbounded-time) unitary $U$ 
over $\mathbf{R}$ and an additional register $\mathbf{Z}$ and any polynomial-size state 
$\ket{\tau}_{\mathbf{Z}}$, it holds that
\[
\left\|
(Q_1 \ket{0}\!\bra{0} Q_1^{\dagger})_{\mathbf{C},\mathbf{R}}
(I_{\mathbf{C}} \otimes U_{\mathbf{R},\mathbf{Z}})
((Q_0 \ket{0})_{\mathbf{C},\mathbf{R}} \ket{\tau}_{\mathbf{Z}})
\right\| = \mathsf{negl}(\lambda).
\]
We say that it is \emph{perfectly binding} if the LHS is $0$ for all unbounded-time unitaries $U$.
\end{definition}

Here is the flavor of conversion lemma from~\cite{HMY23}.
\begin{lemma}[Converting Flavors (restate)~\cite{HMY23}]\label{lemma:com_swtich}
Let $\{Q_0, Q_1\}$ be a canonical quantum bit commitment scheme. 
Let $\{Q_0', Q_1'\}$ be a canonical quantum bit commitment scheme described as follows:

\begin{itemize}
    \item \textbf{Swapping registers:}  
    The roles of the commitment and reveal registers are swapped from $\{Q_0, Q_1\}$, 
    and the new commitment register is augmented by an additional one-qubit register. 
    That is, if $\mathbf{C}$ and $\mathbf{R}$ are the commitment and reveal registers of $\{Q_0, Q_1\}$, 
    then the commitment and reveal registers of $\{Q_0', Q_1'\}$ are defined as 
    $\mathbf{C}' := (\mathbf{R}, \mathbf{D})$ and $\mathbf{R}' := \mathbf{C}$, 
    where $\mathbf{D}$ is a one-qubit register.

    \item \textbf{Construction of flavored-swapped commitment scheme:}  
    For $b \in \{0,1\}$, the unitary $Q_b'$ is defined as follows:
    \[
    Q_b' := (Q_0 \otimes \ket{0}\!\bra{0}_{\mathbf{D}} + Q_1 \otimes \ket{1}\!\bra{1}_{\mathbf{D}})
    \, (I_{\mathbf{R},\mathbf{C}} \otimes Z_{\mathbf{D}}^b H_{\mathbf{D}})
    \]
    where $Z_{\mathbf{D}}$ and $H_{\mathbf{D}}$ denote the Pauli-$Z$ and Hadamard operators on $\mathbf{D}$. Note that we have
    \begin{equation}\label{equ:flavor_swap}
        Q_b' \ket{0}_{\mathbf{C}',\mathbf{R}'} 
        = \frac{1}{\sqrt{2}} 
        \left(
            (Q_0 \ket{0})_{\mathbf{C},\mathbf{R}} \ket{0}_{\mathbf{D}} 
            + (-1)^b (Q_1 \ket{0})_{\mathbf{C},\mathbf{R}} \ket{1}_{\mathbf{D}}
        \right)
    \end{equation}
    for $b \in \{0,1\}$, where $(\mathbf{C}', \mathbf{R}')$ is rearranged as $(\mathbf{C}, \mathbf{R}, \mathbf{D})$.
\end{itemize}

Then, if $\{Q_0, Q_1\}$ is computationally hiding, $\{Q_0', Q_1'\}$ is computational binding.

\noindent

\end{lemma}

\begin{remark}\label{remark:com_switch_reduction}
    Let $U$ be the adversary that breaks the binding property of $\{Q_0', Q_1'\}$. The reduction $\cR$ in the above lemma satisfies the following property:
    \begin{itemize}
        \item \textbf{Uniformity}: If $U$ is uniform (non-uniform), $\cR$ is also uniform (non-uniform).
        \item \textbf{Advantage:} If $U$ breaks the binding property of $\{Q_0', Q_1'\}$ with an advantage $\epsilon$, then $\cR$ breaks the hiding property of $\{Q_0, Q_1\}$ with an advantage $\epsilon^2$.
        \item \textbf{Size and width:} If the circuit size (width) of $U$ is $n$, then the circuit size (width) of $\cR$ is $O(n)$. 
    \end{itemize}
\end{remark}

\begin{remark}
    The original flavor of the conversion lemma in~\cite{HMY23} contains both directions. For our application, we require only one direction. 
\end{remark}

\subsubsection{Algorithmic Uhlmann’s Theorem}

We will require Algorithmic Uhlmann’s Theorem from~\cite{metger2023stateqip}. In order to state their results, we require the following definitions.

\begin{definition}[statePSPACE]
Let $\delta : \mathbb{N} \to [0,1]$ be a function. Then $\text{statePSPACE}_\delta$ 
is the class of all sequences of density matrices $(\rho_n)_{n \in \mathbb{N}}$ 
such that each $\rho_n$ is a state on $n$ qubits, and there exists a space-uniform 
family of general quantum circuits $(C_n)_{n \in \mathbb{N}}$ such that for all 
sufficiently large $n \in \mathbb{N}$, the circuit $C_n$ takes no inputs and 
outputs a density matrix $\sigma_n$ such that
\[
    \text{TD}(\sigma_n, \rho_n) \le \delta(n).
\]
\end{definition}

\begin{definition}[unitaryPSPACE]
Let $\delta : \mathbb{N} \to [0,1]$ be a function. Then $\text{unitaryPSPACE}_\delta$ 
is the class of all sequences $(U_n)_{n \in \mathbb{N}}$ such that each $U_n$ 
is a unitary acting on $n$ qubits, and there exists a space-uniform family of general 
quantum circuits $(C_n)_{n \in \mathbb{N}}$ such that for all sufficiently large 
$n \in \mathbb{N}$, for all $n$-qubit states $\ket{\psi}$, we have
\[
   \text{TD}(C_n(\psi), U_n \psi U_n^{\dagger}) \le \delta(n).
\]

\medskip
\noindent
We define the class $\text{unitaryPSPACE}$ to be
\[
    \text{unitaryPSPACE} = 
    \bigcap_q \text{unitaryPSPACE}_{\exp(-q(n))}
\]
where the intersection is over all polynomials $q : \mathbb{N} \to \mathbb{R}$.
\end{definition}

\begin{lemma}[Algorithmic Uhlmann’s Theorem~\cite{metger2023stateqip}]\label{lemma:pspace_uhl}
Let $\delta(n)$ be a function and let $\\ \{\ket{\psi_n}\}_n, \{\ket{\varphi_n}\}_n \in \text{statePSPACE}_\delta$ 
be pure state families. Suppose that for each $n$, the states $\ket{\psi_n}$ and $\ket{\varphi_n}$ 
have the same number of qubits and that the qubits can be divided into two registers 
$\mathbf{A}_n, \mathbf{B}_n$. Then, for all polynomials $q(n)$, there exists a sequence of unitaries 
$\{K_n\}_n$ in $\text{unitaryPSPACE}$ such that $K_n$ acts on registers $\mathbf{B}_n$ 
and an ancilla register $\mathbf{R}_n$, and satisfies
\begin{equation*}
    \left\| (I_{\mathbf{A}_n} \otimes K_n) \ket{\varphi_n}_{\mathbf{A}_n \mathbf{B}_n} \ket{0}_{\mathbf{R}_n} 
    - \ket{\psi_n}_{\mathbf{A}_n \mathbf{B}_n} \ket{0}_{\mathbf{R}_n} \right\|^2 
    \le 2\left(1 - \mathrm{F}(\rho_n, \sigma_n)\right) + O(\delta(n)) + 2^{-q(n)},
\end{equation*}
where $\rho_n$ and $\sigma_n$ are the reduced density matrices on register $\mathbf{A}_n$ 
of $\ket{\psi_n}$ and $\ket{\varphi_n}$, respectively.

\medskip
Furthermore, the description of the Turing machine that outputs the circuits implementing 
the unitaries $\{K_n\}_n$ can be computed in polynomial time from the descriptions of the 
Turing machines that output the circuits for preparing the states $\{\ket{\psi_n}\}_n$ 
and $\{\ket{\varphi_n}\}_n$.
\end{lemma}

%% file: 3_Definition.tex
\section{Complexity classes for quantum promise problem}\label{section:def}

\input{3-1_Definition}
\input{3-2_Def_oracle}
\input{3-3_Def_reduction}
\input{3-4_Def_condition}

%% file: 3-1_Definition.tex
\subsection{Definition of quantum promise complexity classes}\label{section:defClass}

%In quantum computing, many computational tasks ask to compute certain properties of input quantum states. We can describe this type of decision problem as two sets of quantum states, and the goal is to identify which set the input state belongs to. Along this line, we can define complexity classes for these problems. 
We provide formal definitions for quantum promise problems and quantum promise complexity classes.

\begin{definition}[Pure state quantum promise problem $\mathcal{L}$]\label{Def:problem_p}
A pure state quantum promise problem is defined as a pair of sets of pure quantum states $\mathcal{L}:= (\mathcal{L}_Y, \mathcal{L}_N)$, where $\mathcal{L}_Y, \mathcal{L}_N \subseteq \bigcup_{n=1}^{\infty}\mathcal{H}(n)$. A state $\ket{\psi}$ is called a yes-instance if $\ket{\psi} \in \mathcal{L}_Y$, and a no-instance if $\ket{\psi} \in \mathcal{L}_N$.
\end{definition}

\begin{definition}[Mixed state quantum promise problem $\mathcal{L}$]\label{Def:problem_m}
A mixed state quantum promise problem is defined as a pair of sets of mixed quantum states $\mathcal{L}:= (\mathcal{L}_Y, \mathcal{L}_N)$, where $\mathcal{L}_Y, \mathcal{L}_N \subseteq \bigcup_{n=1}^{\infty}\mathcal{D}\big(\mathcal{H}(n)\big)$. A state $\rho$ is called a yes-instance if $\rho \in \mathcal{L}_Y$, and a no-instance if $\rho \in \mathcal{L}_N$.
\end{definition}
In general, a quantum promise problem can be viewed as either a pair of sets or an input-output problem, where the output is a classical binary bit. The promise on the input represents restrictions on both yes-instances and no-instances.

\begin{remark}
    In the context of mixed-state quantum promise problems, it is sometimes useful to consider the input as a decomposition into pure states with associated probabilities. However, this decomposition is generally not unique. For a mixed-state quantum promise problem to be well-defined, the input state must belong to the same set,  regardless of the chosen decomposition. For instance, the following problem is not well-defined:
    %so the inputs are possibly sampled from different distributions each time (but still the same mixed state $\rho$). 
    %When we define a mixed-state quantum promise problem (especially according to some choices of distribution of inputs), the definition for yes and no instances should be well-defined. For example, the following problem is not well-defined:
     \begin{itemize}
        \item \textbf{Inputs}: $\rho$.
        \item \textbf{Yes instances:} Consider an eigenbasis decomposition $\{\psi_i\}$ of $\rho$, i.e. $\rho= \sum\limits_i \lambda_i \ket{\psi_i}\bra{\psi_i}$. Accept if there exists $i$ such that $\langle 0 |\psi_i \rangle=1$.
        \item  \textbf{No instances:} Consider an eigenbasis decomposition $\{\psi_i\}$ of $\rho$, i.e. $\rho= \sum\limits_i \lambda_i \ket{\psi_i}\bra{\psi_i}$. Reject if for all $i$ the value $\langle 0 |\psi_i \rangle \leq \frac{1}{2}$.
    \end{itemize}
    Let $\rho$ be the maximally mixed state. Clearly, $\rho$ falls into the yes-instance if we choose its decomposition on the standard basis. Still, it falls into the no-instance if we consider its decomposition on the Hadamard basis.
\end{remark}

We define uniform quantum circuits before discussing the various complexity classes used in this paper.

\begin{definition}[P-uniform quantum circuit]\label{def:p_uniform}
    Fix a finite universal gate set. We say that a set of quantum circuits $\{\cV_\lambda\}_{\lambda \in \N}$ is a P-uniform quantum circuit family if there exists a classical polynomial-time Turing machine $\cM$ such that, on input $1^\lambda$, it outputs the polynomial-sized quantum circuit $\cV_\lambda$.
\end{definition}

\HZZ{CHeck}
\begin{definition}[PSPACE-uniform quantum circuit]\label{def:pspace_uniform}
Fix a finite universal gate set. We say that a set of quantum circuits $\{\cV_\lambda\}_{\lambda \in \N}$ is a PSPACE-uniform quantum circuit family if there exists a classical polynomial-space Turing machine $\cM$ such that, on input $(1^\lambda,t)$, it outputs the $t$-th step of the circuit $\cV_\lambda$. Each gate of $\cV_\lambda$ is a gate from the universal gate set or an intermediate measurement. Note that the circuit produced by the classical Turing machine must have a polynomially bounded width but may have an exponentially large depth of quantum gates. These gates will act on the qubits and produce a result when the final measurement is performed.
\end{definition}

We say a uniform circuit has some special operations if its elementary gates are extended to include those operations. For example, in Section \ref{Def:Oracle}, where we discuss oracle access, we allow an operation that applies a CPTP map followed by a measurement.

\begin{definition}[$\pBQP_{c(\lambda),s(\lambda)}^{t(\lambda)}$]\label{Def:pBQP}
Let $t(\cdot)$, $c(\cdot)$, and $s(\cdot)$ be polynomials. $\pBQP_{c(\lambda),s(\lambda)}^{t(\lambda)}$ is a collection of pure state quantum promise problems $\cL:=(\cL_Y, \cL_N)$ such that if there exists a $\text{P}$-uniform quantum circuit family $\{\cV_\lambda\}_{\lambda \in \N}$ and polynomial $p(\cdot)$ satisfying the following properties: For all $\lambda \in \N$, $\mathcal{V}_\lambda$ takes $t(\lambda)\cdot \lambda + p(\lambda)$ qubits as input, consisting of input $\ket{\psi}^{\otimes t(\lambda)} \in \mathcal{H}(\lambda)^{\otimes t(\lambda)}$ in register $I$ and $p(\lambda)$ ancilla qubits initialized to $\ket{0}$ in register $A$. The first qubit of $A$, denoted as $A^{ans}$ is the designated output qubit, a measurement of which in the standard basis after applying $\cV_\lambda$ yields the following (where the outcome $\ket{1}$ denotes ``accept"). For sufficiently large $\lambda$ and $\ket{\phi} \in \cH (\lambda)$ we have,
\begin{itemize}
    \item (Completeness): If $\ket{\phi} \in \mathcal{L}_{Y} $, then $\mathcal{V}_\lambda$ accepts with probability at least $c(\lambda)$.
    \item (Soundness): If $\ket{\phi} \in \mathcal{L}_{N} $, then $\mathcal{V}_\lambda$ accepts with probability at most $s(\lambda)$.
\end{itemize}
We will also interchangeably view $\{\mathcal{V}_\lambda\}_{\lambda \in \N}$ as an algorithm. Then, $\mathcal{V}_\lambda\big(\ket{\phi}^{\otimes t(\lambda)} \otimes \ket{0}^{\otimes p(\lambda)} \big)$ denotes the output of $\mathcal{V}_\lambda$. We also say $\{\mathcal{V}_\lambda\}_{\lambda \in \N}$ is a $\pBQP_{c(\lambda),s(\lambda)}^{t(\lambda)}$ algorithm or a quantum polynomial-time (QPT) algorithm that decides $\mathcal{L}$ with completeness $c(\lambda)$ and soundness $s(\lambda)$.
\end{definition}

\begin{definition}[$\mBQP_{c(\lambda),s(\lambda)}^{t(\lambda)}$]
    The definition of $\mBQP_{c(\lambda),s(\lambda)}^{t(\lambda)}$ is the same as in Definition \ref{Def:pBQP}, except that $\mathcal{H}(\lambda)$ is replaced by $\mathcal{D}\big(\mathcal{H}(\lambda)\big)$, meaning the inputs are mixed-state.
\end{definition}

\begin{definition}[$\pmPSPACE_{c(\lambda),s(\lambda)}^{t(\lambda)}$]\label{Def:pspace}
    Let $t(\cdot),c(\cdot)$, and $s(\cdot)$ be polynomials. The definitions of $\pPSPACE_{c(\lambda),s(\lambda)}^{t(\lambda)}$ and  $\;\mPSPACE_{c(\lambda),s(\lambda)}^{t(\lambda)}$ are the same as in Definition \ref{Def:pBQP}, except that $\{\mathcal{V}_\lambda\}_{\lambda \in \N}$ is a PSPACE-uniform quantum circuit family. Additionally, for mixed-state quantum promise problems, $\mathcal{H}(\lambda)$ is replaced by $\mathcal{D}\big(\mathcal{H}(\lambda)\big)$.\footnote{In our paper, we only consider a single-party algorithm that receives a polynomial number of copies of the input. However, we can still define a scenario where a party receives a superpolynomial number of input copies. For $t(\cdot) = \omega(poly(\cdot))$, the PSPACE-uniform quantum circuit has a special operation. This special operation traces out a register and appends a new register with a fresh instance of the input. Informally, a PSPACE algorithm can obtain a fresh copy of the input whenever it ``presses a button". Even though the total number of copies it receives is $t(\lambda)=\omega(poly(\cdot))$, the algorithm can only use a polynomial number of copies at any given time.}
\end{definition}

%\begin{definition}[$\pmINF_{c(\lambda),s(\lambda)}^{t(\lambda)}$]Let $t(\cdot),c(\cdot)$, and $s(\cdot)$ be polynomials. The definition of $\newline\pINF_{c(\lambda),s(\lambda)}^{t(\lambda)}$ and  $\;\mINF_{c(\lambda),s(\lambda)}^{t(\lambda)}$ are the same as in Definition \ref{Def:pBQP}, except that $\{\mathcal{V}_\lambda\}_{\lambda \in \N}$ is an unbounded uniform quantum circuit family. That is, there exists a classical Turing machine that, on input $1^\lambda$, is only required to halt (i.e., without any restrictions on time or space) and outputs the description of quantum circuit $\cV_\lambda$. We emphasize that the circuit $\cV_\lambda$ uses only a polynomial number of copies of the input. \nai{The current pspace definition allows exponential copies.}\HZZ{OK}\end{definition}

 The $m$-messages (with $m$ even) interactive protocol between a $m$-message $t_v$-copy quantum verifier circuit family $V^{m,t_v} = \{V_1,\cdots,V_{\frac{m}{2} + 1}\}$, and a $m$-message $t_p$-copy quantum prover circuit family $P^{m,t_p} = \{P_1,\cdots,P_{\frac{m}{2}}\}$ is model as follows:
\begin{equation*}
    (V_{\frac{m}{2}+1})_{M,V} (P_\frac{m}{2})_{P,M} (V_\frac{m}{2})_{M,V} \cdots (P_1)_{P,M} (V_1)_{M,V} \big(\ket{\phi}^{\otimes t_v}\ket{0\cdots0}\big)_P  \cdot \ket{0\cdots0}_M \big(\ket{\phi}^{\otimes t_p}\ket{0\cdots0}\big)_V,    
\end{equation*}
where $t_v$ and $t_p$ refer to the number of input state copies available to the verifier and the prover, respectively. Additionally, $M$ denotes the message register, while $P$ and $V$ refer to the prover's and verifier's private registers, respectively. After the interaction, the verifier measures the first bit of register $V$ in the standard basis. If the outcome is $\ket{1}$, the verifier accepts; otherwise, the verifier rejects. Let $\langle P^{m,t_p}, V^{m,t_v} \rangle$ denote the output of the interactive protocol. When the number of messages $m$ is odd, the definition is the same as for even $m$, except that we assume the prover sends the first message instead of the verifier.
\begin{remark}
    When we refer to a prover's quantum circuit family $P$ with $\infty$-copies, it means that the prover can obtain the description of the input state $\ket{\phi} \in \mathcal{H}(\lambda)$ with arbitrary precision. That is, given an arbitrary precision $\epsilon > 0$, the input for the prover is a classical encoding of the unitary $U^{In}$ such that $||U^{In}\ket{0^{\lambda}} - \ket{\phi}||_{tr} \leq \epsilon$. Additionally, we can also view the input as a classical description of $\ket{\psi}$ with all entries having errors within $\epsilon$.
\end{remark}

\begin{definition}[$\pQIP_{c(\lambda),s(\lambda)}^{t_p(\lambda),t_v(\lambda)}\text{[}m(\lambda)\text{]}$]\label{Def:pQIP} 
    Fix polynomials $t_p(\cdot)$, $t_v(\cdot)$, $c(\cdot)$, $s(\cdot)$, and $m(\cdot)$. \\ $\pQIP_{c(\lambda),s(\lambda)}^{t_p(\lambda),t_v(\lambda)}[m(\lambda)]$ is a collection of quantum promise problems $\mathcal{L}:=(\mathcal{L}_{Y}, \mathcal{L}_{N})$ such that if there exists $P$-uniform $m(\lambda)$-messages $t_v(\lambda)$-copy quantum verifier circuit family $V^{m(\lambda),t_v(\lambda)}$  satisfying the following properties. For sufficiently large $\lambda$ and $\ket{\phi} \in \mathcal{H}(\lambda)$ we have,
    \begin{itemize}
       \item (Completeness): If $\ket{\phi} \in \mathcal{L}_{Y} $, then there exists a $m(\lambda)$-messages $t_p(\lambda)$-copy prover's quantum circuit family  
       $P^{m(\lambda),t_p(\lambda)}$ such that
       \begin{equation*}
           \Pr[\langle P^{m(\lambda),t_p(\lambda)}, V^{m(\lambda),t_v(\lambda)} \rangle = 1] \geq c(\lambda).
       \end{equation*}
        \item (Soundness): If $\ket{\phi} \in \mathcal{L}_{N} $, then for all $m(\lambda)$-messages $t_p(\lambda)$-copy prover's quantum circuit family  
       $P^{m(\lambda),t_p(\lambda)}$ such that
        \begin{equation*}
           \Pr[\langle P^{m(\lambda),t_p(\lambda)}, V^{m(\lambda),t_v(\lambda)} \rangle = 1] \leq s(\lambda).
       \end{equation*}
    \end{itemize}
\end{definition}

\begin{definition}[$\mQIP_{c(\lambda),s(\lambda)}^{t_p(\lambda),t_v(\lambda)}\text{[}m(\lambda)\text{]}$]\label{Def:mQIP}
    The definition of $\mQIP_{c(\lambda),s(\lambda)}^{t_p(\lambda),t_v(\lambda)}\text{[}m(\lambda)\text{]}$ is defined the same as Definition \ref{Def:pQIP} but $\mathcal{H}(\lambda)$ is replaced to $\mathcal{D}\big(\mathcal{H}(\lambda)\big)$, i.e., the inputs are mixed state.
\end{definition}

We now define some classes which are special cases of $\pmQIP$.
\begin{definition}[$\pmQMA_{c(\lambda),s(\lambda)}^{t_p(\lambda),t_v(\lambda)}$]\label{Def:QMA}
    The definition of $\pQMA_{c(\lambda),s(\lambda)}^{t_p(\lambda),t_v(\lambda)}$ and $\mQMA_{c(\lambda),s(\lambda)}^{t_p(\lambda),t_v(\lambda)}$ are defined as  $\pQIP_{c(\lambda),s(\lambda)}^{t_p(\lambda),t_v(\lambda)}[1]$ and $\mQIP_{c(\lambda),s(\lambda)}^{t_p(\lambda),t_v(\lambda)}[1]$, respectively.
\end{definition}
\begin{remark}[Remark for Definition \ref{Def:QMA}]
    The state in the message register $\mathcal{M}$ after the prover applies $P_{\lambda,1}$ is called a witness. For a yes instance, we call a witness good if the verifier accepts with probability at least $c(\lambda)$ . By convexity, we can assume a good witness is a pure state without loss of generality.
\end{remark}

\begin{definition}[$\pmQCMA_{c(\lambda),s(\lambda)}^{t_p(\lambda),t_v(\lambda)}$]
    The definition of $\pQCMA_{c(\lambda),s(\lambda)}^{t_p(\lambda),t_v(\lambda)}$ and $\mQCMA_{c(\lambda),s(\lambda)}^{t_p(\lambda),t_v(\lambda)}$ is defined the same as $\pQMA_{c(\lambda),s(\lambda)}^{t_p(\lambda),t_v(\lambda)}$ and $\pQMA_{c(\lambda),s(\lambda)}^{t_p(\lambda),t_v(\lambda)}$, respectively. The only difference is that the verifier always measures the register $M$ in the computational basis, once it receive messages from the prover. Note that $\mathcal{V}_{\lambda,1}$ can still be defined as an unitary.
\end{definition}

\begin{definition}
[$\mathbf{pQSZK}_{\text{hv}, c(\lambda),s(\lambda)}^{t_p(\lambda),t_v(\lambda),t_s(\lambda)}\text{[}m(\lambda)\text{]}$]\label{Def:pQSZKhv}
The completeness and soundness of \\
$\mathbf{pQSZK}_{\text{hv}, c(\lambda),s(\lambda)}^{t_p(\lambda),t_v(\lambda),t_s(\lambda)}\text{[}m(\lambda)\text{]}$ are defined the same as $\pQIP_{c(\lambda),s(\lambda)}^{t_p(\lambda),t_v(\lambda)}\text{[}m(\lambda)\text{]}$. For the honest verifier statistical zero knowledge, we require the following property. For sufficiently large $\lambda$ and $\ket{\phi} \in \mathcal{H}(\lambda)$, there exist a polynomial time simulator on input $(\ket{\phi}^{\otimes t_s},i)$ (for all $i \in [m(\lambda)]$), output a mixed state $\xi_{\ket{\phi},i}$ such that
\begin{center}
    If $\ket{\phi} \in \mathcal{L}_Y, ||\xi_{\ket{\phi},i } -  
    \view_{P,V}(\ket{\phi}, i)||_{tr} \leq \negl(\lambda)$.
\end{center}
where $\view_{P, V}(\ket{\phi}, i)$ is the reduced state after $i$ messages have been sent and tracing out the prover's private qubits.  
\end{definition}

\begin{definition}
[$\mathbf{mQSZK}_{\text{hv}, c(\lambda),s(\lambda)}^{t_p(\lambda),t_v(\lambda),t_s(\lambda)}\text{[}m(\lambda)\text{]}$]\label{Def:mQSZKhv}
    The definition of $\mathbf{mQSZK}_{\text{hv}, c(\lambda),s(\lambda)}^{t_p(\lambda),t_v(\lambda),t_s(\lambda)}\text{[}m(\lambda)\text{]}$ is defined the same as Definition \ref{Def:pQSZKhv} but $\mathcal{H}(\lambda)$ is replaced to $\mathcal{D}\big(\mathcal{H}(\lambda)\big)$, i.e. the inputs are mixed state.
\end{definition}

%The following notation is natural and frequently used in this paper, so we gave it a special name.

In this work, we focus on the setting where a single-party algorithm can access an arbitrary polynomial number of copies of the input states. Additionally, the prover in the interactive proof is allowed to obtain an unbounded number of copies of the input states. For simplicity and ease of notation, we make the following simplifications:

\begin{definition}\label{Def:simplify}
Let $\mathcal{C} \in \{\pBQP,\mBQP,\pPSPACE,\mPSPACE \}$. We define $\mathcal{C}$ as follows:
\begin{equation*}
    \mathcal{C}:= \bigcup\limits_{t(\cdot) \:\in\: \mathcal{P}(\N)} \mathcal{C}^{t(\lambda)}_{\frac{2}{3},\frac{1}{3}}\;\; \text{ and }  \;\;\mathcal{C}_{c(\lambda),s(\lambda)}:= \bigcup\limits_{t(\cdot) \:\in\: \mathcal{P}(\N)} \mathcal{C}^{t(\lambda)}_{c(\lambda),s(\lambda)}.
\end{equation*}
Let $\mathcal{C} \in \{\ \pQMA,\mQMA,\pQCMA,\mQCMA\}$. Define the following classes:
\begin{equation*}
    \mathcal{C}:= \bigcup\limits_{t(\cdot) \:\in\: \mathcal{P}(\N)} \mathcal{C}^{\infty, t(\lambda)}_{\frac{2}{3},\frac{1}{3}}\;\; \text{ and }  \;\;\mathcal{C}_{c(\lambda),s(\lambda)}:= \bigcup\limits_{t(\cdot) \:\in\: \mathcal{P}(\N)} \mathcal{C}^{\infty,t(\lambda)}_{c(\lambda),s(\lambda)}.
\end{equation*}
Let $\mathcal{C} \in \{\pQIP, \mQIP\}$. Define the following classes:
\begin{equation*}
        \mathcal{C}:= \bigcup\limits_{t(\cdot),m(\cdot) \:\in\: \mathcal{P}(\N)} \mathcal{C}^{\infty, t(\lambda)}_{\frac{2}{3},\frac{1}{3}}[m(\lambda)]\;\; \text{ and }  \;\;\mathcal{C}_{c(\lambda),s(\lambda)}:= \bigcup\limits_{t(\cdot) \:\in\: \mathcal{P}(\N)} \mathcal{C}^{\infty,t(\lambda)}_{c(\lambda),s(\lambda)}[m(\lambda)].
\end{equation*}
Let $\mathcal{C} \in \{\pQSZKhv, \mQSZKhv\}$. Define the following classes:
\begin{equation*}
    \mathcal{C}:= \bigcup\limits_{t_1(\cdot),t_2(\cdot),m(\cdot) \:\in\: \mathcal{P}(\N)} \mathcal{C}^{\infty, t_1(\lambda),t_2(\lambda)}_{\frac{2}{3},\frac{1}{3}}[m(\lambda)]\;\; \text{ and }  \;\;\mathcal{C}_{c(\lambda),s(\lambda)}:= \bigcup\limits_{t_1(\cdot),t_2(\cdot),m(\cdot) \:\in\: \mathcal{P}(\N)} \mathcal{C}^{\infty,t_1(\lambda),t_2(\cdot)}_{c(\lambda),s(\lambda)}[m(\lambda)].
\end{equation*}

\end{definition}
\begin{remark}
    In Definition $\ref{Def:simplify}$, we define a natural complexity class for an interactive proof system: the prover has complete knowledge of inputs. Besides, the definition of completeness and soundness of $\pmQMA$ can change to: (Completeness) there exists a witness $\ket{\phi}$ such that the verifier accept with probability at least $\frac{2}{3}$; (soundness) for all witness $\ket{\phi}$ such that the verifier reject with probability at least $\frac{2}{3}$. This is an analog of the classical complexity class, where the prover always gets the complete information it needs.
    However, there are some variants of the definition. For example, the prover gets polynomial copies of input for completeness, and soundness holds for unbounded inputs. See the open problem Section \ref{section:open_problem} for more discussion.
\end{remark} 

If $\mathcal{C}$ is a classical complexity class, we define $\textbf{p}\mathcal{C}$ and $\textbf{m}\mathcal{C}$ as the pure-state and mixed-state quantum promise complexity classes, respectively, both imitating the definition of $\mathcal{C}$. Additionally, for the simplicity of the syntax of statements, we use the following notation: 
\begin{itemize}
    \item $\pmclass\mathcal{C}$ represents $\textbf{p}\mathcal{C}$ or $\textbf{m}\mathcal{C}$.
    \item $\pmclass\mathcal{C}_1 \subseteq \pmclass\mathcal{C}_2$ represents $\textbf{p}\mathcal{C}_1 \subseteq \textbf{p}\mathcal{C}_2$ and $\textbf{m}\mathcal{C}_1 \subseteq \textbf{m}\mathcal{C}_2$.
    %\item $\mathcal{L} \in \pmclass\mathcal{C}_1 \implies \mathcal{L} \in \pmclass\mathcal{C}_2$ represents $\mathcal{L} \in \textbf{p}\mathcal{C}_1 \implies \mathcal{L} \in \textbf{p}\mathcal{C}_2$ and $\mathcal{L} \in \textbf{m}\mathcal{C}_1 \implies \mathcal{L} \in \textbf{m}\mathcal{C}_2$.
\end{itemize}

%The definition of the quantum promise problem has some advantages, which will be explained in section \ref{Def:Oracle} and section \ref{Def:Reduction}. In brief, quantum promise problems preserve the oracle separation of the classical complexity classes. Also, it preserves the simplicity of Karp reduction and amplification.
%\nai{It is not adequate to say ``adavantage''.}
It is natural to ask whether quantum promise problems and complexity classes preserve some fundamental characteristics of classical complexity classes, such as reductions and separations. A proper definition of the quantum promise oracle can help address this question. In Section~\ref{Def:Oracle} and Section~\ref{Def:Reduction}, we formalize the quantum promise oracle, show that quantum promise classes preserve the separation between the classical complexity classes, demonstrate that it preserves the simplicity of Karp and Turing reduction and amplification. 

%% file: 3-2_Def_oracle.tex
\subsection{Quantum promise oracle and relationships to classical complexity classes}
\label{Def:Oracle}

An oracle is a black-box interface that, when queried, typically responds with a quantum state or some other output in a single step. Oracles can be modeled in various ways, such as classical functions, unitary operators, \emph{Completely Positive and Trace-Preserving} (CPTP) maps, or complexity classes. There are two types of oracles: total function oracles and partial function oracles. A total function oracle guarantees a valid response for every possible input, ensuring the output is consistent with a well-defined function. For example, classical function oracles compute $f(x)$ for any input $x$; unitary oracles apply a unitary transformation to an input quantum state; CPTP maps oracles generalize unitary transformations; and complexity class oracles solve decision problems within a particular complexity class. On the other hand, a partial function oracle computes a function that is not defined for every input, meaning that it may return an arbitrary result for some inputs. For instance, promise problem oracles handle decision problems where the input is restricted to a specific subset, known as the ``promise," and the oracle's behavior for inputs outside this subset may be arbitrary. Additionally, the promise complexity class oracles provide solutions to decision promise problems that belong to a specific promise complexity class. However, we must avoid unintended behavior when querying inputs outside the oracle's promise. Hence, the definition of the oracle algorithm should ensure that any responses outside the oracle's promise must not affect the overall result of $\cA$.\HZZ{Check}

Here, we consider an oracle as a decision quantum promise problem stated in Definition \ref{Def:problem_p} and \ref{Def:problem_m}.
Our first goal is to formalize the quantum promise oracle. We recall the original definition of a classical oracle algorithm, denoted as $\cA^\cO$, where $\cA$ is any algorithm and $\cO$ could be viewed as (partial) Boolean functions.\footnote{Boolean functions correspond to languages, which ask to decide whether an input string $x$ is in $L$ ($f(x) = 1$) or not ($f(x)=0$). Partial Boolean functions correspond to promise problems, which allow the oracle’s reply to be arbitrary when $x$ is not in $L_Y\cup L_N$ ($f(x)$ is undefined).} Informally, when the algorithm $\cA$ queries a yes or no instance to the oracle, the oracle replies with the correct answer. The oracle’s responses to queries in the ``don’t care" region should not affect the algorithm's outcome. Hence, the statement \emph{$\cA^\cO$ decides a language $\cL$} can be defined as follows: For all Boolean functions $f$ that implement $\cO$ (i.e., $f$ and $\cO$ share the same input-output pairs for all yes and no instances), $\cA^f$ decides $\cL$. We emphasize that every Boolean function can be constructed by a corresponding circuit, a process called ``physical realization".\HZZ{Check}

Let us come back to formalizing the quantum promise oracle. When the algorithm $\cA$ interacts with a quantum promise oracle, it may provide the oracle with a limited number of copies of a quantum state as input. Two key issues must be addressed. First, similar to the classical case, we do not want the outcome of a ``don't care" instance to affect the result. This issue can be directly addressed by following the definition in the classical setting. 
%However, we must be cautious, as not all oracles are physically realizable; in other words, some oracles cannot be constructed by any CPTP map.\HZZ{Check}
The second issue is more tricky and unique for quantum promise problems. That is, \emph{quantum promise problems require sufficiently many copies of the input states to be solved.} This could lead to the case that some quantum promise problem oracle algorithms $\cA^{\cO}$ are not well-defined. For example, let $\cO$ be the oracle of some quantum promise problems requiring superpolynomial copies of input states. Then, defining $\BQP^{\cO}$ could be problematic: The $\BQP$ machine can only send at most polynomial number of states to $\cO$. If we allow $\cO$ to give answers with that many copies, then $\BQP^{\cO}$ can even solve some information-theoretically hard problem and thus is not ``physically realizable''. On the other hand, if $\cO$ will not give a response unless it obtains sufficient input states, then $\cO$ is useless for the $\BQP$ machine. We conclude that our paper only considers a physically realizable oracle.

Now, we can informally define ``$\cA^\cO$ decides a promise problem $\cL$" as follows: For all CPTP map $\cC$ that instantiate the oracle $\cO$, $\cA^\cC$ should also decides $\cL$. Formally, we define them in Definition \ref{Def:query_oracle1}, \ref{Def:query_oracle2}, \ref{Def:query_oracle3}, and \ref{Def:query_oracle4}.

\begin{definition}[Instantiate a quantum promise oracle]\label{Def:query_oracle3}Consider a family of CPTP maps $\cM := \{\cM_\lambda\}_\lambda$, where $\cM_\lambda$ has input size $\lambda$ and a single qubit output register. Consider a quantum promise problem $\cO:=\{\cL_Y,\cL_N\}$ and a polynomial function $p(\cdot)$. We say that $\cM$ instantiates $\cO$ if there exists a polynomial $p(\cdot)$ such that, given an $n$-qubit quantum state $\rho$, the following holds:
\begin{itemize}
    \item If $\rho \in \cL_Y $, $\Pr\Big[\text{The standard basis measurement outcome of } \cM_{np(n)}\big(\rho^{\otimes p(n)} \big) \text{ is } 1\Big] \geq 1-\negl(n)$.
    \item If $\rho \in \cL_N $, $\Pr\Big[\text{The standard basis measurement outcome of } \cM_{np(n)}\big(\rho^{\otimes p(n)} \big) \text{ is } 0\Big] \geq 1-\negl(n)$.
\end{itemize}
\nai{Why is the sub of $\cM_{np(n)}$ $np(n)$?}
\end{definition}

\begin{definition}[Pyhsically realizable oracle, $\pmINF$]\label{Def:physically_realizable}
    An oracle (or quantum promise problem) $\cO:=\{\cL_Y,\cL_N\}$ is physically realizable if there exists a family of CPTP maps $\cM:=\{\cM_\lambda\}_\lambda$ that instantiates $\cO$ with some polynomial $p(\cdot)$ copies. Otherwise, it is considered physically unrealizable. Let $\pINF$ and $\mINF$ be the collection of physically realizable pure-state and mix-state quantum promise problem $\cL$, respectively.
\end{definition}

\begin{definition}[Quantum oracle circuits]\label{Def:query_oracle1}
Consider a family of CPTP maps $\cM := \{\cM_\lambda\}_\lambda$, where $\cM_\lambda$ has input size $\lambda$ and a single qubit output register. A quantum oracle circuit $\cC$ with oracle access to $\cM$, denoted as $\cC^\cM$, consists of (i) elementary gates and (ii) a CPTP map $\cM_\lambda$ followed by a standard basis measurement on the output register.
\end{definition}

\begin{definition}[Quantum oracle algorithms]\label{Def:query_oracle2}
Consider a family of CPTP maps $\cM := \{\cM_\lambda\}_\lambda$, where $\cM_\lambda$ has input size $\lambda$ and a single qubit output register. An algorithm $\cA$ with oracle access to $\cM$, denoted as $\cA^\cM$, is a family of circuits with oracle access to $\cM$. That is, $\cA^\cM:=\{\cC^M_{1,\lambda},\cC^M_{2,\lambda}\dots\}_\lambda$.
\end{definition}

In the above definition, the algorithm $\cA$ can be specified by a complexity class, e.g. $\pBQP$ or $\pQMA$. We say that $\cA^\cM$ decides a language $\cL$ if it satisfies the condition for that complexity class. For example, let $\cA$ be a $\pBQP$ algorithm and $\cM$ be a family of CPTP maps. Then, $\cA^\cM$ is defined to be a $P$-uniform quantum oracle circuits $\{\cC^\cM_\lambda\}_\lambda$. That is, there exists a polynomial-time classical Turing machine such that, on input $1^\lambda$, it outputs the circuit $\{\cC^\cM_\lambda\}_\lambda$ consists of elementary gates and CPTP maps $\{\cM_\lambda\}_\lambda$ followed by a standard basis measurement on its output.
Furthermore, let $\cB$ be a $\pQMA$ algorithm. Then, $\cB^\cM$ is defined to be $\{\cP^\cM_\lambda, \cV^\cM_\lambda\}_\lambda$, where $\{\cP^\cM_\lambda\}_\lambda$ is a family of arbitrary circuits and $\{\cV^\cM_\lambda\}_\lambda$ is a $P$-uniform quantum oracle circuits.

\begin{definition}[Quantum promise oracle algorithm for deciding $\cL$]\label{Def:query_oracle4}
     Given a physically realizable oracle $\cO$. Let $\cA(\cdot)$ be an algorithm that belongs to some complexity class. We say that $\cA^\cO$ decides a quantum promise problem $\cL$ if the following holds: For all CPTP map $\cM$ such that $\cM$ instantiate $\cO$, $\cA^\cM$ decides $\cL$.\footnote{Suppose $\cM$ instantiates $\cO$ with polynomial $p(\cdot)$ copies. The algorithm $\cA$ obtains $p(\cdot)$ when given access to $\cM$. Additionally, we assume that $p(\cdot)$ is a polynomial-time computable function.}
\end{definition}

%Next, we consider the meaning of a primitive exists under an oracle. The following defines $\textit{cryptographic primitive}$.
%\begin{definition}[Cryptographic primitive \cite{??}]
%A cryptographic primitive $\cP$ is a pair
%\end{definition}

%%%%%%%%%%%%%%%%%%%%%%%%%%%%%%%%%%%%%%%%%%%%%%%%%%%%%%%%%%%%

The following Claims~\ref{Def:prop1}, \ref{Def:prop2}, and \ref{Def:prop3} formally show that quantum promise complexity classes generalize classical promise complexity classes. Note that in those Claims, $\mathcal{C}$ is 
a classical promise class, and hence $\mathcal{L} \in \mathcal{C}$ is a tuple of two sets $(\mathcal{L}_Y, \mathcal{L}_N)$. Besides, we abuse the symbol: Let $a \in \{0,1\}^*$, we view $a$, $\ket{a}$, and $\ket{a}\bra{a}$ as identical elements. We also implicitly use the fact that $\PSPACE=\QPSPACE$. 
\begin{claim}\label{Def:prop1}
    Let $\mathcal{C} \in \{\BQP,\PSPACE,\QIP,\QMA,\QCMA,\QSZKhv\}$. Then,
    \begin{equation*}
        \cC \subseteq \textbf{p}\cC \subseteq \textbf{m}\cC
    \end{equation*}
    Also, for any oracle $\cO$\footnote{\label{footnote:oracle} $\cO$ can be a unitary operator, a function, a CPTP map, a quantum promise problem, or a quantum promise class.},
    \begin{equation*}
        \cC^\cO \subseteq \textbf{p}\cC^\cO \subseteq \textbf{m}\cC^\cO
    \end{equation*}
\end{claim}
\begin{proof}
    Suppose $\mathcal{L} \in \mathcal{C}$. Let the algorithms $\mathcal{A}_1$ and $\mathcal{A}_2$ be an interactive protocol that decides $\mathcal{L}$ (If $\mathcal{C}$ is defined with a single party, then there is no $\mathcal{A}_2$). Then $\mathcal{A}_1$ and $\mathcal{A}_2$ are also a $\textbf{p}\mathcal{C}$ protocol that decides $\mathcal{L}$. The proof is the same for $ \textbf{p}\mathcal{C} \subseteq  \textbf{m}\mathcal{C}$ and the oracle version.
\end{proof}

\begin{claim}\label{Def:prop2}
    Let $\mathcal{C} \in \{\BQP,\PSPACE,\QIP,\QMA,\QCMA,\QSZKhv\}$. Then,
    \begin{equation*}
    \begin{cases}
        \mathrm{if} \; \cL \in \textbf{m}\cC, \; \mathrm{then} \; \cL \cap \cH(\N) \in \textbf{p}\cC \\
        \mathrm{if} \; \cL \in \textbf{p}\cC, \; \mathrm{then} \; \cL \cap \{0,1\}^* \in \cC,
    \end{cases}
    \end{equation*}
    where $\cL \cap \cH(\N)$ refers to the subset of $\cL$ that eliminates mixed-state, and $\cL \cap \{0,1\}^*$ is defined similarly. Additionally, for any oracle $\cO$\fnref{footnote:oracle},
    \begin{equation*}
    \begin{cases}
        \mathrm{if} \; \cL \in \textbf{m}\cC^\cO, \; \mathrm{then} \; \cL \cap \cH(\N) \in \textbf{p}\cC^\cO \\
        \mathrm{if} \; \cL \in \textbf{p}\cC^\cO, \; \mathrm{then} \; \cL \cap \{0,1\}^* \in \cC^\cO.
    \end{cases}
    \end{equation*}
    
\end{claim}
\begin{proof}
    By definition, $\cL \in \textbf{m}\cC$ implies $\cL \cap \cH(\N) \in \textbf{p}\cC$ and $\cL \in \textbf{p}\cC$ implies $\cL \cap \{0,1\}^* \in \textbf{p}\cC$. Next, we want to prove that
    \begin{equation*}
        \mathcal{L} \cap \{0,1\}^* \in \textbf{p}\mathcal{C} \text{ implies } \mathcal{L} \cap \{0,1\}^* \in \mathcal{C}.
    \end{equation*}
    Let the algorithms $\mathcal{A}_1$ and $\mathcal{A}_2$ be an interactive protocol of $\textbf{p}\mathcal{C}$ that decides $\mathcal{L}$ (If $\textbf{p}\mathcal{C}$ is defined with single party, then there is no $\mathcal{A}_2$). Then, the new algorithm $\mathcal{A}_{i^*}$ works as follows: Simulate the original algorithm $\mathcal{A}_{i}$. If $\mathcal{A}_{i}$ uses new inputs, copy the same number of (classical) inputs and use them for the remaining computation. Then, $\mathcal{A}_{i^*}$ is a $\mathcal{C}$ algorithm that decides $\mathcal{L} \cap \{0,1\}^*$. The proof is the same for the oracle version.
\end{proof}

\begin{claim}[Oracle separation]\label{Def:prop3}
    Let $\mathcal{C}_1, \mathcal{C}_2 \in \{\BQP,\PSPACE,\QIP,\QMA,\QCMA,\QSZKhv\}$ be two classical complexity classes.
    Let $\mathcal{O}$ be a (quantum) oracle such that $\mathcal{C}_1^\mathcal{O} \nsubseteq$ $\mathcal{C}_2^\mathcal{O}$. Then, $\textbf{p}\mathcal{C}_1^\mathcal{O} \nsubseteq \textbf{p}\mathcal{C}_2^\mathcal{O}$ and $\textbf{m}\mathcal{C}_1^\mathcal{O} \nsubseteq \textbf{m}\mathcal{C}_2^\mathcal{O}$.\fnref{footnote:oracle}
\end{claim}
\begin{proof}
    Let $\mathcal{L}$ be a classical language such that $\mathcal{L} \in \mathcal{C}_1^\mathcal{O}$ but $\mathcal{L} \notin \mathcal{C}_2^\mathcal{O}$. By Claim \ref{Def:prop1}, we have $\mathcal{L} \in \textbf{p}\mathcal{C}_1^\mathcal{O}$. Suppose (by contradiction) that $\mathcal{L} \in \textbf{p}\mathcal{C}_2^\mathcal{O}$. By Claim $\ref{Def:prop2}$, we have $\mathcal{L} = \mathcal{L} \cap \{0,1\}^*   \in \mathcal{C}_2^\mathcal{O}$. Hence, $\textbf{p}\mathcal{C}_1^\mathcal{O} \nsubseteq \textbf{p}\mathcal{C}_2^\mathcal{O}$. The proof for $\textbf{m}\mathcal{C}_1^\mathcal{O} \nsubseteq \textbf{m}\mathcal{C}_2^\mathcal{O}$ is similarly.
\end{proof}

%% file: 3-3_Def_reduction.tex
\subsection{Reduction and amplification}\label{Def:Reduction}
Reduction and amplification are natural in classical complexity. We first introduce reduction, hardness, and completeness as follows.
\begin{definition}[Karp reduction with quantum polynomial Time and Polynomial Copies of Input] Let $\mathcal{L}^1$ and $\mathcal{L}^2$ be two quantum promise problems. We say that $\mathcal{L}^1$ (Karp) reduces to $\mathcal{L}^2$ with quantum polynomial time and polynomial copies,  if there exists a $P$-uniform quantum circuit family $\{\mathcal{V_\lambda}\}_{\lambda \in \N}$ such that the following holds:
\begin{itemize}
    \item There exists polynomials $t(\cdot)$ and $p(\cdot)$ such that $V_\lambda$ takes $t(\lambda) \cdot \lambda + p(\lambda)$ qubits.
    \item For all $\rho \in \mathcal{L}^1_Y$, $\mathcal{V}_{|\rho|}\big(\rho^{\otimes t(\lambda)} \otimes \ket{0}^{p(\lambda)}\big) \in \mathcal{L}^2_Y$.
    \item For all $\rho \in \mathcal{L}^1_N$, $\mathcal{V}_{|\rho|}\big(\rho^{\otimes t(\lambda)} \otimes \ket{0}^{p(\lambda)}\big) \in \mathcal{L}^2_N$.
\end{itemize}
We denote it as $\mathcal{L}^1 \leq_p \mathcal{L}^2$.
\end{definition}

\begin{definition}[Hardness and Completeness of Quantum Complexity]\label{Def:h/c}
    Given a quantum promise complexity class $\mathcal{C}$ and a quantum promise problem $\mathcal{L}$. We say that $\mathcal{L}$ is $\mathcal{C}$-hard if for all $\mathcal{L}' \in \mathcal{C}$, $\mathcal{L}' \leq_p \mathcal{L}$. We say that $\mathcal{L}$ is $\mathcal{C}$-complete if $\mathcal{L}$ is $\mathcal{C}$-hard and $\mathcal{L} \in \mathcal{C}$.
\end{definition}

\begin{remark}
    The Definition \ref{Def:h/c} is reasonable only for complexity classes that potentially require more computational resources than $\pmBQP$. Indeed, a proper definition of hardness and completeness is that the reduction has less computational power than the class itself.
\end{remark}

Turing reduction can also be defined. Given two quantum promise problems, $\cL^1$ and $\cL^2$. Besides, we view $\cL^2$ as a quantum promise oracle $\cO$. We call $\mathcal{L}^1$ is polynomial-time Turing reducible to $\mathcal{L}^2$ if there exists a $QPT$ oracle algorithm $\mathcal{A}^\mathcal{O}$ that decides $\mathcal{L}^1$.

Next, we define a quantum promise complexity class as an oracle. Informally, suppose we view a quantum promise complexity class $\cC_2$ as an oracle; an algorithm $\cA^{\cC_2}$ can access any oracle $\cO$ such that $\cO \in \cC_2$. Furthermore, if $\cC_2$ has a complete problem $\cO^*$, then without loss of generality, we can assume $\cA$ only accesses the single oracle $\cO^*$. The following formally defines the case where $\cC_2$ has a complete problem.
\begin{definition}[Quantum promise complexity class as an oracle]\label{Def:oracle_class} Let $\cC_1$ and $\cC_2$ be two quantum promise complexity classes such that $\cC_2$ has a complete language with respect to Karp reduction. Let $\cC_1^{\cC_2}$ be the collection of quantum promise problems $\cL$ such that the following holds: there exists a $\cC_1$ algorithm $\cA$, and there exists an $\cC_2$-complete oracle $\cO$ such that $\cA^\cO$ decides $\cL$.  
\end{definition}

\begin{example}
    $\pBQP^\pQCMA \subseteq \pBQP^\pQMA$.
\end{example}
\begin{proof}
    By Definition \ref{Def:query_oracle4}, we need to check whether $\pQCMA$ or $\pQMA$ are physically realizable. Theorem $\ref{thm:QMA_Upperbound}$ gives an affirmative answer.  
    Let $\cL \in \pBQP^\pQCMA$. Then, there exists a $\pBQP$ algorithm $\cA$, and there exists a quantum promise oracle $\cO \in \pQCMA \subseteq \pQMA$ such that $\cA^\cO$ decides $\cL$. Hence, $\cL \in \pBQP^\pQMA$. 
\end{proof}

Next, we extend the classical amplification statement to the quantum promise complexity class.

\begin{lemma}[Amplification lemma] \label{lemma:amplification}
\quad\\
Let $\mathcal{C} \in \{\BQP,\PSPACE,\QIP,\QMA,\QCMA,\QSZKhv\}$. The following are equivalent:
    \begin{enumerate}
        \item $\mathcal{L} \in \pmclass\mathcal{C}_{a,b}$, with some $a,b$, and polynomial $p(\cdot)$ such that $a-b \geq \frac{1}{p(n)}$.

        \item For all exponential $e(\cdot)$, $\mathcal{L} \in \pmclass \mathcal{C}_{1-\frac{1}{e(n)},\frac{1}{e(n)}}$.
    \end{enumerate}
\end{lemma}
We defer the proof to Appendix~\ref{Appendix:amplification}. 

\begin{example}
    $\pBQP^\pBQP = \pBQP$.
\end{example}
\begin{proof}
    Let $\cL \in \pBQP^\pBQP$. Then, there exists a $\pBQP$ algorithm $\cA$, and there exists a quantum promise oracle $\cO \in \pBQP$ such that $\cA^\cO$ decides $\cL$. Since $\cO \in \pBQP$ and by Amplification lemma \ref{lemma:amplification}, there exists a QPT algorithm $\cB$ that instantiates $\cO$. Hence, $\cA^\cB$ decides $\cL$. Clearly, there exists a QPT algorithm, which simulates $\cA^\cB$, and decides $\cL$. Thus, $\cL \in \pBQP$. The other direction is trivial.
\end{proof}

%% file: 3-4_Def_condition.tex
%\subsection{Physically realizable with Polynomial Copies}
\subsection{Condition for a physically realizable quantum promise problem}\nai{Move to appendix?}
\label{sec:physical_realizable}
This section explores the distance condition required for an algorithm to solve a quantum promise problem with a polynomial number of input copies. Let $\cL = (\cL_Y,\cL_N)$ be a quantum promise problem. Suppose a polynomial number of input copies allows us to decide $\cL$. In that case, it is evident that the distance between $\cL_Y$ and $\cL_N$ must be at least inverse polynomial, as no algorithm can distinguish between negligibly close states when equipped with polynomially many copies. A natural question arises: is this condition sufficient? We answer negatively in Theorem \ref{def:condition}. Specifically, we show something even stronger: even if the distance between $\cL_Y$ and $\cL_N$ is almost orthogonal, no algorithm can distinguish them using polynomially many copies. Therefore, the condition that the yes and no instances are almost orthogonal is insufficient for a quantum promise problem to be physically realizable (Definition~\ref{Def:physically_realizable}).

%Let $\cL = (\cL_Y,\cL_N)$ be a quantum promise problem. Given polynomial copies of input, is it possible to distinguish the input with unbounded computational resources? The answer to this question is necessary before using $\cL$ as an oracle. Besides, this question is also asked in the literature of property testing \cite{montanaro2013survey}. 

%Additionally, consider complexity classes defined by two parties. By definition, the distance between $\cL_Y$ and $\cL_N$ is at least inverse polynomial far. Hence, it is intriguing to ask whether these classes are physically realizable. Namely,
%\begin{center}
%    \textit{Are $\pmQCMA$, $\pmQMA$, and $\pmQIP$ decidable with unbounded computational resources but polynomial copies of input?}
%\end{center}
%Section \ref{qma:section_upper} gives an affirmative answer to the first two classes, while Section \ref{sec:mQIP_not_in_mPSPACE} and \ref{sec:pQIP_not_in_pPSPACE} provide a negative answer to the last one.

\begin{theorem}\label{def:condition}
    There exists a pure state quantum promise problem $\cL = (\cL_Y,\cL_N)$ such that the following holds.
    \begin{itemize}
        \item For all $\lambda$ and for all $\ket{\psi}, \ket{\phi} \in \big(\cL_Y \cup \cL_N \big) \cap \cH(\lambda)$, $\ket{\psi} \neq \ket{\phi}$ implies $|\braket{\psi| \phi}|^2 < \frac{1}{\sqrt{2^\lambda}}$.
        \item No CPTP map $\cM:=\{\cM_\lambda\}_\lambda$ can decide $\cL$ with a polynomial number of input copies.
    \end{itemize}
\end{theorem}
\begin{proof}
    For sufficiently large $\lambda$, we will choose $2^{2^{\frac{\lambda}{3}}}$ quantum states of size $\lambda$ that satisfy the following condition: For all distinct quantum states $\ket{\psi}$ and $\ket{\phi}$ that we chose, $|\braket{\psi|\phi}|^2 < \frac{1}{\sqrt{2^\lambda}}$. Namely, we can choose a doubly exponential number of quantum states that are pairwise almost orthogonal, which is counterintuitive given that a basis only contains an exponential many vectors. If we randomly choose $2^{2^{\frac{\lambda}{3}}}$ quantum states (with respect to the  Haar measure), then these states are pairwise almost orthogonal with at least constant probability. Indeed, this is proven by the following inequalities.
    \begin{center}
    \begin{math}
    \begin{aligned}
        &\quad\hspace{0.25em}
        \Pr\left[\text{Any two quantum states } \ket{\psi} \text{ and } \ket{\phi} \text{, we have } |\braket{\psi|\phi}|^2 < \frac{1}{\sqrt{2^\lambda}}\right] \\
        &\geq 
        \left[1-e^{-\sqrt{2^\lambda}}\right] \cdot 
        \left[1-2\cdot e^{-\sqrt{2^\lambda}}\right]\cdots 
        \left[1-\left(2^{2^{\frac{\lambda}{3}}}-1\right)\cdot e^{-\sqrt{2^\lambda}}\right] \\
        &\geq \big(1-2^{2^{\frac{\lambda}{2}-1}}\cdot 2^{-\sqrt{2^\lambda}}\big)^{2^{2^{\frac{\lambda}{2}-1}}} \\
        &= \big(1-2^{-2^{\frac{\lambda}{2}-1}}\big)^{2^{2^{\frac{\lambda}{2}-1}}} \\
        &\geq 0.3
    \end{aligned}
    \end{math}
    \end{center}
    The first inequality follows from Lemma \ref{lemma:Haar} and the union bound. The second inequality follows from $2^{2^{\frac{\lambda}{2}-1}} \geq 2^{2^{\frac{\lambda}{3}}}-1$ for sufficiently large $\lambda$. The last inequality follows from the fact that $\lim\limits_{x \rightarrow \infty}(1-\frac{1}{x})^x = e^{-1} \geq 0.36$, and we only consider sufficiently large $\lambda$.

    We will define a language $\cL$ as follows and show that it is undecidable with polynomial copies. For sufficiently large security parameters $\lambda$, each quantum state (from those we choose) with size $\lambda$ can belong to a yes or no instance. Hence, we can define $n_\lambda := 2 \,\mbox{\textasciicircum}\, 2 \,\mbox{\textasciicircum}\, 2 \,\mbox{\textasciicircum}\,\frac{\lambda}{3}$ many languages $\cL^\lambda_1,\cL^\lambda_2,\cdots, \cL^\lambda_{n_\lambda}$. Let $t^\lambda_i$ be minimal copies of input that (some CPTP maps) could decide $L^\lambda_i$ with completeness $1-4^{-\lambda}$ and soundness $4^{-\lambda}$. Let $\mathsf{maxIdx} := \mathsf{ArgMax}\{t^\lambda_1,\dots,t^\lambda_{n_\lambda}\}$. Finally, we Let $\cL := \bigcup\limits_{\lambda}L^\lambda_\mathsf{maxIdx}$.

    Suppose (by contradiction) that there exists a family of CPTP $\cM:=\{\cM_\lambda\}_\lambda$ that instantiates $\cL$ with some polynomial copies $t(\cdot)$, with completeness $1-4^{-\lambda}$ and soundness $4^{-\lambda}$. Let us fixed an order of the $2^{2^{\frac{\lambda}{3}}}$ chosen quantum pure states. Then, there exists a family of CPTP map $\cM^*:=\{\cM^*_\lambda\}_\lambda$ that decides the following problem with completeness $1-4^{-\lambda}$ and soundness $4^{-\lambda}$.
    \begin{itemize} 
        \item \textbf{Inputs: }Given $t(\lambda)$ copies of a chosen input $\ket{\psi} \in \cH(\lambda)$ and a string $s \in \{0,1\}^{2 \,\mbox{\textasciicircum}\, 2 \,\mbox{\textasciicircum}\, \frac{\lambda}{3}}$.
        \item \textbf{Yes instances: } $\ket{\psi}$ belongs to the $i$-th quantum state and $s_i:=1$. 
        \item \textbf{No instances: } $\ket{\psi}$ belongs to $i$-th quantum state and $s_i:=0$.
    \end{itemize}
    Indeed, for all $\lambda \in \N$ and $i \in [n_\lambda]$, there exists some CPTP map that decides $\cL^\lambda_i$ within $t(\lambda)$ copies of the input. Hence, we can define $\cM^*$ as follows: On input $\ket{\psi}\in\cH(\lambda)$ and $s$, it runs the CPTP map that decides $\cL_s^\lambda$ on $\ket{\psi}^{\otimes t(\lambda)}$. Then, we can construct $U_\lambda$ from $\cM^*_\lambda$ by introducing ancilla registers such that $U_\lambda$ produces the same output distribution as $\cM^*_\lambda$.
    %Indeed, the ancilla of $U_\lambda$ is fixed (or bounded) since $\cM$ will halt on all input, and the basis size is finite.

    Next, we will derive a contradiction using Holevo's theorem, which provides a lower bound on communication complexity. Fix a security parameter $\lambda$. Suppose Alice wants to send $2^{\frac{\lambda}{3}}$ classical bits of information to Bob. It is sufficient for Alice to send $t(\lambda)\cdot \lambda + log\, \lambda$
    qubits, leading to a contradiction. Indeed, suppose Alice wants to send a number $i \in [2 \,\mbox{\textasciicircum}\, 2 \,\mbox{\textasciicircum}\, \frac{\lambda}{3}]$ to Bob. Alice finds the $i$-th quantum state $\ket{\psi}$ with size $\lambda$ and sends $(\lambda, \ket{\psi}^{\otimes t(\lambda)})$ to Bob. Bob then runs the binary search algorithm as follows:

    \begin{algorithm}[H]
        \caption{\bf Binary Search}
        \begin{algorithmic}[1]
            \REQUIRE $\lambda$ and  $\ket{\psi}^{\otimes t(\lambda)}$
            \ENSURE
            Find the order of $\ket{\psi}$
            \STATE
            Let $\mathsf{start} := 1$, $\mathsf{end} := 2 \,\mbox{\textasciicircum}\, 2 \,\mbox{\textasciicircum}\, \frac{\lambda}{3}$, and $\mathsf{mid}:=\frac{1}{2}(\mathsf{start}+\mathsf{end}-1)$.
            \STATE
            Let $s:= 0^{mid} \| 1^{mid}$.
            
            \FOR{$i = 1$ to $2^{\frac{\lambda}{3}}$} 
                \STATE
                Run $U_\lambda$ on input $\big(\lambda, \ket{\psi}^{\otimes t(\lambda)}, s \big)$ with some ancilla $\ket{0^*}$, measure the answer register to get the answer 
                $ans$, and run $U_\lambda^\dag$ to rewind the state.
                
                \IF{$ans = 1$}
                    \STATE
                    $\mathsf{start} = \mathsf{mid} + 1$.
                \ELSE
                    \STATE
                    $\mathsf{end} = \mathsf{mid}$.
                \ENDIF
                \IF{$\mathsf{start} = \mathsf{end}$}
                    \STATE
                    \textbf{break}
                \ENDIF
                \STATE
                $\mathsf{mid}:=\frac{1}{2}(\mathsf{start}+\mathsf{end}-1)$.
                \STATE
                Substitute the substring $s_{\mathsf{start},\, \mathsf{mid}}$ to all 0 and the substring $s_{\mathsf{mid}+1,\, \mathsf{end}}$ to all 1.
            \ENDFOR
            \RETURN{$\mathsf{start}$}
        \end{algorithmic}
    \end{algorithm}
    The correctness is evident if all the measurements in step 4 give the correct answer. It remains to be proven that all measurements are correct with overwhelming probability. We can view the algorithm as performing sequential measurements on the input $(\lambda, \ket{\psi}^{\otimes t(\lambda)})$ with ancillas $\ket{0^*}$. Note that the string $s$ is part of the definition of the sequential measurements. By the quantum union bound $\ref{pre:qub}$, the algorithm returns the correct order of $\ket{\psi}$ with probability at least $1-4 \cdot 2^{\frac{\lambda}{3}} \cdot 4^{-\lambda}= 1-negl(\lambda)$. One might think the measurements are chosen adaptively; however, we can fix the sequential measurements needed when analyzing the algorithm.
\end{proof}

%% file: 4_qma.tex
\section{Structural results for p/mQ(C)MA} 
\label{section:non_interactive}
%
%\lch{add a few sentences to summarize the whole section.}
This section develops the basic structural theory of $\pmQMA$ and $\pmQCMA$: which problems are complete for them, how much power they have, and whether a witness can be found rather than merely verified.

Section~\ref{section:qma} identifies complete problems. In Section \ref{section:QOP_qma}, we demonstrate that the
famous Quantum OR problems are also complete for pure-state and mixed-state Q(C)MA-complete problems. These problems arise in a range of quantum-input settings, 
including shadow tomography \cite{Aar20}, black-box separation of quantum cryptographic primitives \cite{chen2024power}, and quantum property testing \cite{harrow2017sequential}. 
In Sections~\ref{qma:section_p} and~\ref{qcma:section} we introduce quantum-input
analogues of the local Hamiltonian problem \cite{kitaev2002classical} and show that they are complete for $\pmQMA$ and $\pmQCMA$, respectively.

% In section \ref{section:qma}, we show that variants of the local Hamiltonian problem are complete for pure-state and mixed-state Q(C)MA-complete problems. In section \ref{section:QOP_qma}, we demonstrate that the
% famous Quantum OR problems are also complete for pure-state and mixed-state Q(C)MA-complete problems. The Quantum OR problems have various applications in quantum input problems, such as shadow tomography \cite{Aar20}, black-box separation of quantum cryptographic primitives \cite{chen2024power}, and quantum property testing \cite{harrow2017sequential}. 

In Section \ref{qma:section_upper}, we show that $\pmQMA \subseteq \pmPSPACE$, meaning that even if the non-interactive prover receives an unbounded number of input copies, it is no more powerful than a single party algorithm that uses only a polynomial number of input copies. Finally, Section \ref{section:StoD} gives a search-to-decision reduction for $\pmQCMA$-complete problems, mirroring the classical result.
%\nai{What is the main theorem of this section?}
%\nai{Write the problem definitions at the beginning of the section}
%\nai{Write the formal theorem statements for the subsections.}
%\nai{Each subsection is for the proof of the statement}

\input{4-1_qma_complete_problems}

\input{4-4_qma_ubound}

\input{4-5_qcma_StoDreduction}

%% file: 4-1_qma_complete_problems.tex
\subsection{Complete problems in pure- and mixed-state \class{Q(C)MA}}\label{section:qma}
% \lch{combine Hamiltonian and OR; bring OR up first}

% \lch{state the problem definitions + main theorems}

Throughout this section we work with four classes, according to whether the quantum input is a pure or a mixed state and whether the prover's message is a quantum state or a classical string: $\pQMA$, $\mQMA$, $\pQCMA$ and $\mQCMA$. 

\input{4-3_O_qcma_and_qma}

\input{4-1-1_H_qma_complete}
\input{4-1-2_H_qcma_complete}

%% file: 4-3_O_qcma_and_qma.tex
\subsubsection{Quantum OR Problems as \pmQCMMA-complete}\label{section:QOP_qma}

\nai{Need to be polish.}\HZZ{I changed the definition (include $1^m$ in input). Need to check}

%The following problems in Definition $\ref{qcma:qob}$ and $\ref{qma:qob}$ originate from \ref{?}, which .
%\nai{What is the question mark?}
In this section, we first introduce a series of Quantum OR problems and then show that they are complete for $\pmQMA$ and $\pmQCMA$.  

\begin{definition}[Pure- and Mixed-State Standard-Basis Quantum OR Problem (\textbf{pSQO} and \textbf{mSQO})]
\label{qcma:qob}
The Pure-State Standard-Basis Quantum OR Problem (\textbf{pSQO}) is the following promise problem.
    \begin{itemize} 
        % \item \textbf{Inputs: }Given copies of an $n$-qubit input pure state $\ket{\psi}$, $1^m$ with $m \in \N$, and a $2$-outcome efficient projective measurement $\Lambda$ on $\mathscr{H}_A \otimes \mathscr{H}_B$, where $dim\,\mathscr{H}_A = 2^n$ and $dim\,\mathscr{H}_B = 2^m$. The projective measurement $\Lambda$ consists of a sequence of quantum circuits, followed by a single-bit measurement and then the conjugate of the preceding circuit sequence.\footnote{\label{footnote:QOB} The circuit sequence of $\Lambda$ cannot be succinctly described and should instead be presented in a gate-by-gate format.}
        \item \textbf{Inputs:} An instance is specified by a classical description together with copies of an unknown state.
        \begin{itemize}
            \item \textit{Classical part:} $1^m$ with $m\in \N$, and a description of a two-outcome efficient projective measurement $\Lambda$ on $\cH_A \tp \cH_B$, where $\dim(\cH_A) = 2^n$ and $\dim(\cH_B) = 2^m$. 
            The measurement $\Lambda$ is given as a sequence of quantum circuits,\footnote{\label{footnote:QOB} The circuit sequence of $\Lambda$ is not assumed to admit a succinct description and is thus presented explicitly, gate by gate.} followed by a single-qubit measurement, followed by the inverse of that circuit sequence.
            \item \textit{Quantum part:} $\poly(n)$ copies of an unknown pure state $\ket{\psi}\in \cH_A$.
        \end{itemize}
        \item \textbf{Yes instances: } There exists $i \in \{0,1\}^m$ such that 
        $\Tr(\Lambda \,\ketbra{\psi} \otimes \ketbra{i}) \geq \frac{2}{3}$.
        \item \textbf{No instances: } For all $i \in \{0,1\}^m$, 
        $ \Tr(\Lambda \,\ketbra{\psi}\otimes \ketbra{i}) \leq \frac{1}{64\cdot2^m}$.
    \end{itemize}
    
The Mixed-State Standard-Basis Quantum OR Problem (\textbf{mSQO}) is defined in exactly the same way, except that the quantum part of the input consists of $\poly(n)$ copies of an unknown mixed state $\rho \in \cD(\cH_{A})$, and $\ketbra{\psi}$ is replaced by $\rho$ throughout.
\end{definition}

%\begin{itemize}
%        \item Promised input:
%            \begin{itemize}
%                \item An unknown quantum state $\ket{\psi} \in \cH_A$.
%%                \item $N$, the dimension of Hilbert space $\cH_B$.
%                \item $\Lambda$, a 2-outcome efficient projective measurement on a bipartite space $\cH_A \otimes \cH_B$. That is, $\Lambda$ is a sequence of polynomial circuits, followed by a single-bit measurement and the conjugate of the previous sequence of circuits.
%            \end{itemize}
            
%        \item Output:
%            \begin{itemize}
%                \item Accept if there exists $i \in [N-1]_0$ such that $\Tr(\Lambda \,\ket{\psi}\bra{\psi} \otimes \ket{i}\bra{i}) \geq \frac{2}{3}$.
%                \item Reject if for all $i \in [N-1]_0$ such that $ \Tr(\Lambda \,\ket{\psi}\bra{\psi}\otimes \ket{i}\bra{i}) \leq \frac{1}{63N}$.
%            \end{itemize}
%    \end{itemize}

\begin{definition}[Pure- and Mixed-State Quantum OR Problem (\textbf{pQO} and \textbf{mQO})]
\label{qma:qob}
The Pure-State Quantum OR Problem (\textbf{pQO}) is the following promise problem.
    \begin{itemize} 
        % \item \textbf{Inputs: }Given copies of an $n$-qubit input pure state $\ket{\psi}$, $1^m$ with $m \in \N$, and a $2$-outcome efficient projective measurement $\Lambda$ on $\mathscr{H}_A \otimes \mathscr{H}_B$, where $dim\,\mathscr{H}_A = 2^n$ and $dim\,\mathscr{H}_B = 2^m$. The projective measurement $\Lambda$ consists of a sequence of quantum circuits, followed by a single-bit measurement and then the conjugate of the preceding circuit sequence.\fnref{footnote:QOB}
        \item \textbf{Inputs:} An instance is specified by a classical description together with copies of an unknown state.
        \begin{itemize}
            \item \textit{Classical part:} $1^m$ with $m\in \N$, and a description of a two-outcome efficient projective measurement $\Lambda$ on $\cH_A \tp \cH_B$, where $\dim(\cH_A) = 2^n$ and $\dim(\cH_B) = 2^m$. 
            The measurement $\Lambda$ is given as a sequence of quantum circuits,\fnref{footnote:QOB} followed by a single-qubit measurement, followed by the inverse of that circuit sequence.
            \item \textit{Quantum part:} $\poly(n)$ copies of an unknown pure state $\ket{\psi}\in \cH_A$.
        \end{itemize}
         \item \textbf{Yes instances: } There exists $\sigma \in \cH_B$ such that $\Tr(\Lambda \,\ketbra{\psi} \otimes \sigma) \geq \frac{2}{3}$.
        \item \textbf{No instances: } For all $\sigma \in \cH_B$, $ \Tr(\Lambda\,\ketbra{\psi} \otimes \sigma ) \leq \frac{1}{64\cdot 2^m}$.
    \end{itemize}

    The Mixed-State Quantum OR Problem ($\textbf{mQO}$) is defined in exactly the same way, except that the quantum part of the input consists of $\poly(n)$ copies of an unknown mixed state $\rho\in \cD(\cH_A)$, and $\ketbra{\psi}$ is replaced by $\rho$ throughout. 
\end{definition}

%\begin{definition}\label{qma:qob}
%    \textbf{(Quantum Or Problem)} 
%    \newline\newline
%    The pure state quantum or problem \textbf{pQO} is defined as follows.
%    \begin{itemize}
%        \item Promised input:
%            \begin{itemize}
%                \item An unknown quantum state $\ket{\psi} \in \cH_A$.
%%                \item $N$, the dimension of Hilbert space $\cH_B$.
%                \item $\Lambda$, a 2-outcome efficient projective measurement on a bipartite space $\cH_A \otimes \cH_B$. That is, $\Lambda$ is a sequence of polynomial circuits, followed by a single-bit measurement and the conjugate of the previous sequence of circuits.
%            \end{itemize}
            
%        \item Output:
%            \begin{itemize}
%                \item Accept if there exists $\sigma \in \cH_B$ such that $\Tr(\Lambda \,\ket{\psi}\bra{\psi} \otimes \sigma) \geq \frac{2}{3}$.
%                \item Reject if for all $\sigma \in \cH_B$ such that $ \Tr(\Lambda\,\ket{\psi}\bra{\psi} \otimes \sigma ) \leq \frac{1}{63N}$.
%            \end{itemize}
%    \end{itemize}
%    We also define mixed-state quantum or problem, denoted as $\textbf{mQO}$. The problem definition is the same as $\textbf{pQO}$ except that all unknown pure states $\ket{\psi}$ or $\ket{\psi}\bra{\psi}$ are replaced with mixed-state $\rho$.
%\end{definition}
%We are able to show that 
\begin{theorem}
\label{thm:QOB_complete}
Quantum OR problems are complete for $\pmQMA$ and $\pmQCMA$. In particular, 
\begin{itemize}    
    \item \textbf{pSQO} is $\pQCMA$-complete.
    \item \textbf{mSQO} is $\mQCMA$-complete.
    \item \textbf{pQO} is $\pQMA$-complete.
    \item \textbf{mQO} is $\mQMA$-complete.
\end{itemize}
\end{theorem}
\begin{proof}
   The four problems and their corresponding proofs have the same structure. Thus, we only prove the last statement; the other three are analogous. 
   
   \paragraph*{Membership.}
   We first show that $\textbf{mQO} \in \mQMA$. On a yes instance, the prover sends a state $\sigma \in \cH_B$ such that $\Tr(\Lambda(\rho \otimes \sigma)) \geq \frac{2}{3}$. The verifier performs the two-outcome projective measurement $\Lambda$ to $\rho \otimes \sigma$ and outputs the measurement outcome. Completeness is immediate. On a no instance, every witness is accepted with probability at most $\frac{1}{64\cdot 2^m}\leq \frac13$, so soundness holds as well. The verifier runs in time linear in the input size.

   \paragraph*{Hardness.}
    % Next, we show that $\textbf{mQO} \in \mQMA\text{-hard}$. Suppose $\mathcal{L} \in \mQMA_{1-2^{-n}, 2^{-n}}$; there exists a polynomial $p(\cdot)$ such that $(\mathcal{P},\mathcal{V})$ is a $\textbf{mQMA}$ protocol that decides $\mathcal{L}$ with completeness at least $1-2^{-n}$, soundness at most $2^{-n}$, witness size at most $w(n)$, and use $t(n)$ copies of the input. For any function $c(n):\N \rightarrow \N$, we will amplify the soundness of $\cV$ to $2^{-c(n)\cdot n}$ while preserving the witness size. The algorithm is similar to that in Lemma 15 of \cite{aaronson2006qma}, but the proof is shown below for clarity and completeness. We simplify the notation by letting $c(n)$ be denoted as $c$ and $t(n)$ as $t$. For $i \in [c]$, define the unitary $\mathcal{V}_i$ as equivalent to $\mathcal{V}$, but acting on the registers $I_iA_iW$. To get the answer, we measure the register $A^{ans}_i$, the first qubit of register $A_i$. We define $\mathcal{V}'$ as follow:

    Next, we show that $\textbf{mQO} \in \mQMA\text{-hard}$. Let $\mathcal{L} \in \mQMA_{1-2^{-n}, 2^{-n}}$, and let $(\cP, \cV)$ be an $\textbf{mQMA}$ protocol that decides $\cL$ with completeness at least $1-2^{-n}$, soundness at most $2^{-n}$, where $\cV$ uses $t(n)$ copies of the input, a witness of size at most $w(n)$, and $p(n)$ ancilla qubits, for polynomial-time computable functions $t(\cdot), w(\cdot)$ and $p(\cdot)$.
    For any function $c: \N \to \N$, we first amplify the soundness of $\cV$ to $2^{-c(n)\cdot n}$ while preserving the witness size. The algorithm follows Lemma 15 of \cite{aaronson2006qma}; we reproduce the argument here for clarity and completeness. To simplify notation, write $c$ for $c(n)$ and $t$ for $t(n)$. 
    For $i\in [c]$, let $\cV_i$ denote the unitary $\cV$ acting on the registers $I_i A_i W$, and let $\aans$ denote the first qubit of $A_i$, which carries the answer. 
    Write $\Pi_i \coloneqq \ketbra{1}_{\aans} \tp I$ for the projector onto acceptance in round $i$.
    We define $\cV'$ as follows.
    
    \begin{algorithm}[H]
    \caption{\bf QPT $\mathcal{V}'$}
    \begin{algorithmic}[1]
        \REQUIRE $\rho ^{\otimes ct},\sigma$
        \ENSURE
        Amplify soundness error
        \STATE
        Initialize the register to $\rho^{\otimes t}_{I_1}\otimes\cdots\otimes \rho^{\otimes t}_{I_c} \otimes |0\rangle_{A_1}\otimes\cdots\otimes|0\rangle_{A_c} \otimes \sigma_W $
        \FOR{$i\in[c]$}
            \STATE
            Apply $\mathcal{V}_i$ on register $I_iA_iW$.
            \STATE
            Measure register $\aans$ and get outcome $\ans_i$.
            \IF{$\ans_i=0$}
                \RETURN{Reject.}
            \ENDIF
            \STATE
            Apply $\mathcal{V}_i^\dagger$ on register $I_iA_iW$. 
        \ENDFOR
        \RETURN{Accept.}
    \end{algorithmic}
    \end{algorithm}
    Here $\rho$ is the problem input and $\sigma$ is a witness from the prover. Note that all $c$ iterations share a single witness register $W$, so the witness size is indeed unchanged.  
    % Yes instance.
    Suppose first that $\rho \in \cL_Y$. For $i\in [c]$, define the two-outcome measurement $E_i \coloneqq \mathcal{V}_i^\dagger \Pi_i \mathcal{V}_i$. By the completeness of $(\cP, \cV)$, each $E_i$ accepts the state $\rho_{I_i}^{\tp t}\tp \ketbra{0}_{A_i} \tp \sigma_W$ with probability at least $1-2^{-n}$, so by the quantum union bound (Lemma~\ref{pre:qub}) the sequential measurements $E_1, \dots, E_c$ accept in every round, and hence $\cV'$ accepts, with probability at least $1-4c\cdot 2^{-n}$. 
    % No instance.
    Now suppose that $\rho\in \cL_N$. Conditioned on the first $i-1$ iterations accepting, Step 3 applies $\cV$ to a state of the form $\rho^{\tp t} \tp \ket{0^*} \tp \sigma'$ for some state $\sigma'$ on $W$, since each iteration uses a fresh input $I_i$ and a fresh ancilla $A_i$. By soundness, $\Pr[\ans_i=1 | \ans_1 = \cdots = \ans_{i-1} = 1]\leq 2^{-n}$ for every $i\in [c]$, and therefore $\Pr[\cV' \text{ accepts}]\leq 2^{-c\cdot n}$.

    The reduction is now immediate. The witness of $\cV'$ has $w(n)$ qubits, so the \textbf{mQO} instance we build has $m\coloneqq w(n)$, and we also take $c\coloneqq w(n)$. 
    Let $\rho\in \cD(\cH(n))$ be an instance of $\cL$. We map it to the \textbf{mQO} instance with input state $\rho^{\tp c\cdot t(n)} \otimes \ketbra{0}^{\tp c\cdot p(n)}$, unary parameter $1^m = 1^{w(n)}$, and measurement $\Lambda \coloneqq \cV'$.\footnote{The circuit $\cV'$ can, in general, be expressed in the form of a projective measurement.}  This map is computable in polynomial time, and for sufficiently large $n$ we have $1-4c\cdot 2^{-n} > \frac{2}{3}$ and $2^{-c\cdot n} < \frac{1}{64}\cdot 2^{-m}$.

    %
    % Suppose $\rho \in \cL_Y$. For all $i\in [c]$, define the two-outcome measurement $E_i:=\mathcal{V}_i^\dagger \big(\ket{1}\bra{1}_{\aans}\otimes I \big ) \mathcal{V}_i$. By the quantum union bound in Lemma~\ref{pre:qub}, and given that for all $i\in [c]$, $\,Tr[E_i\rho] \geq 1-2^{-n}$, we conclude that $\mathcal{V}'$ accepts with probability at least $1-4c2^{-n}$. Now suppose $|\psi\rangle \in \mathcal{L_N}$. Regardless of the round, step 3 always performs the following: Apply $\mathcal{V}$ to $\rho^{\otimes t}\otimes \ket{0^*} \otimes \sigma'$, for some $\sigma'$. Since $\rho \in \mathcal{L_N}$, for all $i\in[c]$, $\Pr[ans_i=1] \leq 2^{-n}$. Therefore, $\Pr[\mathcal{V}' \text{ accepts}] \leq 2^{-c\cdot n}$. Choose $c(n)$ as the witness size $w(n)$, then we conclude that $\mathcal{L} \in \mQMA_{1-4w(n)\cdot2^{-n}, 2^{-w(n)\cdot n}}$. Let $p(n)$ be the size of the ancilla register of $\cV$. Without loss of generality, we assume that $t(\cdot), p(\cdot)$ and $w(\cdot)$ are polynomial computable.

    % The reduction works as follows: Let $\rho \in \cD(\cH(n))$ be the input instance of $\cL$. The promised input of \textbf{mQO} consists of an unknown state $\rho^{w(n)\cdot t(n)} \otimes \ket{0}^{p(n)\cdot t(n)}$, $1^m$ with $ m:=w(n)$, and $\Lambda = \cV'$.\footnote{The circuit $\cV'$ can, in general, be expressed in the form of a projective measurement.} Indeed, for sufficiently large $n$, $1-4w(n)\cdot 2^{-n} > \frac{2}{3}$ and $2^{-w(n)\cdot n} < \frac{1}{64}2^{-w(n)}$.
\end{proof}

%% file: 4-1-1_H_qma_complete.tex
\subsubsection{Varients of Local Hamiltonian Problems as \pmQMA-complete}\label{qma:section_p}
We define the $k$-Local Hamiltonian problem with pure (respectively, mixed) states (Definitions~\ref{qma:p_problem} and~\ref{qma:m_problem}), 
a quantum-input analogue of the Local Hamiltonian problem~\cite{kitaev2002classical}.
We then show that these two problems are complete for the corresponding pure- and mixed-state complexity classes
(Theorems \ref{thm:pqma_complete} and \ref{thm:mqma_complete}).

%We will show that the $k$-local Hamiltonian with an unknown pure state problem ($k$-\textbf{LHwP}, Definition~\ref{qma:p_problem}) is $\pQMA$-complete. The $k$-\textbf{LHwP} serves as an analog to the local Hamiltonian problem~\cite{kitaev2002classical}. The only difference is that the input to this problem additionally contains a term of the form, $\sum\limits_\ell \ket{\psi}\bra{\psi}^{\otimes poly(n)} \otimes H_\ell$. Specifically, the input consists of a summation of two types of Hamiltonian: one type is a set of local Hamiltonians, and the second type is an unknown state $\ket{\psi}\bra{\psi}^{\otimes poly(n)}$ tensored with local Hamiltonians. The second type is used in proving the $\pQMA$-hardness. We define $k$-\textbf{LHwP} as follows. 

%Pure state
\paragraph{Pure state QMA-complete}

%08/24 comment: 1. S and L overlap? 2. can \ket{\psi} be affected by H_l/H_s?
\begin{definition}[$k$-local Hamiltonian with an unknown pure state problem ($k$-LHwP)]
\label{qma:p_problem}
\leavevmode
    \begin{itemize}
        % \item \textbf{Inputs:} Given the input $1^p$ with $p \in \N$, $a\in \R^+_0$, $b\in\R^+_0$, and
        % \begin{equation*}
        %     H:= \sum\limits_sH_s - \sum\limits_\ell\ket{\psi}\bra{\psi}\otimes H_\ell,
        % \end{equation*}
        % where $\ket{\psi}\in\cH(n)$ is an unknown state, $\{H_s\}_{s\in S}$ and $\{H_\ell\}_{\ell \in L}$ are two sets of local Hamiltonians, with each Hamiltonian acting on at most $k$ qubits. We are given the promise that $|S|+|L| \leq p$, $\;0\preceq H_x \preceq I$ for all $x \in S\cup L$, and $b-a > \frac{2}{p}$.
        \item \textbf{Inputs:} An instance is specified by a classical description together with copies of an unknown state.
        \begin{itemize}
            \item \textit{Classical part:} $1^p$ with $p \in \N$, thresholds $a, b\in\R^+_0$, and classical descriptions of two families of local Hamiltonians $\{H_s\}_{s\in S}$ and $\{H_\ell\}_{\ell \in L}$, each acting on at most $k$ qubits.
            \item \textit{Quantum part:} $\poly(n)$ copies of an unknown state $\ket{\psi} \in \cH(n)$.
        \end{itemize}
        Together these specify the Hamiltonian 
        \begin{equation*}
            H:= \sum\limits_{s\in S}H_s - \sum\limits_{\ell\in L}\ketbra{\psi}\otimes H_\ell.
        \end{equation*}
        We are given the promise that $p=\poly(n)$, $|S|+|L| \leq p$, $\, 0\preceq H_x \preceq I$ for all $x \in S\cup L$, and $b-a > \frac{2}{p}$.
        \item \textbf{Yes instances:} $\lambda_{min}(H) \leq a$.
        \item \textbf{No instances:} $\lambda_{min}(H) \geq b$.
    \end{itemize}

\end{definition}
%We only require $\ket{\psi}\bra{\psi}^{\otimes t}$ and $H_\ell$ act on disjoint registers. The registers, where $H_s$ act on, are not important.\nai{Check.}

% Unlike the traditional local Hamiltonian problem, our problem instance consists of two types of Hamiltonian terms. The first type is the standard collection of local Hamiltonians $\{H_s\}$.
% The second type consists of local Hamiltonians $\{H_\ell\}$ tensored with an unknown state $\ket{\psi}\bra{\psi}^{\otimes poly(n)}$. %\lch{poly copies are not shown in def}
%  As will become clear in the completeness proof, terms of the second type are essential for establishing $\pQMA$-hardness.
The Hamiltonian terms involving the unknown state $\ket{\psi}$ distinguish our problem from the traditional local Hamiltonian problem and, as will become clear in the completeness proof, are essential for establishing $\pQMA$-hardness.
 
 %$\sum\limits_\ell \ket{\psi}\bra{\psi}^{\otimes poly(n)} \otimes H_\ell$. Specifically, the input consists of a summation of two types of Hamiltonian: one type is a set of local Hamiltonians, and the second type is an unknown state $\ket{\psi}\bra{\psi}^{\otimes poly(n)}$ tensored with local Hamiltonians. The second type is used in proving the $\pQMA$-hardness. 
 
%We will show that the $k$-local Hamiltonian with an unknown pure state problem ($k$-\textbf{LHwP}, Definition~\ref{qma:p_problem}) is $\pQMA$-complete.

\begin{theorem}\label{thm:pqma_complete}
    5-$\textbf{LHwP}$ is $\pQMA$-complete.
\end{theorem}
\begin{proof}
   The proof follows immediately from membership (Lemma~\ref{qma:p_in}) and hardness (Lemma~\ref{qma:p_hard}).
   %Lemma~\ref{qma:p_in} and Lemma~\ref{qma:p_hard}, where we will show that 5-$\textbf{LHwP}$ is in $\pQMA$ (Lemma~\ref{qma:p_in}) and that 5-$\textbf{LHwP}$ is hard for $\pQMA$ (Lemma~\ref{qma:p_hard}).
\end{proof}

\begin{lemma}\label{qma:p_in}
    For all constant $k \in \N$, $k$-\textbf{LHwP} $\in$ \pQMA.
\end{lemma}
\begin{proof}
    Throughout, $p$ is the parameter of Definition \ref{qma:p_problem}, with $p=\poly(n)$, $m$ denotes the number of qubits on which $H$ acts,
    and $r\coloneqq p^5$ denotes the number of the witness state that the prover supplies, equivalently the number of registers
    the verifier processes.
    On input $\big(1^p,\ket{\psi}^{\tp \poly(n)},a,b, \{H_s\},\{H_\ell\}\big)$ where $p$ now denotes the input size, we construct an efficient verifier. For every ``yes" instance, there exists a witness state $ \ket{\phi}^{\otimes r}$ that causes the verifier to accept with an overwhelming probability, whereas for every ``no'' instance, the verifier accepts with only negligible probability.

    %Let $m$ be the number of qubits on which $H$ acts. The prover provides a witness $\Tilde{\sigma}$ on $mp^5$ qubits. 
    \subparagraph{Verification.}
    %The verifier proceeds as follows: 
    The prover provides a witness $\Tilde{\sigma}$ on $mr$ qubits, and the verifier proceeds as follows.
    % (i) Partition the witness into $p^5$ registers. Let $\sigma_j$ denote the reduced state of the $j$-th register, i.e., $\sigma_j\coloneq\Tr_{\leq m(j-1), > mj}(\Tilde{\sigma})$. Note that the registers may be entangled. 
    % (ii) For each $j\in[p^5]$, run an unbiased estimator for $\frac{1}{|S|+|L|} \Tr(\sigma_j H)$. 
    % (iii) Compute the average of the resulting estimates and multiply it by $(|S| + |L|)$. 
    % (iv) Accept if the resulting estimate is at most $a + \frac{1}{p}$; otherwise, reject.
    \begin{enumerate}[label=(\roman*),leftmargin=2.2em]
        \item Partition $\Tilde{\sigma}$ into $r$ registers of $m$ qubits each, and let $\sigma_j\coloneq\Tr_{\leq m(j-1), > mj}(\Tilde{\sigma})$ be the reduced state of the $j$-th register; the registers may be entangled.
        \item For each $j\in [r]$, run an unbiased estimator for $\frac{1}{|S|+|L|} \Tr(\sigma_j H)$ on the $j$-th register, consuming one copy of $\ket{\psi}$ when the sampled term comes from $L$; 
        write $M_j$ for the resulting estimate and $M\coloneqq \frac1r \sum_j M_j$ for their average.
        \item Accept if $(|S| + |L|)\cdot M \leq a + \frac1p$, and reject otherwise.
    \end{enumerate}
    It remains to specify the estimator in step (ii). 
    By linearity of the trace, sampling $x$ uniformly from $S \cup L$ and estimating the corresponding term
    yields such an estimator: $\Tr(\sigma_j\cdot H_s \otimes I)$ if $x=s\in S$;
    and $-\Tr\left(\sigma_j \cdot \ketbra{\psi} \otimes H_\ell\otimes I) \right)$ if $x=\ell \in L$.
    %
    %the following algorithm can estimate $\frac{1}{|S|+|L|} \cdot \Tr(\sigma_j H)$: the verifier randomly chooses an $x \in S \cup L$. If $s:=x \in S$, we estimate $\Tr(\sigma_j\cdot H_s \otimes I)$. If $\ell:=x \in L$, we estimate the negation of $\Tr\left(\sigma_j \cdot \ket{\psi}\bra{\psi} \otimes H_\ell\otimes I) \right)$. 
    %
    We describe the estimation procedure for each case below, omitting the subscript $j$ for simplicity.

    \emph{Case 1 ($x=s\in S$). Estimating $\Tr\left(\sigma \cdot (H_s)_A \otimes I_B\right)$.} Since $H_s$ acts on at most $k$ qubits, we can efficiently compute an eigenbasis $\{\ket{v_i}\}$ and the corresponding eigenvalues $\{\lambda_i\}$ of $H_s$,
    so that $H_s= \sum\limits_{i}\lambda_i\ketbra{v_i}$. 
    The verifier measures $\sigma$ on register $A$ in the basis $\{\ket{v_i}\}$. If the measurement outcome is $\ket{v_i}$, the estimator is $X \coloneq \lambda_i$.

    \emph{Case 2 ($x=\ell \in L$). Estimating $-\Tr\left(\sigma \cdot (\ketbra{\psi})_I \otimes (H_\ell)_A\otimes I_B \right)$.}
    Compute an eigenbasis $\{\ket{v_i}\}$ and eigenvalues $\{\lambda_i\}$ of $H_\ell$, so that $H_\ell= \sum\limits_{i}\lambda_i \ketbra{v_i}$, and measure $\sigma$ on register $A$ in the basis $\{\ket{v_i}\}$. If the outcome is $\ket{v_i}$, 
    set $Y\coloneqq \lambda_i$. Then apply the partial swap test of Lemma \ref{pre:sqap} to register $I$ of the post-measurement
    state against a fresh copy of $\ket{\psi}$: if the test accepts, set $Z\coloneqq \lambda_i$; otherwise set $Z\coloneqq 0$.
    The estimate is $Y-2Z$.
    
    % This is the case in which the quantum input makes a difference. In Case 1, the operator to be estimated is fully known to the verifier, so it can simply be measured.
    % Here, by contrast, the operator factorizes into a known part $H_l$ and an unknown state $\ketbra{\psi}$, to which the verifier has access only through copies.
    % The known part is handled exactly as in Case 1; the difficulty is confined to the unknown one.
    % The solution is to replace the projective measurement by a swap test against a fresh copy.
    % A swap test is not a projector, thus suffers with probability $\frac12 + \frac12 |\inner{\psi}{\cdot}|^2$,
    % and is therefore biased by the component of the witness orthogonal to $\ket{\psi}$ that can be controlled by a dishonest prover. 
    % We manage to cancel this bias against the measurement of $H_\ell$ itself.

    % Since $H_\ell$ acts on only $k$ qubits, we can efficiently compute an eigenbasis $\{\ket{v_i}\}$ of $H_\ell$ together with the corresponding eigenvalues $\{\lambda_i\}$. Thus,  $H_\ell = \sum\limits_{i}\lambda_i |v_i\rangle\langle v_i|$.  
    % The verifier measures $\sigma$ on register $A$ in the basis $\{\ket{v_i}\}$. If the measurement outcome is $\ket{v_i}$, let $Y \coloneq \lambda_i$. The verifier then applies the partial swap test (Lemma \ref{pre:sqap}) to the state $\ket{\psi}$ on register $I$. If the test accepts (i.e., outputs $0$), let $Z \coloneq \lambda_i$; otherwise, let $Z \coloneq 0$. 
    % Finally, The estimator outputs $Y-2Z$.

    \subparagraph{Completeness.} 
    Suppose $\lambda_{\min} (H) \leq a$ and the prover sends a witness $\ket{\phi}^{\otimes r}$ with $\bra{\phi} H \ket{\phi} \leq a$. We check that each of the two estimators is unbiased.

    %Consider the case where $x=s \in S$ in the verifier's step 2. 
    For $x=s\in S$, write $\ket{\phi} = \sum\limits_{i}\alpha_i\ket{v_i}_A\ket{G_i}_B$, where the $|G_i\rangle$ are arbitrary states on the remaining registers $I$ and $B$, and are irrelevant to our analysis. Then
    \[
        \Tr\big(\ketbra{\phi} \cdot (H_s)_A \otimes I_B\big) = \bra{\phi} H_s \otimes I \ket{\phi} =
        \sum_i |\alpha_i|^2 \lambda_i,
        \qquad
        \Pr[X = \lambda_i] = \sum_{k \,:\, \lambda_k = \lambda_i} |\alpha_k|^2 ,
    \]
    % The witness state $\ket{\phi}$ can be expressed in the eigenbasis ${\ket{v_i}}$ as follows:
    % \begin{equation*}
    %     \ket{\phi} = \sum\limits_{i}\alpha_i\ket{v_i}\ket{G_i},
    % \end{equation*}
    % where $|G_i\rangle$ is an arbitrary state irrelevant to our analysis. Then,
    
    % \begin{equation*}
    % \begin{cases}
    %     \Tr(\ket{\phi}\bra{\phi}\cdot (H_s)_A \otimes I_B) = \bra{\phi} H_s \otimes I \ket{\phi} = \sum\limits_i|\alpha_i|^2\lambda_i \\
    %     Pr[X=\lambda_i]=       \sum\limits_{k:\lambda_k=\lambda_i}|\alpha_k|^2.
    % \end{cases}
    % \end{equation*}
    \noindent and hence $\E[X] = \Tr(\ketbra{\phi}\cdot (H_s)_A \otimes I_B)$.

    For $x=\ell \in L$, write
    \begin{equation*}
        |\phi\rangle = \sum\limits_{i}\alpha_i|\psi\rangle_I|v_i\rangle_A|G_i\rangle_B + \sum\limits_{i,j}\beta_{i,j}|\psi^{\perp}_j\rangle_I|v_i\rangle_A|G_{ij}\rangle_B,
    \end{equation*}
    %where $|G_i\rangle$ is an arbitrary state irrelevant to our analysis,
    where $\{|\psi^{\perp}_j\rangle\}$ is a basis of the subspace orthogonal to $\ket{\psi}$. Then,
    $
        \Tr\left(\ketbra{\phi}\cdot \ketbra{\psi}_I \otimes (H_l)_A\otimes I_B \right) = \bra{\phi} \cdot\ketbra{\psi} \otimes H_l\otimes I\cdot\ket{\phi} = \sum\limits_i|\alpha_i|^2\lambda_i,
    $
    while Lemma \ref{pre:sqap} gives
    \[
        \Pr[Y = \lambda_i] = \sum_{k \,:\, \lambda_k = \lambda_i}
        \Big( |\alpha_k|^2 + \sum_j |\beta_{k,j}|^2 \Big),
        \qquad
        \Pr[Z = \lambda_i] = \sum_{k \,:\, \lambda_k = \lambda_i}
      \Big( |\alpha_k|^2 + \tfrac12 \sum_j |\beta_{k,j}|^2 \Big)
    \]
    for $\lambda_i \neq 0$, and $\Pr[Z = 0] = 1 - \sum_{i : \lambda_i \ne 0} \Pr[Z = \lambda_i]$. 
    Subtracting,
    \[
        \E[Y - 2Z] = \E[Y] - 2\,\E[Z] = -\sum_i |\alpha_i|^2 \lambda_i
        = -\Tr\big(\ketbra{\phi} \cdot \ketbra{\psi}_I \otimes (H_\ell)_A \otimes I_B\big).
    \]
    
    % Moreover, we have
    % \begin{equation*}
    %     Pr[Y=\lambda_i]=\sum\limits_{k:\:\lambda_k=\lambda_i}\Big(|\alpha_k|^2+\sum\limits_{j}|\beta_{k,j}|^2\Big),
    % \end{equation*}
    % and
    % \begin{equation*}
    %     \begin{cases}
    %         Pr[Z=\lambda_i]=\sum\limits_{k:\:\lambda_k=\lambda_i}\Big(|\alpha_k|^2+\frac{1}{2}\sum\limits_{j}|\beta_{k,j}|^2\Big), & \text{if } \lambda_i\neq 0 \\

    %         \Pr[Z=0]=1-\sum\limits_{i:\;\lambda_i \neq 0}\Pr[Z=\lambda_i].
    % \end{cases}       
    % \end{equation*}
    % Then, $E[Y-2Z]=E[Y]-2E[Z]=- \sum\limits_i|\alpha_i|^2\lambda_i$. Besides, $-1 \leq Y-2Z \leq 1$.

    Averaging over the uniform choice of $x$, step (ii) is therefore an unbiased estimator for $\frac{1}{|S|+|L|}\bra{\phi} H \ket{\phi}$. The $r$ samples are i.i.d. and take values in $[-1, 1]$, so Hoeffding's inequality applied with accuracy $\frac{1}{p^2}$ gives
    \begin{equation*}
        \Pr\Big[\Big| M - \frac{\bra{\phi} H \ket{\phi}}{|S|+|L|} \Big| \geq \frac{1}{p^2}\Big] 
        \leq 2\exp\left(-\frac{2r^2\cdot\frac{1}{p^4}}{4r}\right)
        \leq 2\exp\left(-\frac{p}{2}\right),
     \end{equation*}
     by the choice of $r=p^5$.
    
    % After running step (ii) $p^5$ times, we obtain $p^5$ samples. Let $M$ represent the mean of these samples. Using Chernoff bound, we have
     % \begin{equation*}
     %    \Pr\Big[\Big| M - \frac{1}{|S|+|L|}\bra{\phi} H \ket{\phi} \Big| \geq \frac{1}{p^2}\Big] \leq 2\,\exp(\frac{-2p^{10}\cdot\frac{1}{p^4}}{4p^5}) \leq 2\,\exp(\frac{-p}{2}).
     % \end{equation*}
     Since, $|S|+|L| \leq p$, an error of $\frac{1}{p^2}$ in $M$ becomes an error of at most $\frac1p$ after rescaling:
     \begin{equation*}
        \Pr\Big[\Big| (|S|+|L|) \cdot M -\bra{\phi} H \ket{\phi} \Big| \geq \frac{1}{p}\Big] 
        \leq 2\exp \left(-\frac{p}{2}\right).
     \end{equation*}
     Thus, the verifier estimates $\langle \phi |H| \phi \rangle$ within additive error $\frac1p$, 
     and accepts with probability at least $1-2e^{-p/2}$, %\geq 1-2e^{-n/2}$
     as long as the witness consists of multiple copies of the same low-energy state.
     
     % We conclude that, as long as the witness consists of multiple copies of the same low-energy state, we can estimate $\langle \phi |H| \phi \rangle$ with an additive error of at most $\frac{1}{p}$ (i.e., $\leq a+\frac{1}{p}$), with overwhelming success probability. 

    \subparagraph{Soundness.} Suppose $\lambda_{\min} (H) \geq b$, so that $\Tr(\sigma H)\geq b$ for every state $\sigma$. A register may now carry a mixed state, and the registers may be entangled,
    so the $j$-th round measures the $j$-th register as conditioned on the earlier outcomes.

    Let $\cF_{j-1}$ record the outcomes of the $j-1$ rounds and let $\sigma'_j$ be the state of the $j$-th register conditioned on $\cF_{j-1}$, which agrees with the marginal $\sigma_j$ unless the witness is entangled across registers. Whatever the earlier rounds did, $\sigma'_j$ is again a state on $m$ qubits, and the two cases extend from pure to mixed states by linearity of the trace and of quantum operations. Thus the $j$-th estimate $M_j$ satisfies
    \[
        \E \left[ M_j | \cF_{j-1} \right] = \frac{1}{|S|+|L|} \Tr(\sigma'_jH) \geq \frac{b}{|S|+|L|},
    \]
    the inequality being the promise of a  ``no'' instance. It holds for every state $\sigma_j'$, so no history can push a conditional mean below $\frac{b}{|S|+|L|}$. 
    The rounds are dependent, but a prover cannot exploit that. Thus this uniformity over histories replaces independence. 
    Each $M_j$ lies in $[-1, 1]$ whatever the history, 
    so Lemma~\ref{pre:azuma} applies with $X_j = M_j$, $[a_j, b_j] = [-1,1]$, $\mu = \frac{b}{|S|+|L|}$ and $t = \frac1{p^2}$, and gives
    \begin{equation*}
        \Pr\left[ M < \frac{b}{|S|+|L|} - \frac{1}{p^2}\right] 
        \leq \exp\left(-\frac{r}{2p^4}\right) = \exp \left(-\frac{p}{2}\right).
     \end{equation*}
     Since $|S|+|L|\leq p$, therefore $\Pr[(|S|+|L|)\cdot M < b- \frac1p] \leq \exp (-\frac{p}{2})$,
     and $b-\frac1p > a+ \frac1p$ because $b-a>\frac2p$, so the verifier rejects except with negligible probability.

\end{proof}

The following lemma is similar to Kitaev's quantum Cook-Levin theorem \cite{kitaev2002classical}. A proof sketch is provided in Appendix \ref{Appendix:p_hard} for completeness.
\begin{lemma}[Proof sketch in \ref{Appendix:p_hard}]\label{qma:p_hard}
    5-$\textbf{LHwP}$ is $\pQMA$-hard.
\end{lemma}

%%%%%%%%%%%%%%%%%%%%%%%% Mixed QMA %%%%%%%%%%%%%%%%%%%%%%%%%%%
%%%%%%%%%%%%%%%%%%%%%%%%%%%%%%%%%%%%%%%%%%%%%%%%%%%%%%%%%%%%%%
%%%%%%%%%%%%%%%%%%%%%%%%%%%%%%%%%%%%%%%%%%%%%%%%%%%%%%%%%%%%%%
%%%%%%%%%%%%%%%%%%%%%%%%%%%%%%%%%%%%%%%%%%%%%%%%%%%%%%%%%%%%%%
%%%%%%%%%%%%%%%%%%%%%%%%%%%%%%%%%%%%%%%%%%%%%%%%%%%%%%%%%%%%%%
\paragraph{Mixed state QMA-complete}\label{qma:section_m}
%We will show that $k$-local Hamiltonian with unknown mixed state problem (k-\textbf{LHwM}, Definition \ref{qma:m_problem}) is $\mQMA$-complete. 
In this section, the unknown quantum input is a mixed state $\rho \in \cD(\cH(n))$, which we  view as a distribution over pure states: consider any eigenbasis decomposition $\cD = \{(\ket{\psi_i}, q_i)\}_i \in \cD_\rho$ (Preliminary $\ref{Notation}$, 1.b) such that $\rho = \sum_i q_i \ketbra{\psi_i}$. The reduction maps $\rho$ into a distribution of Hamiltonians $\{(H_{\psi_i}, q_i)\}$, where each pure state $\ket{\psi_i}$ is mapped to a corresponding Hamiltonian $H_{\psi_i}$. The problem asks whether that distribution has a low expected ground energy: for a no instance the expected ground energy is high, while a yes instance must admit a single circuit whose outputs on the $\ket{\psi_i}$ have low energy in expectation.

% We will use the notation of the decomposition set (Preliminary $\ref{Notation}$, 1.b)\nai{Again, revise the ref}\HZZ{Ok} in the context of Definition \ref{qma:m_problem}. Let $\rho$ be an unknown mixed-state.\nai{Input of what?} Without loss of generality, we will view $\rho$ as a distribution over pure states. Let $\cD_\rho$ be a decomposition set of $\rho$ and fix a eigenbasis decomposition $\cD=\{\ket{\psi_i}\}_i \in \cD_{\rho}$. Suppose $\rho:=\sum\limits_i p_i\ket{\psi_i}\bra{\psi_i}$, we can define a distribution of Hamiltonians $\{(H_{\psi_i},p_i)\}$ based on $\mathcal{D}$. Definition \ref{qma:m_problem} will demonstrate how we use $\{(H_{\psi_i},p_i)\}$ to define a $\mQMA$-complete quantum promise problem.

\begin{definition}[$k$-local Hamiltonian with unknown mixed state problem ($k$-LHwM)]
\label{qma:m_problem}
\leavevmode
    % The promise problem $k$-\textbf{LHwM} is defined as follows.
    \begin{itemize}
        % \item \textbf{Inputs:} Given the input $1^p$ with $p \in \N$, $a \in \R^+_0$, $b \in \R^+_0$, an unknown input state $\rho \in \cD(\cH(n))$, and two sets of local Hamiltonians $\{H_s\}_{s\in S}$ and $\{H_\ell\}_{\ell \in L}$, with each Hamiltonian acting on at most $k$ qubits. We are given the promise that $|S|+2^k|L| \leq p$, $\;0\preceq H_x \preceq I$ for all $x \in S\cup L$, and $b-a > \frac{4}{p}$. Additionally, let $\cD_\rho$ denote the decomposition set of $\rho$. For any pure-state $\ket{\psi}$, define 
        % \begin{equation*}
        %     H_\psi:= \sum\limits_sH_s - \sum\limits_\ell\ket{\psi}\bra{\psi}_I \otimes H_\ell.
        % \end{equation*}
        \item \textbf{Inputs:} An instance is specified by a classical description together with copies of an unknown state.
        \begin{itemize}
            \item \emph{Classical part:} $1^p$ with $p\in \N$, thresholds $a, b\in \R^+_0$, and classical descriptions of two families of local Hamiltonians $\{H_s\}_{s\in S}$ and $\{H_\ell\}_{\ell \in L}$, each acting on at most $k$ qubits.
            \item \emph{Quantum part:} $\poly(n)$ copies of an unknown state $\rho\in \cD(\cH(n))$.
        \end{itemize}
        For every pure state $\ket{\psi}$ these specify the Hamiltonian
        \begin{equation*}
            H_\psi\coloneqq \sum\limits_{s\in S}H_s - \sum\limits_{\ell\in L}\ket{\psi}\bra{\psi}_I \otimes H_\ell.
        \end{equation*}
        We are given the promise that $p=\poly(n)$, $|S|+2^k|L| \leq p$, $\;0\preceq H_x \preceq I$ for all $x \in S\cup L$, and $b-a > \frac{4}{p}$. 
        %$\cD_\rho$ denotes the decomposition set of $\rho$.
        
        \item \textbf{Yes instances:} There are a uniform QPT circuit (or unitary) $\mathcal{C}$, a pure state $\ket{\phi}$, and $\alpha \in \R$,  such that, writing 
        \[
            \ket{\eta_{\psi, \phi}} \coloneqq \cC 
            \left(\ket{\psi}_I\ket{\phi}_W\ket{0^*}_A\right)
        \]
        for the state $\cC$ prepares from $\ket{\psi}$, the following hold for every $\cD \in \cD_\rho$, where $\cD_\rho$ denotes the decomposition set of $\rho$. Here $\ket{\eta_{\psi, \phi}}$ carries the same number of qubits as those $H_\psi$ acts on.
        
        % for all $\cD\in\cD_\rho$ the following properties hold: Let $\ket{\eta_{\psi,\phi}}:=\cC\big(\ket{\psi}_I\ket{\phi}_W\ket{0^*}_A\big)$, where $\ket{\eta_{\psi,\phi}}$ has the same number of qubits as those on which $H_\psi$ acts.
        \begin{enumerate}[label = (\arabic*)]
            \item \emph{Expected small eigenvalue:}
            \begin{equation*}
                \E\limits_{\ket{\psi} \leftarrow \mathcal{D}} \big[\langle \eta_{\psi,\phi}| H_{\psi} |\eta_{\psi,\phi} \rangle \big]  \leq a,
            \end{equation*}
            \item \emph{Uniform Initialization:} For all $\ell\in L$, let $\{v_i\}$ be any eigenbasis and $\{ \lambda_i\}$ be the corresponding eigenvalue of $H_\ell$. For all $\ket{\psi} \in \mathcal{D}$,  $|\eta_{\psi,\phi}\rangle$ has the form
            \begin{equation}\label{qma:restrict}
                \sum\limits_{i:\;\lambda_i \neq 0}\Big(\alpha\ket{\psi}_I\ket{\phi}_W\ket{0^*}\ket{v_i} \Big ) +  \sum\limits_{i:\;\lambda_i \neq 0}\Big(\ket{\psi^{\perp}}_I\ket{*} \ket{v_i}\Big )+\sum\limits_{i:\;\lambda_i = 0}|*\rangle|v_i\rangle,
            \end{equation}
            where the unimportant amplitudes of the last two terms are implicitly represented by $\ket{*}$. We highlight that
            \begin{equation*}
                \ket{\psi}_I\left(\ket{\phi} \ket{0^*}\right )^\perp\ket{v_i},
            \end{equation*}
            in Equation ($\ref{qma:restrict}$) has amplitude 0 when $\lambda_i\neq 0$.
            \label{def:klhwm-unif}
            \item \emph{Restriction on $H_\ell$:} For all $
            \ell \in L$, $H_\ell$ acts on the last $k$ qubits, disjoint from registers $I$ and $W$.\label{def:klhwm3}\footnote{\label{footnote:mqma_restriction}In general, $H_\ell$ only needs to act on register $A$, remaining disjoint from registers $I$ and $W$. Moreover, each $H_\ell$ can act on different qubits. However, the quantum state $\ket{v_i}$ in the form of \emph{uniform initialization}, $\ket{\eta_\psi}$, will shift positions corresponding to the qubits on which $H_\ell$ acts.}
            \end{enumerate}

        \item \textbf{No instances:} For all $\cD \in \cD_{\rho}$, the expected minimum eigenvalue is large. Specifically,
        \begin{equation*}
            \E\limits_{\ket{\psi}\leftarrow \mathcal{D}}\Big[\min\limits_{|\eta\rangle} \langle \eta| H_{\psi} |\eta \rangle \Big] \geq b.
        \end{equation*}
    \end{itemize}

\end{definition}

Uniform initialization in Definition~\ref{qma:m_problem} requires $\alpha$ to be a single number, the same for every $\cD\in \cD_{\rho}$ and every $\ket{\psi}\in \cD$. It is needed because copies of $\rho$ do not repeat: $\rho^{\tp r}$ draws its pure states independently, so no $\ket{\psi}$ is available twice and a verifier can only average across \emph{different} draws.  Thus, the quantity measured must be independent of which $\ket{\psi}$ was chosen. 
The restriction on $H_\ell$ fixes where those terms sit, so that the verifier measures the intended observable. 
Both conditions are imposed only so that membership can be proved; the technical details will show where they are used.

% In Definition~\ref{qma:m_problem}, we emphasize that $\textit{uniform initialization}$ ensures that, regardless of the choice of $\mathcal{D} \in \mathcal{D}_{\rho^t}$ or $\ket{\psi} \in \mathcal{D}$, the values of $\alpha \in \R$ remain the same.
% The task is to determine whether $\{(H_{\psi_i},p_i)\}$ has an expected low-energy state. 
% To prove that $k$-\textbf{LHwM} is contained in $\mQMA$, we require two additional promises for the yes instance, allowing us to obtain information about the (small) energy of $H_{\psi}$ with a single copy of $\ket{\psi}$. The reason these additional promises are needed will be explained in the technical details.

\begin{theorem}\label{thm:mqma_complete}
    $5$-$\textbf{LHwM}$ is $\mQMA$-complete.
\end{theorem}
\begin{proof}
    The proof follows immediately from membership (Lemma~\ref{qma:m_in}) and hardness (Lemma~\ref{qma:m_hard}).
    % The proof follows directly from Lemma~\ref{qma:m_in} and Lemma~\ref{qma:m_hard}, where we will show that $5$-$\textbf{LHwM}$ is in $\mQMA$ (Lemma~\ref{qma:m_in}) and that $5$-$\textbf{LHwM}$ is hard for $\mQMA$ (Lemma~\ref{qma:m_hard}).
\end{proof}

Two elementary facts are stated first and used as black boxes in the membership proof. 
\begin{lemma}[Mean-square Inequality]\label{qma:mean_square}
Let $A_1,\dots,A_N$ and $g_1,\dots,g_N$ be complex-valued random variables with
$|g_j| \leq 1$ for all $j$.  Define $\mu_j := \E\big[\mathbf{Re}(A_j g_j)\big]$
and $\pi_j := \E\big[|A_j|^2\big]$.  Then
\begin{equation*}
    \left(\frac{1}{N}\sum\limits_{j=1}^{N}\mu_j\right)^2
    \;\leq\; \frac{1}{N}\sum\limits_{j=1}^{N}\mu_j^2
    \;\leq\; \frac{1}{N}\sum\limits_{j=1}^{N}\pi_j.
\end{equation*}
\end{lemma}
\begin{proof}
The first inequality states that the square of an average is at most the average
of the squares; the difference of the two sides is the variance of the list
$\mu_1, \dots, \mu_N$, which is non-negative.  For the second it suffices that
$\mu_j^2 \leq \pi_j$ for each $j$: since $|\mathbf{Re}(z)| \leq |z|$ and
$|g_j| \leq 1$,
\begin{equation*}
    |\mu_j| \;\leq\; \E\big[|A_j g_j|\big] \;\leq\; \E\big[|A_j|\big]
    \;\leq\; \E\big[|A_j|^2\big]^{1/2} \;=\; \pi_j^{1/2},
\end{equation*}
where the last inequality follows from Jensen's inequality.
\end{proof}

\begin{lemma}[Detection]\label{qma:detection}
Let $X$ be any random variable supported on $[-1,1]$\footnote{The interval $[-1,1]$ is only a normalization; boundedness is what the argument uses. For $X$ supported on an interval of length $\ell$, the same proof gives $\Pr[|X-\alpha|>\epsilon] \geq \frac{\epsilon}{\ell+\epsilon}$.}, and let $\mu := \E[X]$.
For any $\alpha \in [-1,1]$ and any $0 < \epsilon \leq 1$, if
$|\alpha - \mu| \geq 2\epsilon$, then
\begin{equation*}
    \Pr\big[\, |X - \alpha| > \epsilon \,\big] \;\geq\; \frac{\epsilon}{3}.
\end{equation*}
\end{lemma}
\begin{proof}
Assume $\alpha \geq \mu + 2\epsilon$; the other case follows by applying the
argument to $-X$ and $-\alpha$.  If $|X - \alpha| \leq \epsilon$ then
$X \geq \alpha - \epsilon \geq \mu + \epsilon$, so with
$q := \Pr[X \geq \mu + \epsilon]$ we have
$\Pr[|X - \alpha| \leq \epsilon] \leq q$.  Splitting the expectation over the
event $\{X \geq \mu + \epsilon\}$ and its complement, and using $X \geq -1$ on
the complement,
\begin{equation*}
    \mu \;\geq\; q\,(\mu + \epsilon) - (1 - q).
\end{equation*}
Rearranging gives $\mu(1-q) \geq q(\epsilon + 1) - 1$; since $\mu \leq 1$, the
left side is at most $1 - q$, whence $2 \geq q(2 + \epsilon)$ and
\begin{equation*}
    \Pr\big[|X - \alpha| > \epsilon\big] \;\geq\; 1 - q
    \;\geq\; 1 - \frac{2}{2+\epsilon} \;=\; \frac{\epsilon}{2+\epsilon}
    \;\geq\; \frac{\epsilon}{3},
\end{equation*}
the last step because $\epsilon \leq 1$.
\end{proof}

\begin{lemma}\label{qma:m_in}
    For all constant $k \in \N$, \textbf{$k$-LHwM $\in$ mQMA}. 
\end{lemma}
%HERE
% \input{k-LHwM-membership-original}

\input{k-LHwM-membership-new}

The following lemma differs only slightly from the pure-state version.
\begin{lemma}\label{qma:m_hard}
    5-$\textbf{LHwM}$ is $\mQMA$-hard.
\end{lemma}
\begin{proof}
    Given a $\mQMA$ mixed-state quantum promise problem $\cL$. By the amplification Lemma~\ref{App:amplify_QMA}, let $(\cP,\cV)$ be a $\mQMA$ protocol that decides $\cL$ with completeness $1-4^{-n}$ and soundness $4^{-n}$. There exist polynomials $q(\cdot)$ and $q'(\cdot)$ such that $\cV$ uses $q(n)$ qubits of ancilla, requires $c \leq q'(n)$ copies of the input, and runs for $m \leq q'(n)$ steps. Hence, let $\cV$ be represented as a sequence of elementary quantum gates, i.e., $\cV:=\cV_m\cdots\cV_1$, acting on register $I$, $W$, and $A$. The reduction maps the input $\rho \in \cD(\cH(n))$ of the promise problem $\cL$ to the inputs
    \begin{equation*}
        \left(p:=\frac{10(m+1)^3}{d}, a:=a'+\negl(n), b:=\left(1-\negl(n)\right)b', \rho^{\otimes c}, \{H_s\}_{s\in S}, \{H_\ell\}_{\ell\in L},  \right),
    \end{equation*}
    where $a'=\frac{1}{2^{n+1}(m+1)}$, $b'=\frac{d}{2(m+1)^3}$, $\{H_s\}_{s\in S}$, and $\{H_\ell\}_{\ell\in L}$ are defined identically to those in Lemma~\ref{Appendix:p_hard} (with some constant $d$).
    Let $\mathcal{D}_{\rho^{\tp c}}$ denote the decomposition set of $\rho^{\otimes c}$, and consider an arbitrary ensemble $\mathcal{D} \in \mathcal{D}_{\rho^{\tp c}}$.
    
    Suppose $\rho \in \mathcal{L}_Y$. Without loss of generality, consider a ``good" pure-state witness $\ket{\phi}$ for input $\rho$ with respect to the verifier $\mathcal{V}$. That is, define a set $\textbf{Good}$ as follows:
    \begin{equation*}
        \textbf{Good}:=\bigg\{\ket{\psi} \in \mathcal{D} \::\:\Pr\left[\mathcal{V}\left(\ket{\psi},\ket{\phi}, 0^{q(n)}\right)\right] \geq 1-2^{-n} \bigg\}.
    \end{equation*}
    By average argument,
    \begin{equation*}
        \Pr\limits_{\ket{\psi}\leftarrow \mathcal{D}}\big[\ket{\psi} \in \textbf{Good}\big] \geq 1-2^{-n}.
    \end{equation*}
    Define $\cC$ as the unitary that maps $\ket{\psi}\ket{\phi}\ket{0}$ to $\ket{\eta_{\psi,\phi}}$, as stated in Equation (\ref{Appendix:witness}). That is,
    \begin{equation*}
        \cC:\ket{\psi}\ket{\phi}\ket{0} \mapsto \ket{\eta_{\psi,\phi}}:= \frac{1}{\sqrt{m+1}}\sum\limits_{t=0}^m \mathcal{V}_t\cdots\mathcal{V}_1 \big(\ket{\psi}_I\ket{\phi}_W \ket{0}_A \big)\ket{1^t0^{m-t+1}}.
    \end{equation*}
    Additionally, $\cC$ is a uniform QPT unitary. Define the witnesses for a yes instance as 
    \begin{equation*}
        \left(\cC,\ket{\phi}^{\tp \poly(n)}, \alpha:=\frac{1}{\sqrt{m+1}}\right).
    \end{equation*}
    Consider $\ket{\psi} \in \textbf{Good}$. Then, as in Lemma \ref{Appendix:p_hard}, we have $\bra{\eta_{\psi,\phi}}H_{\psi}\ket{\eta_{\psi,\phi}} \leq a'$. Since $0 \preceq H_{\psi} \preceq \poly(n)I$, we obtain $\E\limits_{\ket{\psi} \leftarrow \mathcal{D}}\big[\bra{\eta_{\psi,\phi}}H_{\psi}\ket{\eta_{\psi,\phi}} \big] \leq a' + \negl(n) =: a$. Therefore, $H_\psi$ has an \emph{expected small eigenvalue}. Additionally, \emph{uniform initialization} holds for all inputs. Specifically, $|L|=1$ and $H_\ell = \ket{0}\bra{0}_{T_1}$; hence, for all $\mathcal{D} \in \cD_{\rho^t}$ and $\ket{\psi} \in \mathcal{D}$,
    \begin{equation*}
        \ket{\eta_{\psi}}:=\frac{1}{\sqrt{m+1}}\ket{\psi}_I\ket{\phi}_W\ket{0}_A\ket{0}_{-T_1}\ket{0}_{T_1} + \frac{1}{\sqrt{m+1}}\sum\limits_{t=1}^m \mathcal{V}_t\cdots\mathcal{V}_1 \big(\ket{\psi}_I\ket{\phi}_W \ket{0}_A \big)\otimes \ket{1^{t-1}0^{m-t+1}}_{-T_1}\ket{1}_{T_1}.
    \end{equation*}
    Finally, the \emph{restriction on $H_\ell$} also holds, where the register $T_1$ is disjoint from registers $I$ and $W$. 

    Suppose $\rho \in \mathcal{L}_N$. Define a bad set $\textbf{Bad}$ as follows:
    \begin{equation*}
        \textbf{Bad}:=\bigg\{\ket{\psi} \in \mathcal{D} \::\:\forall \ket{\zeta},\; \Pr\Big[\mathcal{V}\big(\ket{\psi},\ket{\zeta}, 0^{q(n)}\big)\Big] \leq 2^{-n} \bigg\}.
    \end{equation*}
    By the average argument,
    \begin{equation*}
        \text{If } \rho \in \mathcal{L}_N  \implies \Pr\limits_{\ket{\psi}\leftarrow \mathcal{D}}\big[\ket{\psi} \in \textbf{Bad}\big] \geq 1-2^{-n}.
    \end{equation*} 
    Consider $\ket{\psi} \in \textbf{Bad}$. Then as in Lemma~\ref{Appendix:p_hard}, we have $\lambda_{min}(H_{\psi}) \geq b'$. Since $0 \preceq H_{\psi} \preceq \poly(n)I$, we obtain $\E\limits_{\ket{\psi} \leftarrow \mathcal{D}}\big[\min_{\ket{\eta}}\bra{\eta_{\psi}}H_{\psi}\ket{\eta_{\psi}} \big] \geq \big(1-\negl(n)\big) b' =: b$.
\end{proof}

%% file: k-LHwM-membership-new.tex
% \lch{attempt}
\begin{proof}
% Attempt to rewrite the proof for k-LHwM \in mQMA
%
Fix $\cD \in \cD_\rho$ and treat the input as a pure state $\ket{\psi}\gets \cD$.
This is without loss of generality because the yes and the no conditions of Definition \ref{qma:m_problem} are both imposed for every $\cD \in \cD_\rho$. Every quantity below is evaluated inside that single decomposition.
\subparagraph{Witness and verification.}
The witness is $(\cC, \tilde{\sigma}, \aprov)$: a description of a unitary $\cC$ implemented by a uniform QPT circuit, 
a state $\tilde{\sigma}$ on $mp^{10}$ qubits, and a declared $\aprov \in [-1, 1]$. 
Cut $\tilde{\sigma}$ into blocks $\sigma_j$ of $m$ qubits, $j=1, \dots, p^{10}$; each block and each of the $p^{10}$ copies of $\rho$ is used exactly once. 
Write
\[
\eta_{\psi,\sigma}:= \cC(\ketbra{\psi}_I \otimes \sigma_W \otimes \ketbra{0^*}_{DE})\cC^\dagger,
\]
with $E$ the last $k$ qubits. 
% birefly talk about the verification procedure
The verifier reads the register layout off $\cC$ and rejects unless every $H_\ell$
acts on E, which is sound because Definition \ref{qma:m_problem}.\ref{def:klhwm3} holds on every yes instance. 
It then runs \Cref{alg:method2}. In total $p^5$ rounds, each draws one term of $H_{\psi}$ and spends $p^5$ blocks and as many copies of $\rho$ on an estimate $W_t\in [-1,1]$ of it, followed by a threshold test on the average.

Since $x$ is uniform and the trace is linear, what a round should report is the average, over the round's own blocks, of the energy of $H_\psi$: 
\begin{equation}
    R_t \coloneqq \frac{1}{|S| + 2^k |L|} \cdot \frac{1}{p^5} \sum_{j=c+1}^{c+p^5} \E\limits_{\ket{\psi}\gets \cD}
    \left[\Tr(\eta_{\psi, \sigma_j} H_\psi)\right],\qquad c = p^5t.
\label{eq:mixedG}
\end{equation}

%Algorithm 1: k-LHwM verifier
\begin{algorithm}[t]
\caption{$k$-LHwM verifier}
\label{alg:method2}
\begin{algorithmic}[1]
  \REQUIRE $(1^p, \rho^{\otimes p^{10}}, a, b, \{H_s\}, \{H_\ell\})$ and a witness                 $(\cC, \Tilde{\sigma},\alpha_{\prover})$.
  \STATE \textbf{reject} unless every $H_\ell$ acts on $E$ in the layout of C \COMMENT{Definition \ref{qma:m_problem}\ref{def:klhwm3}}
  \STATE partition $\tilde\sigma$ into $\sigma_1, \dots, \sigma_{p^{10}}$,
         each on $m$ qubits
  \FOR{$t = 0, 1, \dots, p^{5}-1$}
    \STATE $c \gets p^{5} t$
    \STATE draw $x$ uniformly from $S \cup (L \times [2^{k}])$
    \IF[\textbf{Case 1}]{$x = s \in S$}
      \FOR{$j=c+1, \dots, c+p^5$}
        \STATE prepare $\eta_{\psi, \sigma_j}$ from one copy of $\rho$ and the block $\sigma_j$
        \STATE measure the register carrying $H_s$ in an eigenbasis of $H_s$
        \STATE $\Lambda_j \gets $ the eigenvalue observed
      \ENDFOR
      \STATE $W_t \gets p^{-5} \sum_j \Lambda_j$
    \ELSIF[\textbf{Case 2:} $H_\ell = \sum_{i\in[2^k]}\lambda_i\ketbra{v_i}$ for the drawn $\ell$]{$x=(\ell, r)$ and $\lambda_r\neq  0$} 
      \FOR{$j=c+1, \dots, c+p^5$}
        \STATE $X_j \gets$ Algorithm \ref{alg:hadamardtest} on $(\cC, \sigma_j, \rho, r, H_\ell)$ \COMMENT{Hadamard test}
      \ENDFOR
      \STATE \textbf{reject} if $|\aprov - p^{-5}\sum_j X_j| > 1/4p^2$ \COMMENT{the check Eq.(\ref{eq:aprov-check})}
      \STATE $W_t \gets -\lambda_r \aprov^2$\COMMENT{Eq.(\ref{eq:mixed-Wt})}
    \ELSE[$\lambda_r = 0$: the term vanishes]
      \STATE $W_t \gets 0$
    \ENDIF
  \ENDFOR
  \STATE $M \gets p^{-5} \sum_{t} W_t$
  \STATE \textbf{accept} if $(|S| + 2^{k}|L|) \cdot M \le a + 2/p$, else
         \textbf{reject}
\end{algorithmic}
\end{algorithm}
%
%Algorithm 2: k-LHwM verifier
\begin{algorithm}[t]
\caption{Hadamard test on one block, at a fixed $(\ell, r)$}
\label{alg:hadamardtest}
\begin{algorithmic}[1]
\renewcommand{\ENSURE}{\item[\textbf{Output:}]}
  \REQUIRE $(\cC, \rho, \sigma, r, H_\ell = \sum_i \lambda_i \ketbra{v_i} \text{ with }\lambda_r\neq 0)$
  \ENSURE $X\in \{-1, 0, 1\}$
  \STATE let $U$ be the unitary on $E$ with $U\ket{v_i} = \ket{i}$
  \STATE prepare $\ketbra{0}_A \tp \rho_I \tp \sigma_W \tp \ketbra{0^*}_{DE}$
  \STATE apply $H$ to $A$
  \STATE apply $\cC$ to $IWDE$, then $U$ to $E$, both controlled on $\ket{1}_A$
  \STATE add $r$ on $E$, controlled on $\ket{0}_A$
  \STATE measure $A\in \{\ket{+}, \ket{-}\}$ and $E$ in the computational basis
  \STATE $Y\gets 1$ if $E$ returned $r$, and $0$ otherwise
  \STATE $Z\gets 1$ if $E$ returned $r$ and $A$ returned $\ket{+}$, and $0$ otherwise
  \STATE \textbf{return} $X\gets 2Z-Y$
\end{algorithmic}
\end{algorithm}

\noindent Each branch of the round below produces the term of Eq.(\ref{eq:mixedG}) that the draw of $x$ selects.

%Case 1
\emph{Case 1: $x=s\in S$}. For each block, prepare $\eta_{\psi, \sigma_j}$ from one copy of $\rho$, measure the register carrying $H_s$ in an eigenbasis of $H_s$, report the eigenvalue, and average over the round's blocks. 
This is unbiased for
\begin{equation}
    \frac{1}{p^5} \sum_{j=c+1}^{c+2p^5} \E \limits_{\ket{\psi}\gets \cD}
    \left[ \Tr (\eta_{\psi, \sigma_j} \cdot H_s \tp I)\right],
\end{equation}
by the computation of Case 1 of Lemma~\ref{qma:p_in} applied to $\eta_{\psi, \sigma_j}$ and averaged over $\ket{\psi}\gets \cD$. 

%Case 2
\emph{Case 2: $x=(\ell, r)\in L\times [2^k]$}. What the round must deliver here is the $(\ell, r)$ term of \Cref{eq:mixedG}, 
\begin{equation}
    T \coloneqq -\frac{\lambda_r}{p^5} \sum_{j=c+1}^{c+p^5} \E_{\ket{\psi}\gets \cD}
    \left[ \Tr \left(
    \eta_{\psi, \sigma_j} \cdot \ketbra{\psi}_I \tp I_{WD} \tp \ketbra{v_r}_E
    \right)\right].
\label{eq:lr-target}
\end{equation}
%Explain the reason to use hadamard test
This quantity is reached by a swap test for pure inputs. However, that approach is not applicable here: a swap test requires two copies of the same state $\ket{\psi}$, which cannot be guaranteed when sampling twice from $\rho^{\tp 2}$. 
The Hadamard test spends one copy only, but what it returns is an amplitude rather than a probability: $\alpha^{\psi,\phi}_r$ rather than $|\alpha^{\psi,\phi}_r|^{2}$.
% It is a signed real number linear in $\alpha^{\psi,\phi}_r$, where the round owes $|\alpha^{\psi,\phi}_r|^{2}$.
Recovering the square would mean multiplying two such outcomes, whose expectation is $\E[X_j X_{j'}]$ across two blocks. The blocks are marginals of the single state $\tilde\sigma$ the prover chooses and may be correlated arbitrarily.

% briefly talk about Hadamard test
Algorithm \ref{alg:hadamardtest} runs a Hadamard test on each of the round's $p^5$ blocks. Let $U$ be the $k$-qubit unitary carrying each $\ket{v_i}$ to the computational basis state $\ket{i}$ of $E$, so that the eigenbasis measurement becomes one the verifier can perform. Each block then yields two bits: $Y_j$ is $1$ if the measurement of $E$ returns $r$ and $0$ otherwise, and $Z_j$ is $1$ if the measurement of $E$ returns $r$ and that of $A$ returns $\ket{+}$, and 0 otherwise; the block's outcome is $X_j \coloneqq 2Z_j-Y_j$, which is $0$ unless $E$ returns $r$, and otherwise $+1$ or $-1$ according as $A$ returns $\ket{+}$ or $\ket{-}$. All $p^5$ outcomes go into a single consistency check, 
\begin{equation}
    \left| \aprov - \frac{1}{p^5} \sum_{j=c+1}^{c+p^5} X_j\right| \leq \frac{1}{4p^2},  \ \ \text{the verifier rejecting if it fails,}
\label{eq:aprov-check}
\end{equation}
and the round's output is the declared number squared,
\begin{equation}
    W_t \coloneqq -\lambda_r \aprov^2.
\label{eq:mixed-Wt}
\end{equation}
Both are actions of the verifier: $\aprov$ is a number the prover claims, and Eq.(\ref{eq:aprov-check}) confronts that claim with the verifier's own measurements.
When $\lambda_r = 0$ the term (\ref{eq:lr-target}) vanishes whatever the state, so the verifier simply sets $W_t\gets 0$ without running the test or the check. 

To see what Algorithm \ref{alg:hadamardtest} returns, fix a pure $\ket{\psi}$ and a pure $\ket{\phi}$ drawn from the block $\sigma_j$. Expand
\begin{equation}
    \ket{\eta_{\psi, \phi}} \coloneqq \cC \ket{\psi}_I \ket{\phi}_W \ket{0^*}_{DE} = 
    \sum_i \alpha_i^{\psi, \phi} \ket{\psi}_I \ket{G_i}_{WD}\ket{v_i}_E + 
    \sum_{i,j}\beta_{i,j}^{\psi, \phi} \ket{\psi_j^\bot}_I \ket{G_{i,j}}_{WD} \ket{v_i}_E,
\label{eq:eta-expansion-b}
\end{equation}
so that the summand of (\ref{eq:lr-target}) is $\E[|\alpha_r^{\psi, \phi}|^2]$. 
Lines~3--5 then evolve the state prepared in line~2 as follows, the last line being the state just before the measurement in line~6:
\begin{center}
    \begin{math}
    \begin{aligned}
        |0\rangle_A|\psi\rangle_I|\phi\rangle_W|0^*\rangle_{DE} &\xmapsto{H_A} \tfrac{1}{\sqrt{2}}|0\rangle_A|\psi\rangle_I|\phi\rangle_W|0^*\rangle_{DE} + \tfrac{1}{\sqrt{2}}|1\rangle_A|\psi\rangle_I|\phi\rangle_W|0^*\rangle_{DE} \\     
        &\xmapsto{c-\mathcal{C}_{A,IWDE}} \tfrac{1}{\sqrt{2}}|0\rangle_A|\psi\rangle_I|\phi\rangle_W|0^*\rangle_{DE} + \tfrac{1}{\sqrt{2}}|1\rangle_A|\eta_{\psi,\phi}\rangle_{IWDE} \\
        &= \tfrac{1}{\sqrt{2}}|0\rangle_A|\psi\rangle_I|\phi\rangle_W|0^*\rangle_{DE}  \\ 
        & \quad\hspace{0.3em} + \tfrac{1}{\sqrt{2}}|1\rangle_A\left(\sum\limits_i\alpha_i^{\psi,\phi}|\psi\rangle_I|G_i\rangle_{WD}|v_i\rangle_E+\sum\limits_{i,j}\beta_{i,j}^{\psi,\phi}|\psi^{\perp}_j\rangle_I|G_{i,j}\rangle_{WD}|v_i\rangle_{E}\right) \\
        &\xmapsto{c-U_{A,E}} \tfrac{1}{\sqrt{2}}|0\rangle_A|\psi\rangle_I|\phi\rangle_W|0^*\rangle_{DE} \\ 
        &\qquad\qquad + \tfrac{1}{\sqrt{2}}|1\rangle_A\left(\sum\limits_i\alpha_i^{\psi,\phi}|\psi\rangle_I|G_i\rangle_{WD}|i\rangle_E+\sum\limits_{i,j}\beta_{i,j}^{\psi,\phi}|\psi^{\perp}_j\rangle_I|G_{i,j}\rangle_{WD}|i\rangle_{E}\right) \\  
        &\xmapsto{(\Bar{c}-\text{add } r)_{A,E}} \tfrac{1}{\sqrt{2}}\ket{0}_A\ket{\psi}_I\ket{\phi}_W\ket{0^*}_D\ket{r}_E \\
        &\qquad\qquad\quad\hspace{0.75em} + \tfrac{1}{\sqrt{2}}|1\rangle_A\left(\sum\limits_i\alpha_i^{\psi,\phi}|\psi\rangle_I|G_i\rangle_{WD}|i\rangle_E+\sum\limits_{i,j}\beta_{i,j}^{\psi,\phi}|\psi^{\perp}_j\rangle_I|G_{i,j}\rangle_{WD}|i\rangle_{E}\right).
    \end{aligned}
    \end{math}
    \end{center}

% Just before the measurement (line 6 of \Cref{alg:hadamardtest}) the state is 
% \[
%   \tfrac{1}{\sqrt2} \ket{0}_A \ket{\psi}_I \ket{\phi}_W \ket{0^{*}}_D \ket{r}_E
%   \;+\;
%   \tfrac{1}{\sqrt2} \ket{1}_A\, U_E \ket{\eta_{\psi,\phi}} ,
% \]
Its two branches interfere only through the $\alpha$ terms, since the $\beta$ terms being orthogonal to $\ket{\psi}$ on $I$. Writing $S_r\coloneqq |\alpha_r^{\psi, \phi}|^2 + \sum_j |\beta_{r,j}^{\psi, \phi}|^2$,
\[
  \Pr[Y_j = 1] = \tfrac12 + \tfrac12 S_r ,
  \qquad
  \Pr[Z_j = 1] = \tfrac14 + \tfrac14 S_r
    + \tfrac12 \operatorname{Re}\big(\alpha^{\psi,\phi}_r \braket{\phi,0^{*}|G_r}\big) ,
\]
and therefore
\begin{equation}\label{qma:completeness_x}
\E[X_j | \psi, \phi] = \operatorname{Re}\left(\alpha^{\psi,\phi}_r \braket{\phi,0^{*}|G_r}\right),
\end{equation}
which is conditional on the pair $(\ket{\psi}, \ket{\phi})$. The input $\rho$ is a mixed state by definition, and a cheating prover's witness need not be pure either. Averaging over both sources, 
\begin{equation*}
    \E[X_j] := \E\limits_{\substack{\psi \leftarrow \cD \\ \phi \leftarrow \sigma_j}}
    \big[\E[X_j|\psi,\phi]\big]
    = \E\limits_{\substack{\psi \leftarrow \cD \\ \phi \leftarrow \sigma_j}}
    \left[\textbf{Re}\left(\alpha_r^{\psi,\phi}\cdot\braket{\phi,0^*|G_r}\right)\right],
\end{equation*}
where each $X_j$ has support $\{-1,0,1\}$.

\subparagraph{Completeness.} Suppose the prover sends a witness $\cC, \ket{\phi}^{\tp p^{10}}$, and $\aprov$ satisfying all the conditions for a yes instance. The estimation result for Case 2 remains to be provided. For simplicity, we often omit the subscript $j$.

By the promise of \textit{uniform initialization}, for every $\ket{\psi}\in \cD$ and every $i$ with $\lambda_i \neq 0$ we have $\alpha_i^{\psi, \phi} = \alpha$ and $\ket{G_i} = \ket{\phi}\ket{0^*}$. Hence, for every block, $\E[X_j] = \alpha$ and $\E[|\alpha_i^{\psi, \phi}|^2] = \alpha^2$.
The honest prover sends $\aprov = \alpha$, so the round's output satisfies
\begin{equation}\label{qma:m_hard_est_case2}
\begin{aligned}
    T &= -\frac{\lambda_r}{p^5} \cdot \sum\limits_{j=c+1}^{c+p^5}\E\limits_{\substack{\ket{\psi_j} \leftarrow \cD \\ \ket{\phi} \leftarrow \sigma_j }}\left[\bra{\eta_{\psi_j,\phi}} \cdot(\ket{\psi_j}\bra{\psi_j})_I \otimes  I_{WD}\otimes (\ket{v_r}\bra{v_r})_E \cdot\ket{\eta_{\psi_j,\phi}}\right] \\
    &=-\frac{\lambda_r}{p^5} \cdot \sum\limits_{j=c+1}^{c+p^5}\E\limits_{\substack{\ket{\psi_j} \leftarrow \cD \\ \ket{\phi} \leftarrow \sigma_j }}\left[ |\alpha_r^{\psi_j,\phi}|^2\right] \\
    &= -\lambda_r \alpha^2 \;=\; -\lambda_r \aprov^2 \;=\; W_t,
\end{aligned}
\end{equation}
where the first equality writes each block state $\sigma_j$ in a pure-state
decomposition, and the second is Eq.(\ref{eq:eta-expansion-b}), and the third is \textit{uniform initialization}. Note that this is an identity rather than an unbiased estimate.

Finally, the verifier must not reject a yes instance. Recall that $\E[X_j]=\alpha$ for every block. Since the witness is a product state and the copies of $\rho$ are drawn independently, the $X_j$ within a round are independent. Hoeffding's inequality therefore gives
\begin{equation}\label{qma:completeness_check}
    \Pr\left[\left|\alpha- \frac{1}{p^5}\sum\limits_{j=c+1}^{c+p^5} X_{j}\right| \geq \frac{1}{4p^2} \right] \leq 2\,\exp\left(\frac{-p^{10}p^{-4}}{32p^5}\right).
\end{equation}

The $W_t$ are likewise independent across rounds and supported on $[-1,1]$, so a second application of Hoeffding's inequality, together with a union bound over the at most $p^5$ checks, gives
\begin{equation*}
\begin{aligned}
    &\pad\Pr\left[ \left|M-\frac{1}{|S|+2^k|L|}\cdot\E\limits_{\ket{\psi} \leftarrow \cD}[\bra{\eta_{\psi,\phi}}H_{\psi} \ket{ \eta_{\psi,\phi}}]\right|  \geq \frac{1}{p^{2}} \;\lor\; \text{some check } (\ref{eq:aprov-check}) \text{ fails}\right]\\
    &\leq 2\cdot \exp\left(\frac{-2p^{10}p^{-4}}{4p^5}\right)+ 2p^5\cdot \exp\left(\frac{-p^{10}p^{-4}}{32p^5}\right).
\end{aligned}
\end{equation*}
On the complement, Definition \ref{qma:m_problem}.(1) gives
$(|S|+2^k|L|)\cdot M \leq a + \frac{1}{p} \leq a+\frac{2}{p}$, so
the verifier accepts, and a yes instance is rejected with negligible probability.

\subparagraph{Soundness.} Suppose the prover sends a  witness $\cC$, $\tilde{\sigma} = \sigma_1,\dots, \sigma_{p^{10}}$, and $\aprov$.  The estimation
result for Case 2 remains to be provided. 
We will show that the output of a Case~2 round is at least Eq.(\ref{eq:lr-target}) minus a small error with high probability, by the check in Eq.(\ref{eq:aprov-check}).
Consequently, the expected estimated outcome of Algorithm \ref{alg:method2} is also at least Eq.(\ref{eq:mixedG}) up to the same error.

Verifying Eq.$(\ref{eq:aprov-check})$ allows us to ensure that the prover's declared $\aprov$ satisfies the following condition:
\begin{equation}\label{qma:condition_expect}
    \left|\aprov- \frac{1}{p^5}\sum\limits_{j=c+1}^{c+p^5} \E[X_{j}]\right| < \frac{1}{2p^2}.
\end{equation}
For convenience, define $X_{\sample}:=\frac{1}{p^5}\sum\limits_{j=c+1}^{c+p^5} X_{j}$. We will show that a dishonest declaration by the prover is detected with inverse polynomial probability.  
Unfortunately, since the random variables $X_{j}$ are not independent, we cannot simply use Hoeffding's inequality to argue that $X_{\sample}$ will concentrate around $\E[X_{\sample}]$.  
However, $X_{\sample}$ has support $[-1,1]$, and $\E[X_{\sample}] = \frac{1}{p^5}\sum_{j}\E[X_j]$ by linearity of expectation alone, whatever the correlations between the blocks.  Suppose $\aprov$ does not satisfy Eq.$(\ref{qma:condition_expect})$.  Applying Lemma \ref{qma:detection} with $X = X_{\sample}$, $\alpha = \aprov$ and $\epsilon = \frac{1}{4p^2}$ shows that the check in Eq.(\ref{eq:aprov-check}) fails, and thus the verifier rejects, with probability at least $\frac{\epsilon}{3} = \frac{1}{12p^2} \geq \frac{1}{p^3}$.

Suppose now that Eq.(\ref{qma:condition_expect}) does hold, and consider the first round $t=0$. Since $0\leq \lambda_r \leq 1$, we obtain the following 
    \begin{equation*}
    \begin{aligned}
        T&=-\frac{\lambda_r}{p^5}\sum\limits_{j=c+1}^{c+p^5}\E\limits_{\substack{\ket{\psi_j} \leftarrow \cD \\ \ket{\phi} \leftarrow \sigma_j }}\left[|\alpha_r^{\psi_j, \phi}|^2\right] \\
        &\leq -\lambda_r\E[X_{\sample}]^2 \\
        & \leq -\lambda_r\left(\aprov^2 - \frac{1}{p^2}-\frac{1}{4p^4}\right)\\
        &\leq W_t+\frac{1}{p^2}+\frac{1}{4p^4}.
    \end{aligned}
    \end{equation*}
The equality combines the pure-state decomposition of each $\sigma_j$ with
Eq.(\ref{eq:eta-expansion-b}), as in the first two steps of
Eq.(\ref{qma:m_hard_est_case2}).  The first inequality is
Lemma \ref{qma:mean_square}, applied with $A_j = \alpha_r^{\psi_j,\phi}$ and
$g_j = \braket{\phi,0^*|G_r}$, so that $\mu_j = \E[X_j]$ by
Eq.(\ref{qma:completeness_x}).  The second follows by squaring
Eq.(\ref{qma:condition_expect}): it gives
$|\aprov| \leq |\E[X_{\sample}]| + 1/2p^2$, and $|\E[X_{\sample}]| \leq 1$, so
$\aprov^2 \leq \E[X_{\sample}]^2 + 1/p^2 + 1/4p^4$.  The last uses
$W_t = -\lambda_r\aprov^2$.  That is $W_t \geq T - \frac{1}{p^2} - \frac{1}{4p^4}$.

Combining Case 1 and 2, and averaging over the draw of $x$,
\begin{equation*}
    \E[W_t| \cF_{t-1}] \;\geq\; R_t - \frac{1}{p^2} - \frac{1}{4p^4}
    \;\geq\; \frac{b}{|S|+2^k|L|} - \frac{1}{p^2} - \frac{1}{4p^4},
\end{equation*}
where $\cF_{t-1}$ records the outcomes of the earlier rounds and 
the second inequality follows from the no-instance condition.  
Starting from the second round, the blocks $\sigma_{c+1}, \dots, \sigma_{c+p^5}$ may be post-measurement states resulting from the earlier rounds. Since the per-round estimate derived above holds for an arbitrary block state, the bound holds after every history. Thus, as in the pure-state case, we apply Lemma~\ref{pre:azuma} with $X_t = W_t$, $[a_t,b_t] = [-1,1]$, $\mu = \frac{b}{|S|+2^k|L|} - \frac{1}{p^2} - \frac{1}{4p^4}$ and deviation $\frac{1}{2p^2}$; since $|S|+2^k|L| \leq p$, the event below implies $M \leq \mu - \frac{1}{2p^2}$, and we obtain

% Thus, as in the completeness case, Hoeffding's inequality gives
\begin{equation*}
    \Pr\Big[ (|S|+2^k|L|) \cdot M < b- \frac{2}{p}\Big] \leq \exp\Big(\frac{-p}{8}\Big).
\end{equation*}
 However, if $\aprov$ fails Eq.$(\ref{qma:condition_expect})$ for some pair $(\ell, r)$ after some history $\cF_{t-1}$, then a round draws that pair with probability at least $\frac{1}{|S|+2^k|L|} \geq \frac1p$, and conditioned on the draw the verifier rejects at line~15 of \Cref{alg:method2} with probability at least $\frac{1}{12p^2}$ by Lemma~\ref{qma:detection}; a single round therefore rejects with probability at least $\frac{1}{12p^3} \geq \frac{1}{p^4}$. Hence the soundness is at most $1-\frac{1}{p^4}+\exp(\frac{-p}{8})$, which is bounded above by $1-\frac{1}{\poly(p)}$.
 Finally, we apply the Amplification Lemma $\ref{lemma:amplification}$ to further reduce the soundness error.
\end{proof}

%% file: 4-1-2_H_qcma_complete.tex
\subsubsection{Varients of Local Hamiltonian Problems as \pmQCMA-complete}\label{qcma:section}
In the following section, we fix a universal quantum gates set $\,\mathcal{U}$. The following definition is an analog of Definition \ref{qma:p_problem} and \cite{wocjan2003two}.
\begin{definition} \label{qcma:p_problem}
    \textbf{(Low complexity low energy states for $k$-local Hamiltonian with unknown pure state problem ($k$-LLHwP))} 
    \newline\newline
    The promise problem $k$-\textbf{LLHwP} is defined as follows.
    \begin{itemize}
        \item \textbf{Inputs:} Given the input $1^p$ with $p \in \N$, $a\in \R^+_0$, $b\in\R^+_0$, and
        \begin{equation*}
            H:= \sum\limits_sH_s - \sum\limits_\ell\ket{\psi}\bra{\psi} \otimes H_\ell,
        \end{equation*}
        where $\ket{\psi}\in\cH(n)$ is an unknown state, $\{H_s\}_{s\in S}$ and $\{H_\ell\}_{\ell \in L}$ are two sets of local Hamiltonians, with each Hamiltonian acting on at most $k$ qubits. We are given the promise that $|S|+|L| \leq p$, $\;0\preceq H_x \preceq I$ for all $x \in S\cup L$, and $b-a > \frac{2}{p}$.

        \item \textbf{Yes instances:} If there exist $m \leq p$ and a sequence of 2-qubit elementary gates $(U_i)_{i=1}^{m} \in \mathcal{U}^{\times m}$ such that 
        \begin{equation*}
            \ket{\eta}:= U_{m}\cdots U_1 \big( \ket{\psi}_I\otimes \ket{0^*}_{WA}\big )
        \end{equation*}
        is a state with energy less than $a$, i.e.
        \begin{equation*}
            \bra{\eta} H \ket{\eta} \leq a.
        \end{equation*}
        \item \textbf{No instances:}
            Let \textbf{Bad} denote the set of all quantum states $\ket{\eta}$ that can be constructed using at most $p(n)$ elementary gates and a single copy of the input state $\ket{\psi}$. We are promised that $\ket{\eta}$ is a state with high energy, i.e.,
            \begin{equation*}
                \min\limits_{\ket{\eta} \in \textbf{Bad} } \bra{\eta} H \ket{\eta} \geq b.
            \end{equation*}
    \end{itemize}
\end{definition}
\begin{theorem}\label{qcma:p_complete}
    5-$\textbf{LLHwP}$ is $\pQCMA$-complete.
\end{theorem}
The proof follows directly from Lemma~\ref{qcma:p_c_in} and Lemma~\ref{qcma:p_hard}, where we will show that 5-$\textbf{LLHwP}$ is in $\pQCMA$ (Lemma~\ref{qcma:p_c_in}) and that 5-$\textbf{LLHwP}$ is hard for $\pQCMA$ (Lemma~\ref{qcma:p_hard}).

\begin{lemma}\label{qcma:p_c_in}
    For all constant $k \in \N$, $k$-\textbf{LLHwP} $\in$ \pQCMA.
\end{lemma}
\begin{proof}
    The prover sends $(U_i)_{i=1}^m \in \mathcal{U}^{\times m}$ to the verifier. Consequently, the verifier can efficiently generate polynomially many copies of $\ket{\eta}$ using a polynomial number of inputs. By Lemma~\ref{qma:p_in}, with a polynomial number of inputs, we can compute $\bra{\eta} H \ket{\eta}$ with an additive error of $\frac{1}{p}$ with overwhelming probability. 
\end{proof}

The following lemma is similar to that in \cite{wocjan2003two}. 
\begin{lemma}[Sketch proof in Appendix \ref{Appendix:p_c_hard}]\label{qcma:p_hard}
    5-$\textbf{LLHwP}$ is $\pQCMA$-hard.
\end{lemma}

The following definition is analogous to Definition~\ref{qma:m_problem} and Definition~\ref{qcma:p_problem}.
\begin{definition}\label{qcma:m_problem}
    \textbf{(Low complexity low energy states for $k$-local Hamiltonian with unknown mixed sate problem ($k$-LLHwM))} 
    \newline\newline
    The promise problem $k$-\textbf{LLHwM} is defined as follows.
    \begin{itemize}
        \item \textbf{Inputs:} Given the input $1^p$ with $p \in \N$, $a \in \R^+_0$, $b \in \R^+_0$, an unknown input state $\rho \in \cD(\cH(n))$, and two sets of local Hamiltonians $\{H_s\}_{s\in S}$ and $\{H_\ell\}_{\ell \in L}$, with each Hamiltonian acting on at most $k$ qubits. We are given the promise that $|S|+2^k|L| \leq p$, $\;0\preceq H_x \preceq I$ for all $x \in S\cup L$, and $b-a > \frac{4}{p}$. Additionally, let $\cD_\rho$ denote the decomposition set of $\rho$. For any pure-state $\ket{\psi}$, define 
        \begin{equation*}
            H_\psi:= \sum\limits_sH_s - \sum\limits_\ell\ket{\psi}\bra{\psi}_I \otimes H_\ell.
        \end{equation*}
        \item \textbf{Yes instances:} There exists $m\leq p$, a sequence of 2-qubit elementary gates $(U_i)_{i=1}^{m} \in \mathcal{U}^{\times m}$, $\{\alpha \in \R\}$, for all $\cD\in\cD_{\rho}$ such that the following holds: Let $\ket{\eta_{\psi}}:= U_{m}\cdots U_1 \left( \ket{\psi}_I\otimes \ket{0}_{WA}\right)$, where $\ket{\eta_{\psi}}$ has the same number of qubits as those on which $H_\psi$ acts.
        \begin{enumerate}
            \item (Expected small eigenvalue):
            \begin{equation*}
                \E\limits_{\ket{\psi} \leftarrow \mathcal{D}} \big[\langle \eta_{\psi}| H_{\psi} |\eta_{\psi} \rangle \big]  \leq a,
            \end{equation*}
            \item (Uniform Initialization): For all $\ell\in L$, let $\{v_i\}$ be any eigenbasis and $\{ \lambda_i\}$ be the corresponding eigenvalue of $H_\ell$. For all $\ket{\psi} \in \mathcal{D}$,  $|\eta_{\psi}\rangle$ has the form
            \begin{equation}\label{qcma:restrict}
                \sum\limits_{i:\;\lambda_i \neq 0}\Big(\alpha\ket{\psi}_I\ket{\phi}_W\ket{0^*}\ket{v_i} \Big ) +  \sum\limits_{i:\;\lambda_i \neq 0}\Big(\ket{\psi^{\perp}}_I\ket{*} \ket{v_i}\Big )+\sum\limits_{i:\;\lambda_i = 0}|*\rangle|v_i\rangle,
            \end{equation}
            where the unimportant amplitudes of the last two terms are implicitly represented by $\ket{*}$. Additionally,
            \begin{equation*}
                \ket{\psi}_I\ket{\phi\hspace{1pt} 0^*}^\perp\ket{v_i},
            \end{equation*}
            in Equation ($\ref{qcma:restrict}$) has amplitude 0 when $\lambda_i\neq 0$.
            \item (Restriction on $H_\ell$): For all $
            \ell \in L$, $H_\ell$ acts on the last $k$ qubits, disjoint from registers $I$ and $W$\fnref{footnote:mqma_restriction}.
            \end{enumerate}
        \item \textbf{No instances:} For all $\cD \in \cD_{\rho}$, the expected minimum eigenvalue is large. Specifically, let \textbf{Bad} denote the set of all quantum states $\ket{\eta}$ that can be constructed using at most $p(n)$ elementary gates and a single copy of the input state $\ket{\psi}$. We are promised that $\ket{\eta}$ is a state with high energy, i.e.,
        \begin{equation}\label{qcma:m_problem_no}
            \E\limits_{\ket{\psi}\leftarrow \mathcal{D}}\Big[\min\limits_{\ket{\eta} \in \textbf{Bad} } \langle \eta| H_{\psi} |\eta \rangle \Big] \geq b.
        \end{equation}
    \end{itemize}
\end{definition}
\begin{theorem}\label{qcma:m_complete}
    5-$\textbf{LLHwM}$ is \mQCMA-complete.
\end{theorem}
\begin{proof}
    First, we aim to prove that for all constants $k \in \N$, $k$-\textbf{LLHwM} $\in$ $\mQCMA$. The proof utilizes the same techniques from  Lemma~\ref{qma:m_in} and Lemma~\ref{qcma:p_c_in}. Second, we aim to prove that 5-\textbf{LLHwM} is \mQCMA-hard, using the same techniques from Lemma~\ref{qma:m_hard} and Lemma~\ref{qcma:p_hard}.
\end{proof}

%% file: 4-4_qma_ubound.tex
\subsection{Upper bound of \pmQMA}\label{qma:section_upper}

\begin{theorem}\label{thm:QMA_Upperbound}
    $\pmQMA \subseteq \pmPSPACE$.
\end{theorem}

\begin{proof}
    For simplicity, we considered only the mixed-state version as the proof for the pure-state version is similar. It suffices to show $\cL:= \textbf{mQO} \in \mPSPACE$ since \textbf{mQO} is $\mQMA$-complete. Consider an instance $\rho_{in}:= (\rho, 1^m, \Lambda)$ of $\cL$. By Theorem \ref{thm:QOR}, there exist an algorithm that accepts $\rho_{in}\in \mathcal{L}_{Y}$ with probability at least $(\frac{2}{3})^2 \cdot \frac{1}{7} = \frac{4}{63}$, and accepts $\rho_{in}\in \mathcal{L}_{N}$ with probability at most $4\cdot2^m\cdot\frac{1}{64\cdot2^m}=\frac{4}{64}$. Suppose the algorithm uses polynomial space. Then we have $\textbf{mQO} \in \mPSPACE_{\frac{4}{63}, \frac{4}{64}}$. By the amplification Lemma \ref{lemma:amplification}, we have  $\mPSPACE_{\frac{4}{63}, \frac{4}{64}} = \mPSPACE$. As a result, $\textbf{mQO} \in \mPSPACE$, and thus we have $\mQMA \subseteq \mPSPACE$.
    
    Now, we show that the algorithm uses polynomial space. Let us recall the notation from Theorem \ref{thm:QOR}. For $i \in [N-1]_0$, define a 2-outcome projective measurement $\Lambda_i$ on the Hibert space $\mathscr{H}_A \otimes \mathscr{H}_B$ as follow: (i) apply $I_A \otimes {X_i}_B$, where $X_i: \ket{a} \rightarrow \ket{a \oplus i}$; (ii) apply $\Lambda$; and (iii) apply $I_A \otimes {X_i}^\dagger_B$. Define $\Pi:= \sum\limits_{i=0}^{N-1} (\Lambda_i)_{AB} \otimes \ket{\hat{i}}\bra{\hat{i}}_C$, where $\{\ket{\hat{i}}\}$ is the Fourier basis. Also, define $\Delta := I_{AB} \otimes \ket{0}\bra{0}_C$. We emphasize that $\Lambda$ is a polynomial-time operator. We will show that the projective measurements $\{\Pi,I-\Pi\}$ and  $\{\Delta,I-\Delta\}$ run in polynomial time. Since these projective measurements will repeat $\left\lceil \frac{3}{2}\cdot 2^m \right\rceil$ times, the algorithm uses polynomial space. The latter case is straightforward, so we focus only on $\{\Pi,I-\Pi\}$. The measurement $\{\Pi,I-\Pi\}$ is equivalent to the following algorithm, which is efficient: 
    \begin{enumerate}
        \item Apply the Fourier transform $Q$ on register $C$.
        \item Apply CNOT on registers $CB$ (Conditioned on $C$ and apply NOT gate on $B$).
        \item Apply $\Lambda$ on registers $AB$ (i.e., apply a polynomial sequence of unitaries, followed by a computational basis measurement, and then the conjugate of the previous sequence of unitaries.).
        \item Apply CNOT on registers $CB$ .
        \item Apply the conjugate Fourier transform $Q^*$ on register $C$. 
    \end{enumerate}
    Indeed, consider an arbitrary pure state $\sum\limits_{x,i} \ket{x}_{AB}\ket{\hat{i}}_C$. Then,
    \begin{equation}\label{qma:ubound_algo}
        \Pi \cdot \sum\limits_{x,i} \ket{x}_{AB}\ket{\hat{i}}_C = \sum\limits_{x,i} \Lambda_i \ket{x} \otimes \ket{\hat{i}}.
    \end{equation}
    Running the above algorithm (without normalization) to $\sum\limits_{x,i} \ket{x}_{AB}\ket{\hat{i}}_C$ yields the same result as in Equation $(\ref{qma:ubound_algo})$. Therefore, the measurement is equivalent to the algorithm.
    
\end{proof}

%% file: 4-5_qcma_StoDreduction.tex
\subsection{Search-to-decision reductions for \pmQCMA}\label{section:StoD}

We show how to use a $\pmQCMA$ oracle to find a good witness of yes instance in $\pmQCMA$. The proof is essentially the same for pure-state or mixed-state promise problems. Hence, we consider pure-state promise problems only. The $\mathcal{L}_{StoD}$ structure in Equation $(\ref{qma:StoDLan})$ is similar to classical language design for $\NP$-type search-to-decision reductions. However, $\NP$-type search-to-decision reduction deals with languages, not promise problems. Therefore, input instances never ``drop" outside the promise, but this is not true for $\pmQCMA$. To address this issue, we modify the size of ``yes" region and ``no" region for queries with different lengths of prefixes. Precisely, the ``yes" region of longer prefix lengths always encompasses the complement of ``no" region of shorter prefix lengths.\footnote{\cite{Aar20} also use the similar technique.}

\begin{lemma}[Finding the witness for $\pQCMA$ problems] \label{lem:searchToDecision_version1}
Consider a pure-state quantum promise problem $\cL:=(\cL_Y,\cL_N)$. Suppose there exists a $\pQCMA$ protocol $\braket{\cP, \cV}$ that decides $\cL$ with completeness $a$ and soundness $a-\frac{1}{\poly(\cdot)}$. Then there exists a quantum promise problem $\cL_{StoD} \in \pQCMA$, which may depend on $\cV$, and an efficient oracle algorithm $\cA$ such that, for all $\lambda$-qubit $\ket{\psi}\in\mathcal{L}_Y$, $\cA^{\cL_{StoD}}$ finds a witness of $\ket{\psi}$ corresponding to the verifier $\cV$. Specifically, for all $c\in\N$, there exists polynomials $t(\cdot)$ and $p(\cdot)$ such that the following holds:
\begin{equation*}
    \text{for sufficiently large } \lambda,\, \Pr\left[\cV\left(w, \ket{\psi}^{\otimes t(\lambda)}\right)=1: w \leftarrow \mathcal{A}^{\mathcal{L}_{StoD}}\left(\ket{\psi}^{\otimes p(\lambda)}\right)\right] \geq a-\frac{1}{\lambda^c}.
\end{equation*}

\end{lemma}
\begin{proof}
Suppose that $\cV$ uses $t(\lambda)$ copies of the input and the witness register have size $n(\lambda)$. Since $\cV$ runs in polynomial time, $t(\lambda)$ and $n(\lambda)$ are polynomial bounded. We denote $t(\lambda)$ and $n(\lambda)$ to $t$ and $n$, respectively. Define
$\mathcal{L}_{StoD}:= \big(\mathcal{L}^*_{Y}, \mathcal{L}^*_{N}\big)$ as follow:
\begin{equation}\label{qma:StoDLan}
    \begin{aligned}
        &\mathcal{L}^*_{Y}:=\left\{\ket{\psi},x,\#^{n - |x|} : \;\exists y \in \{0,1\}^{n - |x|} \;s.t \;\Pr \left[ \cV(x\|y, \ket{\psi}^{t(\lambda)}) = 1 \right] \geq a - \frac{|x|} {2n\lambda^{c}}\right\} \\
        &\mathcal{L}^*_{N}:=\left\{\ket{\psi},x,\#^{n - |x|} : \; \forall y \in \{0,1\}^{n - |x|} \;s.t \;\Pr \left[ \cV(x||y, \ket{\psi}^{t(\lambda)}) = 1 \right] < a - \frac{|x|+1}{2n\lambda^{c}}\right\},
    \end{aligned}
\end{equation}
where $\cL^*_Y$ collects all $\ket{\psi}\in \cL_Y$ and $x\in \{0,1\}^*$ with $1 \leq |x|\leq n$. Here, $x$ represents the prefix of the witness, and the symbol $\#$ pads different lengths of prefixes to the length $n$.

First, we show that $\mathcal{L}_{StoD} \in \pQCMA$. The proof follows a similar approach to the amplification  Lemma~\ref{lemma:amplification}. The prover sends a witness $y$ to the verifier, and the verifier runs $\mathcal{V}$ with the inputs and the witness $\alpha n^2\lambda^{2c}$ times (for some sufficiently large constant $\alpha$). If the fraction of acceptances, $\frac{\# \text{ of accepts}}{\alpha n^2\lambda^{2c}}$, is at least $a-\frac{1}{2n\lambda^{c}}(|x|+\frac{1}{2})$, the verifier accepts; otherwise, it rejects. By the Chernoff bound, this shows that $\mathcal{L}_{StoD} \in \pQCMA$.

Next, we define the oracle algorithm $\mathcal{A}^{\mathcal{L}_{StoD}}$ in Algorithm~\ref{alg:searchToDecision}. This algorithm is well-defined because, by Therorem~\ref{thm:QMA_Upperbound}, we know that $\cL_{StoD}$ is physically realizable (see Section~\ref{Def:Oracle}).
\begin{algorithm}[H]
    \caption{$\mathcal{A}^{\mathcal{L}_{StoD}}$}
    \label{alg:searchToDecision}
    \begin{algorithmic}[1]
        \REQUIRE $\ket{\psi}^{\otimes p(\lambda)}$, where $p(\lambda) = nt'$, with $t'$ representing the number of copies required for querying the oracle $\mathcal{L}_{StoD}$.
        \ENSURE
        Find $w_n \in \{0,1\}^{n}\;$ s.t $\;\Pr[\mathcal{V}(w_n, \ket{\psi}^{t(\lambda)}) = 1] \geq a - \frac{n+1}{2n\lambda^{c}} \geq a-\frac{1}{\lambda^{c}}$.
        \STATE
        Initialize $w_0 = \phi$ (an empty string).
        \FOR{$i \in [n]$}
            \STATE $w^{test}_i \leftarrow w_{i-1}\|0$.
            \STATE
            Query $\cL_{StoD}$ with input $(\ket{\psi},w^{test}_i, \#^{n - i})^{\otimes t'}$ and receive outcome $b$.
            \IF{$b = 0$ (i.e., the oracle return ``no") }
                \STATE $w_i \leftarrow w_{i-1}\|1$.
            \ELSE
                \STATE $w_i \leftarrow w^{test}_i$.
            \ENDIF
        \ENDFOR
        \RETURN{$w_n$}
    \end{algorithmic}
\end{algorithm}

Each query to $\mathcal{L}_{StoD} \in \pQCMA$ requires a polynomial number of copies; therefore, both $t'$ and $p$ are polynomial. Now, we show that the algorithm will output $w_n$ such that $\Pr[\mathcal{V}(w_n, \ket{\psi}^{t(\lambda)}) = 1] \geq a - \frac{n+1}{2n\lambda^{c}}$. Fix an arbitrary $\ket{\psi} \in \mathcal{L}_Y$. We define a good set $\mathbf{Good}$ as follows:
\begin{equation*}
    \mathbf{Good}:=\Big\{x \in \{0,1\}^n \text{ with } 0\leq |x|\leq n:\exists y \in \{0,1\}^{n - |x|}\; s.t \; \Pr[\mathcal{V}(x\|y, \ket{\psi}^{t(\lambda)}) = 1] \geq a - \frac{|x|+1}{2n\lambda^{c}} \Big\}.
\end{equation*}
    
Then, $\mathbf{Good}$ satisfies following properties:
\begin{enumerate}
    \item \label{Good_1} The empty string $\phi \in \mathbf{Good}$ follows from the fact that $\ket{\psi} \in \mathcal{L}_Y$.
    \item \label{Good_2} For $|x| < n$, 
    \begin{equation*}
        x \in \mathbf{Good} \implies (\ket{\psi},x\|0,\#^{n - |x|-1}) \in \mathcal{L}^*_{Y} \;\lor\; (\ket{\psi},x\|1,\#^{n - |x|-1}) \in \mathcal{L}^*_{Y}.
    \end{equation*}
    \item \label{Good_3} If $(\ket{\psi},x,\#^{n - |x|}) \nin \mathcal{L}^*_{N}$, then $x \in \mathbf{Good}$.
\end{enumerate}

It is sufficient to show that for all $i \in [n]_0$, $w_i \in \mathbf{Good}$. By Property $\ref{Good_1}$, $w_0 \in \mathbf{Good}$, which is a partial witness $\cA$ determine so far. Suppose $w_i \in \mathbf{Good}$; we will now show that $w_{i+1} \in \mathbf{Good}$. This implies that $w_n \in \mathbf{Good}$, and the proof is complete. Suppose that when we query $\mathcal{L}_{StoD}$ with the input $(\ket{\psi},w^{test}_{i+1}, \#^{n - i-1})^{\otimes t'}$, we receive the outcome $b=0$. By Property \ref{Good_2}, $(\ket{\psi},w_{i}\|1, \#^{n - i-1})^{\otimes t'} \in \mathcal{L}^*_{Y}$. Hence, by Property \ref{Good_3}, $w_{i+1}=w_i\|1 \in \mathbf{Good}$. On the other hand, if we receive $b = 1$, then $(\ket{\psi},w^{test}_{i+1}, \#^{n - i-1})^{\otimes t'} \nin \mathcal{L}^*_{N}$. Again, by Property \ref{Good_3}, we conclude that $w_{i+1}=w^{test}_{i+1}\in\mathbf{Good}$.
\end{proof}

\begin{corollary}[Search-to-decision reduction for $\pQCMA$]\label{lem:searchToDecision_version2}
    Consider a $\pQCMA$-complete quantum promise problem $\cL$, corresponding to a protocol $\braket{\cP, \cV}$. There exists an efficient oracle algorithm $\cA^\cL$ that finds a witness, corresponding to the verifier $\cV$, for any ``yes" instance in $\cL$.
\end{corollary}
\begin{proof}
    By Lemma~\ref{lem:searchToDecision_version1}, there exists a quantum promise problem $\cL_{StoD}$ and an efficient oracle algorithm $\cA^{\cL_{StoD}}$ such that $\cA^{\cL_{StoD}}$ finds a witness, corresponding to the verifier $\cV$, for any ``yes" instance in $\cL$. Since $\cL$ is $\pQCMA$-complete, any oracle query to $\cL_{StoD}$ can be Karp-reduced to $\cL$. Therefore, we conclude that $\cA^{\cL}$ also finds a witness for any ``yes" instance in $\cL$.
\end{proof}

In fact, the prover can find a witness for $\pmQCMA$ (with some loss) using only a polynomial number of copies of the input. This implies that the following variants of $\pmQCMA$ are equivalent.
\begin{definition}
    Let $\mathcal{C} \in \{\pQCMA,\mQCMA\}$. Define the classes $\mathcal{C}^{\poly}$ and $\mathcal{C}^{\poly}_{c(\lambda),s(\lambda)}$ similarly to $\mathcal{C}$ and $\mathcal{C}_{c(\lambda),s(\lambda)}$, respectively. The only difference is that, in the case of yes instances, the prover can find a witness using only a polynomial number of copies. Note that soundness still holds when the prover can access an unbounded number of copies. Consequently, we obtain $\mathcal{C}^{\poly} \subseteq \mathcal{C}$.
\end{definition}

\begin{corollary}\label{thm:qcmaR}
    $\pmQCMA^{\poly} = \pmQCMA$.
\end{corollary}
\begin{proof}
    The proofs for pure-state and mixed-state versions are equivalent. Therefore, we consider only the pure-state promise problem. By definition, $\pQCMA^{\poly} \subseteq \pQCMA$. Suppose $\mathcal{L} \in \pQCMA= \pQCMA_{\frac{3}{4},\frac{1}{4}}$, and let $\braket{\cP,\cV}$ denote a $\pQCMA_{\frac{3}{4},\frac{1}{4}}$ protocol that decides $\cL$. By Lemma~\ref{lem:searchToDecision_version1} with respect to the verifier $\cV$ and Theorem~\ref{thm:QMA_Upperbound}, the promise problem $\cL_{StoD}$ (as defined in Lemma~\ref{lem:searchToDecision_version1}) is decidable using a polynomial number of input copies.
    Furthermore, by Lemma~\ref{lem:searchToDecision_version1}, there exists a polynomial-time algorithm $\cA$ such that $\cA^{\cL_{StoD}}$ finds a witness for any ``yes" instance in $\cL$, corresponding to the verifier $\cV$. The prover then simulates the algorithm $\cA^{\cL_{StoD}}$ and sends the output to the verifier. We conclude that the prover finds a witness using a polynomial number of input copies, allowing the verifier to accept with probability $\frac{2}{3}$ in a yes instance. Therefore, $\mathcal{L} \in \pQCMA^{\poly}_{\frac{2}{3},\frac{1}{4}+\negl(\cdot)}= \pQCMA^{\poly}$.
\end{proof}

%% file: 7_Uncond_separation.tex
\section{Unconditional Separation}

In this section, we prove several separation results. We first review two types of complexity classes, which were defined in the previous section. The first type consists of $\pINF$ and $\mINF$ (Definition~\ref{Def:physically_realizable}), which contain all pure and mixed QPPs that can be decided by a computationally unbounded single-party algorithm given only a polynomial number of copies of the input state. The second type consists of $\pQSZKhv$, $\mQSZKhv$, $\pQIP$, and $\mQIP$ (Definition~\ref{Def:pQIP},~\ref{Def:pQSZKhv},~\ref{Def:mQIP},~\ref{Def:mQSZKhv} and~\ref{Def:simplify}), which capture interactive proofs and zero knowledge for pure and mixed QPPs. One important point to emphasize is that throughout this section we only focus on the cases in which the prover has a full description of the input state (equivalently, access to an unbounded number of copies of the input state).

Next, we outline the results established in each subsection. In Sections~\ref{subsec:sep_mQSZK_and_mINF},~\ref{subsec:sep_pQSZK_and_pINF}, and~\ref{subsec:sep_pBQPpoly_and_mBQPpoly}, we prove the following three separations: (i) $\mQSZKhv \not\subseteq \mINF$. (ii) $\pQSZKhv \not\subseteq \pINF$ (iii) $\pBQPqpoly \not\subseteq \pBQPpoly$.In Section~\ref{subsec:sep_mQIP_and_mcoQIP}, we show that neither $\mQIP$ nor $\mQSZKhv$ is closed under complement. In Section~\ref{subsec:sep_property_testing}, we extend the separations between $\pQSZKhv$ and $\pINF$, and between $\mQSZKhv$ and $\mINF$, to some well-studied quantum property testing problems. Finally, in Section~\ref{subsec:imposs_unitary_synth}, we establish an interesting connection between separations of $\pINF$ from $\pQIP$, or of $\mINF$ from $\mQIP$, and the impossibility of quantum-input unitary synthesis problems.

Finally, in addition to the separation results, we establish several structural results for $\pQSZKhv$. For example, we show that $\pQSZKhv$ is closed under complement, in contrast to the behavior of $\mQSZKhv$. We also identify a natural complete problem for $\pQSZKhv$ and show that every QPP in $\pQSZKhv$ admits a public-coin protocol. Interested readers may refer to Appendix~\ref{Appendix:pQSZK_hv} for formal statements and proofs.

,

\subsection{Separation between $\mQIP$ and $\mPSPACE$} \label{subsec:sep_mQSZK_and_mINF}

This subsection will show that $\mQIPsec \not\subseteq \mPSPACE$. Specifically, we show a stronger theorem, which is stated as follows:
\begin{theorem}\label{thm:mQSZKhv_not_inside_mINF}
    $\mQSZKhvsec \not \subseteq\mINF$. That is, $\mQSZKhvsec$ cannot be regarded as a physically realizable oracle (Definition~\ref{Def:physically_realizable}).
\end{theorem}
Indeed, $\mQSZKhvsec \subseteq \mQIPsec$ and $\mPSPACE \subseteq \mINF$ give the following corollary.
\begin{corollary}
    $\mQIPsec \not \subseteq \mPSPACE$.
\end{corollary}

To prove theorem \ref{thm:mQSZKhv_not_inside_mINF}, we use the sample complexity lower bound from testing the mixedness property (testing a state whether it is close to or far from a totally mixed state)~\cite{CHW07,MdeW16, o2021quantum}. For our purposes, we only need specific parameters, which we state as follows.
\begin{theorem}[\cite{o2021quantum}, restate]\label{thm:totally_mix_state_hardess_half}
    For any polynomial $q(\cdot)$ and all sufficiently large $\lambda$, any algorithm give $q(\lambda)$ copies of state $\rho$. The state $\rho$ is either $\rho = \frac{I}{2^{\lambda}}$ or $\rho$ is maximally mixed on a uniform random subspace of dimension $2^{\lambda-1}$. The algorithm’s advantage in distinguishing these two cases is at most $\frac{q(\lambda)}{2^\lambda}$. \footnote{The upper boud $q(\lambda)/2^{\lambda}$ is derived from the proof of \cite{o2021quantum} Corollary 4.3 by taking $\epsilon = \frac{1}{2},n = q(\lambda), d= 2^{\lambda}$, where $q(\cdot)$ is some polynomial.}
\end{theorem}

% We state one of the results as follows.
% \begin{theorem}[\cite{o2021quantum}, restate]\label{thm:totally_mix_state_hardess}
%     Any algorithm that distinguishes, with a probability of success at least $\frac{2}{3}$, between two cases that $\rho = \frac{I}{2^{\lambda}}$ or $\rho$ is maximally mixed on a uniform random subspace of dimension $2^r$ ($1 \leq r \leq \lambda-1$), must use ${\theta}(2^r)$ copies of $\rho$.
% \end{theorem}

We define the following notation\(\rho_{\halfs_\lambda}:= \frac{1}{2^{\lambda-1}} \sum_{i \in \{0,1\}^{\lambda-1}}\ket{i||0}\bra{i||0}.\) The mixed state $\rho_{\halfs_\lambda}$ is equivalent to the classical distribution over $\lambda$-bit strings obtained by first sampling $i \leftarrow \{0,1\}^{\lambda-1}$ uniformly at random and then outputting $i||0$. Then, we define the quantum promise problem of the hard instance mentioned above.
\begin{definition}[$\mathcal{L}_{mix^*}$]\label{def:L_mix_constant}
Define a quantum promise problem $\cL_{mix^*}:=(\cL_Y:= \bigcup\limits_{\lambda \in \N} \cL_{mix^*,Y}^{\lambda}, \cL_N:= \bigcup\limits_{\lambda \in \N}\cL_{mix^*,N}^{\lambda})$, where 
\begin{itemize}
    \item  $\mathcal{L}_{mix^*,Y}^{\lambda}:=\{ T\rho_{\halfs_\lambda}T^{\dagger}: T \text{ is arbitrary $\lambda$-qubit untaries} \}$.
    \item $\mathcal{L}_{mix^*,N}^{\lambda}:=\{ \frac{I}{2^\lambda} \}$.
\end{itemize}
\end{definition}

% We define the quantum promise problem of the hard instance mentioned above.
% \begin{definition}[$\mathcal{L}_{mix,\epsilon(\cdot)}$]\label{def:L_mix_constant}
% Define a quantum promise problem $\cL_{mix,\epsilon(\cdot)}:=(\cL_Y:= \bigcup\limits_{\lambda \in \N} \cL_{mix,Y}^{\lambda}, \cL_N:= \bigcup\limits_{\lambda \in \N}\cL_{mix,N}^{\lambda})$, where 
% \begin{itemize}
%     \item  $\mathcal{L}_{mix,Y}^{\lambda}:=\{\rho : \text{TD}(\rho,I_{H(\lambda)}) = \epsilon(\lambda) \}$.
%     \item $\mathcal{L}_{mix,N}^{\lambda}:=\{ I_{\cH(\lambda)} \}$.
% \end{itemize}
% \end{definition}
% \begin{definition}[$\mathcal{L}_{mix,\geq\epsilon(\cdot)}$]\label{def:L_mix_range}
%     The quantum promise problem $\mathcal{L}_{mix,\geq\epsilon(\cdot)}$ follows the same definition as $\mathcal{L}_{mix,\epsilon(\cdot)}$, except that $\mathcal{L}_{mix,Y}^\lambda$ is defined as follows:
%     \begin{equation*}%\label{equ:language_mix}
%         \mathcal{L}_{mix,Y}^{\lambda}:=\{\rho : \text{TD}(\rho,I_{H(\lambda)}) \geq \epsilon(\lambda) \}.
%     \end{equation*}
% \end{definition}

% It is easy to see Theorem~\ref{thm:totally_mix_state_hardess_half} can be restate as an average case hardness of $\cL_{mix^*}$. We immediately get the worst case hardness of $\cL_{mix^*}$ as follow.

% \begin{lemma}\label{lemma:mix_notin_mINF}
%    $\cL_{mix^*} \notin \mINF$. 
% \end{lemma}

It is easy to see that Theorem~\ref{thm:totally_mix_state_hardess_half} can be restate as an average case hardness of $\cL_{mix^*}$, but we show that $\cL_{mix^*}$ can be solved in the worst case with the help of the prover. We state the results as follows. 
\begin{lemma}\label{lem:mix_in_mqszkhv}
    $\mathcal{L}_{mix^*} \in \mQSZKhvsec$.
\end{lemma}
\begin{proof}
    %The following protocol is similar to Lemma \ref{lemma:UNQSK_in_pQSZK}.
    We present the $\mQSZKhvsec$ protocol for $\mathcal{L}_{mix^*}$ as follows.  
\begin{protocal}{
 Mixed State Honest Verifier Statistical Zero-Knowledge Protocal for $\mathcal{L}_{mix^*}$
}
\begin{description}

\item[Verifier's step 1:]\quad\\
Let $\lambda$ be the number of qubits of input state $\rho_{in}$. Receive $\lambda^2$ copies of input state $\rho_{in}$. Sample a uniformly random string $m \in \{0,1\}^\lambda$. Consider $\lambda$ registers $A_1\cdots A_\lambda$, each of size $\lambda^2$. For $i \in [\lambda]$,
if $m_i = 0$, let $A_i = \rho_{in}^{\otimes \lambda}$. Otherwise, let $A_i = (\frac{I}{2^\lambda})^{\otimes\lambda}$. Sends the registers $A_1\cdots A_\lambda$ to the prover.

\item [Prover's step 1:]\quad\\
For $i \in [\lambda]$, apply optimal measurement for $(\rho_{in}^{\otimes\lambda} , (\frac{I}{2^\lambda})^{\otimes\lambda})$ on register $A_i$ and get result $m'_i$. Send $m'_1\cdots m'_\lambda$ back to the verifier.

\item [Verifier's step 2:]\quad\\
The verifier accepts if for all $i \in [\lambda], m_i = m_i'$. Otherwise, the verifier rejects it.
\end{description}
\end{protocal}

Suppose $\rho_{in} \in \cL_Y$, we know that $\text{TD}(\rho_{in}, \frac{I}{2^\lambda}) = \frac{1}{2}.$ By standard calculation, we get $\text{TD}(\rho_{in}^{\otimes \lambda}, ( \frac{I}{2^\lambda})^{\otimes \lambda}) = 1 - \negl(\lambda)$. Since the prover gets unbounded copies of input, the optimal measure obtains correct $m'_i$ with probability at least $1 - \negl(\lambda)$. By union bound, the verifier accepts with probability at least $1- \lambda\cdot\negl(\lambda) = 1 - \negl(\lambda)$. Suppose $\rho_{in} \in \cL_N$, then the value of $A_i$ is independent to $b$. Then, the verifier accepts with probability at most $(\frac{1}{2})^\lambda =\negl(\lambda)$. 

For the statistical zero-knowledge property, we allow the simulator to receive $\lambda^2$ copies of the input, just like the verifier. The simulator can perfectly simulate the first message by executing the verifier's first step. For the second message, the simulator runs the verifier's first step and let for all $i \in [\lambda]$, $m_i' = m_i$. In the honest case, the prover will send the correct $m_i'$ with probability $1-\negl(\lambda)$, so the error at most $\negl(\lambda)$ for each $i$. Lastly, by applying the union bound over all $i\in [t]$, the total simulation error remains at most $\negl(\lambda)$. 
\end{proof}

\begin{proof}[Proof of Theorem \ref{thm:mQSZKhv_not_inside_mINF}]
 To separate $\pINF$ and $\pQSZKhvsec$, we need to show that there exists some quantum promise problem $\cL$ such that it is easy for $\pQSZKhvsec$ and hard for $\pINF$. Here, we let $\cL := \cL_{mix^*}$. The easiness result follows by Lemma~\ref{lem:mix_in_mqszkhv} and the hardness result comes from Theorem~\ref{thm:totally_mix_state_hardess_half} since average case hardness trivially imply worst case hardness of $\cL_{mix^*}$. Then we get the separation. 
\end{proof}

\subsection{Separation between $\pQIP$ and $\pPSPACE$} \label{subsec:sep_pQSZK_and_pINF}

In this subsection, we will extend $\mQIPsec \not \subseteq \mPSPACE$ to $\pQIPsec \not \subseteq \pPSPACE$. In particular, we show Theorem \ref{thm:mQSZKhv_not_inside_mINF} can be extended to the pure state version. Here is our main theorem.
\begin{theorem}\label{thm:pQSZKhv_not_inside_pINF}
    $\pQSZKhvsec \not \subseteq \pINF$. That is, $\pQSZKhvsec$ cannot be regarded as a physically realizable oracle (Definition~\ref{Def:physically_realizable}).
\end{theorem}
The fact that $\pQSZKhvsec \subseteq \pQIPsec$ and $\pPSPACE \subseteq \pINF$ give the following corollary.
\begin{corollary}
    $\pQIPsec \not \subseteq \pPSPACE$.
\end{corollary}
\begin{remark}
Theorem~\ref{thm:pQSZKhv_not_inside_pINF} also implies that $\pQSZKhvsec$ and $\pQIPsec$ are not physically realizable oracles (Definition~\ref{Def:physically_realizable}). Section~\ref{sec:physical_realizable} provides a detailed discussion of the condition for a physically realizable oracle.
\end{remark}
Before we define our hard instance, we define two additional notations $\ket{\epr_\lambda}:=\frac{1}{\sqrt{2^\lambda}}\sum\limits_{i \in \{0,1\}^\lambda}\ket{i}\ket{i}$ and $\ket{\halfs_\lambda}:=\frac{1}{\sqrt{2^{\lambda-1}}}\sum\limits_{i \in \{0,1\}^{\lambda - 1}}\ket{i\|0}\ket{i\|0}$, which will also be useful in the later subsections. Now, we introduce our hard instance the purified version of $\mathcal{L}_{mix^*}$. We formally state as follows.
\begin{definition}[$\mathcal{L}_{purifyMix^*}$]\label{def:L_purifymix_constant}
A quantum promise problem $\mathcal{L}_{purifyMix^*}:=\bigcup\limits_{\lambda \in \N}(\cL_Y:=\mathcal{L}_{purifyMix^*,Y}^{\lambda}, \cL_N:=\mathcal{L}_{purifyMix^*,N}^{\lambda})$
\begin{itemize}
    \item $\mathcal{L}_{purifyMix^*,Y}^{\lambda}:=\left\{ (T_1 \otimes T_2)\ket{\halfs_\lambda}:T_1,T_2 \text{ are arbitrary $\lambda$-qubit unitaries} \right\}$.
    \item $\mathcal{L}_{purifyMix^*,N}^{\lambda}:=\left\{(I \otimes T)\ket{\epr_\lambda)} : T \text{ is arbitrary $\lambda$-qubit unitary}\right\}$.

\end{itemize}
\end{definition}

\begin{lemma}\label{lem:L_purifymix_in_pQSZKhv}
    $\mathcal{L}_{purifyMix^*} \in \pQSZKhvsec$.
\end{lemma}
\begin{proof}
    We adopt the protocol in Lemma \ref{lem:mix_in_mqszkhv}. The protocol $\rho_{in}$ is a mixed state, but the input state $\ket{\phi_{in}} \in \cL_{purifyMix^*}$ is the purify version of $\cL_{mix^*}$. To run the protocol, we just replace $\rho_{in}$ by tracing out the purification part of $\ket{\phi_{in}}$ and the proof is identical to Lemma~\ref{lem:mix_in_mqszkhv}.  
\end{proof}

We show that if the purification part of $\cL_{purifyMix^*}$ is sampled from the Haar random distribution, then the sample complexity of $\cL_{purifyMix^*}$ and average-case hardness of $\cL_{mix^*}$ (Theorem~\ref{thm:totally_mix_state_hardess_half}) are identical. To show the result, we rely on the optimal average-case local tester Theorem 3.7 in \cite{CWZ24}, which we state in Theorem~\ref{thm:optimal-average-case-local-tester}. We state our results as follows.
\begin{lemma}\label{lemma:avg_purify_mixed_hardness_half}
For all polynomials $q(\cdot)$, for sufficiently large $\lambda$, and for all algorithms $\cA$ , the following hold:
\begin{equation}\label{equ:commitment_random}
    \begin{aligned}
        &\left| \Pr\limits_{T \leftarrow Haar_\lambda}\left[ \cA ((I \otimes T)\ket{\epr_\lambda})^{\otimes q(\lambda)}=1 \right] - \Pr\limits_{T_1,T_2 \leftarrow Haar_\lambda}\left[ \cA ((T_1 \otimes T_2)\ket{\halfs_\lambda})^{\otimes q(\lambda)}=1 \right] \right|  \\
        \leq &\,\frac{q(\lambda)}{2^\lambda},
    \end{aligned}
    \end{equation}
    where $Haar_\lambda$ is the Haar measure on $\lambda$-qubit unitaries. 
\end{lemma}
\begin{proof}
We will show, by contradiction, that if there exists an algorithm $\cA$ that breaks Equation~(\ref{equ:commitment_random}) using a polynomial number $q(\lambda)$ of copies, then there exists another quantum circuit $\hat{\cT_\lambda}$ that breaks Theorem~\ref{thm:totally_mix_state_hardess_half} using the same number of copies and with advantage at least $\frac{q(\lambda)}{2^\lambda}$, leading to a contradiction. Indeed, this follows as a consequence of Theorem~\ref{thm:optimal-average-case-local-tester}, with
    \begin{itemize}
        \item $\cP^{\mathrm{yes}}:= \cL_{purifyMix^*,Y}$
        \item $\cP^{\mathrm{no}}:= \cL_{purifyMix^*,N} $
        \item $\cD^{\mathrm{yes}}:$ A family of distributions are defines as follow: for every $\lambda \in \N$, samples $T_1,T_2\leftarrow Haar_{\lambda}$, and outputs $(T_1\otimes T_2) \ket{\halfs_\lambda}$
        \item $\cD^{\mathrm{no}}$: A family of distributions are defines as follow: for every $\lambda \in \N$, samples $T\leftarrow Haar_\lambda$, and outputs $(I\otimes T) \ket{\epr_\lambda}$.
.
        \item Let $\cT = \cA$ be an algorithm with input sample complexity $q(\lambda)$. The advantage $c - s$ is at least $ \frac{q(\lambda)}{2^\lambda}$.
    \end{itemize}
    For any fix $\lambda \in \N$, let $\ket{\phi_{in}}$ be any $2\lambda$ qubits input state, $A$ register be the first $\lambda$ of qubits and $B$ register be the last $\lambda$ of qubits.  It is easy to see that $\cP^{\mathrm{yes}}$ and $\cP^{\mathrm{no}}$ are unitarily invariant on the register $B$, as defined in Definition~\ref{def:unitarily-invariant}, and that $\cD^{\mathrm{yes}}$ and $\cD^{\mathrm{no}}$ are unitarily invariant distributions on the register $B$, as defined in Definition~\ref{def:unitarily-invariant-dist}. Then, by Theorem~\ref{thm:optimal-average-case-local-tester}, we obtain another quantum circuit $\hat{\cT_\lambda}$ that acts only on the $A$ register of the input state $\ket{\phi_{in}}_{AB}$, with sample complexity $q(\lambda)$ and advantage $\frac{q(\lambda)}{2^\lambda}$. Since the quantum circuit $\hat{\cT}$ acts only on the $A$ register, the result remains the same even when $\cD^{\mathrm{yes}}$ and $\cD^{\mathrm{no}}$ only send the $A$ register to $\hat{\cT}$. Then the distribution becomes the maximally mixed state (in the case of $\cD^{\mathrm{yes}}$) or the maximally mixed state over a random subspace of dimension $2^{\lambda-1}$ (in the case of $\cD^{\mathrm{no}}$) when only the $A$ register is sent, which matches the setting of Theorem~\ref{thm:totally_mix_state_hardess_half}. This implies that $\hat{\cT}$, with $q(\lambda)$ copies of the input state and advantage $\frac{q(\lambda)}{2^\lambda}$, contradicts Theorem~\ref{thm:totally_mix_state_hardess_half}.
\end{proof}

% It is easy to see that Lemma~\ref{lemma:avg_purify_mixed_hardness_half} can be restated as an average case hardness of $\cL_{purify\_mix^*}$. We immediately get the worst case hardness of $\cL_{purify\_mix^*}$ as follows.
% \begin{theorem}\label{thm:L_purfiymx_not_in_pINV}
%      $\cL_{purify\_mix^*} \not \in \pINF$.\footnote{In~\cite{CWZ24}, they already obtain the same results, we restate the result in terms of our framework.}
% \end{theorem}

%Combine Lemma \ref{lem:L_purifymix_in_pQSZKhv} and Theorem \ref{thm:L_purfiymx_not_in_pINV}, we get Theorem \ref{thm:pQSZKhv_not_inside_pINF}.

\begin{proof}[Proof of Theorem \ref{thm:pQSZKhv_not_inside_pINF}]
 To separate $\pINF$ and $\pQSZKhvsec$, we need to show that there exists some pure quantum promise problem $\cL$ such that it is easy for $\pQSZKhvsec$ and hard for $\pINF$. Here, we let $\cL:=\cL_{purifyMix^*}$. The easiness result follows by Lemma \ref{lem:L_purifymix_in_pQSZKhv}, and the hardness result comes from Lemma~\ref{lemma:avg_purify_mixed_hardness_half} since average case hardness trivially imply worst case hardness \footnote{In~\cite{CWZ24}, they also obtain the worst case hardness of $\cL_{purifyMix^*}$ for $\pINF$.}. Then we get the separation. 
\end{proof}

\subsection{Separation between $\pBQPpoly$ and $\pBQPqpoly$}
\label{subsec:sep_pBQPpoly_and_mBQPpoly}

We first define the QPP complexity class that is similar to $\BQPpoly$ and $\BQPqpoly$.

\begin{definition}[$\pBQPqpoly$ and $\mBQPqpoly$]\label{Def:mBQPqpoly}
    The definition of $\pBQPqpoly$ (respectively, $\mBQPqpoly$) is the same as that of $\pBQP$ (respectively, $\mBQP$), expect that the $P$-uniform quantum circuit family $\{\cV_\lambda\}_{\lambda \in \N}$ is additional given a polynomial-size family of quantum advice $\{\ket{\eta_\lambda}\}_{\lambda \in \N}$ as input.
\end{definition}

\begin{definition}[$\pBQPpoly$ and $\mBQPpoly$]\label{Def:mBQPqpoly}
    The definition of $\pBQPpoly$ (respectively, $\mBQPpoly$) is the same as that of $\pBQP$ (respectively, $\mBQP$), expect that the $P$-uniform quantum circuit family $\{\cV_\lambda\}_{\lambda \in \N}$ is additional given a polynomial-size family of classical advice $\{\eta_\lambda\}_{\lambda \in \N}$ as input.
\end{definition}

Here is our main theorem in this section. 

\begin{theorem}\label{thm:sep_pure_BQPpoly_BQPqpoly}
    $\pBQPpoly \subsetneq \pBQPqpoly$.
\end{theorem}

The pure QPP is a special case of the mixed QPP. As a result, we immediately obtain the following.
\begin{theorem}\label{thm:sep_mix_BQPpoly_BQPqpoly}
    $\mBQPpoly \subsetneq \mBQPqpoly$. 
\end{theorem}

Next, we define a parameterize version of $\cL_{purifyMix^*}$.
\begin{definition}[$\cL_{purifyMix^*}(\{T^*_\lambda\})$] Let $\{T^*_\lambda\}$ be any family of unitary such that for each $T^*_\lambda$ is a $\lambda$-qubit unitary. Define a quantum promise problem $\cL_{purifyMix^*}(\{T^*_\lambda\})=\bigcup\limits_{\lambda \in \N}(\cL_Y:=\mathcal{L}_{purifyMix^*,Y}^{\lambda}, \cL_N:=\mathcal{L}_{purifyMix^*,N}^{\lambda})$, where 
\begin{itemize}
    \item $\mathcal{L}_{purifyMix^*,Y}^{\lambda}:=\left\{ (T_1 \otimes T_2)\ket{\halfs_\lambda}:T_1,T_2 \text{ are arbitrary $\lambda$-qubit unitaries} \right\}$.
    \item $\mathcal{L}_{purifyMix^*,N}^{\lambda}:=\left\{(I \otimes T^*_\lambda)\ket{\epr_\lambda)} \right\}$.
\end{itemize}
\end{definition}

\begin{lemma}\label{lemma:in_pBQPqpoly}
    For any family of unitaries $\{T^*_\lambda\}$, 
    $\cL_{purifyMix^*}(\{T^*_\lambda\}) \in \pBQPqpoly$.
\end{lemma}
\begin{proof}
 Fix the security parameter to $\lambda$. Let the quantum advice to be $(I \otimes T^*_\lambda) \ket{\epr_\lambda}$. The algorithm performs a swap-test between the input instance and the quantum advice. Then output $0$ if the swap-test passes; otherwise output $1$. In the no case, the probability passes the swap test is $1$ because the input and the advice state are the same. The algorithm output $1$ with probability $0$. In the yes case, the probability passes the swap-test with at most 
 \begin{equation*}
 \begin{aligned}
          \frac{1}{2} + \frac{1}{2}F((I\otimes T_\lambda )\ket{\epr_\lambda}, \ket{\phi_{in}})^2 &\leq           \frac{1}{2} + \frac{1}{2}F((\Tr_{>\lambda}((I\otimes T_\lambda )\ket{\epr_\lambda}), \Tr_{>\lambda}(\ket{\phi_{in}}))^2 \\ &=  \frac{1}{2} + \frac{1}{2} F(\frac{I}{2^\lambda},\sigma)^2 \\ &\leq \frac{1}{2} + \frac{1}{2} (1 - \text{TD}(\frac{I}{2^\lambda},\sigma)^2)\\ &= \frac{1}{2} + \frac{1}{2}\cdot\frac{3}{4} = \frac{7}{8},
 \end{aligned}    
 \end{equation*}
 where $\sigma:= \Tr_{>\lambda}(\ket{\phi_{in}})$. The first inequality follows from a property of fidelity, which increases under any CPTP map. The second inequality follow from $F(\rho,\sigma)^2 \leq 1 - \text{TD}(\rho,\sigma)^2$. The second equality follow from $\text{TD}(\frac{I}{2^\lambda},\sigma) = \frac{1}{2}$ because $\sigma$ is maximally mixed state on some subspace with dimension $2^{\lambda-1}$. Then. the algorithm outputs 1 with probability at least $\frac{1}{8}$.  Hence, for any family of unitary $\{T^*_\lambda\}$, $\cL_{purifyMix^*}(\{T^*_\lambda\}) \in \pBQPqpoly$.
\end{proof}

\begin{lemma}\label{lemma:notin_pBQPpoly}
 There exists a family of unitaries $\{T^*_\lambda\}$, 
    $\cL_{purifyMix^*}(\{T^*_\lambda\}) \notin \pBQPpoly$. 
\end{lemma}

In the rest of this section, we are going to prove Lemma~\ref{lemma:notin_pBQPpoly}.
\JW{I rewrite the original proof as follows.} 
\begin{proof}[Proof of Lemma~\ref{lemma:notin_pBQPpoly}]
    We first define the following notation. Let 
    \begin{equation*}
        \beta_{\lambda,\cA,q(\cdot)}:= \Pr_{T_1,T_2 \leftarrow Haar_\lambda} [\cA(((T_1\otimes T_2)\ket{\halfs_\lambda})^{\otimes q(\lambda)}) = 1],
    \end{equation*}  and 
    \begin{equation*}
        f_{\lambda,\cA, q(\cdot)}(T) = \Pr[\cA(((I\otimes T)\ket{\epr_\lambda})^{\otimes q(\lambda)}) = 1] - \beta_{\lambda,\cA, q(\cdot)}.
    \end{equation*} Using the above notation, we can first restate the average hardness of $\cL_{purfyMix^*}$ with distribution $T,T_1,T_2 \leftarrow Haar_\lambda$ (Lemma~\ref{lemma:avg_purify_mixed_hardness_half}) as follows: for any polynomial $q(\cdot)$, any sufficiently large $\lambda$ and any adversaries $\cA$ the following hold:
    \begin{equation*}
        |E_{T\leftarrow Haar_\lambda}[f_{\lambda,\cA,p(\cdot)}(T)]| \leq \frac{q(\lambda)}{2^\lambda}.
    \end{equation*} It is easy to see that the value $|f_{\lambda,\cA, q(\cdot)}(T^*_\lambda)|$ is exactly the advantage of $\cA$ of distinguishing the average case of $\cL_{purifyMix^*}(\{T^*_\lambda\})$ with the distribution $T_1$ and $T_2$ are sampled from $Haar_\lambda$. We can also restate our goal as follows: for any sufficiently large $\lambda$, there exist $T^*_\lambda$ such that for all polynomial size quantum circuits $C_{\lambda}$\footnote{It is sufficient to consider all polynomial-size quantum circuits because polynomial-size advice is equivalent to a polynomial-size classical circuit.} the following hold: 
    \begin{equation*}
        |f_{\lambda,C_\lambda,p(\cdot)} (T^*_\lambda)| = \negl(\lambda).
    \end{equation*}

     We state our key lemma. \begin{lemma}\label{lem:haar_concen_for_average_purify_hard}
       For any polynomial $q(\cdot)$, any sufficiently large $\lambda$ and any adversaries $\cA$ if 
        \begin{equation*}
            |E_{T\leftarrow Haar_\lambda}[f_{\lambda,\cA,p(\cdot)}(T)]| \leq \frac{q(\lambda)}{2^\lambda},
        \end{equation*} Then 
        \begin{equation*}
            \Pr_{T\leftarrow Haar_\lambda} [|f_{\lambda,\cA,q(\cdot)}(T)| \geq 2^{\frac{-\lambda}{3}} +\frac{q(\lambda)}{2^\lambda}] \leq  \exp(-2^{\frac{\lambda}{4}}).
        \end{equation*}
    \end{lemma}
    Lemma~\ref{lem:haar_concen_for_average_purify_hard} tells us that, for every algorithm $\cA$, the fraction of choices of $T$ that $\cA$ can distinguish with non-negligible advantage is at most doubly exponentially small. We can then apply a standard counting argument to obtain the desired result, since the number of distinct polynomial-size quantum circuits is at most exponential.

    To make the argument formal, we consider the set of all quantum circuits of size less than $1.1^\lambda$. The number of distinct quantum circuits of size less than $1.1^\lambda$ is at most $2^{1.1^\lambda}$.\footnote{We assume a universal gate set consisting of the Toffoli and Hadamard gates. Under this assumption, the number of circuits of size less than $1.1^\lambda$ is at most $2^{1.1^\lambda}$.} Therefore, by Lemma~\ref{lem:haar_concen_for_average_purify_hard} and union bound, the fraction of choices of $T$ that can be distinguished with non-negligible advantage by some quantum circuit of size less than $1.1^\lambda$ is at most $\exp(-2^{^{\lambda/4}})\cdot2^{1.1^{\lambda}} = \negl(\lambda)$. Hence, 
     there must exist $T^*_\lambda$ such that for all polynomial-size quantum circuit $C_{\lambda}$ the following hold: 
    \begin{equation*}
        |f_{\lambda,C_\lambda,p(\cdot)} (T^*_\lambda)| = \negl(\lambda).
    \end{equation*}
    
    In the rest of the proof, we focus on proving Lemma~\ref{lem:haar_concen_for_average_purify_hard}.
    \begin{proof}[Proof of Lemma~\ref{lem:haar_concen_for_average_purify_hard}]

    First, we show that the function $f_{\lambda,\cA,q(\cdot)}$ is $q(\lambda)$-Lipschitz with respect to the Frobenius norm. Since the value $\beta_{\lambda,\cA,q(\cdot)}$ is constant (independent of $T$), it suffices to consider the term
\[
\Pr\!\left[\cA\!\left(((I\otimes T)\ket{\epr_\lambda})^{\otimes q(\lambda)}\right)=1\right].
\]
By Lemma~\ref{lemma:query_Lipschitz}, it is sufficient to show that $\cA(((I\otimes T)\ket{\epr_\lambda})^{\otimes q(\lambda)})$ can be simulated by a $q(\lambda)$-query quantum algorithm $\cA^{T}$. The simulation of $\cA^{T}$ proceeds as follows.
\begin{enumerate}
    \item On input $\ket{0}$, prepare the state $\ket{\epr_\lambda}^{\otimes q(\lambda)}$.
    \item For each copy of $\ket{\epr_\lambda}$, apply the oracle $T$ to the second register. The resulting state is
        \[
        ((I \otimes T)\ket{\epr_\lambda})^{\otimes q(\lambda)}.
        \]
    \item Run $\cA$ on the resulting state and output whatever $\cA$ outputs.
\end{enumerate}

It is easy to see that this simulation is perfect and makes exactly $q(\lambda)$ queries to the oracle $T$. Therefore, the function $f_{\lambda,\cA,q(\cdot)}$ is $q(\lambda)$-Lipschitz with respect to the Frobenius norm.

Second, we apply haar-random concentration (Theorem~\ref{thm:concentration_on_unitary}) to $f_{\lambda,\cA,p(\cdot)}$ with $t = 2^{-\frac{\lambda}{3}}$, we get 
    \begin{equation*}
                \Pr_{T \leftarrow Haar_\lambda}\left[ \left |f_{\lambda,\cA,p(\cdot)}(T) - \E_{V \leftarrow Harr_\lambda}[f_{\lambda,\cA,p(\cdot)}(V)] \right| \geq 2^{-\frac{\lambda}{3}}  \right] \leq 2\cdot \exp\left(-\frac{(2^\lambda-2) 2^{-\frac{2\lambda}{3}}}{24q(\lambda)^2}\right) .
    \end{equation*}
    Since $q(\cdot)$ is polynomial, we get the right hand side is at most $\exp(-2^{\frac{\lambda}{4}})$ for sufficiently large $\lambda$. We can rewrite above as follow:
      \begin{equation*}
                \Pr_{T \leftarrow Haar_\lambda}\left[ \left |f_{\lambda,\cA,p(\cdot)}(T) - \E_{V \leftarrow Harr_\lambda}[f_{\lambda,\cA,p(\cdot)}(V)] \right| \leq 2^{-\frac{\lambda}{3}}  \right] > 1- \exp(-2^\frac{\lambda}{4}).
    \end{equation*}
    Then we use triangle-inequality, we get  
          \begin{equation*}
                \Pr_{T \leftarrow Haar_\lambda}\left[ \left |f_{\lambda,\cA,p(\cdot)}(T) \right| \leq 2^{-\frac{\lambda}{3}} + \left|\E_{V \leftarrow Harr_\lambda}[f_{\lambda,\cA,p(\cdot)}(V)]\right|  \right] > 1- \exp(-2^\frac{\lambda}{4}).
    \end{equation*}
    Last, by the condition $|\E_{V \leftarrow Harr_\lambda}[f_{\lambda,\cA,p(\cdot)}(V)]|\leq \frac{q(\lambda)}{2^\lambda}$, we get our results.
    \end{proof}
\end{proof}

\begin{proof}[Proof of Theorem~\ref{thm:sep_pure_BQPpoly_BQPqpoly}]
Combine Lemma~\ref{lemma:in_pBQPqpoly} and Lemma~\ref{lemma:notin_pBQPpoly}, we obtain the results.    
\end{proof}

\subsection{Separation between $\mQIP (\mQSZKhv)$ and $\mathbf{mcoQIP} $($\mathbf{mcoQSZK}_{\mathrm{hv}}$)} \label{subsec:sep_mQIP_and_mcoQIP}

This subsection will show that $\mQIP$ and $\mQSZKhv$ are not closed under complement, in contrast to $\pQSZKhv$ and classical complexity class $\QSZKhv$, which are closed under complement. In conclusion, we highlight a distinction between mixed-state quantum promise problems and other types of complexity classes.

\begin{theorem}\label{thm:mQSZKhv_and_mQIP_not_close_under_complement}
    $\mQIP$ and $\mQSZKhv$ are not closed under complement.
\end{theorem}
\begin{proof}
    It remains to show that $\overline{\cL_{mix^*}} \notin \mQIP$  $(\mQSZKhv)$, where $\cL_{mix^*}$ is defined in Definition~\ref{def:L_mix_constant}. Indeed, combining that with Lemma~\ref{lem:mix_in_mqszkhv} will complete the proof. We apply the same proof strategy in~\cite{HR22}.
    Suppose (by contradiction) that there exists an interactive protocol for $\overline{\mathcal{L}_{mix^*}}$ with completeness $\frac{2}{3}$ and soundness $\frac{1}{3}$. We will construct a single-party algorithm $\cD$ that decides $\overline{\mathcal{L}_{mix^*}}$. The algorithm $\cD$ simulates the interactive protocol by itself but replaces the input state with a totally mixed state only for the prover strategy. For the yes case, $\cD$ accepts with probability at least $\frac{2}{3}$ because the input can only be totally mixed states. For the no case, $\cD$ accepts with probability at most $\frac{1}{3}$ because the soundness of interactive proof holds for arbitrary malicious prover and the algorithm $\cD$ is just one of them. Then we get $\overline{\mathcal{L}_{mix^*}}$ is in $\mINF$, which contradicts to Theorem~\ref{thm:totally_mix_state_hardess_half}.
\end{proof}

\subsection{Separation between single-party algorithm and interactive proof in quantum property testing} \label{subsec:sep_property_testing}

Quantum property testing is a special case of quantum promise problems, where the no case includes all instances which are $\epsilon$-far from the yes case. For example, the hard QPP problems \(\cL_{mix^*}\) and \(\cL_{purifyMix^*}\) studied in the previous sections are not valid quantum property testing problems. In previous sections, we already showed the separation of single-party algorithm and interactive proof for both pure and mixed QPP problems (Theorem~\ref{thm:mQSZKhv_not_inside_mINF} and Theorem~\ref{thm:pQSZKhv_not_inside_pINF}). In this subsection, we will extend the above results to quantum property testing.

Formally, we present the following three results: (i) there exists a mixed property \(\cL\) such that \(\cL \notin \mQIP \cup \mINF\); (ii) there exists another mixed property \(\cL\) such that \(\cL \notin \mINF\) and \(\cL \in \mQSZKhv\); (iii) there exists a pure property \(\cL\) such that \(\cL \notin \pINF\) and \(\cL \in \pQSZKhv\).

Previously, work in quantum property testing focused mainly on the sample complexity of single-party algorithms. Here, we consider an interactive model, viewed as a two-party algorithm, for quantum property testing. To the best of our knowledge, such a model has not been previously studied in the quantum setting.

We can interpret results (ii) and (iii) as follows: there exist both pure and mixed properties that can only be decided by single-party algorithms using nearly exponentially many samples, but, with the help of a prover, the verifier can verify the property using only polynomially many samples, communication, and verification time. This demonstrates the power of interactive proofs for quantum property testing.

On the other hand, result (i) show that if the verifier in the interactive model is limited to polynomially many samples, then some properties still cannot even be verified. It is an interesting open question to explore other natural properties in both the positive and negative directions.

\paragraph{Mixed quantum property testing problems}

Here we try to modify $\cL_{mix^*}$ to a valid quantum property testing problem.
A natural way to modify $\cL_{mix^*}$ to be a valid quantum property testing problem is to define the property of testing whether a state is a maximally mixed state or $\epsilon$-far from it. Here we state the property as a QPP as follows. 
\begin{definition}[$\mathcal{L}_{mix,\epsilon(\cdot)}$]\label{def:L_maximall_mixed_property_problem}
Define a quantum promise problem $\cL_{mix,\epsilon(\cdot)}:=(\cL_Y:= \bigcup\limits_{\lambda \in \N} \cL_{mix,Y}^{\lambda}, \cL_N:= \bigcup\limits_{\lambda \in \N}\cL_{mix,N}^{\lambda})$, where 
\begin{itemize}
    \item $\mathcal{L}_{mix,Y}^{\lambda}:=\{ \frac{I}{2^{\lambda}} \}$.
    \item  $\mathcal{L}_{mix,N}^{\lambda}:=\{\rho : \text{TD}(\rho,\frac{I}{2^\lambda}) \geq \epsilon(\lambda) \}$.
\end{itemize}
\end{definition}

We first obtain the following corollary.

\begin{corollary}\label{cor:mix_property_is_hard_for_single_party}
    Fix any polynomial $p(\cdot) \geq 2$.  Then the property $\cL_{mix,\frac{1}{p(\cdot)}} \notin \mINF$ and the complement property $\overline{\cL_{mix,\frac{1}{p(\cdot)}}} \notin \mINF$.  
\end{corollary}

\begin{proof}
Since $p(\cdot)\geq 2$, it is easy to see $\overline{\cL_{mix,\frac{1}{p(\cdot)}}} \supseteq \cL_{mix^*}$. Then, the lower bound of $\cL_{mix^*}$ directly implies the lower bound of $\overline{\cL_{mix,\frac{1}{p(\cdot)}}}$. We get $\overline{\cL_{mix,\frac{1}{p(\cdot)}}} \notin \mINF$ by Theorem~\ref{thm:totally_mix_state_hardess_half}. Since the single party algorithm is close under complement, we also get $\cL_{mix,\frac{1}{p(\cdot)}} \notin \mINF$ 
\end{proof}

Then we find that the property of testing a state is maximally mixed state or $\epsilon$-far from it cannot be verified even with the help of the prover. 
\begin{corollary}\label{cor:mixed_prpoery_hard_for_single_and_interactive}
 Fix any polynomial $p(\cdot) \geq 2$.  Then the property $\mathcal{L}_{mix,\frac{1}{p(\cdot)}} \notin \mQIP \;\cup\; \mINF $.
\end{corollary}
\begin{proof}
Combined with Corollary~\ref{cor:mix_property_is_hard_for_single_party} and Theorem~\ref{thm:mQSZKhv_and_mQIP_not_close_under_complement}, we get the results.
\end{proof}

Next, we show that the complement of $\cL_{mix,\epsilon(\cdot)}$ can be verified in polynomial samples, times and communication with the help of the prover.  
\begin{corollary} \label{corollary:mixness_cpmlement_in_mQSZK}
 Fix any polynomial $p(\cdot) \geq 2$. Then the property $\overline{\mathcal{L}_{mix,\frac{1}{p(\cdot)}}} \notin\mINF $. However $\overline{\mathcal{L}_{mix,\frac{1}{p(\cdot)}}} \in \mQSZKhvsec$   
\end{corollary}
\begin{proof}By Corollary~\ref{cor:mix_property_is_hard_for_single_party}, it remains to show $\overline{\mathcal{L}_{mix,\frac{1}{p(\cdot)}}} \in \mQSZKhvsec$. We adopt the protocol from Lemma~\ref{lem:mix_in_mqszkhv}. In the previous case, we knew that $\text{TD}(\rho_{in},\frac{I}{2^\lambda}) = \frac{1}{2}$. Thus, we only need $\lambda$ copies of the state to amplify the distance to $1-\negl(\lambda)$. In our case, we have $\text{TD}(\rho_{in},\frac{I}{2^\lambda}) \geq \frac{1}{p(\lambda)}$. We know that there exists another polynomial $q(\cdot)$ such that $\text{TD}(\rho_{in}^{\otimes q(\lambda)},(\frac{I}{2^\lambda})^{\otimes q(\lambda)}) \geq 1 -\negl(\lambda)$. We let the verifier receive $\lambda q(\lambda)$ copies of the input state. Each $A_i$ contains $q(\lambda)$ copies of $\rho_{in}$ or $\frac{I}{2^\lambda}$, and the rest of the protocol remains the same. The analysis of completeness, soundness, and statistical zero-knowledge are the same as Lemma~\ref{lem:mix_in_mqszkhv}.
\end{proof}

\paragraph{Pure quantum property testing problems}

Here we try to modify $\cL_{purify\_mix^*}$ to a valid quantum property testing problem.
A natural way to modify $\cL_{pureifyMix*}$ to be a valid quantum property testing problem is to define the property of testing whether a state is a maximally entangle (ME) state or $\epsilon$-far from it. Here we state the property as a QPP as follows. 

\begin{definition}\label{def:L_ME_range}[$\mathcal{L}_{ME,\epsilon(\cdot)}$]\label{def:purifyMix_property}
Define a quantum promise problem $\mathcal{L}_{ME,\epsilon(\cdot)}:= \bigcup\limits_{\lambda \in \N}(\mathcal{L}_{ME,Y}^{\lambda}, \mathcal{L}_{ME,N}^{\lambda})$, where 
\begin{itemize}
    \item $\mathcal{L}_{ME,Y}^{\lambda}:=\left\{(I \otimes T)\ket{\epr_\lambda)} : T \text{ is arbitrary $\lambda$-qubit unitary}\right\}$.
    
    \item $\mathcal{L}_{ME,N}^{\lambda}:=\big\{ \ket{\psi}:   \forall \; \ket{\phi}\in \mathcal{L}_{ME,Y}^{\lambda} \,,\,\;\text{TD}(\ket{\phi}, \ket{\psi}) \geq \epsilon(\lambda) 
    \big\}$.

\end{itemize}
\end{definition}

We first obtain the following corollary which is similar to the corollary~\ref{cor:mix_property_is_hard_for_single_party}.

\begin{corollary}\label{cor:pure_property_is_hard_for_single_party}
    Fix any polynomial $p(\cdot) \geq 2$.  Then the property $\cL_{ME,\frac{1}{p(\cdot)}} \notin \mINF$ and the complement property $\overline{\cL_{ME,\frac{1}{p(\cdot)}}} \notin \mINF$.  
\end{corollary}
\begin{proof}
The proof is almost identical to that of Corollary~\ref{cor:pure_property_is_hard_for_single_party}, except that we replace the containment \(\overline{\cL_{mix,\frac{1}{p(\cdot)}}} \supseteq \cL_{mix^*}\) with \(\overline{\cL_{ME,\frac{1}{p(\cdot)}}} \supseteq \cL_{purifyMix^*}\), and replace the hardness results from Theorem~\ref{thm:totally_mix_state_hardess_half} with Lemma~\ref{lemma:avg_purify_mixed_hardness_half}.
\end{proof}

Next, we state our main theorem.
\begin{theorem} \label{thm:purifymixPorperty_in_pQSZKhv}
    Fix any polynomial $p(\cdot) \geq 2$.  The property $\cL_{ME, \frac{1}{p(\cdot)}} \not \in \pINF$ and $\overline{\cL_{ME, \frac{1}{p(\cdot)}}} \not \in \pINF$. However, $\cL_{ME, \frac{1}{p(\cdot)}}  \in \pQSZKhvsec$ and $\overline{\cL_{ME, \frac{1}{p(\cdot)}}}  \in \pQSZKhvsec$. 
\end{theorem}

Unlike the case in the mixed property, $\cL_{mix,\epsilon(\cdot)}$ and $\overline{\cL_{mix,\epsilon(\cdot)}}$ cannot be both in $\mQSZKhvsec$. Here, $\cL_{ME,\epsilon(\cdot)}$ and $\overline{\cL_{ME,\epsilon(\cdot)}}$ are both inside $\pQSZKhvsec$. Before we prove our main theorem~\ref{thm:purifymixPorperty_in_pQSZKhv}, we require the following technical Lemma. 

\begin{lemma} \label{lem:fidelity_of_L_purify_n} 
Fix any function $\epsilon(\cdot)$. Let $\ket{\psi}$ be any $2\lambda$-qubit state in $\cL^{\lambda}_{ME,N}$ ( dependend on $\epsilon$). Then $F(\sigma,\frac{I}{2^\lambda})^2 \leq 1 - \epsilon(\lambda)^2$, where $\sigma$ is the mixed state of tracing out the last $\lambda$-qubit of $\ket{\psi}$.
\end{lemma}
\begin{proof}
   Suppose (by contradiction) that there exist a state $\ket{\psi} \in \cL^\lambda_{ME,N}$ such that $F(\sigma, \frac{I}{2^\lambda})^2 > 1 - \epsilon(\lambda)^2$, where $\sigma$ is the mixed state of tracing out the last $\lambda$-qubit of $\ket{\psi}$. We will show that there must exist another state $\ket{\phi^*}$ such that $\ket{\phi^*} \in \cL^{\lambda}_{ME,Y}$ and $\text{TD}(\ket{\psi},\ket{\phi^*}) < \epsilon(\lambda)$. This leads to the contradiction because we know $\ket{\psi}$ must be at least $\epsilon(\lambda)$-far from all state in $\cL^\lambda_{ME,Y}$.
   
   Given that $F(\sigma,\frac{I}{2^\lambda})^2 > 1 - \epsilon(\lambda)^2$, then by Uhlmann's Theorem, we know that there exist two pure states $\ket{\psi'}$ and $\ket{\phi'}$ such that $F(\ket{\psi'}_,\ket{\phi'})^2 > 1 - \epsilon(\lambda)^2$, where $\ket{\psi'}$ and $\ket{\phi'}$ are purifications of $\sigma$ and $\frac{I}{2^\lambda}$, respectively. Since $\ket{\psi'}$ is a purification of $\sigma$, we know that there exists a $\lambda$-qubit unitary $U$ act on last $\lambda$-qubit of $\ket{\psi}$ such that $(I\otimes U)\ket{\psi} = \ket{\psi'}$. We let $\ket{\phi^*}:=(I\otimes U^{\dagger})\ket{\phi'}$. Because $\ket{\phi^*}$ is just another purification of $\frac{I}{2^\lambda}$, then we know that $\ket{\phi^*}\in \cL^\lambda_{ME,Y}$. Hence, it remains to show that $\text{TD}(\ket{\psi},\ket{\phi^*}) < \epsilon$.
   We observe the following: \begin{equation*}
   F(\ket{\psi},\ket{\phi^*})^2= F(\ket{\psi}, (I \otimes U^{\dagger})\ket{\phi'})^2 = F((I \otimes U)\ket{\psi},\ket{\phi'})^2 =  F (\ket{\psi'},\ket{\phi'})^2  > 1 - \epsilon(\lambda)^2.
   \end{equation*}
   The second equality follows from that fact that fidelity value is preserved under unitary operation. Lastly, applying the relation of $F(\ket{\phi},\ket{\psi})^2 = 1 - \text{TD}(\ket{\phi}, \ket{\psi})^2$, we get our desired result
   \begin{equation*}
       \text{TD}(\ket{\psi},\ket{\phi^*}) < \epsilon(\lambda).
   \end{equation*}
\end{proof}

\begin{proof}[Proof of Theorem~\ref{thm:purifymixPorperty_in_pQSZKhv}]
By Corollary~\ref{cor:pure_property_is_hard_for_single_party}, it remains to show $\overline{\cL_{ME,\frac{1}{p(\cdot)}}}$ and $\cL_{ME,\frac{1}{p(\cdot)}}$ is in $\pQSZKhvsec$. We first show $\overline{\cL_{ME,\frac{1}{p(\cdot)}}} \in \pQSZKhvsec$. Let $\ket{\phi_{in}}$ be the input state in $\overline{\cL_{ME,\frac{1}{p(\cdot)}}}$ and $\sigma$ be the mixed state by tracing out the last $\lambda$-qubit of $\ket{\phi_{in}}$. We adopt the same protocol in Corollary~\ref{corollary:mixness_cpmlement_in_mQSZK} by letting $\rho_{in}:=\sigma$. In the no case, $\rho_{in}$ is exactly equal to maximally mixed state. If in the yes case, we get $\text{TD}(\rho_{in},\frac{I}{2^\lambda}) \geq \frac{1}{2p(\lambda)^2}$. Then we complete the proof because we reduce the case from $\overline{\cL_{ME,\frac{1}{p(\cdot)}}}$ to $\overline{\cL_{mix,\frac{1}{p^2(\cdot)}}}$. It remains to show in the yes case, $\text{TD}(\rho_{in},\frac{I}{2^\lambda}) \geq \frac{1}{2p(\lambda)^2}$.The results is derived as follows:
\begin{equation*}
    \text{TD}(\rho_{in},\frac{I}{2^\lambda}) \geq 1 - \sqrt{F(\rho_{in},\frac{I}
    {2^\lambda})^2} = 1 - \sqrt{F(\sigma,\frac{I}
    {2^\lambda})^2} \geq 1 - \sqrt{1 - \frac{1}{p(\lambda)^2}} \geq \frac{1}{2p(\lambda)^2}.
\end{equation*}
The first inequality is followed by $\text{TD}(\rho,\sigma)\ge 1 - F(\sigma,\rho)$. The second inequality is followed by lemma~\ref{lem:fidelity_of_L_purify_n}. The last inequality is followed by $(1-x)^{\frac{1}{2}} \leq 1 - \frac{1}{2}x$ for all $|x|\leq1$.

Next, we show $\cL_{ME,\frac{1}{p(\cdot)}} \in \pQSZKhvsec$.We present a $\pQSZKhvsec$ protocol as follows.
    
\begin{protocal}{ Pure State Honest Verifier Statistical Zero-Knowledge Protocal for $\mathcal{L}_{ME, \frac{1}{p(\epsilon)}}$}
    \begin{description}
    \item[Verifier's step 1:]\quad\\
    Let \(\ket{\phi_{in}}\) be a \(2\lambda\)-qubit input state and  $m := \lambda p^2(\lambda)$. Receive $m$ copies of the input $\ket{\phi_{in}}$. Let register \(A\) be the collection of the first \(\lambda\) qubits of all \(m\) copies of \(\ket{\phi_{in}}\), and let register \(B\) be the collection of the last \(\lambda\) qubits of all \(m\) copies of \(\ket{\phi_{in}}\). Then send register \(B\) to the prover. 
    
    %Let $t := \lambda p^4(\lambda)$. 
    % For each $i \in [t]$, construct a new state $\ket{\epr_\lambda}$ and let the register $A_i$ ($B_i$) be the first (last) $\lambda$-qubit of $\ket{\epr_\lambda}$. Then, send $B_1\cdots B_t$ to the prover.
    
    \item [Prover's step 1:]\quad\\
    If \(\ket{\phi_{in}} \in \cL^\lambda_{ME,Y}\), then, by the definition of \(\cL^\lambda_{ME,Y}\), there must exist a \(\lambda\)-qubit unitary \(U\) acting on the last \(\lambda\) qubits of \(\ket{\phi_{in}}\) such that
    \[
    (I \otimes U) \ket{\phi_{in}} =  \ket{\epr_\lambda}.
    \] Then, applying this $U^{\otimes m}$ to $B$, and then send $B$ back to the verifier.
    
    \item [Verifier's step 2:]\quad\\
    Performing binary projective measurement $\{\ket{\epr_\lambda}\bra{\epr_\lambda}^{\otimes m}, I - \ket{\epr_\lambda}\bra{\epr_\lambda}^{\otimes m}\}$ on register $AB$. Here, the ordering of the \(AB\)-registers in \(\ket{\epr_\lambda}^{\otimes m}\) is the same as the ordering of the \(AB\)-registers in \(\ket{\phi_{in}}^{\otimes m}\). The verifier accepts if the measurement outcome refer to the first projector. Otherwise, the verifier rejects.  
     % For all $i \in [t]$, apply a swap-test between the registers $A_iB_i$ and $\ket{\phi_{in}}$. The verifier accepts if all the swap tests pass (i.e., the measurement outcome of each swap test is $0$). Otherwise, the verifier rejects.
        
    \end{description}
    \end{protocal}
    For completeness, by the definition of \(\cL_{ME,Y}\), every state \(\ket{\phi_{in}} \in \cL_{ME,Y}\) can be written as \((I\otimes T)\ket{\epr_\lambda}\) for some \(\lambda\)-qubit unitary $T$. The prover simply let \(U := T^{\dagger}\). Then the register $AB$ in the verifier's step 2 is exactly $\ket{\epr_\lambda}^{\otimes m}$, the probability that the verifier accepts is equal to $1$.
    
    For soundness, suppose the registers $AB$ contain states $\rho_{AB}$ at the beginning of the verifier's step 2. Let $\sigma$ be the mixed state of last $\lambda$-qubit of $\ket{\phi_{in}}$
    Since the prover cannot manipulate the $A$ register, we have $\Tr_{B}(\rho_{AB}) = \sigma^{\otimes m}$. The probability of the verifier accept is equal to the following:
    \begin{equation*}
    \begin{aligned}
              F(\rho_{AB}, \ket{\epr_\lambda}^{\otimes m})^2 &\leq F( \Tr_B(\rho_{AB}), \Tr_B(\ket{\epr_\lambda}^{\otimes m}))^2 \\ &= F(\sigma^{\otimes m}, (\frac{I}{2^\lambda})^{\otimes m})^2 \\ &= F(\sigma,\frac{I}{2^\lambda})^{2m} \\
              &\leq (1-\frac{1}{p^2(\lambda)})^m \\&\leq e^{-\frac{m}{p^2(\lambda)}} = \negl(\lambda)   
    \end{aligned}
    \end{equation*}
    The first inequality follows from a property of fidelity, which increases under any CPTP map. The second equality follows from $F(\sigma_1 \otimes \sigma_2,\rho_1\otimes\rho_2)^2 = F(\sigma_1,\sigma_2)^2 \cdot F(\rho_1, \rho_2)^2$. The second inequality follows from Lemma \ref{lem:fidelity_of_L_purify_n}. The last equality follows from $1-x \leq e^{-x}$ for all real $x$.
    
    For the statistical zero-knowledge property, let the simulator receive $t$ copies of the inputs, the same number as the verifier. The simulator can perfectly simulate the first message. For the second message, we know that the $AB$ register will be the same as $\ket{\epr_\lambda}^{\otimes m}$, so the simulator just simulate the registers $AB$ by directly compute the state $\ket{\epr_\lambda}^{\otimes m}$. Then apply the projective measurement on the register $AB$. Notice that this simulation is perfect.
\end{proof}

Last, consider a property testing problem called \emph{entanglement testing}, which determines whether a state is entangled or far from the entangled state. Montanaro and de Wolf\cite{MdeW16}
demonstrate how Theorem~\ref{thm:totally_mix_state_hardess_half} can be used to show that the \emph{entanglement testing} does not belong to $\mINF$. Combine this result with $\mQMA \subseteq \mPSPACE$ (Theorem \ref{thm:QMA_Upperbound}), we obtain the following Corollary: 
\begin{corollary}\label{cor:ka04_contradict}
    Entanglement testing is not in $\QMA$.
\end{corollary}
The above corollary appears to contradict the statement in \cite{kashefi2004complexity}, which asserts that the \emph{entanglement testing} problem belongs to $\QMA$. We provide a more detailed discussion of the potential proof gap in~\cite{kashefi2004complexity} in Appendix~\ref{Appendix:proof_gaps_in_KA04}.

\subsection{Impossibility of quantum input unitary synthesis problems}\label{subsec:imposs_unitary_synth}

In the previous subsections, we established several separation results between single-party algorithms and interactive proofs for quantum promise problems, and even for quantum property testing problems. Interestingly, as a byproduct of these separations, we are able to derive some impossibility results for quantum-input unitary synthesis problems. We first review some background on the unitary synthesis problem.

Bostanci et al. \cite{bostanci2023unitary} define the notion of a unitary synthesis problem, which involves a sequence $\mathcal{U}=(U_x)_{x\in\{0,1\}^*}$ of unitary or even partial isometries. The unitary synthesis problem is asking that given a classical string $x$ and determining how much computational resources is sufficient to synthesize the desired unitary. We generalize their problem as follows: Given a quantum input $\ket{\psi}$ (or a mixed-state $\rho$) with multiple copies, the goal is to implement a unitary $U_{\ket{\psi}}$ ($U_\rho$) defined by $\ket{{\psi}}$ ($\rho$). We focus on the sample complexity required to synthesize $U_{\ket{\psi}}$ ($U_\rho$). We present negative results related to state discrimination unitary (e.g pretty good measurement) and Uhlmann's transform. We provide the formal definition of above problems. 

\begin{definition}[\textbf{State Discrimination unitary synthesis problem}]\label{Def:unitary_dis} Fix three polynomials, $t(\cdot)$, $p(\cdot)$, and $\alpha(\cdot) \geq 1$. The task of the state discrimination unitary unitary synthesis problem associated with those parameters is defined as follows.
\begin{itemize}
 \item \textbf{Inputs}: Given $t(\lambda)$ copies of two unknown $\lambda$-qubit mixed states $\rho$ and $\sigma$, where it is promised that $0 \leq F(\rho,\sigma) \leq \frac{1}{\alpha(\lambda)}$. 
 \item \textbf{Goal}: Compute a CPTP map $R$ with a single bit output such that 
     \begin{equation}\label{def:succ_prob}
         \frac{1}{2}\Pr[R(\rho, \rho^{t(\lambda)}, \sigma^{t(\lambda)})=0] + \frac{1}{2} \Pr[R(\sigma, \rho^{t(\lambda)}, \sigma^{t(\lambda)})=1] \geq \frac{1}{2} + \frac{1}{p(\lambda)}.
     \end{equation}
\end{itemize}
We abbreviate this problem as $(t(\cdot),p(\cdot),\alpha(\cdot))$-\textbf{qDIS}. We can also consider the problem where $R$ has access to quantum advice that depends only on the security parameter.
\end{definition}

\begin{definition}[\textbf{Uhlmann's transform unitary synthesis problem}]\label{def:unitary_Uhl}
$\newline$
Let us fix two polynomials, $t(\cdot)$ and $0 \leq \epsilon(\cdot) \leq 1$. The task of quantum input Uhlmann's transform unitary synthesis problem associated with $t(\cdot)$ and $\epsilon(\cdot)$ is defined as follows.
 \begin{itemize}
     \item \textbf{Inputs}: Given $t(\lambda)$ copies of two unknown $\lambda$-qubit bipartite pure states $\ket{\phi}_{AB}$ and $\ket{\psi}_{AB}$.
     \item \textbf{Goal}: Let $\sigma:=\Tr_{B}(\ket{\phi})$ and $\rho:=\Tr_{B}(\ket{\psi})$. We say that a unitary $R^{Uhl}_{\ket{\phi}, \ket{\psi}}:=I_\lambda\otimes U $ is Uhlmann if it satisfy the following 
     \begin{equation*}
         |\bra{\psi}I_A\otimes U_B \ket{\phi}| = \max\limits_{U^*}|\bra{\psi}I_A\otimes U_B^*\ket{\phi}| = F(\sigma,\rho).
     \end{equation*}
     Compute a CPTP map $R$ such that there exists a Uhlmann unitary $R^{Uhl}_{\ket{\phi}, \ket{\psi}}:=I_A\otimes U_B$ for which the following holds:
    \begin{equation*}
         \frac{1}{2}\|R(\cdot, \ket{\phi}^{t(\lambda)}, \ket{\psi}^{t(\lambda)}) - R^{Uhl}_{\ket{\phi}, \ket{\psi}}(\cdot)\|_{\diamondsuit} \leq 
         \epsilon(\lambda).
    \end{equation*}
 \end{itemize}
 We abbreviate this problem as $(t(\cdot),\epsilon(\cdot))$-\textbf{qUHL}. We can also consider the problem where $R$ has access to quantum advice that depends only on the security parameter.
\end{definition}

We prove the following theorems: It is impossible to approximate Uhlmann's transform unitary and the state discrimination distinguisher unitary using a polynomial number of copies of unknown quantum states.
\begin{theorem}\label{thm:lower_bound_dis}
    For all polynomials $t(\cdot)$, $p(\cdot)$, and $\alpha(\cdot) \geq 1$, there is no algorithm that can compute $(t(\cdot),p(\cdot), \alpha(\cdot))$-\textbf{qDIS} for sufficiently large $\lambda$, even when given advice of arbitrary size.
\end{theorem}
\begin{theorem}\label{thm:lower_bound_uhl}
    For all polynomial $t(\cdot)$ and $q(\cdot)\geq 1$ , there is no algorithm that can compute $(t(\cdot),1 - \frac{1}{q(\cdot)})$-\textbf{qUHL} for sufficiently large $\lambda$, even when given advice of arbitrary size.
\end{theorem}

\begin{remark}
    \cite{Qia23, MNY24} also implicitly gave the same result as Theorem~\ref{thm:lower_bound_dis}. However, we provide an alternative proof: while they rely on the multi-instance games technique, we do not. We also emphasize that the quantum advice cannot depend on the input; otherwise, there exists a trivial unbounded algorithm that achieves the goal.
 \end{remark}
 Before we present the formal proof, we provide some high level idea why the above impossibility result is connected to the separation between single-party algorithm and interactive proof. Fixed some QPPs or quantum property testing problems $\cL$ that are hard for any single-party algorithms, but there exist some protocols that can decide $\cL$. The key observation is the following: if any single-party algorithm can simulate the behavior of the honest prover, then such single-party algorithm can decide $\cL$ with roughly the same completeness and soundness in the interactive protocol. The single-party algorithm just run the interactive protocol itself with the simulate honest prover and verifier. The completeness is almost the same if the simulation of honest prover is not too bad, while the soundness remains exactly the same because the soundness of the interactive proof must hold against any prover, and our simulation of honest prover is just one possible prover. 
 
 Since we obtain a separation, there must exist some step, namely some unitary operation, of the honest prover that cannot be simulated by a single-party algorithm; otherwise, we would obtain a contradiction by the observation mentioned above. The only difference between the single-party algorithm and the honest prover is that the single-party algorithm can obtain only polynomially many copies of the input state, whereas the honest prover can obtain the full description of the state. Therefore, there must exist some unitary operation that cannot be simulated using only polynomially many samples of the input state. In our previous separation results (Theorem~\ref{thm:mQSZKhv_not_inside_mINF} and Theorem~\ref{thm:purifymixPorperty_in_pQSZKhv}), the only unitary operations that depend on the input state and are performed by the honest prover are the state-discrimination unitary and the Uhlmann's transformation unitary. Consequently, we obtain two impossibility results for these two quantum-input unitary synthesis problems (Theorem~\ref{thm:lower_bound_dis} and Theorem~\ref{thm:lower_bound_uhl}).  

In summary, once we obtain an advantage from the interactive model for any particular quantum property testing problem or QPP that is hard for single-party algorithms, we can establish a sample-complexity lower bound for certain unitary synthesis problems. This relationship has a flavor similar to the well-known algorithmic approach used to establish circuit lower bounds in complexity theory~\cite{Wil21}.

\begin{proof}[Proof of Theorem \ref{thm:lower_bound_dis}]
Suppose (by contradiction) that there exists polynomials $t(\cdot)$, $p(\cdot)$, and $\alpha(\cdot) \geq 1$ such that there exists an algorithm $\mathcal{A}$ that computes $(t(\cdot),p(\cdot), \alpha(\cdot))$-\textbf{qDIS}. Consider the quantum promise problem $\cL_{mix^*}$, defined in Definition~\ref{def:L_mix_constant}. We will construct a distinguisher $\mathcal{D}$ that decides $\cL_{mix^*}$ with overwhelming probability and polynomial copies of input. This contradict to Theorem~$\ref{thm:totally_mix_state_hardess_half}$. The distinguisher $\cD$ simulates the protocol in Lemma \ref{lem:mix_in_mqszkhv} (i.e., $\cD$ acts as both verifier and prover). We state the distinguisher \(\cD\) below and mark the differences in red.
\begin{protocal}{ Single-Party Algorithm $\cD$ for $\mathcal{L}_{mix*}$}
\begin{description}
\item[Simulated Verifier's step 1:]\quad\\
Let $\lambda$ be the number of qubits of input state $\rho_{in}$. Let {\color{red} $\ell=\lambda p^2(\lambda^2)$}.
Receive ${\color{red}\lambda \cdot \ell}$ copies of input state $\rho_{in}$. Sample a uniformly random string {\color{red}$m \in \{0,1\}^{{\ell}}$}. Consider ${\color{red}\ell}$ registers {\color{red}$A_1\cdots A_{\ell}$}, each of size $\lambda^2$. For {\color{red}$i \in [\ell]$},
if $m_i = 0$, let $A_i = \rho_{in}^{\otimes \lambda}$. Otherwise, let $A_i = (\frac{I}{2^\lambda})^{\otimes\lambda}$. Sends the registers {\color{red}$A_1\cdots A_{\ell}$} to the prover.

\item [Simulated Honest Prover's step 1:]\quad\\
 Receive {\color{red}$\ell\cdot\lambda\cdot t(\lambda^2)$} copies of input state.
For {\color{red}$i \in [{\color{red}\ell}]$}, apply {\color{red}$\cA$} with the input 
\[{\color{red}\big(\cdot, (\rho_{in}^{\otimes \lambda})^{t(\lambda^2)}, ((\frac{I}{2^\lambda})^{\otimes \lambda})^{t(\lambda^2)}\big)}
\] where register {\color{red}$A_i$} is placed in the first slot, and get the output $m'_i$.
Send {\color{red}$m'_1\cdots m'_\ell$} back to the verifier.

\item [Simulated Verifier's step 2:]\quad\\
 The verifier accepts if at least {\color{red}$\big(\frac{1}{2}+\frac{1}{2p(\lambda^2)}\big)$} fraction of $m'_i$ equals to $m_i$.
\end{description}
\end{protocal}
The number of copies of the input state used by \(\cD\) is
\[
\lambda \cdot \ell + \ell \cdot \lambda \cdot t(\lambda^2) = \lambda^2p^2(\lambda^2)(t(\lambda^2) + 1)
\]
which is at most polynomial because \(t(\cdot)\) and \(p(\cdot)\) are also polynomial. When \(\rho_{in} \in \cL_{mix^*,Y}\), we know that
\[
\begin{aligned}
&\text{TD}\left(\rho_{in}^{\otimes \lambda}, \left(\frac{I}{2^\lambda}\right)^{\otimes \lambda}\right)
\geq 1-\negl(\lambda) \\
\implies &F\left(\rho_{in}^{\otimes \lambda}, \left(\frac{I}{2^\lambda}\right)^{\otimes \lambda}\right) \leq \negl(\lambda) \leq \frac{1}{\alpha(\lambda^2)},
\end{aligned}
\]
which satisfies the promise of \(\cA\). Therefore, \(\cA\) can distinguish \(\rho_{in}^{\otimes \lambda}\) from \(\left(\frac{I}{2^\lambda}\right)^{\otimes \lambda}\) with advantage \(\frac{1}{p(\lambda^2)}\) when $\rho_{in} \in \cL_{mix^*,Y}$. We also know  no algorithm can distinguish between these two cases when  $\rho_{in} \in \cL_{mix^*,N}$ because $\rho_{in}$ is equal to $\frac{I}{2^\lambda}$. Then, by the Chernoff--Hoeffding bound, \(\cD\) decides \(\mathcal{L}_{mix^*}\) with overwhelming probability.
\end{proof}

\begin{proof}[Proof of Theorem \ref{thm:lower_bound_uhl}]
Suppose (by contradiction) that there exists a polynomial $t(\cdot)$ and $q(\cdot) \geq 1$ such that there exists an algorithm $\mathcal{A}$ that computes $(t(\cdot),1 - \frac{1}{q(\cdot)})-\textbf{qUHL}$. Fix any polynomial $p(\cdot) \geq 1$.  Consider the quantum promise problem $\cL_{ME,\frac{1}{p(\cdot)}}$, defined in Definition~\ref{def:L_ME_range}. We will construct a distinguisher $\mathcal{D}$ that decides $\cL_{ME,\frac{1}{p(\cdot)}}$ with overwhelming probability and polynomial copies of input. This contradict to Corollary~\ref{cor:pure_property_is_hard_for_single_party}. The distinguisher $\cD$ simulates the protocol in Theorem~\ref{thm:purifymixPorperty_in_pQSZKhv} (i.e., $\cD$ acts as both verifier and prover).  We state the distinguisher \(\cD\) below and mark the differences in red. 

\begin{protocal}{ Single-Party Algorithm $\cD$ for $\mathcal{L}_{ME,\frac{1}{p()}}$}
    \begin{description}
    \item[Simulated Verifier's step 1:]\quad\\
    Let \(\ket{\phi_{in}}\) be a \(2\lambda\)-qubit input state, $m := \lambda p^2(\lambda)$ and {\color{red}$\ell = \lambda \cdot q^2(2m\lambda)$}.  Receive {\color{red}$\ell\cdot m$} copies of the input $\ket{\phi_{in}}$. For each {\color{red}$i \in [\ell]$}, we pick fresh $m$ fresh copies of input state \(\ket{\phi_{in}}\). Then we let register {\color{red}\(A_{i}\)} be the collection of all the first \(\lambda\) qubits of these fresh \(m\) copies of \(\ket{\phi_{in}}\), and let register {\color{red}\(B_{i}\)} be the collection of all the last \(\lambda\) qubits of these fresh \(m\) copies of \(\ket{\phi_{in}}\). Send register {\color{red}\(B_1,\cdots B_\ell\)} to the prover. 
     
    \item [Simulated Honest Prover's step 1:]\quad\\
    Receive {\color{red}$m\cdot \ell \cdot t(2m\lambda)$} copies of input state. For each {\color{red}$i \in [\ell]$}, let the register $A_iB_i$ ordering of $\ket{\epr_\lambda}^{\otimes m}$ is the same as the ordering of the $A_iB_i$ register in $\ket{\phi_{in}}^{\otimes m}$. Applying {\color{red}$\cA$} with the input 
    \[
        \color{red}(\cdot, (\ket{\phi_{in}}^{\otimes m})^{t(2m\lambda)}, (\ket{\epr_\lambda}^{\otimes m})^{t(2m\lambda)})
    \] where register {\color{red}$B_i$} is place in the first slot. Then send $B_1, \cdots B_{\ell}$ back to the verifier.

    \item [Simulated Verifier's step 2:]\quad\\
    For each {\color{red}$i\in[\ell]$}, performing binary projective measurement 
    \[\{\ket{\epr_\lambda}\bra{\epr_\lambda}^{\otimes m}, I - \ket{\epr_\lambda}\bra{\epr_\lambda}^{\otimes  m}\}
    \] on register {\color{red}$A_iB_i$}. Here, the ordering of the \(A_iB_i\)-registers in \(\ket{\epr_\lambda}^{\otimes m}\) is the same as the ordering of the {\color{red}\(A_iB_i\)}-registers in \(\ket{\phi_{in}}^{\otimes m}\). The verifier accepts if the at least {\color{red}$\frac{1}{2q(2m\lambda)}$} fraction of measurement outcome refer to the first projector. Otherwise, the verifier rejects.  
     % For all $i \in [t]$, apply a swap-test between the registers $A_iB_i$ and $\ket{\phi_{in}}$. The verifier accepts if all the swap tests pass (i.e., the measurement outcome of each swap test is $0$). Otherwise, the verifier rejects.
        
    \end{description}
    \end{protocal}

The number of copies of the input state used by \(\cD\) is
\[
 m\cdot\ell\cdot(t(2m\ell) + 1) = \lambda^2 \cdot p^2(\lambda) \cdot q^2\!\bigl(2\lambda^2 p^2(\lambda)\bigr)
\cdot \left(
t\!\left(2\lambda^2 \cdot p^2(\lambda) \cdot q^2\!\bigl(2\lambda^2 p^2(\lambda)\bigr)\right)
+1
\right),
\]
which is at most polynomial because \(t(\cdot)\), \(p(\cdot)\) and \(q(\cdot)\) are also polynomial. When $\ket{\phi_{in}} \in \cL_{ME,Y}$, \(\ket{\phi_{in}}\) can be written as \((I\otimes T)\ket{\epr_\lambda}\) for some \(\lambda\)-qubit unitary $T$. We know $R^{Uhl}_{\ket{\phi}^{\otimes m},\ket{\epr_\lambda}^{\otimes m}}$ is equal to $(T^{\dagger})^{\otimes m}$ if the register $A_iB_i$ ordering are the same as above protocol. Hence, if the $\cA$ simulate $R^{Uhl}_{\ket{\phi}^{\otimes m},\ket{\epr_\lambda}^{\otimes m}}$ perfectly, for each $A_iB_i$ measure to the first projector is equal to $1$. Since we know the simulation error of $\cA$ is at most $1 - \frac{1}{q(2m\lambda)}$, by the definition of the diamond norm, for each $A_iB_i$ measure to the first projector is at least $\frac{1}{q(2m\lambda)}= \frac{1}{q(2\lambda^2p^2(\lambda))}$ which is still non-negligible. 

When $\ket{\phi_{in}} \in \cL_{ME,N}$, for each $A_iB_i$, the probability of measuring to the first projector is $\negl(\lambda)$. The analysis is the same as the soundness proof in Theorem~\ref{thm:purifymixPorperty_in_pQSZKhv}. Then, by the Chernoff--Hoeffding bound, \(\cD\) decides \(\mathcal{L}_{ME,\frac{1}{p(\cdot)}}\) for all polynomial $p(\cdot)\ge 1$ with overwhelming probability.  
\end{proof}

Theorem \ref{thm:lower_bound_dis} also implies that it is impossible to approximate pretty-good-measurement unitary using a polynomial number of copies of unknown quantum states.

\begin{definition}[\textbf{Pretty good measurement unitary synthesis problem}]\label{def:unitary_PGM}
$\newline$
Let us fix a polynomial $t(\cdot)$ and a constant $0 \leq \epsilon \leq 1$. The task of the pretty good measurement unitary synthesis problem associated with $t(\cdot)$ and $\epsilon$ is defined as follows.
 \begin{itemize}
     \item \textbf{Inputs}: Given $t(\lambda)$ copies of two unknown $\lambda$-qubit mixed states $\rho$ and $\sigma$.
     \item \textbf{Goal}: Let $R^{PGM}_{\rho, \sigma}$ denote the pretty good measurement (PGM) for $\rho$ and $\sigma$. We briefly recall the algorithm for the PGM: define $S:=\frac{1}{2}(\rho+\sigma)$. The PGM is the measurement $\{\frac{1}{2}S^{-\frac{1}{2}} \rho S^{-\frac{1}{2}}, \frac{1}{2}S^{-\frac{1}{2}} \sigma S^{-\frac{1}{2}}\}$ applied to the input. Specifically, $R^{PGM}_{\rho, \sigma}$ is a CPTP map that, given either $\rho$ or $\sigma$ as inputs, outputs a single bit to distinguish them. Compute a CPTP map $R$ such that 
     \begin{equation*}
         \frac{1}{2}\|R(\cdot, \rho^{\otimes t(n)}, \sigma^{\otimes t(n)}) - R^{PGM}_{{\rho, \sigma}}(\cdot)\|_{\diamond} \leq \epsilon.
     \end{equation*}
 \end{itemize}
 We abbreviate this problem as $(t(\cdot),\epsilon(\cdot))$-\textbf{qPGM}.
\end{definition}
\begin{corollary}\label{thm:PGM}
    For all polynomials $t(\cdot)$ and for all constant $0 \leq \epsilon < \frac{1}{2}$, there is no algorithm that can compute $(t(\cdot),\epsilon)$-\textbf{qPGM} for sufficiently large $\lambda$.
\end{corollary}
\begin{proof}
    Suppose (by contradiction) there exists a polynomial $t(\cdot)$ and constant $\frac{1}{2}>\epsilon\geq0$ such that there exists an algorithm $\cA$ that computes $(t(\cdot),\epsilon)$-\textbf{qPGM}, we will show the same $\cA$ can also compute $(t(\cdot),\frac{1}{2}(\frac{1}{2}-\epsilon),\frac{2}{1-2\epsilon})$-\textbf{qDIS}. This contradicts to Theorem~\ref{thm:lower_bound_dis}. 
    
    To see why this is true. First, we use the condition in $\alpha(\cdot) = \frac{2}{1-2\epsilon}$, this imply $F(\sigma,\rho) \leq \frac{1}{2} - \epsilon$. Second, it is known that the pretty good measurement successfully distinguish $\sigma$ and $\rho$ with probability at least $1 - \frac{1}{2}F(\rho,\sigma) \geq \frac{3}{4}+\frac{\epsilon}{2}$. Third, we know that the simulation error of $\cA$ at most $\epsilon$, then by the definition of diamond norm $\cA$ with $t(\lambda)$ copies of $\rho$ and $\sigma$ can successfully distinguish $\sigma$ and $\rho$ with probability
    \[
    \geq\frac{3}{4} + \frac{\epsilon}{2} - \epsilon = \frac{3}{4} - \frac{\epsilon}{2} = \frac{1}{2} + \frac{1}{2} (\frac{1}{2} - \epsilon).
    \]

    % Suppose we are distinguishing $\rho$ from $\sigma$. Let $R^{PGM}$ be the pretty good measurement distinguisher. Also, define the following values:
    % \begin{equation*}
    % \begin{cases}
    %     p^{PGM}_{succ}:= \frac{1}{2}\Pr[R^{PGM}(\rho)=0] +\frac{1}{2}\Pr[R^{PFM}(\sigma)=1] \\
    %     p^{PGM}_{err}:= 1-p^{PGM}_{succ}.
    % \end{cases}       
    % \end{equation*}
    % It is known by the result of the pretty good measurement that $p^{PGM}_{err} \leq \frac{1}{2}F(\rho, \sigma) \leq  \frac{1}{2}(\frac{1}{2}-\epsilon)$. By the definition of diamond norm, the goal of the distinguisher $R$ should have an error smaller than or equal to $\frac{1}{2}(\frac{1}{2}-\epsilon) + \epsilon \leq c$, where $c < \frac{1}{2}$ is a constant. That is, \begin{equation*}
    % \begin{cases}
    %     p_{succ}:= \frac{1}{2}\Pr[R(\rho, \rho^{t(n)}, \sigma^{t(n)})=0] +\frac{1}{2}\Pr[R(\sigma, \rho^{t(n)}, \sigma^{t(n)})=1] \\
    %     p_{err}:= 1-p_{succ} \leq c < \frac{1}{2}.
    % \end{cases}       
    % \end{equation*}
    % Let $\frac{1}{p(\cdot)}:= \frac{1}{2}(\frac{1}{2}-c)$. Then, this contradict to Theorem $\ref{thm:lower_bound_dis}$ that any CPTP map will have infinitely many $n$ such that for some states $\rho$ and $\sigma$, $p_{err} \geq \frac{1}{2}-\frac{1}{p(n)} = \frac{1}{4} +\frac{c}{2} > c$.
    
\end{proof}

%% file: 6_App.tex
\section{Applications}\label{section:app}

\input{6-3_Crypto}

%\subsection{Cryptography}
%\input{6-3-1_prs}
%\input{6-3-2_owsg}
%\input{6-3-3_efi}
%\input{6-3-4_average_case_imply_EFI}

%% file: 6-3_Crypto.tex
\subsection{Cryptography}
\begin{figure}
\centering
\includegraphics[width=0.9\textwidth]{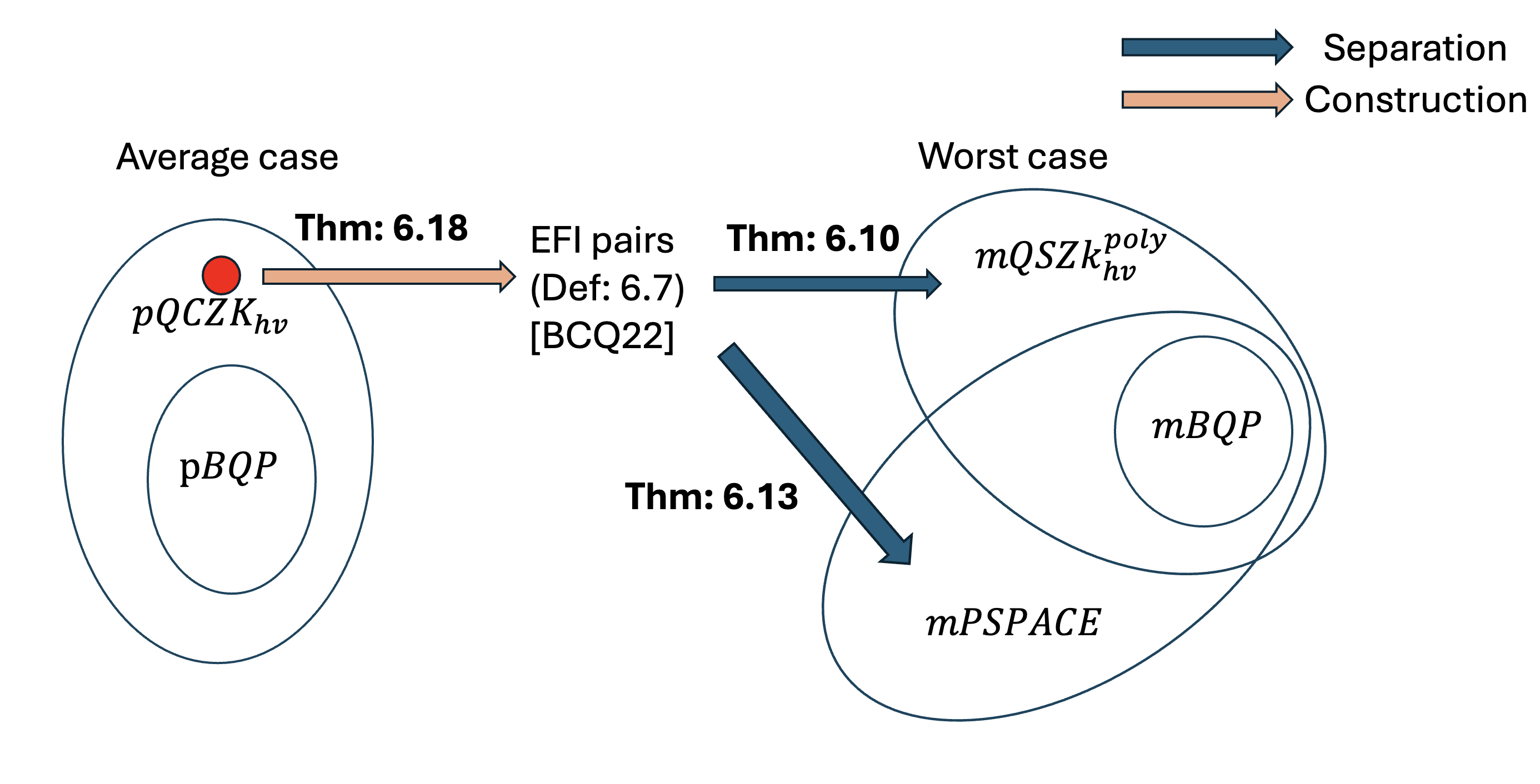}
\caption{\label{fig:crypto_result} The separation results of  EFI pairs comes from Theorem \ref{thm:EFI_PSPACE} and Theorem \ref{thm:EFI}. The construction arrow comes from Theorem \ref{Thm:pQSZK_hv_imply_EFI}.} 
\end{figure}

%In Section~\ref{sec:un_shcb_commit}, we construct an unconditional secure perfectly hiding and computationally binding commitment scheme in the auxiliary input model. 
In Section~\ref{sec:PRS_lower}, we show that if pseudorandom states (PRS) exist, then $\pBQP$ is not equal to $\pQCMA$. In Section~\ref{sec:OWSG_lower}, we demonstrate that a one-way state generator (OWSG) exists, then $\pmBQP$ is not equal to $\pmQCMA$. The above results can be seen as a quantum analog of the statement that if one-way functions (OWFs) or pseudorandom generators (PRGs) exist, then $\P$ is not equal to $\NP$. In Section~\ref{sec:EFI_lower}, we show that if EFI pairs exist, then $\mBQP$ is not equal to some variant of $\mQSZKhv$ and $\mPSPACE$. In Section~\ref{sec:EFI_const}, we present that EFI pairs can be constructed from the average case hardness of $\pQCZKhv$. We summarize the applications of EFI pairs in Figure \ref{fig:crypto_result}.

\input{6-3-2_prs}

\input{6-3-3_owsg}
\input{6-3-4_efi}

\input{6-3-5_average_case_imply_EFI}

%% file: 6-3-2_prs.tex
\subsubsection{Pseudorandom states}
\label{sec:PRS_lower}
First, we recall the definition of pseudorandom states (PRS) as given in \cite{pqs}.

\begin{definition}[Pseudorandom states (PRS), \cite{pqs}]
A pseudorandom state (PRS) is a QPT algorithm $\StateGen$ that, on input $k\in\{0,1\}^\lambda$, outputs an $m(\lambda)$-qubit quantum state $\ket{\phi_k}$ for some $m(\lambda) \geq \mathsf{log}_2\,\lambda$. We require the following condition for security: for any polynomial $t$ and any QPT adversary $\mathcal{A}$, such that for all $\lambda$,
\begin{equation*}
       \big|\Pr[\mathcal{A}(\ket{\phi_k}^{\otimes t(\lambda)})\rightarrow 1: k\leftarrow\{0,1\}^{\lambda}] - \Pr[\mathcal{A}(\ket{\phi}^{\otimes t(\lambda)})\rightarrow 1: \ket{\phi}\leftarrow \mu_{m(\lambda)}]\big| \leq \negl(\lambda),
\end{equation*}
where $\mu_{m(\lambda)}$ is the Haar measure on $m(\lambda)$-qubit states.
\end{definition}

Our first application gives an upper bound on the complexity of breaking PRS. 

\begin{theorem}\label{thm:PRS}
If $\mathsf{PRS}$ exist, then $\pBQP \subsetneq \pQCMA$.
\end{theorem}

Let $\lambda$ be the security parameter of the PRS, and let $\lambda$, $m(\lambda)$, and $\ket{\phi_k}$ represent the input length, output length, and output state of $\StateGen$, respectively. We first define a quantum promise problem $\mathcal{L}_{PRS}$, which essentially asks whether a given input state is an image of the PRS or far from all images of the PRS.

\begin{definition}[$\mathcal{L}_{PRS}$]
Define a quantum promise problem $\mathcal{L}_{PRS}:=\bigcup\limits_{\lambda \in \N}(\mathcal{L}_{PRS,Y}^{\lambda}, \mathcal{L}_{PRS,N}^{\lambda})$, where
\begin{itemize}
    \item $\mathcal{L}_{PRS,Y}^{\lambda}:=\{\ket{\phi_k}: k \in \{0,1\}^{\lambda} \}$

    \item $\mathcal{L}_{PRS,N}^{\lambda}:=\{\ket{\psi} \in \mathcal{H}(m(\lambda)) : \forall k \in \{0,1\}^{\lambda},\; |\braket{\psi|\phi_k}|^2 \leq 0.7\}$.
\end{itemize}  
\end{definition}

\begin{lemma}\label{lemma:L_prs_QCMA}
    $\mathcal{L}_{PRS}\in \pQCMA$.
\end{lemma}
\begin{proof}
    The verifier, given an input $\ket{\phi}$ and a witness $k$, applies a swap test between $\ket{\phi}$ and $\ket{\phi_k}$. If the swap test passes, outputs 1; otherwise, outputs 0. By Lemma \ref{lemma:swap-test}, the protocol achieves perfect completeness and a soundness value of $0.85$. Using the  amplification Lemma \ref{lemma:amplification}, we conclude that $\mathcal{L}_{PRS}$ is in $\pQCMA$. 
\end{proof}

We are ready to prove Theorem \ref{thm:PRS}.
\begin{proof}[Proof of Theorem \ref{thm:PRS}]
    Suppose $\pQCMA = \pBQP$. Define the oracle algorithm $\mathcal{A}^{\mathcal{L}_{PRS}}$ that queries the $\mathcal{L}_{PRS}$ oracle once with the given input $\ket{\phi}$ and outputs the oracle's results. By Lemma \ref{lemma:Haar} and the union bound, $\mathcal{A}^{\mathcal{L}_{PRS}}$ breaks the security of the PRS. Indeed, $\ket{\phi}$ is either an image of the PRS or a random state sampled from the Haar measure, with equal probability. Moreover, a random Haar state lies in $\mathcal{L}_{PRS,N}^{\lambda}$ with probability $1-2^\lambda\cdot e^{-0.7\cdot2^{log\,\lambda}}$, which is overwhelming. \nai{How?}\HZZ{OK} By the hypothesis that $\pQCMA = \pBQP$ and Lemma~\ref{lemma:L_prs_QCMA}, we can simulate the oracle query in quantum polynomial time and with polynomial many copies of $\ket{\phi}$, with a simulation error of $\negl(\lambda)$. Hence, there exists a QPT algorithm $\mathcal{A}'$ such that
    \begin{equation*}
        \big|\Pr[\mathcal{A'}(\ket{\phi_k}^{\otimes \poly(\lambda)})\rightarrow 1: k\leftarrow\{0,1\}^{\lambda}] - \Pr[\mathcal{A'}(\ket{\psi}^{\otimes \poly(\lambda)})\rightarrow 1: \ket{\psi}\leftarrow \mu_{m(\lambda)}]\big| \geq 1 - \negl(\lambda).
    \end{equation*}
\end{proof}

%% file: 6-3-3_owsg.tex
\subsubsection{One-way state generators}
\label{sec:OWSG_lower}

In this subsection, we recall the definition of a one-way state generator (OWSG) in \cite{morimae2022one}
\begin{definition} [Pure / Mixed one-way state generators (p/mOWSGs), \cite{morimae2022one}]\label{Def:owsg}

A pure/mixed one-way state generator (p/mOWSG) is a set of algorithms $(\KeyGen, \StateGen, \Ver)$, where
\begin{itemize}
    \item $\KeyGen(1^\lambda)\rightarrow k$: It is a QPT algorithm that, on input of the security parameter $\lambda$, outputs a classical key $k \in \{0,1\}^{n(\lambda)}$,
    
    \item $\StateGen(k)\rightarrow \phi_k$ :It is a QPT algorithm that, on input $k$, outputs an $m(\lambda)$-qubit pure or mixed quantum state $\phi_k$, and
    
    \item  $\Ver(k', \phi_k) \rightarrow \top /\ \bot$ : It is a QPT algorithm that, on input $\phi_k$ and a bit string $k'$, outputs $\top$ or $\bot$,
\end{itemize}
such that the following holds:
\begin{itemize}
    \item (Correctness): $\Pr[\top \leftarrow \Ver(k, \phi_k): k \leftarrow \KeyGen(1^\lambda), \phi_k \leftarrow \StateGen(k)] \geq 1 - \negl(\lambda)$,
    \item (Security): For any QPT adversary $\mathcal{A}$ and any polynomial $t(\cdot)$,
    \begin{center}
        $\Pr[\top \leftarrow \Ver(k', \phi_k):k \leftarrow \KeyGen(1^\lambda), \phi_k \leftarrow \StateGen(k), k' \leftarrow \mathcal{A}(\phi_k^{\otimes t(\lambda)})] \leq \negl(\lambda)$.
    \end{center}

\end{itemize}
\end{definition}

Our second application provides an upper bound on the complexity of breaking $\mathsf{p/mOWSG}$s. This result can be seen as a classical analog to the existing result that one-way functions imply $\P \neq \NP$.

\begin{theorem}[]\label{thm:OWSG}
If $\mathsf{p/mOWSG}$ exist, then $\pmBQP \subsetneq \pmQCMA$.
\end{theorem}

\begin{proof}
    First, we consider the pure-state version. Let us define a set of algorithms  $(\KeyGen, \StateGen, \Ver)$ that satisfy the input/output requirements and the correctness condition in Definition $\ref{Def:owsg}$.  We define a quantum promise problem $\mathcal{L}_{\StateGen}:=\bigcup\limits_{\lambda \in \N}(\mathcal{L}_{\StateGen,Y}^{\lambda}, \mathcal{L}_{\StateGen,N}^{\lambda})$, where
    \begin{itemize}
        \item $\mathcal{L}_{\StateGen,Y}^{\lambda}:=\{\ket{\phi_k}: k \in \{0,1\}^{n(\lambda)} \text{ such that } \Pr[\Ver(k, \ket{\phi_k} )=\top] \geq 1-\negl'(\lambda) \}$
    
        \item $\mathcal{L}_{\StateGen,N}^{\lambda}:=\{\}$, the empty set.
    \end{itemize}
    Note that, by the average argument, an overwhelming fraction of $\{\ket{\phi_k}\}$ are ``yes" instances. Although deciding $\mathcal{L}_{\StateGen}$ is trivial, we will construct a non-trivial $\pQCMA$ protocol $\braket{\cP,\cV}$ that decides $\mathcal{L}_{\StateGen}$ and is useful for constructing the search-to-decision reduction. The protocol $\braket{\cP,\cV}$ works as follows: The prover $\cP$, given the state description of input $\ket{\phi_{in}}$, finds a $k'$ such that  $\StateGen(k')$ is the closest match to the given state description. The prover then sends $k'$ to the verifier $\cV$. The verifier $\cV$, given the input $\ket{\phi_{in}}$ and the witness $k'$, runs $\Ver(\ket{\phi_{in}}, \StateGen(k')$, and outputs whatever $\Ver$ outputs. It is clear that the protocol $\braket{\cP,\cV}$ decides $\cL_{\StateGen}$. Since we can find the witness for $\cL_{\StateGen}$ (Lemma \ref{lem:searchToDecision_version1}) with respect to the verifier $\cV$, there exists an efficient oracle algorithm $\mathcal{A}^{\pQCMA}$ that breaks pOWSG security with probability at least $\frac{2}{3}$.

    Suppose $\pQCMA = \pBQP$. Since all oracle queries can be efficiently generated, we can simulate the $\pQCMA$ oracle in quantum polynomial time using polynomial copies of $\ket{\phi_k}$. Additionally, the simulation error is $\negl(\lambda)$.  Therefore, we obtain a QPT algorithm that breaks the security of the pOWSG with probability at least $\frac{2}{3}-\negl(\lambda)$.

    Note that we only prove our result in pOWSG. For mOWSG, we must replace every instance of pure-state syntax with the corresponding mixed-state syntax. For example, replace $\ket{\phi_k}$ with $\sigma_k$, $\ket{\phi_{in}}$ with $\sigma_{in}$, $\pQCMA$ with $\mQCMA$, and $\pBQP$ with $\mBQP$.
\end{proof}

%% file: 6-3-4_efi.tex
\subsubsection{EFI Pairs}
\label{sec:EFI_lower}

\paragraph{Separate mixed-state honest-verifier QSZK from BQP} In the first part, we show that the existence of EFI pairs implies that mixed-state honest-verifier $\QSZK$ is strictly larger than $\BQP$. We recall the definition of EFI pairs from \cite{brakerski2022computational}.

\begin{definition}[EFI pairs\cite{brakerski2022computational}]\label{def:EFI}
An EFI is a family of QPT circuits $\{Q_0(\lambda),Q_1(\lambda)\}_{\lambda \in \N}$ acting on two registers: $A$ (the output register) and $B$ (the trace-out register). For $b \in \{0,1\}$, let $\rho_b^\lambda = \Tr_{B}(Q_b(\lambda)\ket{0}_{AB})$. We require that $\rho_0^\lambda$ and $\rho_1^\lambda$ satisfy the following two conditions:
\begin{itemize}
    \item $\rho_0^\lambda$ and $\rho_1^\lambda$ are computationally indistinguishable. That is, for any QPT adversary $\mathcal{A}$, 
    \begin{equation*}
        \left|\Pr\left[1 \leftarrow \mathcal{A}(\rho_0^\lambda)\right] -\Pr\left[1 \leftarrow \mathcal{A}(\rho_1^\lambda)\right]\right| \leq \negl(\lambda).
    \end{equation*}
    
    \item $\rho_0^\lambda$ and $\rho_1^\lambda$ are statistically distinguishable. That is, $\left\| \rho_0^\lambda - \rho_1^\lambda\right\|_{tr} \geq \frac{1}{\poly(\lambda)}$.
\end{itemize}

Note that in the above definition, the trace distance can be amplified to $1 - \negl(\lambda)$ by repeatedly running the same construction, i.e., $Q_0^{\otimes n}(\lambda)$ and $Q_1^{\otimes n}(\lambda)$, for some polynomial $n = \poly(\lambda)$. Moreover, the computationally indistinguishable property is still preserved, as shown in  \cite{brakerski2022computational}. We first define a slightly modified version of $\mQSZKhv$. 
\begin{definition}[$\mpolyQSZKhv$]\label{def:mpolyQSZKhv}
The soundness and statistical zero-knowledge properties of $\mpolyQSZKhv$ are the same as $\mQSZKhv$. The only difference lies in the completeness: the honest prover is limited to obtaining only a polynomial number of input copies to run the protocol.
\end{definition}
\begin{remark}
In Definition $\ref{def:mpolyQSZKhv}$, we emphasize that soundness still holds even when the malicious prover is given an unbounded number of input copies. Consequently, $\mpolyQSZKhv \subseteq \mQSZKhv$. Additionally, by Theorem \ref{thm:mQSZKhv_not_inside_mINF}, we know that $\mQSZKhv$ is not in $\mINF$. Therefore, $\mQSZKhv$ is not a physically realizable oracle (Definition~\ref{Def:physically_realizable}). To avoid this issue,
we define $\mpolyQSZKhv$. Since $\mpolyQSZKhv \in \mINF$,  $\mpolyQSZKhv$ is a physically realizable oracle.
\end{remark}

Our third application provides an upper bound on the complexity of breaking EFI pairs.

\begin{theorem}\label{thm:EFI}
    If EFI pairs exist, then $\mBQP \subsetneq \mpolyQSZKhv$.
\end{theorem}

Let $\lambda$ be the security parameter, and let $Q_0(\lambda)$ and $Q_1(\lambda)$ be the corresponding circuits, with  $\rho_0^\lambda$ and $\rho_1^\lambda$ as the mixed states defined in Definition \ref{def:EFI}. For simplicity, we omit the parameter $\lambda$ and rewrite them as $Q_0$, $Q_1$, $\rho_0$, and $\rho_1$. Additionally, we assume that  $ \|\rho_0 - \rho_1\|_{tr} \geq 1 - negl(\lambda)$. We now define a quantum promise problem $\mathcal{L}_{EFI}$.

\begin{definition}[$\mathcal{L}_{EFI}$]
\label{def:EFI_QPP}
Define a quantum promise problem $\mathcal{L}_{EFI}:=\bigcup\limits_{\lambda \in \N}(\mathcal{L}_{EFI,Y}^{\lambda}, \mathcal{L}_{EFI,N}^{\lambda})$ as follows. 

\begin{itemize}
    \item $\mathcal{L}_{EFI,Y}^{\lambda}:=\{(\rho_0,\rho_1, Q_0, Q_1),(\rho_1,\rho_0, Q_0, Q_1)\}$

    \item $\mathcal{L}_{EFI,N}^{\lambda}:=\{(\rho_0,\rho_0, Q_0, Q_1),(\rho_1,\rho_1, Q_0, Q_1)\}$,
\end{itemize}
where $Q_0$ and $Q_1$ are QPT circuits that generate EFI pairs, encoded as classical strings; $\rho_0$ and $\rho_1$ are their mixed-state outputs.
\end{definition}

\begin{lemma}\label{lem:L_efiInQSZKHV}
$\mathcal{L}_{EFI} \in \mpolyQSZKhv$.  
\end{lemma}

\begin{proof}

The $\mpolyQSZKhv$ protocol for $\mathcal{L}_{EFI}$ is described as follows.
\begin{protocal}{
  The $\mpolyQSZKhv$ protocol for $\mathcal{L}_{EFI}$ }
\begin{description}
\item[Verifier's step 1:]\quad\\
Receive $t = \poly(\lambda)$ copies of the input $(\rho_a, \rho_b, Q_0, Q_1)$. Define registers $A_1,\cdots A_t$ to contain the state $\rho_a^{\otimes t}$, and let registers $B_1, \cdots B_t$ contain the state $\rho_b^{\otimes t}$. Let register $C$ contain a uniformly sampled string $n \leftarrow \{0,1\}^{t}$, where each bit serves as a control bit. For all $i \in [t]$, control on $i$-th bit in register $C$ and apply the SWAP operation on registers $A_i, B_i$. Finally, for all $i \in [t]$, send registers $A_i$ and $B_i$ to the prover.

\item [Prover's step 1:]\quad\\
 First, apply optimal measurement for $\rho_0, \rho_1$ on input $\rho_a, \rho_b$ and obtain the results $ans_a$ and $ans_b$. Let $m = 0^t$. For every $i \in [t]$, apply the optimal measurement on registers $A_i$ and $B_i$, and obtain the results $ans_{a_i}$ and $ans_{b_i}$. If $ans_{a_i} \neq ans_a$ or $ans_{b_i} \neq ans_b$, set $m_i = 1$. Finally, send $m$ to the verifier. 

\item [Verifier's step 2:]\quad\\
Check if $m = n$. If they are equal, the verifier accepts; otherwise, the verifier rejects.
\end{description}
\end{protocal}

The above protocol decides $\mathcal{L}_{EFI}$. First, we check that the prover's first step can be executed with a polynomial number of input copies. The prover only performs the optimal measurement between $\rho_0$ and $\rho_1$ in this step, which is feasible because the promise problem itself includes the construction of $\rho_0$ and $\rho_1$. Therefore, the prover can determine the optimal measurement for $\rho_0$ and $\rho_1$ without any input state. Next, we show completeness. When the input is a ``yes" instance, $\rho_a$ and $\rho_b$ are either $\rho_0, \rho_1$ or $\rho_1, \rho_0$. Since $ \|\rho_0 - \rho_1\|_{tr} \geq 1 - \negl(\lambda)$, there exists an optimal POVM measurement $\{\Pi_0, \Pi_1\}$ such that $\Tr(\Pi_0\rho_0) \geq 1 - \negl(\lambda)$ and $\Tr(\Pi_1\rho_1) \geq 1 - \negl(\lambda)$. By the union bound, the verifier accepts with probability at least $1 - (2t+1)\negl(\lambda)$.

Suppose the input is a ``no" instance. In this case, $\rho_a$ and  $\rho_b$ are either $\rho_0, \rho_0$ or $\rho_1, \rho_1$. Regardless of the control string in register $C$, the registers $A_i$, $B_i$, $\cdots$, $A_t$, and $B_t$ will contain identical states. This implies that $n$ is independent of those registers. Since the prover has no information about $n$, the verifier accepts with probability at most $2^{-t} = 2^{-\poly(\lambda)} = \negl(\lambda)$. 

Lastly, we will show the statistical zero-knowledge property. For the first message, the simulator receives $t$ copies of the input and follows the verifier's step 1 to produce a distribution identical to the first message. For the second message, $m$, the simulator outputs $n$ in the final step. Since the protocol accepts with probability at least $1 - \negl(\lambda)$, the simulator's error is $\negl(\lambda)$. 
\end{proof}

We are ready to prove Theorem \ref{thm:EFI}.
\begin{proof}[Proof of Theorem \ref{thm:EFI}]

Suppose $\mpolyQSZKhv = \mBQP$ and that $\rho_0$ and $\rho_1$ are statistically far. We will show that $\rho_0$ and $\rho_1$ are computationally distinguishable, thereby breaking the security of EFI. The following $\mpolyQSZKhv$ oracle algorithm demonstrates how to distinguish $\rho_0$ and $\rho_1$.
\begin{algorithm}[H]
    \caption{$\mathcal{A}^{\mpolyQSZKhv}$}
    \label{alg:efibreaker}
    \begin{algorithmic}[1]
        \REQUIRE $\rho_b^{\otimes t}, Q_0, Q_1$, where $t$ is the number of copies required to solve the $\mpolyQSZKhv$ promise problem $\mathcal{L}_{EFI}$).
        \ENSURE
        Find $b$.
        \STATE
        Compute  $\rho_0 = \Tr_{\mathbf{B}}(Q_0\ket{0}_{\mathbf{A,B}})$ and repeat this process $t$ times to obtain $\rho_0^{\otimes t}$.
        \STATE
        Query $\mathcal{L}_{EFI}$ with input $(\rho_0, \rho_b, Q_0, Q_1)^{\otimes t}$ and obtain the result $b'$. 
        \RETURN{$b'$}
    \end{algorithmic}
\end{algorithm}
By the definition of $\mathcal{L}_{EFI}$, the above algorithm uses polynomial copies of $\rho_b$ and outputs $b$ with probability $1$. Since $\mpolyQSZKhv = \mBQP$ and all input queries of $\mathcal{L}_{EFI}$ are generated efficiently, we can simulate the query to $\mathcal{L}_{EFI}$ in quantum polynomial time with negligible error. Hence, there exists a QPT algorithm $\mathcal{A'}$ that distinguishes $\rho_0$ and $\rho_1$. That is,
\begin{equation*}
    \Pr\left[0 \leftarrow \mathcal{A'}(\rho_0^{\otimes t})\right] - \Pr\left[0 \leftarrow \mathcal{A'}(\rho_1^{\otimes t})\right] \geq 1 - \negl(\lambda).
\end{equation*}
\end{proof}
\end{definition}

\paragraph{Separate mixed-state PSPACE from BQP} 
In the second part, we show that the existence of EFI pairs implies that $\PSPACE$ is strictly larger than $\BQP$. Morimae and Yamada observe that the combination of the flavored-swapped commitment scheme and the algorithmic Uhlmann’s theorem enables one to break EFI~\cite{MS}. We formalize this using QPP complexity class.

\begin{theorem}\label{thm:EFI_PSPACE}
    If EFI pairs exist, then $\mBQP \subsetneq \mPSPACE$.
\end{theorem}

We now provide a high-level overview of the proof. The key idea is to interpret the EFI scheme $\{Q_0, Q_1\}$ as a computationally binding canonical quantum bit-commitment scheme (see Definition~\ref{definiton:can_com}).
For each $b \in \{0,1\}$, the state $\rho_b$ corresponds precisely to the commitment state $\Com(b)$, obtained by tracing out the reveal register from $Q_b\ket{0}$.
Thus, distinguishing $\rho_0$ from $\rho_1$ is equivalent to breaking the binding property of $\{Q_0, Q_1\}$. By the flavor-conversion theorem~\cite{HMY23} (see Lemma~\ref{lemma:com_swtich}), breaking the hiding property of $\{Q_0, Q_1\}$ can be reduced to breaking the binding property of another canonical quantum bit-commitment scheme $\{Q'_0, Q'_1\}$. %If $\{Q_0, Q_1\}$ is efficient, then $\{Q'_0, Q'_1\}$ is also efficient. 
We then apply the algorithmic Uhlmann theorem~\cite{metger2023stateqip} (see Lemma~\ref{lemma:pspace_uhl}) to break the binding property of $\{Q'_0,Q'_1\}$, and hence to break the hiding property of $\{Q_0,Q_1\}$.

\begin{proof} [Proof of Theorem~\ref{thm:EFI_PSPACE}]
It is sufficient to show that $\cL_{EFI} \in \mPSPACE$ (Definition~\ref{def:EFI_QPP}). Then, by the same argument as in the proof of Theorem~\ref{thm:EFI}, showing $\cL_{EFI} \in \mPSPACE$ would imply breaking EFI. 

The $\mPSPACE$ algorithm for $\cL_{EFI}$ is described as follows.
\begin{algorithm}[H]
    \caption{$\mathcal{A}$ for solving $\cL_{EFI}$}
    \label{alg:efibreaker}
    \begin{algorithmic}[1]
        \REQUIRE $(\rho_a,\rho_b,Q_0, Q_1)$, where $a,b\in\{0,1\}$
        \ENSURE
        Determine $\rho_a$ and $\rho_b$ are different or the same. 
        \STATE
        Construct two unitaries $\{Q'_0,Q'_1\}$ from $\{Q_0,Q_1\}$ using Lemma~\ref{lemma:com_swtich}. 
        \STATE
         Construct the unitary $K_n$ in Lemma~\ref{lemma:pspace_uhl} by defining  $\ket{\psi_n}_{A_n,B_n}$ to be $Q'_0\ket{0}_{R',C'}$ and defining $\ket{\varphi_n}_{A_n,B_n}$ to be $Q'_1\ket{0}_{R',C'}$. Besides, the registers $R'$ and $C'$ are rename $A_n$ and $B_n$, respectively.
        \STATE
         Let $K_n$ be the unitary that breaks the computational binding of $\{Q'_0, Q'_1\}$. Let the reduction $\cR$, defined using $K_n$, be an algorithm with properties described in Lemma~\ref{lemma:com_swtich} (and Remark~\ref{remark:com_switch_reduction}). Apply the reduction $R$ on input $\rho_a$ and $\rho_b$. 
         If the results are different, outputs 1.
         Otherwise output 0.
    \end{algorithmic}
\end{algorithm}

The complexity of $\cA$ is polynomial space because the only inefficient component of $\cA$ is to construct $K_n$ which requires polynomial-space computation.

Next, we analyze the completeness and soundness of $\cA$. We view $\{Q_0,Q_1\}$ as a canonical quantum bit-commitment protocol. Moreover $\forall a\in\{0,1\}$, $\rho_a$ is precisely the states of $\Com(a)$, i.e., tracing out the revealing register of $Q_a \ket{0}$. The same argument also hold for $\rho_b, \forall b\in\{0,1\}$. 

The completeness of $\cA$ are the same as the advantage of breaking the hiding property of $\{Q_0, Q_1\}$, since in the yes-instances of $\cL_{EFI}$, the states $\rho_a$ and $\rho_b$ are always distinct. From Lemma~\ref{lemma:com_swtich} (Remark~\ref{remark:com_switch_reduction}), we know reduction $\cR$ break the hiding property of $\{Q_0, Q_1\}$ with advantage $\epsilon^2$ if $K_n$ break the binding property of $\{Q'_0, Q'_1\}$ with advantage $\epsilon$.
The completeness of $\cA$ further reduce to analyzing the advantage of $K_n$ in breaking the binding property of $\{Q'_0,Q'_1\}$. By lemma~\ref{lemma:pspace_uhl}, we know that $K_n$ breaks the binding property of $\{Q'_0,Q'_1\}$ with advantage $\geq 1-\negl(\lambda)$. Indeed,
\begin{equation*}
    \Tr_{B_n}(\ket{\psi_n})=\Tr_{C'}(Q'_0\ket{0})=\frac{1}{2}(\rho_0+\rho_1)=\Tr_{C'}(Q'_1\ket{0})= \Tr_{B_n}(\ket{\varphi_n}),
\end{equation*}
where the second equality follows from Equation~(\ref{equ:flavor_swap}) in Lemma~\ref{lemma:com_swtich}. Hence $F(\rho_n,\sigma_n) = 1$. Therefore, the completeness of $\cA$ is also $1 - \negl(\lambda)$.

The soundness of $\cA$ is $0$, since in the no-instances of $\cL_{EFI}$, the states $\rho_a$ and $\rho_b$ are always identical. Consequently, no algorithm can distinguish between $\rho_a$ and $\rho_b$.

Finally, if EFI pairs exist, it follows from the same argument as in the proof of Theorem~\ref{thm:EFI} that $\cL_{EFI} \notin \mBQP$. Together with Theorem~\ref{thm:EFI_PSPACE}, we have $\mBQP \subsetneq \mPSPACE$.
\end{proof}

%% file: 6-3-5_average_case_imply_EFI.tex
\subsubsection{EFI from average case hardness of $\pQCZKhv$}
\label{sec:EFI_const}

We first define the complexity class $\pQCZKhv$. The definition is almost identical to $\pQSZKhv$, except for the change from statistical zero-knowledge to computational zero-knowledge.

\begin{definition}[$\pQCZKhv$]
The completeness and soundness of $\pQCZKhv$ are the same as $\pQSZKhv$. Let the polynomial $m(\cdot)$ be the function of the number of messages in $\pQSZKhv$. We say $\cL=(\cL_Y,\cL_N)$ satisfies honest verifier computational zero-knowledge property if the following conditional hold. For sufficiently large $\lambda$ and $\ket{\phi} \in \mathcal{H}(\lambda)$, there exist a polynomial $q(\cdot)$ and a polynomial time simulator on input $(\ket{\phi}^{\otimes q(\lambda)},i)$ (for $i \in [m(\lambda)]$), output a mixed state $\xi_{\ket{\phi},i}$ such that for all QPT $\cA$, and polynomial $t(\cdot)$
\begin{equation*}
    |\Pr[\cA(\ket{\phi}^{\otimes t(\lambda)},\xi_{\ket{\phi},i}) = 1] - \Pr[\cA(\ket{\phi}^{\otimes t(\lambda)},\view_{P, V}(\ket{\phi}, i)) = 1]| \leq \negl(\lambda) \;\;\mathbf{if}\; \ket{\phi} \in \cL_Y.
\end{equation*}
where $\view_{P, V}(\ket{\phi}, i)$ is the reduced state after $i$ messages have been sent and tracing out the prover's private qubits.   
\end{definition}

We aim to define the average-case quantum promise problem. In the worst-case quantum promise problem, the problem input is polynomial identical copies of input states. A natural definition of the average-case quantum promise problem is to consider the distribution over identical copies of the input state. Therefore, the distribution in the average-case quantum promise problem must be able to sample any number of identical states. We call such a distribution a strongly samplable distribution.
\begin{definition}[Strongly samplable distribution]
Let $D_\lambda:=\{(p_i^\lambda, \sigma_i^\lambda)\}_{i \in \N}$ be a distribution on $\lambda$-qubits state. We say a collection of distributions $\mathcal{D}:=\{D_\lambda\}_{\lambda \in \N}$ is a strongly samplable distribution if there exist an algorithm $\mathsf{Samp}$ that, given input $(1^{t},1^{\lambda})$, outputs the following mixed state
\begin{equation*}
     \rho_\lambda^t:=\sum_{i\in \N} p_i^\lambda \sigma_i^{\lambda\otimes{t}}.
\end{equation*} 
When the running time of $\mathsf{Samp}$ is polynomial in $1^t$ and $1^\lambda$, we call $\mathcal{D}$ a $\class{BQP}$ strongly samplable distribution.  
\end{definition}

We also define the following indicator function and the notion of hard on average for pure quantum promise problem.
\begin{definition}
Let $\cL=(\cL_Y,\cL_N)$ be a pure-state quantum promise problem, and we define the following function 
\begin{equation*}
    \mathcal{L}(\ket{\phi}):=\begin{cases}
        \{1\} &\mathbf{if}\; \ket{\phi} \in \mathcal{L}_Y \\
        \{0\} &\mathbf{if}\; \ket{\phi} \in \mathcal{L}_N \\
        \{0,1\} &\mathbf{if}\; \ket{\phi} \notin \mathcal{L}_Y \cup \mathcal{L}_N.
    \end{cases}
\end{equation*}
    
\end{definition}

\begin{definition}
We say a pure-state quantum promise problem $\mathcal{L}=(\mathcal{L}_Y,\mathcal{L}_N)$ is hard on average for $\pBQP$ with respect to some strongly sampler of distribution $\mathcal{D}$ if for all polynomial $t(\cdot)$ and QPT $\mathcal{A}$
\begin{equation*}
    \Pr[\mathcal{A}(\ket{\phi}^{\otimes{t(\lambda)}}) \in \mathcal{L}(\ket{\phi}): \ket{\phi}^{\otimes{t(\lambda)}} \leftarrow \mathsf{Samp}(1^\lambda,1^{t(\lambda)}) ] \leq \frac{1}{2} + \negl(\lambda),
\end{equation*}
where $\mathsf{Samp}$ is the algorithm that generates the strongly samplable distribution $\mathcal{D}$.  
\end{definition}

We are ready to state our main theorem. 
\begin{theorem}\label{Thm:pQSZK_hv_imply_EFI}
    If there is a quantum promise problem $\mathcal{L}=(\mathcal{L}_Y,\mathcal{L}_N)$ in $\pQCZKhv$ that is hard on average for $\pBQP$ with respect to some $\class{BQP}$ strongly samplable distribution, then EFI pairs exist. 
\end{theorem}

Brakerski et al. \cite{brakerski2022computational}  showed how to construct EFI from the average-case hardness of $\QCZKhv$. Since the input of $\QCZKhv$ is classical, they can construct EFI based on classical input problems. In Theorem \ref{Thm:pQSZK_hv_imply_EFI},  we extend the possibility of constructing EFI from not just a classical input problem but also a quantum input problem. Our approach is similar to \cite{brakerski2022computational} where we first construct an instance-dependent EFI under the condition that $\mathcal{L}$ is in $\pQCZKhv$. We then apply the average-case hardness condition to upgrade the instance-dependent EFI to EFI.

\begin{lemma}\label{lem:ins_dep_EFI} If a pure-state quantum promise problem $\mathcal{L}=(\mathcal{L}_Y,\mathcal{L}_N)$ is in $\pQCZKhv$, then there are instance-dependent EFI $\{\gamma_{b,\ket{\phi}}\}_{b,\ket{\phi}}$ for $\mathcal{L}$ where $b = 0, 1$ and $\ket{\phi}\in \mathcal{H(\lambda)}$ such that
\begin{enumerate}
    \item There is a polynomial $q(\cdot)$ and QPT that on input $b$ and $q(\lambda)$ copies of $\ket{\phi}$ generate $\gamma_{b,\ket{\phi}}$
    \item For all polynomial $t(\cdot)$, for every QPT distinguisher $\mathcal{A}$, for all $\ket{\phi} \in \mathcal{L}_Y$,
    \begin{equation*}
        |\Pr[\mathcal{A}(\ket{\phi}^{\otimes{t(\lambda)}},\gamma_{0,\ket{\phi}}) = 1] - \Pr[\mathcal{A}(\ket{\phi}^{\otimes{t(\lambda)}},\gamma_{1,\ket{\phi}}) = 1]| \leq \negl(\lambda)
    \end{equation*}
    \item There is some constant $c$ such that $\ket{\phi} \in \mathcal{L}_N, ||\gamma_{0,\ket{\phi}} - \gamma_{1,\ket{\phi}}||_{tr} \geq c$.
\end{enumerate}
\end{lemma}

\begin{proof}[Proof sketch]
The construction of Lemma \ref{lem:ins_dep_EFI} is the same as the hardness reduction in Lemma~\ref{Applemma:pQSZK_hard} ($\gamma'_0,\gamma'_1$ in Equation~\ref{equation:EFI_construction}). The only difference in the proof of Lemma~\ref{Applemma:pQSZK_hard}  is that the zero-knowledge property is statistical. Then, by replacing the statistical zero-knowledge property in Lemma~\ref{Applemma:pQSZK_hard} to the computational zero-knowledge property, we get Lemma \ref{lem:ins_dep_EFI}.
\end{proof}

\begin{proof}[Proof of Theorem \ref{Thm:pQSZK_hv_imply_EFI}]
    Assume that there exists a pure-state quantum promise problem $\mathcal{L}=(\mathcal{L}_Y,\mathcal{L}_N)$ that is in $\pQCZKhv$ but is hard on average for $\pBQP$ with respect to some $\class{BQP}$ strongly samplable distribution $
    \cD$ \nai{D}. Let $\mathsf{Samp}$ be the efficient algorithm that generates the strongly samplable distribution $\mathcal{D}$, and $q(\cdot)$ be the polynomial used in the construction of instance-dependent EFI in Lemma \ref{lem:ins_dep_EFI}.
    The EFI construction is as follows: first run $\mathsf{Samp}(1^\lambda,1^{q(\lambda)+1 )})$ and obtain $\ket{\phi}^{\otimes q(\lambda)+1}$. Use the first $q(\lambda)$ of $\ket{\phi}$ to construct instance-dependent EFI $\gamma_{b,\ket{\phi}}$ in Lemma \ref{lem:ins_dep_EFI}. Then, output $\rho_b:=\ket{\phi}\bra{\phi}\otimes \gamma_{b,\ket{\phi}}$.

    First, we show that $\rho_0$ and $\rho_1$ are statistically far. Since $\cL$ is hard on average, the fraction of ``yes" and ``no" instances must lie within the range $[\frac{1}{2}-\negl(\lambda),\frac{1}{2} + \negl(\lambda)]$. Otherwise, a Turing machine that always outputs $0$ or $1$ will already have a noticeable advantage. Then, by the property 3 of instance-dependent EFI in Lemma \ref{lem:ins_dep_EFI} and ``no" instances must have a fraction at least $\frac{1}{2}-\negl(\lambda)$, we conclude the $\rho_0$ and $\rho_1$ are statistical far. 

    Next, we show that $\rho_0$ and $\rho_1$ are computationally indistinguishable. Suppose there exists a QPT $\mathcal{A}$ that can distinguish between $\rho_0$ and $\rho_1$ with noticeable advantage. We can use $\mathcal{A}$ to break the average-case hardness of $\cL$. The reduction $\cR$ work as follows: The challenger runs the $\mathsf{Samp}(1^\lambda,1^{q(\lambda) + 1})$ and sends $\ket{\phi}^{\otimes q(\lambda)+1}$ to the reduction $\cR$. The reduction $\cR$ samples a uniformly random bit $b$, and constructs $\rho_b$ from $\ket{\phi}^{\otimes q(\lambda)+1}$. Then, the reduction $\cR$ run $\cA$ on input $\rho_b$ and obtains the result $b'$. If $b'= b$, the reduction $\cR$ returns 0; otherwise, it returns $1$ to the challenger. When $\ket{\phi}$ is in $\mathcal{L}_Y$, the advantage of $\mathcal{M}$ can only be $[\frac{1}{2}-\negl(\lambda),\frac{1}{2} + \negl(\lambda)]$ by the property 2 of instance-dependent EFI in Lemma~\ref{lem:ins_dep_EFI} and $\cA$ is QPT. The noticeable advantage of $\cA$ all come form $\ket{\phi}$ in $\mathcal{L}_N$. Therefore, for a ``no" instance, reduction $\cR$ correctly rejects with noticeable probability. This breaks the average-case hardness of $\cL$.
\end{proof}

%% file: Appendix.tex
\section{Amplification Proofs in Section 3}
To reduce the error of the $\pmQMA$ protocol, the new protocol requires the prover to send many copies of the (original) witness. The verifier will run the (original) verification algorithm many times. However, the prover can be dishonest, and the witness can be entangled. Hence, in no instance, each round of accept (or reject) probability is not independent. To analyze the maximum accept probability in the no instance, we have to upper bound the accept probability by some independent random variables.

\begin{lemma} (Parallel amplification for \pmQMA) \label{App:amplify_QMA}
    Consider a quantum promise problem $\mathcal{L}$. The following are equivalent:
    \begin{enumerate}
        \item[(1)] There exists a polynomial $p(\cdot)$ and $\frac{1}{p(n)} \leq a \leq 1$ such that $\mathcal{L} \in \pmQMA_{a,a-\frac{1}{p(n)}}$.
        \item[(2)] For all exponential $e(\cdot)$, $\mathcal{L} \in \pmQMA_{1-\frac{1}{e(n)},\frac{1}{e(n)}}$.
    \end{enumerate}
\end{lemma}
\begin{proof}
    The proof is essentially the same for pure state promise problem or mixed state promise problem. Hence, we considered pure state promise problem only. 
    \newline
    
    $(1) \implies (2):$
    Let $(\mathcal{P},\mathcal{V})$ be a $\textbf{pQMA}_{a,b}$ protocol that decides $\mathcal{L}$.
    We view $\mathcal{V}$ as a unitary acting on register $I_i$ (input), $W_i$ (witness), $A_i$ (ancilla), for some $i$. Suppose $\mathcal{V}$ uses $t$ copies of inputs. To get the answer, we measure the register $A^{ans}_i$, the first qubit of register $A_i$. Choose a polynomial $s(\cdot)$ such that $e^{-\frac{s(n)}{2p^2(n)}} \leq \frac{1}{e(n)}$ for sufficiently large $n$. We define $\mathcal{V}'$ as follow:
    \begin{algorithm}[H]
    \caption{\bf QPT $\mathcal{V}'$}
    \begin{algorithmic}[1]
        \REQUIRE $|\psi\rangle^{ts(n)},\rho$
        \ENSURE
        Amplify error
        \STATE
        Initial the register to $\ket{\psi}^{ts(n)}_{I_1I_2,\dots,I_{s(n)}} \otimes \rho_{W_1W_2\dots W_{s(n)}}\otimes\ket{0}_{A_1A_2,\dots,A_{s(n)}}$.
        \FOR{$i \in [s(n)]$}
            \STATE
            Apply $\mathcal{V}$ on register $I_iW_iA_i$.
            \STATE
            Measure register $A^{ans}_i$ and get outcome $ans_i$.
        \ENDFOR
        \IF{$\frac{1}{s(n)}\sum\limits_{i=1}^{s(n)}ans_i \geq \frac{1
        }{2} (2a-\frac{1}{p(n)})$}
            \RETURN{Accept.}
        \ELSE
            \RETURN{Reject.}
        \ENDIF
    \end{algorithmic}
    \end{algorithm}
    Suppose $|\psi\rangle \in \mathcal{L}_N$. Define random variables $X_i$ such that for all $v\in \{0,1\}$, $\Pr[X_i=v]=\Pr[ans_i=v]$. Also, define i.i.d. random variables $Y_i$ (also independent to $X_i's$) such that $\Pr[Y_i=1]=a-\frac{1}{p(n)}$ and $\Pr[Y_i=0]=1-\Pr[Y_i=1]$. Then, for all $c \in [s(n)]$, $\Pr[\sum\limits_{i=1}^{s(n)} X_i \geq c] \leq \Pr[\sum\limits_{i=1}^{s(n)} Y_i \geq c]$. Indeed, it follows from:
    \begin{center}
    \begin{math}
    \begin{aligned}
        &\quad\hspace{0.25em}
        \Pr\Big[\sum\limits_{i=1}^{s(n)} X_i \geq c\Big] \\
        &= \Pr\Big[\sum\limits_{i=1}^{s(n)-1} X_i \geq c\Big]+\Pr\Big[\sum\limits_{i=1}^{s(n)-1} X_i = c-1\Big]\cdot \Pr\Big[X_{s(n)}=1 \Big| \sum\limits_{i=1}^{s(n)-1} X_i = c-1\Big] \\
        &\leq  \Pr\Big[\sum\limits_{i=1}^{s(n)-1} X_i \geq c\Big]+\Pr\Big[\sum\limits_{i=1}^{s(n)-1} X_i = c-1\Big]\cdot (a-\frac{1}{p(n)}) \\
        &=\Pr\Big[\sum\limits_{i=1}^{s(n)-1} X_i \geq c\Big]+\Pr\Big[\sum\limits_{i=1}^{s(n)-1} X_i = c-1\Big]\cdot \Pr[Y_{s(n)}=1]\\
        &= \Pr\Big[\sum\limits_{i=1}^{s(n)-1} X_i + Y_{s(n)} \geq c\Big].
    \end{aligned}
    \end{math}
    \end{center}
    By Chernoff bound, we have
    \begin{center}
    \begin{math}
    \begin{aligned}
        &\quad\hspace{0.25em}
        \Pr\Big[\frac{1}{s(n)}\sum\limits_{i=1}^{s(n)} X_i\geq \frac{1}{2} (2a-\frac{1}{p(n)})\Big] \\
        &\leq  \Pr\Big[\frac{1}{s(n)}\sum\limits_{i=1}^{s(n)} Y_i\geq \frac{1}{2} (2a-\frac{1}{p(n)})\Big] \\
        &\leq \frac{1}{e(n)}.
    \end{aligned}
    \end{math}
    \end{center}
    Suppose $|\psi\rangle \in \mathcal{L}_Y$. Then, there exists a (separable) witness such that $X_i's$ are independent. Again by Chernoff-Hoeffding bound (i), we have $\Pr\Big[\frac{1}{s(n)}\sum\limits_{i=0}^{s(n)-1} X_i < \frac{a+b}{2}\Big] \leq \frac{1}{e(n)}$.
    \newline

    $(2)\implies (1):$ Trivially.
\end{proof}

\begin{lemma}[Amplification Lemma] \label{Appendix:amplification}
Let $\mathcal{C} \in \{\BQP,\PSPACE,\QIP, \QMA,\QCMA,\QSZKhv\}$. The following are equivalent:
    \begin{enumerate}
        \item[(1)] $\mathcal{L} \in \pmclass\mathcal{C}_{a,b}$, with some $a,b$, and polynomial $p(\cdot)$ such that $a-b \geq \frac{1}{p(n)}$.
        \item[(2)] For all exponential $e(\cdot)$, $\mathcal{L} \in \pmclass \mathcal{C}_{1-\frac{1}{e(n)},\frac{1}{e(n)}}$.
    \end{enumerate}
\end{lemma}
 
\begin{proof}
    Since the other direction is trivial, we will only prove $(1) \implies (2)$. The proof follows a similar approach as classical error reduction, achieved by repeatedly running the same input many times. Let $s(n)$ be a polynomial such that $e^{-\frac{s(n)}{2p^2(n)}} \leq \frac{1}{e(n)}$. Then choose the threshold to be $s(n) \cdot \frac{1}{2}(2a-\frac{1}{p(n)})$. If the number of 1s is greater than or equal to this threshold, output 1; otherwise, output 0.

    For $\mathcal{C}\in \{ \BQP,\PSPACE,\QCMA \}$. Let $t(n)$ be the number of copies required by deciding $\mathcal{L} \in \mathcal{C}$. Then, we require a total of $t(n) \cdot s(n)$ (polynomial) copies of inputs to run the same algorithm $s(n)$ times. Each time, the accept (or reject) probability is independent and identical, and hence, by Chernoff bound, we achieve the desired completeness and soundness.

    For $\mathcal{C}=\QMA$. The new protocol requires the prover to send $s(n)$ copies of the (original) witness. However, the prover can be dishonest, and the witness can be entangled. Lemma \ref{App:amplify_QMA} gives an analysis that this protocol still works for amplification.

    For $\mathcal{C} \in \{ \QIP, \QSZKhv\}$. Let $t(n)$ be the number of copies required by deciding $\mathcal{L} \in \mathcal{C}$. We sequentially and repeatedly run the (original) protocol $s(n)$ times with $t(n) \cdot s(n)$ (polynomial) copies of inputs. For yes instances, the accept probability of each round is at least $a$. For no instances, the accept probability of each round is at most $a-\frac{1}{p(n)}$. We can use the same technique as Lemma $\ref{App:amplify_QMA}$ to analyze the completeness and the soundness. To see that zero-knowledge property also holds. Let $P$, $M$, and $V$ be the prover's private register, message register, and verifier's private register (of the original protocol), respectively. Let the new protocol has register $P_1P_2\dots P_{s(n)}$, $M_1M_2\dots M_{s(n)}$, and $V_1V_2\dots V_{s(n)}$, where $qubit(P_i)=qubit(P)$, $qubit(M_i)=qubit(M)$, and $qubit(V_i)=qubit(V)$. The new protocol runs as follows: For round $i\in[s(n)]$, runs the original protocol with register $P_i$, $M_i$, and $V_i$. To simulate the view of round $i$, message $j$. The new simulator runs as follows: For all $k\in [i-1]$, runs the original simulator on register $M_k$ and $V_k$. Then, run the original simulator to simulate the $j$th message on register $M_i$ and $V_i$.
\end{proof}

\section{Completeness Proofs in Section 4}

\begin{lemma}[Partial swap test]\label{App:swap}
    Consider two states $\ket{\phi}_{BC}$ and $\ket{\psi}_D$, where $qubit(B)=qubit(D)$. 
    We write $\ket{\phi}$ as the following.
    \begin{equation*}
        \ket{\phi} = \alpha\ket{\psi}_B\ket{G}_C + \sum\limits_j \beta_j \ket{\psi_j^\bot}_B\ket{G_j}_C,
    \end{equation*}
    where $\ket{G}$ and $\{\ket{G_j}\}$ are unimportant garbage state and $\{|\psi^{\perp}_j\rangle\}$ is a basis of subspace $\big\{\ket{\eta} \::\: \bra{\eta}\cdot \ket{\psi}=0\big\}$. Consider the following state
    \begin{equation*}
        H_A(c_A-\mathbf{SWAP}_{BD})H_A \ket{0}_A\ket{\phi}_{BC}\ket{\psi}_D.
    \end{equation*}
    Measuring the register $A$ gives outcome 0 (which we call accept) with probability $|\alpha|^2 + \frac{1}{2}\sum\limits_j |\beta_j|^2$.
\end{lemma}
\begin{proof}
    $\quad$
    \begin{center}
    \begin{math}
    \begin{aligned}
        &\qquad\hspace{6pt}\ket{0}_A\ket{\phi}_{BC}\ket{\psi}_D \\ 
        &\xmapsto{H_A} \frac{1}{\sqrt{2}}(\ket{0}+\ket{1})_A\ket{\phi}_{BC}\ket{\psi}_D \\
        &\xmapsto{(c_A-\mathbf{SWAP}_{BD})} 
        \frac{1}{\sqrt{2}} \ket{0}_A\Big(\alpha\ket{\psi}\ket{G}\ket{\psi} +   \sum\limits_j \beta_j \ket{\psi_j^\bot}\ket{G_j}\ket{\psi} \Big) + 
        \\&\hspace{7em} \frac{1}{\sqrt{2}} \ket{1}_A\Big(\alpha\ket{\psi}\ket{G}\ket{\psi} + \sum\limits_j \beta_j \ket{\psi_j}\ket{G_j}\ket{\psi^\bot} \Big) \\
        &\xmapsto{H_A} \frac{1}{2} \ket{0}_A\Big(2\alpha\ket{\psi}\ket{G}\ket{\psi} +   \sum\limits_j \beta_j \ket{\psi_j^\bot}\ket{G_j}\ket{\psi} + \sum\limits_j \beta_j \ket{\psi_j}\ket{G_j}\ket{\psi^\bot}  \Big) + \frac{1}{2}\ket{1}\ket{*}.
    \end{aligned}
    \end{math}
    \end{center}
    Measuring the register $A$ gives outcome 0 with probability $|\alpha|^2 + \frac{1}{2}\sum\limits_j |\beta_j|^2$.
\end{proof}

\begin{lemma}[Geometric Lemma \cite{kitaev2002classical}\label{Appendix:geometric}]
    Let $\mathcal{X}$ and $\mathcal{Y}$ be two (inner product) vector spaces. Define the angle (or closeness) between spaces $\mathcal{X}$ and $\mathcal{Y}$ is defined as
    \begin{center} 
        $\angle(\mathcal{X},\mathcal{Y}):=\text{arccos}\Bigg[\max_{\substack{\ket{x}\in\mathcal{X},\:\ket{y}\in\mathcal{Y} \\ \|\ket{x}\|_2=\|\ket{y}\|_2=1}}\big|\langle x\ket{y}\big|\Bigg]$.
    \end{center}
    \begin{enumerate}
        \item     
            Let $H_1, H_2 \succeq 0$, and let $v$ lower bound the minimum non-zero eigenvalues of both $H_1$ and $H_2$. Then,
            \begin{equation*}
                \lambda_{min}(H_1+H_2) \geq 2v\cdot sin^2(\frac{\angle \big(\text{Null}(H_1),\:\text{Null}(H_2)\big)}{2}).
            \end{equation*}
        \item Define $\Pi_{\mathcal{X}}$ be a projector onto the space $\mathcal{X}$. For spaces $\mathcal{X}$ and $\mathcal{Y}$,
        \begin{equation*}
            \lambda_{max}(\Pi_{\mathcal{X}}+\Pi_{\mathcal{Y}}) \leq 1 + cos\big(\angle(\mathcal{X},\:\mathcal{Y})\big).
        \end{equation*}
    \end{enumerate}

\end{lemma}

The following lemma is a slight modification of the proof that a local Hamiltonian problem is $\QMA$-hard in \cite{kitaev2002classical,kempe20033}.
\begin{lemma}\label{Appendix:p_hard}
    5-$\textbf{LHwP} \in \pQMA$-hard.
\end{lemma}
\begin{proof}
    Given a $\pQMA$ pure-state promise problem $\mathcal{L}$. By the amplification Lemma~ \ref{App:amplify_QMA}, let $(\mathcal{P},\mathcal{V})$ be a $\pQMA$ protocol that decides $\mathcal{L}$ with completness $1-2^{-n}$ and soundness $2^{-n}$. There exist a polynomial $m$ such that $\mathcal{V}$ uses at most $m$ many qubits ancilla, requires $c \leq m$ copies of the input, and runs for at most $m$ steps. Thus, let $\cV$ be represented as a sequence of elementary quantum gates, i.e., $\cV:=\cV_m\cdots\cV_1$ acting on registers $I$, $W$, and $A$. For the input $\ket{\psi}$, we define the Hamiltonians as follow:
    \begin{equation*}
    \begin{cases}
        H'_{in} := (I-\ket{\psi}\bra{\psi}^c)_I \otimes I_W \otimes I_A \otimes \ket{0}\bra{0}_T + I_I \otimes I_W \otimes (I-\ket{0}\bra{0}_A)\otimes \ket{0}\bra{0}_T \\
        H'_{out} := I_I \otimes I_W \otimes \ket{0}\bra{0}_{A^{ans}} \otimes I_{-A^{ans}} \otimes \ket{m}\bra{m}_T \\
        H''_{prop} := \sum\limits_{t=0}^{m-1} -\mathcal{V}_{t+1} \otimes \ket{t+1}\bra{t}_T -\mathcal{V}^\dagger_{t+1} \otimes \ket{t}\bra{t+1}_T + I \otimes \ket{t}\bra{t}_T + I \otimes \ket{t+1}\bra{t+1}_T.
    \end{cases}       
    \end{equation*}
    Next, we make these Hamiltonians 5-local by introducing new Hamiltonians.
    \begin{equation*}
    \begin{cases}
        H_{in}:= (I-\ket{\psi}\bra{\psi}^c)_I \otimes I_W \otimes I_A \otimes \ket{0}\bra{0}_{T_1} \otimes I_{-T_{1}} + I_I \otimes I_W \otimes \sum\limits_{i=1}^{q(n)}\ket{1}\bra{1}_{A_i}\otimes \ket{0}\bra{0}_{T_1} \otimes I_{-T_{1}} \\
        H_{out} = I_I \otimes I_W \otimes \ket{0}\bra{0}_{A^{ans}} \otimes I_{-A^{ans}} \otimes \ket{1}\bra{1}_{T_m} \otimes I_{-T_m} \\
        H_{prop} := \sum\limits_{t=0}^{m-1} -\mathcal{V}_{t+1} \otimes \ket{110}\bra{100}_{T_{[t,t+2]}}\otimes I_{-T_{[t,t+2]}} -\mathcal{V}^\dagger_{t+1} \otimes \ket{100}\bra{110}_{T_{[t,t+2]}}\otimes I_{-T_{[t,t+2]}} + \\
        \qquad\qquad I \otimes \ket{100}\bra{100}_{T_{[t,t+2]}}\otimes I_{-T_{[t,t+2]}} + I \otimes \ket{110}\bra{110}_{T_{[t,t+2]}}\otimes I_{-T_{[t,t+2]}} \\
        H_{stab} := m^{12} \cdot I \otimes \sum\limits_{i=1}^{m-1} \ket{0}\bra{0}_{T_i} \otimes \ket{1}\bra{1}_{T_{i+1}}.
    \end{cases}       
    \end{equation*}

    Define $H:=H_{in}+H_{out}+H_{prop}+H_{stab}$. Suppose $\ket{\psi}\in\mathcal{L}_Y$ and let $\ket{\phi}$ be a witness. Then, the witness (also called the history state) of Hamiltonian $H$ is 
    \begin{equation}\label{Appendix:witness}
        \ket{\eta}:= \frac{1}{\sqrt{m+1}}\sum\limits_{t=0}^m \mathcal{V}_t\cdots\mathcal{V}_1 \big(\ket{\psi}^c_I \otimes \ket{\phi}_W \otimes \ket{0}_A \big)\otimes \ket{1^t0^{m-t+1}}.
    \end{equation}
    Following a calculation similar to that in \cite{kitaev2002classical}, we obtain $\bra{\eta} H \ket{\eta} \leq \frac{1}{2^n(m+1)}$; hence we set $a=\frac{1}{2^n(m+1)}$. 
    
    Suppose $\ket{\psi}\in\mathcal{L}_N$. We claim that for all witness $\ket{\phi}$, $\bra{\eta} H \ket{\eta} \geq \frac{d}{(m+1)^3}$ for some constant $d$; therefore we set $b=\frac{d}{(m+1)^3}$. 
    
   %%%%%%%%%%%%%%%%%New proof starting here%%%%%%%%%%%%%%%%%%%%%%%%
   Let $\mathcal{H}_{unstab}$ denote the subspace whose clock register represents a legal unary encoding, i.e. it is in the form of $1^i0^{m+1-i}$ for $i=0,1,\dots, m$. We also denote the $\mathcal{H}_{stab}$ to be the subspace orthogonal to $\mathcal{H}_{unstab}$. Note that states in $\mathcal{H}_{unstab}$ are orthogonal to $H_{stab}$. 

   Let $H_{circ} = H_{in} + H_{out} + H_{prop}$. We can obtain the following upper bound on the operator norm of $H_{circ}$: 
   \begin{align*}
       \|H_{circ}\| \leq &\|H_{in}\| + \|H_{out}\| + \|H_{prop}\| \\
       \leq&m+1 + 1 + 4m \\
       \leq&7m
   \end{align*}

    Then, we write $\ket{\eta} = \alpha_1\ket{\eta_1} + \alpha_2\ket{\eta_2}$, where $\ket{\eta_1} \in \mathcal{H}_{unstab}$ and $\ket{\eta_2} \in \mathcal{H}_{stab}$. In case that $\alpha_2\geq \frac{1}{m^5}$, 
    \begin{align*}
        \bra{\eta}H\ket{\eta} \geq \bra{\eta}H_{stab}\ket{\eta} - \|H_{circ}\| \geq \alpha_2^2 m^{12} - 7m >1.
    \end{align*}

    Otherwise, in case that $\alpha_2< \frac{1}{m^5}$, we get 
    \begin{align*}
        \bra{\eta}H\ket{\eta} &= \bra{\eta}H_{circ}\ket{\eta} + \bra{\eta}H_{stab}\ket{\eta} \\
        & \geq \bra{\eta}H_{circ}\ket{\eta} = (1-\alpha_2^2)\bra{\eta_1}H_{circ}\ket{\eta_1} + 2\alpha_1\alpha_2Re(\bra{\eta_1}H_{circ}\ket{\eta_2}) + \alpha_2^2 \bra{\eta_2}H_{circ}\ket{\eta_2}\\
        &\geq \bra{\eta_1}H_{circ}\ket{\eta_1} - (\frac{2}{m^{10}}+\frac{2}{m^5})\|H_{circ}\| \\
        &\geq \bra{\eta_1}H_{circ}\ket{\eta_1} - \frac{15}{m^4}.
    \end{align*}
    The last inequality uses the upper bound on $\|H_{circ}\|$. Then, it remains to show that $\bra{\eta_1}H_{circ}\ket{\eta_1}$ for $\ket{\eta_1} \in \mathcal{H}_{unstab}$. Let $U$ be the following operator changing the basis in the subspace $\mathcal{H}_{unstab}$:
    \begin{equation*}
        U:=\sum\limits_{t=0}^{m}V_1^\dagger\cdots V_t^\dagger  \otimes \ket{1^t0^{m-t+1}}\bra{1^t0^{m-t+1}}_T.
    \end{equation*}
    We aim to show that for $\ket{\eta_1}\in \mathcal{H}_{unstab}$, $\bra{\eta_1} H_{circ} \ket{\eta_1} = \bra{\eta_1}U^{\dagger} H_{circ} U\ket{\eta_1}.$ First, we apply $U$ to change the basis of $H_{in}$, $H_{out}$, and $H_{prop}$ in the subspace of $\mathcal{H}_{unstab}$.  
    \begin{equation*}
    \begin{cases}
        UH_{in}U^\dagger = H'_{in} \\
        UH_{out}U^\dagger = H'_{out} \\
        UH_{prop}U^\dagger = \sum\limits_{t=0}^{m-1} -I \otimes \ket{t+1}\bra{t}_T -I \otimes \ket{t}\bra{t+1}_T + I \otimes \ket{t}\bra{t}_T + I \otimes \ket{t+1}\bra{t+1}_T := H'_{prop}.\\
        UH_{stab}U^\dagger = 0.
    \end{cases}    
 \end{equation*}
 The following describe the null spaces of these Hamiltonians.
    \begin{equation*}
    \begin{cases}
        \text{NULL}(H'_{prop}) = \mathscr{H}_I \otimes \mathscr{H}_W \otimes \mathscr{H}_A \otimes \frac{1}{\sqrt{m+1}}\sum\limits_{t=0}^m\ket{1^t0^{m-t+1}} \\
        \text{NULL}(H'_{in}+H'_{out}) = N_1 \oplus N_2 \oplus N_3,
    \end{cases}    
    \end{equation*}
    where
    \begin{equation}
    \begin{cases}
        N_1 = \ket{\psi}^c \otimes \mathscr{H}_W \otimes \ket{0}_A \otimes \ket{0}_T \\
        N_2 = \mathscr{H}_I \otimes \mathscr{H}_W \otimes \ket{0}_A \otimes \text{Span}(\ket{1},\dots,\ket{m-1})_T \\
        N_3 = \{\ket{\eta}:\mathcal{V}\ket{\eta} \text{ has qubit } A^{ans} \text{ set to } \ket{1} \}_{IWA} \otimes \ket{m}_T.    \end{cases}       
    \end{equation}
    Let $N=N_1 \oplus N_2 \oplus N_3$. By the geometric Lemma~\ref{Appendix:geometric}, we need to show two things.
    First, the minimum non-zero eigenvalue of $H'_{in}+H'_{out}$ is 1 since these operators commute and are all projectors. The minimum non-zero eigenvalue of $H'_{prop}$ is at least $\frac{\pi^2}{(m+1)^2}$. Second,
    \begin{equation*}
        \cos^2\angle\big(N,\text{NULL}(H'_{prop})\big) \leq \max\limits_{\ket{y} \in \text{Null}(H'_{prop}) \:\land\: \|\ket{y}\|_2=1} \bra{y}\Pi_N\ket{y} \leq 1- \frac{1-\sqrt{\epsilon}}{m+1}.
    \end{equation*}
    By the geometric lemma \ref{Appendix:geometric},
    \begin{equation*}
        \lambda_{min}\big((H'_{in}+H'_{out}) + H'_{prop} \big) \geq 2\big(\frac{\pi^2}{(m+1)^2}\big)\big(\frac{1-2^{-n}}{4(m+1)}\big) \geq \frac{\pi^2(1-\sqrt{2^{-n}})}{2(m+1)^3}.
    \end{equation*}
    This shows that when $\alpha_2< \frac{1}{m^5}$, $\bra{\eta} H \ket{\eta} \geq \frac{d}{(m+1)^3}$ for some constant $d$.

    It is straightforward to find  $\{H_s\}_{s\in S}$ and $\{H_\ell\}_{\ell\in L}$ such that $H_{in}+H_{out}+H_{prop}+H_{stab}=\sum\limits_{s\in S}H_s+\sum\limits_{\ell\in L}(I-\ket{\psi}\bra{\psi}^c)\otimes H_\ell$. Additionally, for all $i\in S\bigcup L $, we have $0 \preceq H_i \preceq 2I$. Thus, we divide all $H_i$ by 2, resulting in the threshold $a$ and $b$ being divided by 2.
    We conclude that the reduction maps the input $\ket{\psi}$ of the promise problem $\cL$ to the inputs $\\ \left(p:=\frac{10(m+1)^3}{\pi^2(1-\sqrt{2^{-n}})}, a:=\frac{1}{2^{n+1}(m+1)}, b:=\frac{d}{2(m+1)^3}, \ket{\psi}^{\otimes c},  \{H_s\}_{s\in S}, \{H_\ell\}_{\ell\in L}\right)$ of the promise problem $5$-\textbf{LHwP}.
\end{proof}

The following lemma combines the technique of \cite{wocjan2003two} and Lemma~\ref{Appendix:p_hard}.
\begin{lemma}\label{Appendix:p_c_hard}
    5-$\textbf{LLHwP} \in \pQCMA$-hard.
\end{lemma}
\begin{proof}
    Let $\cL \in \pQCMA$. Let $(\cP,\cV)$ be a $\pQCMA$ protocol that decides $\cL$ with completeness $1-2^{-n}$ and soundness $2^{-n}$. Suppose that $\cV$ uses $c:=\poly(n)$ copies of the input. Define a new verifier $\mathcal{V}'$ as follows: (i) Apply $n$ C-NOT gates to copy the witness register to an additional $n$-qubit ancilla register $A_{new}$; (ii) Apply $\cV$ on the original register (excluding $A_{new}$). Then $(\cP,\cV')$ is a $\pQMA$ protocol that also decides $\cL$ with completeness $1-2^{-n}$ and soundness $2^{-n}$. Indeed, a quantum witness exists such that $\mathcal{V}'$ accepts with probability $pr$ if and only if a classical witness exists such that $\mathcal{V}$ accepts with probability $pr$. Additionally, in the $(\cP,\cV')$ protocol, it is sufficient for the prover to send a classical witness for a ``yes" instance. Let $\cV$ be represented as a sequence of $m$ elementary quantum gates, i.e., $\cV:=\cV_m\cdots\cV_1$ acting on registers $I$, $W$, and $A$. Then, the reduction maps the input $\ket{\psi} \in \cH(n)$ of the promise problem $\cL$ to the inputs
    \begin{equation*}
        \left(p, a, b, \ket{\psi}^{c}, \{H_s\}_S, \{H_\ell\}_L\right)
    \end{equation*}
    of 5-\textbf{LLHwP}, where these parameters are defined identically as in Lemma~\ref{Appendix:p_hard} with respect to the verifier $\cV'$. Let $H_{\psi}:= \sum\limits_sH_s+\sum\limits_\ell\big((I- \ket{\psi}\bra{\psi}^{c}) \otimes H_\ell\big)$.
    
    Suppose $\ket{\psi} \in \mathcal{L}_Y$ and the witness is a classical string $s$. By Lemma~\ref{Appendix:p_hard}, the low-energy vector $\ket{\eta}$ is defined as follows:
    \begin{equation}
        \ket{\eta}:= \frac{1}{\sqrt{m+1}}\sum\limits_{t=0}^m \mathcal{V}_t\cdots\mathcal{V}_1 \big(\ket{\psi}^c_I \otimes \ket{s}_W \otimes \ket{0}_A \big)\otimes \ket{1^t0^{m-t+1}},
    \end{equation}
    which is identical to Equation~ (\ref{Appendix:witness}), except that the register $W$ only contains a classical string. Therefore, $\ket{\eta}$ can be constructed using a polynomial number of elementary gates on $c$ copies of $\ket{\psi}$, and we have $\bra{\eta} H_\psi \ket{\eta} \leq a$. We conclude that the reduction maps $\ket{\psi}$ to a ``yes" instance of 5-\textbf{LLHwP}.
    
    Suppose $\ket{\psi} \in \mathcal{L}_N$. Then, by Lemma~\ref{Appendix:p_hard} with respect to the verifier $\cV'$, we obtain that $\lambda_{min}(H) \geq b$, which is a stricter condition than that promised for ``no" instances.
\end{proof}

%% file: Appendix_pQSZK.tex
\section{Other Structure results of $\pQSZKhv$} \label{Appendix:pQSZK_hv}

The definition of $\pQSZKhv$ can refer to Definition~\ref{Def:pQIP},~\ref{Def:pQSZKhv}, and ~\ref{Def:simplify}.

\subsection{Complete problem for $\pQSZKhv$ and $\pQSZKhv$ is closed under complement.}\label{section:pQSZK_hv}

In this section, we will show that there exists a complete problem for $\pQSZKhv$ and prove that $\pQSZKhv$ is closed under complement. 

We define the $\pQSZKhv$-complete problem, analogous to the quantum state distinguishability problem introduced in~\cite{Wat02}.
\begin{definition} \label{definition:QSDwP}[(\textbf{$\alpha,\beta$)-Quantum State Distinguishability with Unknown Pure State ($\\(\alpha,\beta$)-QSDwP)}]
Fix efficient computable polynomials $p(\cdot)$, $q(\cdot)$, and $k(\cdot)$. $(\alpha,\beta)$-\textbf{QSDwP} is defined as follows:
\begin{itemize}
    %\item Input: $n$ qubits unknow state $\ket{\phi}$, two efficient quantum circuits (unitary) $Q_0$, $Q_1$ size at most $p(n)$ acting on acting on $n+q(n)$ qubits and having $k(n)$ specified output qubits.  
    \item \textbf{Inputs: } An $n$-qubit quantum state $\ket{\phi}$ and $p(n)$-size quantum circuits $Q_0$, $Q_1$ acting on $n+q(n)$ qubits and having $k(n)$ specified output qubits. Let $\rho_b$ denote the mixed state by running $Q_b$ on $\ket{\phi}\ket{0^{q(n)}}$ and trace out the non-output qubits.
    \item \textbf{Yes instances:} $|| \rho_0 - \rho_1||_{tr} \geq \beta$
    \item \textbf{No instances:} $||\rho_0 - \rho_1||_{tr} \leq \alpha$
    %\item \textbf{Promise}: Let $\rho_b$ denote the mixed stated by running $Q_b$ on $\ket{\phi}\ket{0^{q(n)}}$ and trace out the non-output qubits. Then either $|| \rho_0 - \rho_1||_{tr} \geq \beta$ or $||\rho_0 - \rho_1||_{tr} \leq \alpha$.
    %\item \textbf{Output}: Accept when $|| \rho_0 - \rho_1||_{tr} \geq \beta$, and reject when $||\rho_0 - \rho_1||_{tr} \leq \alpha$.
\end{itemize}
\end{definition}

%% compare the difference in original problem
%We can prove the following two theorems.

The main theorems we are going to prove in this section are the following: 
\begin{theorem}\label{thm:UNQSD_QSZKcomplete}
    Let $\alpha$ and $\beta$ satisfy $0 < \alpha < \beta^2 < 1$. Then $(\alpha, \beta)$-\textbf{QSDwP} is complete for $\pQSZKhv$.
\end{theorem}

\begin{theorem}\label{thm:pQSZKhv_close}
    $\pQSZKhv$ is closed under complement.
\end{theorem}

%The way we prove Theorem \ref{thm:pQSZKhv_close} and Theorem \ref{thm:UNQSD_QSZKcomplete} is similar to \cite{Wat02}, instead of directly showing that $(\alpha, \beta)$-\textbf{QSDwP} is complete for $\pQSZKhv$. \nai{This grammar is incorrect.}First, we show that co-$(\alpha, \beta)$-\textbf{QSDwP} is complete for $\pQSZKhv$. Then, we show that $(\alpha, \beta)$-\textbf{QSDwP} is also in $\pQSZKhv$. Consequently, we obtain Theorem \ref{thm:pQSZKhv_close} and Theorem \ref{thm:UNQSD_QSZKcomplete} immediately. \nai{Revise this paragraph or remove it.}
%\nai{We prove these two theorems by starting with showing that}

%First, we need the polarization lemma from \cite{Wat02}, increasing the gap between $\alpha$ and $\beta$ to a value close to 1. \nai{To prove Theorem \ref{thm:pQSZKhv_close} and Theorem \ref{thm:UNQSD_QSZKcomplete}, we show the following polarization lemma that reduces $(\alpha, \beta)$-\textbf{QSDwP} to $(1-negl(n), negl(n))$-\textbf{QSDwP}}.

To prove Theorem \ref{thm:pQSZKhv_close} and Theorem \ref{thm:UNQSD_QSZKcomplete}, we show the following polarization lemma that reduces $(\alpha, \beta)$-\textbf{QSDwP} to $(\negl(n), 1 - \negl(n))$-\textbf{QSDwP}.
%We restate it as follows: \nai{This lemma is different from Watrous's lemma. Why do you say this is from Watrous?}
\begin{lemma}[Polarization Lemma]\label{lemma:polar}
For any $(\ket{\phi}, Q_0, Q_1) \in (\alpha, \beta)$-$\mathbf{QSDwP}$, satisfying $0 < \alpha < \beta^2 < 1$. Let $n$ be the number of qubits of $\ket{\phi}$ and $\rho_b$ be the output state of $Q_b$ apply on input $\ket{\phi}$ and trace out non-output bit. Then there exists  an polynomial time deterministic algorithm $\cA$ and polynomials $t(\cdot), q'(\cdot), p'(\cdot), k'(\cdot)$ such that 
\begin{itemize}
        \item On input $(1^n, Q_0, Q_1)$, $\cA$ outputs two $p'(n)$-size quantum circuits, $R_0$ and $R_1$, both acting on input state $\ket{\phi}^{\otimes t(n)}$ and ancilla $\ket{0^{q'(n)}}$ with $k'(n)$ specified output qubits.
        \item If $\| \rho_0 - \rho_1\|_{tr} \geq \beta$, then $ \|\sigma_0 - \sigma_1\|_{tr} \geq 1 - \negl(n)$.
       \item If $\| \rho_0 - \rho_1\|_{tr} \leq \alpha$, then $ \|\sigma_0 - \sigma_1\|_{tr} \leq \negl(n)$,
    \end{itemize}
    where $\sigma_b$  is the mixed state by running $R_b$ on $\ket{\phi}^{\otimes t(n)}\ket{0^{q'(n)}}$ and then trace out the non-output qubits. 
\end{lemma}

%\begin{lemma}[Polarization lemma]\label{lemma:polar}
%    Given the input $(\ket{\phi}, Q_0, Q_1)$ \nai{$(\alpha, \beta)$-$\mathbf{QSDwP}$'s input only has $\ket{\phi}, Q_0, Q_1$} from $(\alpha, \beta)$-$\mathbf{QSDwP}$, satisfying $0 < \alpha < \beta^2 < 1$. Let $n$ be the number of qubits of $\ket{\phi}$. There exist polynomials $t(\cdot), q'(\cdot), p'(\cdot), k'(\cdot)$, an $nt(n)$-qubit quantum state  and $p'(n)$-size quantum circuits $R_0, R_1$ , acting on $n \cdot t(n) + q'(n)$ qubits, and having $k'(n)$ specified output qubits, satisfying the following conditions:\nai{You shall also specify the input states here.}
%    \begin{itemize}
%        \item If $\| \rho_0 - \rho_1\|_{tr} \geq \beta$ then $ ||\sigma_0 - \sigma_1||_{tr} \geq 1 - negl(n)$
%        \item If $\| \rho_0 - \rho_1\|_{tr} \leq \alpha$ then $ ||\sigma_0 - \sigma_1||_{tr} \leq negl(n)$,
%    \end{itemize}
%    where $\sigma_b$ is the mixed state by running $R_b$ on $\ket{\phi}^{\otimes t(n)}\ket{0^{q'(n)}}$ and then trace out the non-output qubits.
%\end{lemma}
\begin{proof}[Proof sketch]
The deterministic algorithm $\cA$ is the same as in \cite{Wat02} because the proof relies solely on the trace distance between $\rho_0$ and $\rho_1$, which is the same in our case. The only difference is that in \cite{Wat02}, $R_b$ is applied only to the all-$\ket{0}$ state, whereas in our case, it is applied to $t(n)$ copies of the input state $\ket{\phi}$. This difference comes from in \cite{Wat02} $Q_0$ and $Q_1$ are only applied to the all-$\ket{0}$ state. However, in the $(\alpha, \beta)$-$\mathbf{QSDwP}$ problem, $Q_0$ and $Q_1$ are applied not just to $\ket{0}$ but to an unknown n-qubit quantum state $\ket{\phi}$.
\end{proof}

\begin{lemma}\label{lemma:coUNQSD_in_pQSZK}
    Let $\alpha$ and $\beta$ satisfy $0 < \alpha < \beta^2 < 1$. Then co-($\alpha, \beta$)-\textbf{QSDwP} $\in \pQSZKhv$. 
\end{lemma}
\begin{proof}
We describe the $\pQSZKhv$ protocol for co-($\alpha, \beta$)-\textbf{QSDwP} as follows.
\begin{protocal}{
 Pure State Honest Verifier Statistical Zero-Knowledge Protocal for co-($\alpha, \beta$)-\textbf{QSDwP}
}
\begin{description}
\item[Notation:]\quad\\
Let the instance of the co-($\alpha, \beta$)-\textbf{QSDwP} problem be ($\ket{\phi}, Q_0, Q_1$) and $n$ be the number of qubit of $\ket{\phi}$. We let $\ket{\phi}^{\otimes t(n)}$, $R_0$, $R_1$, $q'(\cdot)$, $t(\cdot)$, $\sigma_0$, and $\sigma_1$ represent the inputs, circuits, polynomials, and the output mixed states obtained by applying Lemma \ref{lemma:polar} to the instance ($\ket{\phi}, Q_0, Q_1$). To simplify the notation, we will write $t, q'$ instead of $t(n), q'(n)$. 
\item[Verifier's step 1:]\quad\\
Receive $2t$ copies of input $(\ket{\phi}, Q_0, Q_1)$ and compute $R_0\ket{\phi}^{\otimes t}\ket{0^{q'}}$. Let $A$ and $B$ be the output and trace-out registers, respectively. Sends $B$ register to the prover.

\item [Prover's step 1:]\quad\\
Let $U$ be a unitary only operate on register $B$ such that (exist by Uhlmann's Theorem (Theorem \ref{theorem:Ulman}))
\begin{center}
$ \big|\bra{\phi}^{\otimes t}\bra{0^{q'}}R_1^{\dagger}(I_{A}\otimes U_{B})R_0\ket{\phi}^{\otimes t}\ket{0^{q'}}\big| = F(\sigma_0, \sigma_1)$
\end{center}

Apply such $U$ to register $B$, then send register $B$ back to the verifier.

\item [Verifier's step 2:]\quad\\
 Apply swap-test between the state in register $A$ and $B$ and $R_1\ket{\phi}^{\otimes t}\ket{0^{q'}}$. The verifier accepts if the swap test passes (result $0$). Otherwise, the verifier rejects.
    
\end{description}
\end{protocal}

To show completeness, we know that $\|\sigma_0 - \sigma_1\|_{tr} \leq \negl(n)$, which implies that $F(\sigma_0, \sigma_1) \geq 1 - negl(n)$. The probability that the verifier accepts is equal to $\frac{1}{2} + \frac{1}{2}F((I_{A}\otimes U_{B})R_0\ket{\phi}^{\otimes t}\ket{0^{q'}},R_1\ket{\phi}^{\otimes t}\ket{0^{q'}})^2$ by Lemma \ref{lemma:swap-test}. This value is equal to $\frac{1}{2} + \frac{1}{2}F(\sigma_0,\sigma_1)^2 \geq 1 - \negl(n)$ by Uhlmann's theorem. 

To show soundess, we know that $\|\sigma_0 - \sigma_1\|_{tr} \geq 1 - \negl(n).$ Since the prover can act arbitrarily, we let $\hat{\rho}_{AB}$ represent the state in registers $AB$ at the verifier step 2.  The probability that the verifier accepts is equal to 

\begin{equation*}
    \begin{aligned}
        \frac{1}{2} + \frac{1}{2}F(\hat{\rho}_{AB},R_1\ket{\phi}^{\otimes t}\ket{0^{q'}})^2 &\leq \frac{1}{2} + \frac{1}{2}F(\Tr_{B}(\hat{\rho}_{AB}),\Tr_{B}(R_1\ket{\phi}^{\otimes t}\ket{0^{q'}}))^2 \\
    &= \frac{1}{2} + \frac{1}{2}F(\sigma_0,\sigma_1)^2 \leq \frac{1}{2} +\negl(n).      
    \end{aligned}
\end{equation*}

The first inequality comes from the trace operator, which can only increase the fidelity. The equality follows from the fact that the prover cannot touch the register $A$, so we have $\Tr_{B}(\hat{\rho}_{AB}) = \sigma_0$. The last inequality comes from $\|\sigma_0 - \sigma_1\|_{tr} \geq 1 - negl(n).$, which implies $F(\sigma_0, \sigma_1) \leq \negl(n)$.  Last, we apply error amplification (Lemma \ref{lemma:amplification}) to reduce soundness error to $\negl(n)$.

For the statistical zero-knowledge property, we let the simulator get $2t$ copies of input, the same as the verifier. For the first message, the simulator can simulate perfectly. For the second message, the simulator simulates registers $A$ and $B$ by simply computing $R_1\ket{\phi}^{\otimes t}\ket{0^{q'}}$, causing only a negligible error compared to the actual view. Then, the simulator does the same as the verifier's step 2.  We conclude that the simulation error is at most $\negl(n)$.
\end{proof}
%The hardness part is similar to the \cite{Wat02}. We provide a proof sketch in Appendix \ref{Applemma:pQSZK_hard}.

We then show that co-($\alpha, \beta$)-\textbf{QSDwP} is $\pQSZKhv$-complete.

\begin{lemma}\label{lemma:pQSZK_complete}
    Let $\alpha$ and $\beta$ satisfy $0 < \alpha < \beta^2 < 1$, Then co-($\alpha, \beta$)-\textbf{QSDwP} is $\pQSZKhv$-complete. 
\end{lemma}

The proof of Lemma~\ref{lemma:pQSZK_complete} follows a similar approach to that in~\cite{Wat02}. We defer the proof to Appendix~\ref{subsec:pQSZK_complete}.

\begin{lemma}\label{lemma:UNQSK_in_pQSZK}
        Let $\alpha$ and $\beta$ satisfy $0 < \alpha < \beta^2 < 1$, Then ($\alpha, \beta$)-\textbf{QSDwP} is $\in \pQSZKhv$.
\end{lemma}
\begin{proof}
We describe the $\pQSZKhv$-protocol as follows.
\begin{protocal}{
 Pure State Honest Verifier Statistical Zero-Knowledge Protocal for ($\alpha, \beta$)-\textbf{QSDwP}
}
\begin{description}
\item[Notation:]\quad\\
Let the instance of the ($\alpha, \beta$)-\textbf{QSDwP} problem be ($\ket{\phi}, Q_0, Q_1$) and $n$ be the number of qubit of $\ket{\phi}$. We let $\ket{\phi}^{\otimes t(n)}$, $R_0$, $R_1$, $q'(\cdot)$, $t(\cdot)$, $\sigma_0$, and $\sigma_1$ represent the inputs, circuits, polynomials, and the output mixed states obtained by applying Lemma \ref{lemma:polar} to the instance ($\ket{\phi}, Q_0, Q_1$). To simplify the notation, we will write $t, q'$ instead of $t(n), q'(n)$.
\item[Verifier's step 1:]\quad\\
Receive $t$ copies of input $(\ket{\phi}, Q_0, Q_1)$, sample uniform random bit $b$ and compute $R_b\ket{\phi}^{\otimes t}\ket{0^{q'}}$. Let $A$ and $B$ be the output and trace out registers, respectively. Send $A$ to the prover.

\item [Prover's step 1:]\quad\\
Apply optimal measure for $(\sigma_0, \sigma_1)$ to $A$ and get result $b'$. Then, send $b'$ back to the verifier.

\item [Verifier's step 2:]\quad\\
If $b = b'$, then the verifier accepts. Otherwise, the verifier rejects.
\end{description}
\end{protocal}

To show completeness, we know $\|\sigma_0 - \sigma_1\|_{tr} \geq 1 - \negl(n).$ This implies that the optimal measure can cause an error at most $\negl(n)$. The verifier will accept with probability $\geq 1 - \negl(n)$.

To show soundness, we know $\|\sigma_0 - \sigma_1\|_{tr} \leq \negl(n).$ This implies that the best way to distinguish $\sigma_0$ and  $\sigma_1$ is at most $\frac{1}{2} + \negl(n)$. Then we apply error amplification (Lemma \ref{lemma:amplification}) to reduce soundness error to $\negl(n)$.

For the statistical zero-knowledge property, we let the simulator get $t$ copies of input, the same as the verifier. For the first message, the simulator can simulate perfectly. The simulator simulates $b'$, identical to $b$, for the second message—the simulation error is at most $\negl(n)$.

\end{proof}

%It is easy to see that Theorem \ref{thm:pQSZKhv_close} followed from Lemma \ref{lemma:pQSZK_complete} and Lemma \ref{lemma:UNQSK_in_pQSZK}. And Theorem \ref{thm:UNQSD_QSZKcomplete} follows from Theorem \ref{thm:pQSZKhv_close} and Lemma \ref{lemma:pQSZK_complete}.

\begin{proof}[Proof of Theorem \ref{thm:UNQSD_QSZKcomplete}]
    Let $\alpha$ and $\beta$ satisfy $0 < \alpha < \beta^2 < 1$. By Lemma \ref{lemma:pQSZK_complete}, we know co-($\alpha, \beta$)-\textbf{QSDwP} is $\pQSZKhv$ complete. With the same reduction, ($\alpha, \beta$)-\textbf{QSDwP} is complete for the complement of $\pQSZKhv$. Therefore, it suffices to show that co-($\alpha, \beta$)-\textbf{QSDwP} is in the complement of $\pQSZKhv$, or equivalently, that ($\alpha, \beta$)-\textbf{QSDwP} is in $\pQSZKhv$. In Lemma \ref{lemma:UNQSK_in_pQSZK}, we have already shown that ($\alpha, \beta$)-\textbf{QSDwP} is in $\pQSZKhv$. This completes the proof. 

    %\nai{By Lemma~\ref{lemma:pQSZK_complete} and Lemma~\ref{lemma:UNQSK_in_pQSZK}, we know that co-($\alpha, \beta$)-\textbf{QSDwP} is $\pQSZKhv$-complete and that ($\alpha, \beta$)-\textbf{QSDwP} belongs to $\pQSZKhv$. Therefore, it suffices to prove that ($\alpha, \beta$)-\textbf{QSDwP} is hard for $\pQSZKhv$. This can be accomplished by reducing any problem in $\pQSZKhv$ to a co-($\alpha, \beta$)-\textbf{QSDwP} instance and then complementing the output of the co-($\alpha, \beta$)-\textbf{QSDwP} solver.}
\end{proof}

\begin{proof}[Proof of Theorem \ref{thm:pQSZKhv_close}]
It suffices to show that the complement of any complete problem $\cL$ of $\pQSZKhv$ is in $\pQSZKhv$ because the complement of $\cL$ is also a complete problem for the complement of  $\pQSZKhv$. Let $\alpha$ and $\beta$ satisfy $0 < \alpha < \beta^2 < 1$. By Lemma \ref{lemma:pQSZK_complete} and Lemma \ref{lemma:UNQSK_in_pQSZK}, we know that co-($\alpha, \beta$)-\textbf{QSDwP} is $\pQSZKhv$ complete and ($\alpha, \beta$)-\textbf{QSDwP} is in $\pQSZKhv$. This completes the proof.

\end{proof}

\subsection{Proofs of Lemma~\ref{lemma:pQSZK_complete}} \label{subsec:pQSZK_complete}

In this subsection, we provide the proof of Lemma \ref{lemma:pQSZK_complete}. 
Since we already obtain co-($\alpha, \beta$)-\textbf{QSDwP} is in $\pQSZKhv$ (Lemma~\ref{lemma:coUNQSD_in_pQSZK}), it is sufficient to show the following lemma:

\begin{lemma}\label{Applemma:pQSZK_hard}
    Let $\alpha$ and $\beta$ satisfy $0 < \alpha < \beta^2 < 1$, Then, co-($\alpha, \beta$)-\textbf{QSDwP} is $\pQSZKhv$ hard. 
\end{lemma}

We require the following two lemmas to prove Lemma \ref{Applemma:pQSZK_hard}.
\begin{lemma}\label{Applemma:td_amp}(\cite{Wat02})
  Let $\sigma, \rho$ be two mixed states satisfy $||\rho - \sigma||_{tr} = \epsilon$. Then,
  \begin{equation*}
      1 - e^{-l\epsilon^2} < ||\rho^{\otimes l } - \sigma^{\otimes l}||_{tr} \leq l\epsilon.
  \end{equation*}
\end{lemma}

\begin{lemma}\label{Applemma:pQSZK_hard_core}
    Consider an $m$-message pure-state quantum interactive proof system for even $m$ message of  $\ket{\phi}$ and the verifier receives $t$ copies of input $\ket{\phi}$. Let the maximum acceptance probability of the above protocol be $\epsilon$, let $V_1,\cdots ,V_k$ denote the verifier's circuits (where $k = \frac{m}{2} + 1 $), let $\rho_0,\cdots,\rho_{k-1}$ be any mixed quantum states over $\mathcal{V}\otimes\mathcal{M}$, let $\sigma_j = V_j\rho_{j-1}V_j^*$ for $j = 1,\cdots ,k$ and assume that $\Tr(\Pi_{init}\rho_0)$ = $\Tr(\Pi_{acc}\sigma_k) = 1$ Then,
    \begin{equation*}
        \big|\big|\Tr_{\mathcal{M}}\sigma_1 \otimes \cdots \otimes \Tr_{\mathcal{M}}\sigma_{k-1} - \Tr_{\mathcal{M}}\rho_1 \otimes \cdots \otimes \Tr_{\mathcal{M}}\rho_{k-1}\big|\big|_{tr} \geq \frac{(1-\sqrt{\epsilon})^2}{4(k-1)},
    \end{equation*}
    where $\Pi_{init}$ denote projection of the initial state of the verifier, which is equal to $\ket{\phi}^{t}\ket{0}$, and $\Pi_{acc}$ denote the projection onto states for which verifier's output qubit is set to 1 (accept). 
\end{lemma}
\begin{proof}
    The proof is the same as Lemma 15 in \cite{Wat02}.
\end{proof}

We are ready to show Lemma \ref{Applemma:pQSZK_hard}.
\begin{proof}
    Let $\mathcal{L} \in \pQSZKhv$, $\ket{\phi}$ be the n-qubit input state, and $(V,P)$ be a $\pQSZKhv$ protocal decides $\mathcal{L}$ with completeness $1 - 2^{-n}$ and soundness $2^{-n}$.  Let $m = m(n)$ be the number of message exchanges by $P$ and $V$, and $t = t(n)$ be the number of copies of input $\ket{\phi}$ used for $V$. Without loss of generality, we assume the number of messages $m$ is even for all input by adding an initial move for the verifier. Then, the verifier will apply transform $V_1,\cdots V_k$ for $k = m/2 + 1$ and will send the first message in the protocol. We let output of simulator for $(V,P)$ on input ($\ket{\phi}^{\otimes t'(n)},i$) be $\xi_{\ket{\phi},i }$, where $t'$ is some polynomial. Note that for every $i = 1 \cdots m$, $\xi_{\ket{\phi},i }$ can be written as apply a size $r(n)$ unitary $U^i$ on input $\ket{\phi}^{t'(n)}\ket{0^{q(n)}}$ then trace out some non-output qubits, where $q, r$ are some polynomial.

    We are ready to show the reduction. For any fixed input $\ket{\phi}$, we define $\rho_i$ for all $i \in [k - 1]_0$ and $\sigma_i$ for all $i \in [k]$ as follows.
    \begin{enumerate}
        \item Let $\rho_0$ be the state which verifier get $t$ copies of $\ket{\phi}$, and all other qubits initial as $\ket{0}$.
        \item Let $\sigma_k$ denote the state by applying $V_k$ to $\xi_{\ket{\phi},m}$, then discarding the output qubit, and replacing it with a qubit in state $\ket{1}$.
        \item Let $\rho_i = \xi_{\ket{\phi}, 2i}$ for $i = 1, \cdots, k-2$ and let $\rho_{k-1} = V^{\dagger}_k\sigma_k V_k$.
        \item Let $\sigma_i = V_i\rho_{i-1}V_i^{\dagger}$ for $i = 1, \cdots k - 1$.
    \end{enumerate}

    Let $Q_0$ and $Q_1$ be the unitary apply on input  $\ket{\phi}^{\frac{m(n)t'(n)}{2}}\ket{0^{q'(n)}}$ for some polynomial $q'$, then trace out some non-output qubits and get following two state
    \begin{equation}
    \begin{cases}
        \gamma_0 = \Tr_{\mathcal{M}}\sigma_1 \otimes \cdots \otimes \Tr_{\mathcal{M}}\sigma_{k-1} \\
        \gamma_1 = \Tr_{\mathcal{M}}\rho_1 \otimes \cdots \otimes \Tr_{\mathcal{M}}\rho_{k-1},
    \end{cases}
    \end{equation}
    respectively. We can observe that the size of $Q_0,Q_1$ at most $r'(n)$, where $r'$ is some polynomial, because the simulator of $(P,V)$ is efficient.  

    We claim that the following hold
    \begin{itemize}
        \item If $\ket{\phi} \in \mathcal{L}_{Y}$, then $||\gamma_0 - \gamma_1||_{tr} < negl(n)$.
        \item If $\ket{\phi} \in \mathcal{L}_{N}$, then $||\gamma_0 - \gamma_1||_{tr} \geq c/k$, where $c > 0$ is some constant.
    \end{itemize}
    The second one follows from Lemma \ref{Applemma:pQSZK_hard_core}. The first one follows from the pure state statistical zero knowledge. Since $\ket{\phi} \in \mathcal{L}_{Y}$, we can replace $\xi_{\ket{\phi},j}$ with actual view of the verifier $V$ interacting with $P$, for input $\ket{\phi}$ and message $j$. Then we can see $\sigma_i$ and $\rho_i$ are the verifier's private and message registers just before and after the prover replies. Because prover does not touch the private register of the verifier, we get $||\Tr_{\mathcal{M}}\sigma_i -\Tr_{\mathcal{M}}\rho_i||_{tr} \leq negl(n)$ for $i = 1,\cdots ,k - 2$. For the last message, completeness is at least $1 - negl(n)$. This implies replacing the output bit to state $\ket{1}$ has a negligible effect on the state. Then, we get the first result.  

    Let $Q'_0$ and $Q'_1$ be the unitary by applying Lemma \ref{Applemma:td_amp} to $Q_0$ and $Q_1$ with $l = n(k/c)^2$, and the following two states are the output of $Q'_0$ and $Q'_1$ respectively.
    \begin{equation}\label{equation:EFI_construction}
    \begin{cases}
        \gamma'_0 = \gamma_0^{\otimes l} \\
        \gamma'_1 = \gamma_1^{\otimes l}.
    \end{cases}
    \end{equation}
    The output state $\gamma'_b$ can be expressed as the result of applying  $Q'_b$ to the input $\ket{\phi}^{\frac{n\cdot m(n)t'(n)k(n)^2 }{2c^2}}\ket{0^{q''(n)}}$ for some polynomial $q''$, and then trace out the non-output qubits. Moreover, the sizes of $Q'_0$ and $Q'_1$ are at most $r''(n)$ for some polynomial $r''$. By Lemma \ref{Applemma:td_amp}, $\gamma'_0$ and $\gamma'_1$ satisfy the following:
        \begin{itemize}
        \item If $\ket{\phi} \in \mathcal{L}_{Y}$, then $||\gamma'_0 - \gamma'_1||_{tr} < n(k/c)^2negl(n) < \alpha$
        \item If $\ket{\phi} \in \mathcal{L}_{N}$, then $||\gamma'_0 - \gamma'_1||_{tr} > 1 - e^{-n} > \beta$,
    \end{itemize}
    for any constant $\alpha,\beta$ with $\beta > \alpha$. We complete the proof because
        \begin{itemize}
        \item If $\ket{\phi} \in \mathcal{L}_{Y}$, then $
        (\ket{\phi}^{\frac{n\cdot m(n)t'(n)k(n)^2}{2c^2}}, Q'_0,Q'_1) \in$ co-($\alpha, \beta$)-$\mathbf{QSDwP}_{Y}$
        \item If $\ket{\phi} \in \mathcal{L}_{N}$, then       $(\ket{\phi}^{\frac{n\cdot m(n)t'(n)k(n)^2}{2c^2}}, Q'_0,Q'_1) \in$ co-($\alpha, \beta$)-$\mathbf{QSDwP}_{N}$.
    \end{itemize}
\end{proof}

\subsection{co-QSDwP has a public coin pure state honest verifier QSZK}

In this subsection, we provide proof of the following theorem.
\begin{theorem}\label{appthm:pQSZKhv_public_coin}
    For any $\mathcal{L} \in \pQSZKhv$, there exists a three-message public coin honest verifier pure-state quantum statistical zero-knowledge proof system with completeness $1 - \negl(n)$ and soundness $\frac{3}{4} + \negl(n)$.
\end{theorem}

We show the following lemma first.
\begin{lemma}\label{Applemma:coUNQSD_public_coin}
    Let $\alpha$ and $\beta$ satisfy $0 < \alpha < \beta^2 < 1$, Then co-($\alpha, \beta$)-\textbf{QSDwP} has a public coin pure state statistical zero-knowledge protocol with completeness $1 - \negl(n)$ and soundness $\frac{3}{4} + \negl(n)$.
\end{lemma}

We require the following lemma:
\begin{lemma}\label{Applemma:fidelity_iequality}
For any $n$-qubtis mixed states $\sigma, \rho, \xi$, then $F(\rho,\sigma)^2 +F(\sigma,\xi)^2 \leq 1 + F(\rho, \xi)$. 
\end{lemma}

We are ready to show Lemma \ref{Applemma:coUNQSD_public_coin}.
\begin{proof}
We define the protocol as follows.
\begin{protocal}{
 Public Coin Pure State Honest Verifier Statistical Zero-Knowledge Protocol for co-($\alpha, \beta$)-\textbf{QSDwP}
}
\begin{description}
\item[Notation:]\quad\\
Let the instance of the co-($\alpha, \beta$)-\textbf{QSDwP} problem be ($\ket{\phi}, Q_0, Q_1$) and $n$ be the number of qubit of $\ket{\phi}$. We let $\ket{\phi}^{\otimes t(n)}$, $R_0$, $R_1$, $q'(\cdot)$, $t(\cdot)$, $\sigma_0$, and $\sigma_1$ represent the inputs, circuits, polynomials, and the output mixed states obtained by applying Lemma \ref{lemma:polar} to the instance ($\ket{\phi}, Q_0, Q_1$). To simplify the notation, we will write $t, q'$ instead of $t(n), q'(n)$.

\item[Prover's step 1:]\quad\\
Receive $t$ copies of input $(\ket{\phi}, Q_0, Q_1)$ and compute $R_0\ket{\phi}^{\otimes t}\ket{0^{q'}}$, and let $A$,$B$ be the output register and trace out register, and send $A$ to the verifier.

\item [Verifier's step 1:]\quad\\
Receive $t$ copies of input $(\ket{\phi}, Q_0, Q_1)$. Then, sample uniform random bit $b$, and send b to Prover's.

\item [Prover's step 2:]\quad\\
Let $U$ be a unitary only operate on $B$ such that (exist by Uhlmann's Theorem (Theorem \ref{theorem:Ulman}))
\begin{equation*}
 \big|\bra{\phi}^{\otimes t}\bra{0^{q'}}R_1^{\dagger}(I_{A}\otimes U_{B})R_0\ket{\phi}^{\otimes t}\ket{0^{q'}}\big| = F(\sigma_0, \sigma_1).   
\end{equation*}

If $b = 1$, then apply such $U$ to $B$,
Otherwise, do nothing.
Send $B$ back to the verifier.
\item [Verifier's step 2:]\quad\\
Perform swap-test on $AB$ and $R_b\ket{\phi}^{\otimes t}\ket{0^{q'}}$. The verifier accepts if the swap test passes (result $0$). Otherwise, the verifier rejects
\end{description}
\end{protocal}

We show that the above protocol satisfies the completeness property. When $b$ is $0$, the verifier state of register of $AB$ is $R_0\ket{\phi}^{\otimes t}\ket{0^{q'}}$. Then, the verifier accepts with probability 1. When $b$ is $1$, the verifier accepts with the probability at least $1 - \negl(n)$ by completeness analysis in Lemma \ref{lemma:coUNQSD_in_pQSZK}. Combining both cases, we conclude that the verifier accepts with probability at least $1 - \negl(n)$.

To show soundess, we know $\|\sigma_0 - \sigma_1\|_{tr} \geq 1 - \negl(n).$ This imply $F(\sigma_0, \sigma_1) \leq \negl(n)$. Since prover can do anything, we let $\hat{\rho}_{AB}$ be the state of register $AB$ in the verifier's step 2. The verifier's accepting probability is the following.
\begin{equation*}
    \frac{1}{2}\big(\frac{1}{2}+\frac{1}{2}F(\hat{\rho}_{AB},R_0\ket{\phi}^{\otimes t}\ket{0^{q'}})^2\big)+\frac{1}{2}\big(\frac{1}{2}+\frac{1}{2}F(\hat{\rho}_{AB},R_1\ket{\phi}^{\otimes t}\ket{0^{q'}})^2\big).
\end{equation*}
By simple calculation, we can simplify above equation to
\begin{equation*}
    \frac{1}{2}+\frac{1}{4}\big(F(\hat{\rho}_{AB},R_0\ket{\phi}^{\otimes t}\ket{0^{q'}})^2+ F(\hat{\rho}_{AB},R_1\ket{\phi}^{\otimes t}\ket{0^{q'}})^2\big).
\end{equation*}
This value can be upper bound by
\begin{equation*}
    \frac{1}{2}+\frac{1}{4}\big(F(\Tr_{B'}(\hat{\rho}_{AB}),\sigma_0)^2+ F(\Tr_{B'}(\hat{\rho}_{AB}),\sigma_1)^2\big).
\end{equation*}
Next, we apply Lemma \ref{Applemma:fidelity_iequality}, then the accept probability can be upper bound by
\begin{equation*}
    \frac{1}{2}+\frac{1}{4}\big(1 + F(\sigma_0,\sigma_1)\big) = \frac{3}{4} + \negl(n).
\end{equation*}
Last, we show the statistical zero-knowledge property. We can simulate the first and second messages perfectly. For the third message, we look at the simulated value $b$. If $b = 0$, then simulate register $AB$ by $R_0\ket{\phi}^{\otimes t}\ket{0^{q'}}$. If $b = 1$, simulate register $AB$ by $R_1\ket{\phi}^{\otimes t}\ket{0^{q'}}$. Then, the error at most $\negl(n)$.
\end{proof}

Theorem \ref{appthm:pQSZKhv_public_coin} simply follows from Lemma \ref{lemma:pQSZK_complete} and Lemma \ref{Applemma:coUNQSD_public_coin}.

%% file: Appendix_rep_KA04.tex
\section{Potential Proof gaps  in~\cite{kashefi2004complexity}.} \label{Appendix:proof_gaps_in_KA04}

\emph{Entanglement testing} is a quantum property testing problem that determines whether a state is entangled or far from the entangled state. In~\cite{kashefi2004complexity}, they show that \emph{entanglement testing} problem belongs to $\QMA$ The above result contradicts our Corollary~\ref{cor:ka04_contradict}. In the following, we provide a more detailed discussion of the potential proof gap in~\cite{kashefi2004complexity}.

First, we will recall the concept of an entanglement witness. Second, we will state their $\QMA$ protocol for the \emph{entanglement testing} problem. Finally, we will point out some gaps in the proof. The entanglement witness is defined as follows: given an entangled state $\rho_Y$, an entanglement witness is a Hermitian operator $A$ such that $\Tr(A\rho_Y) < 0$, while all separable states $\rho_N$, we have $\Tr(A\rho_N) \geq 0$. We state their $\QMA$ protocol for the \emph{entanglement testing} problem: the honest prover sends real numbers $c_1, \cdots, c_l$ and $\lambda$-qbits mixed states $\rho_1, \cdots, \rho_l$ to the verifier, where $A:=\sum_{i=1}^{l} c_i\rho_i$ is an entanglement witness for $\rho_{in} \in \rho_Y$, $l$ is at most polynomial, and $c_1, \cdots c_l$ are polynomial-time computable real numbers. Then, the verifier estimates $\Tr(A\rho_{in})$ by computing $\Tr(\rho_{in}\rho_i)$ for all $i\in [l]$. The verifier accepts if $\Tr(A\rho_{in})$ is negative and $A$ is indeed an entanglement witness. To ensure $A$ is an entanglement witness, the verifier prepares the basis of separable states (with polynomial size) and checks that the expectation value on these states is non-negative.

We identify the following gaps in their proof. First, the entanglement witness does not provide any guarantee on the gap of $\Tr(A\rho_{\text{in}})$ between the entangled state and all separable states. It is possible that the gap could be negligible. It is unclear how to estimate $\Tr(A\rho_{in})$ with a polynomial number of copies of the unknown operator $A$ and the input state $\rho_{in}$ to the desired precision. Second, even when the gap is polynomial, the verifier estimates $\Tr(A\rho_{\text{in}})$ by computing $\Tr(\rho_{\text{in}}\rho_i)$. The verifier requires identical copies of $\rho_i$ in order to correctly compute $\Tr(\rho_{\text{in}}\rho_i)$. How to ensure the witness contains identical copies of $\rho_i$ is unclear. Third, even if the verifier can ensure that the witness contains identical copies of $\rho_i$, it is unclear how to verify that the operator $A$ is indeed an entanglement witness. Checking only a polynomial-sized basis of separable states is insufficient, as the possible basis of separable states could be exponentially large. Specifically, checking whether all separable states $\rho_N$ satisfy $\Tr(A\rho_N) \geq 0$ would require identifying the entire basis of separable states, which could be computationally inefficient.